\documentclass[11pt]{article}

\usepackage{jheppub}
\pdfoutput=1

\usepackage{epstopdf}
\usepackage{jheppub}
\usepackage{mathrsfs}
\usepackage{psfrag}
\usepackage{color}
\usepackage{hyperref}

\usepackage{slashed}

\usepackage{simplewick}
\usepackage{cancel}

\usepackage{todonotes}
\usepackage{tikz, pict2e}

\usepackage{lmodern}
\usepackage{graphicx}
\usepackage{subcaption}
\usepackage{tkz-euclide}
\usepackage[utf8]{inputenc}

\usetikzlibrary{calc,backgrounds,intersections}
\usetikzlibrary{arrows,shapes,positioning,shadows,trees}
\usetikzlibrary{mindmap}
\usetikzlibrary{decorations.pathreplacing}
\usetikzlibrary{decorations.pathmorphing}
\usetikzlibrary{decorations.markings}
\usetikzlibrary{matrix}

\tikzstyle arrowstyle=[scale=1]
\tikzstyle directed=[postaction={decorate,decoration={markings,
    mark=at position .5 with {\arrow[arrowstyle]{stealth}}}}]
\tikzstyle reverse directed=[postaction={decorate,decoration={markings,
    mark=at position .5 with {\arrowreversed[arrowstyle]{stealth};}}}]

\usepackage{mathtools,bbm,array,amsfonts,graphicx,wrapfig,arydshln,lscape,float,multirow,longtable,rotating,makecell}
\usepackage{url}

\definecolor{redb}{rgb}{0.700, 0.000, 0.300}
\DeclareMathAlphabet\mathbfcal{OMS}{cmsy}{b}{n}

\newcommand{\eq}{\begin{equation}}

\newcommand{\eqe}{\end{equation}}
\newcommand{\eqa}{\begin{eqnarray}}
\newcommand{\eqae}{\end{eqnarray}}

\DeclareMathOperator*{\Det}{\mbox{det}}

\newcommand{\xdownarrow}[1]{%
  {\left\downarrow\vbox to #1{}\right.\kern-\nulldelimiterspace}
}

\title{Cosmological Polytopes and the Wavefunction of the Universe}

\author{Nima Arkani-Hamed,$^1$}
\author{Paolo Benincasa,$^{2}$}
\author{Alexander Postnikov$^{3}$}

\affiliation{$^1$ School of Natural Sciences, Institute for Advanced Study, Princeton, NJ 08540, USA}
\affiliation{$^2$ Perimeter Institute for Theoretical Physics, Waterloo, Canada}
\affiliation{$^3$ Department of Mathematics, Massachusetts Institute of Technology, Cambridge, MA, USA}

\emailAdd{arkani@ias.edu, pablowellinhouse@anche.no, apost@math.mit.edu}

\abstract{We present a connection between the physics of cosmological time evolution and the wavefunction of the universe, and the mathematics of positive geometries, roughly analogous to similar connections seen in the context of scattering amplitudes. We consider the late-time wavefunction of the universe in  a class of toy models of scalars with time-dependent coupling constants, including conformally coupled scalars (with non-conformal interactions) in FRW cosmologies as a special case. The contribution of each Feynman diagram to the wavefunction of the universe is associated with a certain universal rational integrand. We show that this integrand can be identified with the  canonical form of a ``cosmological polytope''. These polytopes have an intrinsic definition making no reference to physics, and the connection to ``time'', along with familiar  properties of the wavefunction, arises from this definition. In particular the singularity structure of the wavefunction for this toy model of scalars  is common to all theories, and is geometrized by the cosmological polytope. A natural triangulation of the polytope is associated with the time-integral representation of the wavefunction; another natural triangulation of the dual polytope reproduces ``old-fashioned perturbation theory''. Other triangulations are associated with efficient recursive computations of the wavefunction, while recently discovered new representations for the canonical forms of general polytopes give new representations of the wavefunction with no extant physical interpretation. We show in some examples how symmetries of the cosmological polytope descend to symmetries of the wavefunction, (such as conformal invariance of deSitter wavefunctions). In cases such as $\phi^3$ theory in $dS_4$, the final wavefunction obtained from integration of the rational functions  gives rise to polylogarithms associated with every graph. We give an explicit expression for the symbol of these polylogs, which record the geometry of sequential projections of the cosmological polytope.}

\begin{document}

\maketitle

\section{Cosmological Time and The Wavefunction of the Universe}

Cosmology is the ultimate historical science. Arguably the central aim of cosmology is to answer the question --``what happened in the early universe?'', which might be a stepping stone to the (still vague and ill-defined) question of ``what was the origin the universe?''. Like any historical science, cosmology has an interesting relationship with the concept of time. After all, no observers were present in the early universe to record what was happening. Instead, from measurements of spatial correlations in the late universe  we infer the existence of a cosmological history that gives a simple and logically consistent account of these correlations, following from time evolution. Cosmological history is then something that is ``integrated in'' to explain patterns we see in the present day. A paleontologist does the same thing, ``integrating in" the existence of dinosaurs roaming the earth to (amongst other things) account for giant fossilized bones and teeth found in the ground today. 

Said more formally, in cosmology we can talk about the ``wavefunction of the universe''. For cosmologies that end in flat space (not ours due to our current accelerated expansion), this is the only quantum-mechanical observable we can talk about -- where an infinite number of measurements can be done with infinitely large measuring apparatuses. The wavefunction of the universe gives answers to questions like ``what fractions of stars are red, what fraction blue?'', or $n$-point correlations in the spectrum of density perturbations. 

What are the rules for the wavefunction of the universe? If someone were to produce such a wavefunction, how could we check if it is right or wrong? The most obvious property is that is ``normalizable'', but this is clearly a far too weak condition. What properties does the wavefunction have to have in order to be at least approximately compatible with ``unitary evolution in cosmological space-time''? We do not yet have the answer to this question. The situation can be compared with that of the $S$-matrix, where things are somewhat clearer -- at least the requirements of Lorentz invariance and Unitarity are well-defined, though still not constraining enough: it is trivial to give examples of exactly unitary $S$-matrices that do not correspond to any sensible theory. The really non-trivial issue is how {\it causality} is imprinted in amplitudes. This is famously  related to the analytic structure of scattering amplitudes, but we don't  yet know what this precisely means non-perturbatively.  Things are better in perturbation theory, though even here we don't exactly know how to characterize the class of ``appropriately analytic functions" for the actual amplitudes beyond one-loop. However, at least when the notion of a ``loop integrand" is available we have the cutting rules that give us highly non-trivial consistency conditions at all loop orders. 

It is interesting to contrast our relatively meager understanding of the rules for cosmology or even the $S$-matrix, with that of  the rules for correlation functions in Euclidean CFT's, which are perfectly well-defined. The OPE gives us the non-perturbative rules and conformal symmetry places powerful constraints on the class of functions that can make an appearance. The difference between these two situation is not a technical one but is deeply tied to the physics of time: we do not have an understanding, for either the $S$-matrix and the wavefunction of the universe, of how ``causal time evolution'' is reflected in ``boundary observables at infinity''. 

It is fascinating that the physics of the $S$-matrix is actually {\it contained} in that of the cosmological correlation functions in a simple way, by looking at the ``total energy singularities", where $(E_1 + \cdots + E_n)\,\longrightarrow\,0$, with each $E_i = |\vec{p}_i|$ being the magnitude of spatial momenta  \cite{Maldacena:2011nz, Arkani-Hamed:2015bza}\footnote{A similar phenomenon occurs for $AdS$ correlators, \cite{Raju:2012zr, Raju:2012zs}.}: dynamical amplitudes are contained in analytic continuations of static (spatial) correlators!. So at least some of the singularities of cosmological correlators are related to objects we are familiar with. It is  likely that the rules for cosmological wavefunction, at least in perturbation theory, might be accessible at least at the same level as for amplitudes, for instance we should be able to sharply understand the analog of the cutting rules for cosmological integrands.  There is of course an independent excellent motivation for studying weakly coupled theories for cosmological correlators, since the physics giving rise to inflationary density perturbations is famously very weakly coupled, with interaction strengths given by powers of $\delta \rho/\rho \sim 10^{-5}$. 

Looking for the rules prescribing how  ``consistent time evolution'' is encoded in the late-time wavefunctions/correlation functions in this way is the cosmological analog of ``how do we look for dinosaurs in the fossil record?'', and some aspects of this question are potentially relevant for experiments: we can for instance ask how the presence of heavy particles with a masses comparable to the inflationary Hubble scale are reflected in non-Gaussian correlation functions \cite{Maldacena:2002vr} in the spectrum of density perturbations \cite{Chen:2009we, Chen:2009zp, Byrnes:2010xd, Suyama:2010uj, Baumann:2011nk, Assassi:2012zq, Noumi:2012vr, Arkani-Hamed:2015bza}. 

While understanding the way in which  consistent causal dynamics is imprinted on boundary observables in various examples is interesting, we can not escape the deeper question of what fixes these properties in a more fundamental way to begin with. The failure to have an a-priori answer to this question was the fundamental flaw at the heart of the $S$-matrix program in the 1960's. However the myriad of advances in our understanding of scattering amplitudes over the past decade are beginning to suggest a different and more radical line of attack on this question. Instead of slavishly ``checking", by hook or by crook, that the $S$-matrix is compatible with causal, unitary evolution in space-time, we are beginning to see that scattering amplitudes should be thought of as the answer to entirely different sorts of natural mathematical questions. As an example, in the context of ${\cal N}=4$ SYM amplitudes in the planar limit, at any loop order, the (integrand of) amplitudes are thought of as canonical differential forms associated with a new geometric structure -- the amplituhedron \cite{Arkani-Hamed:2013jha, Arkani-Hamed:2013kca, Arkani-Hamed:2017vfh} -- generalizing the notion of triangles and polygons to higher-dimensional spaces. There is no reference to standard physical notions here -- neither Space-time nor Hilbert spaces make an appearance.  Locality and unitarity are not primary in this story, but arise instead as derivative notions from the ``positive geometry" of the amplituhedron. 

It is tempting to try and extend these ideas to deal with the wavefunction of the universe, partially because the rewards for  success  are much higher in cosmology. After all seeing ``emergent quantum mechanics and space-time" in non-gravitational scattering amplitudes is a luxury, not a necessity, but the situation is entirely different in cosmology, where especially the notion of  ``emergent time" finds its most fundamental and pressing setting. Here there {\it must} be some picture for the wavefunction of the universe making no reference to time evolution, if for no other reason than that the notion of time breaks down at the big bang.  This motivates looking for a picture of the wavefunction of the universe as an answer to a different question making no reference to the cosmological time evolution, starting with toy models of cosmological evolution and  wavefunction of the universe which do (of course) have a conventional time evolution description. 

With these various questions and goals in mind, in this paper, we will initiate an exploration of the detailed structure of cosmological wavefunctions in perturbation theory. At the same time we will also begin to look for ``a different question to which the cosmological wavefunction is the answer'', leading to a new  connection between physics and ``positive geometries'', roughly analogous to what has been seen in the context of scattering amplitudes.  

We will work in the the context of a class of toy models of scalar fields with time-dependent coupling constants, which  include including conformally coupled scalars (with non-conformal interactions) in FRW cosmologies as a special case. 

In Section \ref{sec:model}, we will study perturbative computation of the wavefunction of the universe for these models;  as we will see the contribution of each Feynman graph to the wavefunction of the universe is associated with a certain universal integrand. Consider e.g. $\phi^3$ theory, some representative Feynman diagrams contributing to the wavefunction at asymptotic future conformal time $\eta=0$ are schematically 
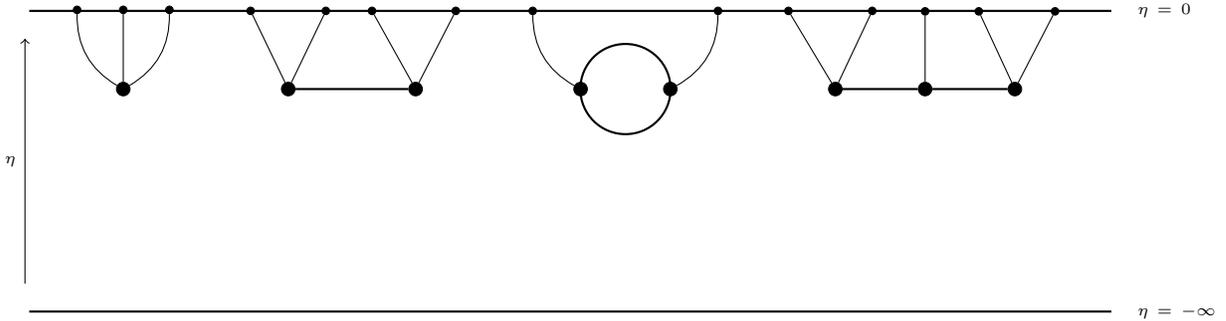
\begin{equation*}
  \begin{tikzpicture}[ball/.style = {circle, draw, align=center, anchor=north, inner sep=0}]
    \node[ball,text width=.1cm,fill,color=white] at (-2.5,0) (d1) {};
    \node[ball,text width=.1cm,fill,color=white] at (12,0) (d2) {};
    \node[right=.1cm of d2.east] (tl) {\tiny $\eta\,=\,0$};
    \draw[thick,color=black] (d1) -- (d2); 
    \node[ball,text width=.1cm,fill,color=white] at (-2.5,-4) (d3) {};
    \node[ball,text width=.1cm,fill,color=white] at (12,-4) (d4) {};
    \node[right=.1cm of d4.east] (te) {\tiny $\eta\,=\,-\infty$};    
    \draw[thick,color=black] (d3) -- (d4);  
    \node[ball,text width=.1cm,fill,color=white,above=.2cm of d3.north] (tinf) {};
    \node[ball,text width=.1cm,fill,color=white,below=.2cm of d1.south] (t0) {}; 
    \draw[->,color=black] (tinf) edge node [text width=.18cm,left=-0.05cm,midway] {\tiny $\eta$} (t0);

    \node[ball,text width=.18cm,fill,color=black] at (1,-1) (x1) {};    
    \node[ball,text width=.18cm,fill,color=black, right=1.5cm of x1] (x2) {};
    \draw[-,thick,color=black] (x1) -- (x2);
    \node[ball,text width=.1cm,fill,color=black] at (.5,0) (p1) {};
    \node[ball,text width=.1cm,fill,color=black] at (1.5,0) (p2) {};
    \draw[-,color=black] (p1) -- (x1) -- (p2);
    \node[ball,text width=.1cm,fill,color=black, right=.5cm of p2] (p3) {};
    \node[ball,text width=.1cm,fill,color=black, right=1cm of p3] (p4) {};
    \draw[-,color=black] (p3) -- (x2) -- (p4);

    \node[ball,text width=.18cm,fill,color=black, left=2cm of x1] (z1) {};
    \node[ball,text width=.1cm,fill,color=black, above=.9cm of z1] (r2) {};    
    \node[ball,text width=.1cm,fill,color=black, left=.5cm of r2] (r1) {};
    \node[ball,text width=.1cm,fill,color=black, right=.5cm of r2] (r3) {};
    \draw[-,color=black] (z1) -- (r2);
    \draw[-,color=black] (z1) edge[bend left] (r1);
    \draw[-,color=black] (z1) edge[bend right] (r3);    

    \node[ball,text width=.18cm,fill,color=black, right=2cm of x2] (v1) {};
    \node[ball,text width=.18cm,fill,color=black, right=1cm of v1] (v2) {};    
    \draw[thick] ($(v1)!0.5!(v2)$) circle (.6cm);   
    \node[ball,text width=.1cm,fill,color=black] at (4.25,0) (q1) {};
    \draw[-,color=black] (q1) edge[bend right] (v1);    
    \node[ball,text width=.1cm,fill,color=black, right=2.35cm of q1] (q2) {};
    \draw[-,color=black] (q2) edge[bend left] (v2);      

    \node[ball,text width=.18cm,fill,color=black, right=2cm of v2] (y1) {};
    \node[ball,text width=.18cm,fill,color=black, right=1cm of y1] (y2) {};
    \node[ball,text width=.18cm,fill,color=black, right=1cm of y2] (y3) {};    
    \draw[-,thick,color=black] (y1) -- (y2) -- (y3);    
    \node[ball,text width=.1cm,fill,color=black] at (7.65,0) (k1) {};    
    \node[ball,text width=.1cm,fill,color=black,right=1cm of k1] (k2) {};        
    \draw[-,color=black] (k1) -- (y1) -- (k2);        
    \node[ball,text width=.1cm,fill,color=black,above=.88cm of y2] (k3) {};

    \draw[-,color=black] (k3) -- (y2);
    \node[ball,text width=.1cm,fill,color=black,right=.6cm of k3] (k4) {};    
    \node[ball,text width=.1cm,fill,color=black,right=.9cm of k4] (k5) {};        
    \draw[-,color=black] (k4) -- (y3) -- (k5);            
  \end{tikzpicture} 
\end{equation*}

As we will see, it is natural to associate these Feynman diagrams with closely related graphs, where all the lines propagating out the boundary at $\eta=0$ are truncated, and each vertex $v$ and edge $e$ of the graph are accompanied by 
``energy'' variables $x_v$ and $y_e$. The universal integrand is a rational function of these $(x_v,y_e)$ variables. The graphs and integrands associated with the Feynman diagrams in the above examples are 
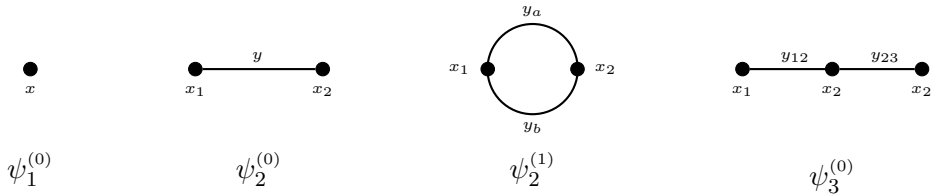
\begin{equation*}
  \begin{tikzpicture}[ball/.style = {circle, draw, align=center, anchor=north, inner sep=0}]
   \node[ball,text width=.18cm,fill,color=black, label=below:{\tiny $x_1$}] at (1,-1) (x1) {};    
   \node[ball,text width=.18cm,fill,color=black, right=1.5cm of x1, label=below:{\tiny $x_2$}] (x2) {};
   \draw[-,thick,color=black] (x1) edge node [text width=.18cm,above=-0.05cm,midway] {\tiny $y$}  (x2);
   \coordinate (c2) at ($(x1)!0.5!(x2)$);
   \node[text width=.18cm,below=.7cm of c2,label=below:{$\psi_2^{\mbox{\tiny $(0)$}}$}]{};

   \node[ball,text width=.18cm,fill,color=black,left=2cm of x1, label=below:{\tiny $x$}] (z1) {};
   \node[text width=.18cm,below=.6cm of z1,label=below:{$\psi_1^{\mbox{\tiny $(0)$}}$}]{};

   \node[ball,text width=.18cm,fill,color=black, right=2cm of x2, label=left:{\tiny $x_1$}] (v1) {};
   \node[ball,text width=.18cm,fill,color=black, right=1cm of v1, label=right:{\tiny $x_2$}] (v2) {};    
   \coordinate (c1) at ($(v1)!0.5!(v2)$);
   \draw[thick] ($(v1)!0.5!(v2)$) circle (.6cm);
   \node[text width=.18cm,above=.3cm of c1,label=above:{\tiny $y_a$}] {};
   \node[text width=.18cm,below=.3cm of c1,label=below:{\tiny $y_b$}] {};   
   \node[text width=.18cm,below=.7cm of c1,label=below:{$\psi_2^{\mbox{\tiny $(1)$}}$}]{};

   \node[ball,text width=.18cm,fill,color=black, right=2cm of v2, label=below:{\tiny $x_1$}] (y1) {};
   \node[ball,text width=.18cm,fill,color=black, right=1cm of y1, label=below:{\tiny $x_2$}] (y2) {};
   \node[ball,text width=.18cm,fill,color=black, right=1cm of y2, label=below:{\tiny $x_2$}] (y3) {};    
   \draw[-,thick,color=black] (y1) edge node [text width=.18cm,above=-0.05cm,midway] {\tiny $y_{12}$}  (y2); 
   \draw[-,thick,color=black] (y2) edge node [text width=.18cm,above=-0.05cm,midway] {\tiny $y_{23}$}  (y3);    
   \node[text width=.18cm,below=.7cm of y2,label=below:{$\psi_3^{\mbox{\tiny $(0)$}}$}]{};   
  \end{tikzpicture}  
\end{equation*}
where
\begin{equation*}
 \begin{split}
  &\psi_1^{\mbox{\tiny $(0)$}}\:=\:\frac{1}{x},\\
  &\psi_2^{\mbox{\tiny $(0)$}}\:=\:\frac{1}{(x_1+x_2)(x_1+y)(y+x_2)},\\
  &\psi_2^{\mbox{\tiny $(1)$}}\:=\:\frac{2(x_1+x_2+y_a+y_b)}{(x_1+x_2)(x_1+y_a+y_b)(x_2+y_a+y_b)(x_1+x_2+2y_a)(x_1+x_2+2y_b)},\\
  &\psi_3^{\mbox{\tiny $(0)$}}\:=\:\frac{x_1+2x_2+x_3+y_{12}+y_{23}}{(x_1+x_2+x_3)(x_1+y_{12})(y_{12}+x_2+y_{23})(y_{23}+x_3)(x_1+x_2+y_{23})(y_{12}+x_2+x_3)}.
 \end{split}
\end{equation*}

We will discuss various natural approaches to computing this integrand, from the ``bulk" path integral perspective involving time integrations, to the ``boundary" perspective of old-fashioned perturbation theory, as well as new techniques exploiting efficient recursion relations. 

In Section \ref{sec:CosmP} we switch gears, and describe a new class of polytopes -- ``cosmological polytopes'' -- which have a natural definition entirely in their own setting with no reference to physics. We will see that this definition is  naturally associated with a graph. The vertices and edges of a graph with $V$ vertices and $E$ edges are associated with basis vectors $\bf{x}_v,\bf{y}_e$ in an $(E+V)$ dimensional space. The vertices of the cosmological polytope can be read off from the graph in the following way. Each edge $e$ of the graph is associated with three vertices of the polytope:
\begin{equation*} 
 \begin{tikzpicture}[ball/.style = {circle, draw, align=center, anchor=north, inner sep=0}]
  \node[ball,text width=.18cm,fill,color=black, label=above:{\tiny $v$}, label=below:{\tiny ${\bf x}_v$}] at (0,0) (x1) {};       
  \node[ball,text width=.18cm,fill,color=black, label=above:{\tiny $v'$}, label=below:{\tiny ${\bf x}_{v'}$}, right=1.5cm of x1.east] (x2) {};         
  \draw[-,thick,color=black] (x1) edge node [text width=.18cm,above=-0.05cm, midway] {\tiny $e$} node[text width=.18cm,below=-0.05cm, midway]{\tiny ${\bf y}_e$} (x2);
  \node[right=1cm of x2.east] {$\displaystyle\xrightarrow{\hspace{2cm}}\qquad
                                \left\{
                                \begin{array}{l}
                                 {\bf x}_v + {\bf x}_{v'}-{\bf y}_e\\
                                 {\bf x}_v + {\bf y}_{e}-{\bf x}_{v'}\\
                                 {\bf x}_{v'} + {\bf y}_{e}-{\bf x}_v
                                \end{array}
                                \right.
                               $};
 \end{tikzpicture}    
\end{equation*}

The cosmological polytope is (projectively) the convex hull of all these $3 E$ vertices ${\bf V}_i$, i.e. all points ${\bf Y}$ of the form ${\bf Y}= c_i {\bf V}_i$ with $c_i>0$. This gives us a polytope that lives in 
$\mathbb{P}^{E+V-1}$. As an example, the polytope associated with the graph
$$
 
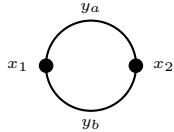
\begin{tikzpicture}[ball/.style = {circle, draw, align=center, anchor=north, inner sep=0}]
   \node[ball,text width=.18cm,fill,color=black, label=left:{\tiny $x_1$}]  at (0,0) (v1) {};
   \node[ball,text width=.18cm,fill,color=black, right=1cm of v1, label=right:{\tiny $x_2$}] (v2) {};    
   \coordinate (c1) at ($(v1)!0.5!(v2)$);
   \draw[thick] ($(v1)!0.5!(v2)$) circle (.6cm);
   \node[text width=.18cm,above=.3cm of c1,label=above:{\tiny $y_a$}] {};
   \node[text width=.18cm,below=.3cm of c1,label=below:{\tiny $y_b$}] {};   
 \end{tikzpicture}
$$
is the convex hull of the six points $\{{\bf x}_1 + {\bf x}_2 - {\bf y}_a, {\bf x}_1 + {\bf y}_a - {\bf x}_2, {\bf x}_2 + {\bf y}_a - {\bf x}_1; {\bf x}_1 + {\bf x}_2 - {\bf y}_b, {\bf x}_1 + {\bf y}_b - {\bf x}_2, {\bf x}_2 + {\bf y}_b - {\bf x}_1\}$, which is a three-dimensional polytope: 
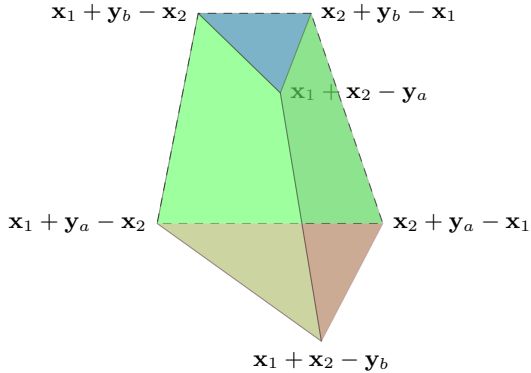
\begin{equation*}
 \begin{tikzpicture}[line join = round, line cap = round, ball/.style = {circle, draw, align=center, anchor=north, inner sep=0}]
 \begin{scope}
  \pgfmathsetmacro{\factor}{1/sqrt(2)};
  \coordinate [label=right:{\footnotesize ${\bf x}_2 + {\bf y}_b - {\bf x}_1$}] (c1a) at (9.75,0,-.75*\factor);
  \coordinate [label=left:{\footnotesize ${\bf x}_1 + {\bf y}_b - {\bf x}_2$}] (b1a) at (8.25,0,-.75*\factor);
  \coordinate [label=right:{\footnotesize ${\bf x}_1 + {\bf x}_2 - {\bf y}_a$}] (a2a) at (9.75,-.65,.75*\factor);

  \coordinate [label=right:{\footnotesize ${\bf x}_2 + {\bf y}_a - {\bf x}_1$}] (c2a) at (10.5,-3,-1.5*\factor);
  \coordinate [label=left:{\footnotesize ${\bf x}_1 + {\bf y}_a - {\bf x}_2$}] (b2a) at (7.5,-3,-1.5*\factor);
  \coordinate [label=below:{\footnotesize ${\bf x}_1 + {\bf x}_2 - {\bf y}_b$}] (a1a) at (10.5,-3.75,1.5*\factor);

  \draw[-,dashed,fill=green!50,opacity=.6] (c1a) -- (b1a) -- (b2a) -- (c2a) -- cycle;
  \draw[draw=none,fill=red!60, opacity=.45] (c2a) -- (b2a) -- (a1a) -- cycle;
  \draw[-,fill=blue!,opacity=.3] (c1a) -- (b1a) -- (a2a) -- cycle; 
  \draw[-,fill=green!50,opacity=.4] (b1a) -- (a2a) -- (a1a) -- (b2a) -- cycle;
  \draw[-,fill=green!45!black,opacity=.2] (c1a) -- (a2a) -- (a1a) -- (c2a) -- cycle; 
 \end{scope}                                              
 \end{tikzpicture}
\end{equation*}

As with scattering amplitudes, the connection to physics is via the ``canonical form" with logarithmic singularities on (and only on) all the facets of polytope, which computes the corresponding (integrand of the graph associated with the) cosmological wavefunction.  In Section \ref{sec:PP}, we illustrate this connection in more detail, showing how different natural triangulations of the polytope (and the dual of the polytope) correspond to the ``bulk" time integral and ``boundary'' old-fashioned perturbation theory computations of the wavefunction. Recent investigations of canonical forms and positive geometries \cite{Arkani-Hamed:2017abh} have also revealed new representations of canonical forms not associated  with any sort of triangulation, in terms of contour-integrals and ``push-forwards", some of which are especially well-suited for application to the cosmological polytope as we describe in Section \ref{sec:NRWF}. 

In Section \ref{sec:SymPol}, we take a first look at symmetries of the canonical forms associated with cosmological polytopes, seeing how these ultimately geometrical symmetries of the polytope descend to more familiar physical symmetries of the wavefunction (such as conformal invariance for deSitter correlators). 

Finally in Section \ref{sec:TT}, we move beyond characterizing the integrands and also consider performing the integrals that give rise to the final wavefunction,  which are especially simple in the context of e.g. $\phi^3$ theory in $dS_4$. Here the final result associated with any graph is seen to be polylogarithmic, and we give an explicit expression for the symbol of these polylogs, which can be seen as a record of the geometry of the cosmological polytope. 

We end in Section \ref{sec:Concl} with some speculations and comments on directions for future work. 


\section{Computations of the Wavefunction}\label{sec:model}

\subsection{The Model}

We will be considering a toy model of a massless scalar field $\phi$ in $(d+1)$-dimensional flat space, with time-dependent polynomial interactions; the action is
\begin{equation}\label{eq:model2}
 S\:=\:\int\,d^dx\,d\eta\,
  \left[
   \frac{1}{2}\left(\partial\phi\right)^2-\sum_{k\,\ge\,3}\frac{\lambda_k(\eta)}{k!}\phi^k
  \right].
\end{equation}
For a specific form of the time-dependence for $\lambda_k(\eta)$, this is equivalent to a conformally-coupled scalar, with non-conformal polynomial interactions, in a general FRW cosmology \footnote{The late-time behavior of the perturbative wavefunction for purely conformally coupled case in four-dimensional de Sitter space-time has been discussed in \cite{Anninos:2014lwa}.} Indeed we can begin with 
\begin{equation}\label{eq:model}
 \begin{split}
  &
   S\:=\:\int d^dx\,d\eta\,\sqrt{-g}\,
    \left[
     \frac{1}{2}g^{\mu\nu}\left(\partial_{\mu}\phi\right)\left(\partial_{\nu}\phi\right)- \xi R\phi^2-\sum_{k\,\ge\,3}\frac{\lambda_k}{k!}\phi^k
    \right],
   \\
  &\hspace{1.5cm}
   ds^2\:\equiv\:g_{\mu\nu}dx^{\mu}dx^{\nu}\:=\:
    a^2(\eta)
    \left[
     -d\eta^2\:+\:dx^idx_i
    \right],
    \qquad
    \xi \:=\: \frac{(d-1)}{4 d}
 \end{split}
\end{equation}
where the metric has been written in comoving coordinates, with conformal time $\eta$, the index $i$ runs over the spatial directions ($i\,=\,1,\,\ldots,\,d$). The  conformal transformation 

\begin{equation}\label{eq:ct}
 g_{\mu\nu}\,\longrightarrow\,a^2(\eta)g_{\mu\nu},\qquad\phi\,\longrightarrow\,a^{-\Delta}(\eta)\phi,\quad\Delta\:=\:\frac{d-1}{2},
\end{equation}
allows to rewrite \eqref{eq:model} as an action of the general form we are considering, with 

\begin{equation}
 \lambda_k(\eta)\,\equiv\,\lambda_k\,\left[a(\eta)\right]^{(2-k)\Delta+2}.
\end{equation}

Our analysis will be focused on the structure and computation of the universe wavefunction, which we will write as
\begin{equation}\label{eq:wvu}
 \Psi\:=\:\mbox{exp}
  \left\{
   \sum_{n\,\ge\,2}\frac{1}{n!}\int\prod_{v=1}^n\,\left[d^d\,z_v\,\phi(z_v)\right]\hat{\psi}_n(z)
  \right\}.
\end{equation}
More precisely, we will analyze and (perturbatively) compute the $\psi_n$'s in \eqref{eq:wvu} in momentum space\footnote{It is important to stress that, while the $\psi_n$'s in \eqref{eq:wvu} are written in position space $z$, we will instead study the $\psi_n$ in momentum space.}. Spatial translational invariance allows us to pull out an overall delta function corresponding to spatial momentum conservation: 
\begin{equation}\label{eq:GenWvF1}
 \hat{\psi}_n = \delta^{\mbox{\tiny $(d)$}}\left(\sum_{i=1}^n\overrightarrow{p}^{\mbox{\tiny $(i)$}}\right)\tilde{\psi}_n
\end{equation}

The wavefunction at $\eta=0$ can be computed by the Feynman path integral, by integrating over all field configurations $\phi(z, \eta)$ with the boundary conditions that $\phi(z,\eta \to 0) = \phi(z)$. This motivates expanding (now working in momentum space) $\phi(p,\eta) = \phi(p) e^{+i E_p \eta} + \delta \phi(p, \eta)$ where $\delta \phi(p, \eta \to 0) = 0$. The first term satisfies the boundary condition and solves the free equation of motion;  the choice of the solution that oscillates as $e^{+ i E_{p} \eta}$ ensures the correct adiabatic/Bunch-Davies/Hartle-Hawking vacuum in the deep past as $\eta \to -\infty$. We then perform the path integral over $\delta \phi$. 

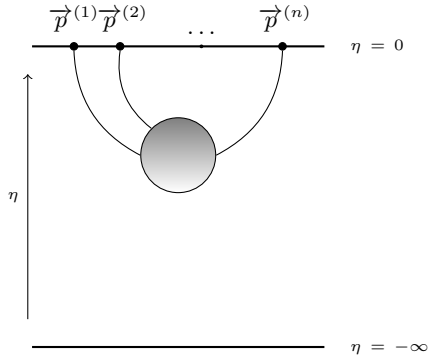
\begin{equation}\label{eq:wvn}
  \begin{tikzpicture}[ball/.style = {circle, draw, align=center, anchor=north, inner sep=0}]
    \node[ball,text width=.1cm,fill,color=white] at (0,0) (d1) {};
    \node[ball,text width=.1cm,fill,color=white] at (4,0) (d2) {};
    \node[right=.1cm of d2.east] (tl) {\tiny $\eta\,=\,0$};
    \draw[thick,color=black] (d1) -- (d2);    
    \node[ball,text width=.1cm,fill,color=white] at (0,-4) (d3) {};
    \node[ball,text width=.1cm,fill,color=white] at (4,-4) (d4) {};
    \node[right=.1cm of d4.east] (te) {\tiny $\eta\,=\,-\infty$};    
    \draw[thick,color=black] (d3) -- (d4);  
    \node[ball,text width=.1cm,fill,color=white,above=.2cm of d3.north] (tinf) {};
    \node[ball,text width=.1cm,fill,color=white,below=.2cm of d1.south] (t0) {}; 
    \draw[->,color=black] (tinf) edge node [text width=.18cm,left=-0.05cm,midway] {\tiny $\eta$} (t0);
    \node[ball,text width=.1cm,fill,color=black,right=.5cm of d1, label=above:{\footnotesize $\overrightarrow{p}^{(1)}$}] (p1) {};
    \node[ball,text width=.1cm,fill,color=black,right=.5cm of p1, label=above:{\footnotesize $\:\overrightarrow{p}^{(2)}$}] (p2) {};    
    \node[ball,text width=.0cm,fill,color=black,right=1cm of p2, label=above:{\footnotesize $\ldots$}] (d5) {};        
    \node[ball,text width=.1cm,fill,color=black,left=.5cm of d2, label=above:{\footnotesize $\:\overrightarrow{p}^{(n)}$}] (pn) {};        
    \node[ball,text width=1cm,shade] (S1) at (2,-1) {};  
    \draw[color=black] (S1.west) edge[bend left] (p1.south);
    \draw[color=black] (S1.north west) edge[bend left] (p2.south);    
    \draw[color=black] (S1.east) edge[bend right] (pn.south);    
    \node[right=3cm of S1.east, color=black] (form) {$\displaystyle \tilde{\psi}_n\:=\:\int_{-\infty}^0\prod_{v\in\mathcal{V}}\left[d\eta_v\,V_v\,H_v\right]\,\prod_{e\in\mathcal{E}}G_e(\eta_{v_e},\eta_{v'_e})$};
  \end{tikzpicture}
\end{equation}

\begin{figure}
 \begin{tikzpicture}[ball/.style = {circle, draw, align=center, anchor=north, inner sep=0}]
  \node[ball,text width=.18cm,fill,color=black,label=below:{\tiny $x_1$}] at (0,0) (x1) {};
  \node[ball,text width=.18cm,fill,color=black,right=1.5cm of x1.east,label=below:{\tiny $x_2$}] (x2) {};
  \draw[-,thick,color=black] (x1.east) edge node [text width=.18cm,above=-0.05cm,midway] {\tiny {$y$}} (x2.west);

  \node[ball,text width=.18cm,fill,color=black,label=below:$x_1$,right=1.75cm of x2.west] (z1) {};
  \node[ball,text width=.18cm,fill,color=black,right=1.5cm of z1.east,label=below:{\tiny $x_2$}] (z2) {};
  \node[ball,text width=.18cm,fill,color=black,right=1.5cm of z2.east,label=below:{\tiny $x_3$}] (z3) {};  
  \draw[-,thick,color=black] (z1.east) edge node [text width=.18cm,above=-0.05cm,midway] {{\tiny $y_{12}$}} (z2.west);
  \draw[-,thick,color=black] (z2.east) edge node [text width=.18cm,above=-0.05cm,midway] {{\tiny $y_{23}$}} (z3.west);

  \node[ball,text width=.18cm,fill,color=black,label=left:{\tiny $x_4$},right=2.25cm of z3.east] (w4) {};    
  \node[ball,text width=.18cm,fill,color=black,label=below:{\tiny $x_3$}] at ({1.5*cos(0)+9.5},{1.5*sin(0)}) (w3) {};    
  \node[ball,text width=.18cm,fill,color=black,label=left:{\tiny $x_1$}] at ({1.5*cos(120)+8.5},{1.5*sin(120)}) (w1) {};
  \node[ball,text width=.18cm,fill,color=black,label=left:{\tiny $x_2$}] at ({1.5*cos(240)+8.5},{1.5*sin(240)}) (w2) {};    
  \draw[-,thick,color=black] (w1) edge node [text width=.18cm,left=.3cm,midway] {\tiny $y_{14}$} (w4);    
  \draw[-,thick,color=black] (w2) edge node [text width=.18cm,right=-.1cm,midway] {\tiny $y_{24}$} (w4);        
  \draw[-,thick,color=black] (w3) edge node [text width=.18cm,above=-.05cm,midway] {\tiny $y_{34}$} (w4);

  \node[ball,text width=.18cm,fill,color=black,label=left:{\tiny $x_1$},right=1.5cm of w3.east] (y1) {};    
  \node[ball,text width=.18cm,fill,color=black,right=1.5cm of y1.east,label=right:{\tiny $x_2$}] (y2) {};   
  \draw[thick] ($(y1)!0.5!(y2)$) circle (.88cm);
 \end{tikzpicture}
 \caption{Two-site chain, three-site chain, four-site star and two-site one-loop graphs.}
 \label{fig:Graphs}
\end{figure}
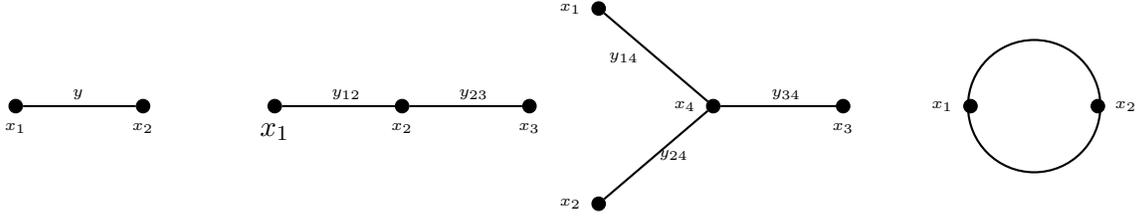

The path integral can be performed diagrammatically in the usual way with Feynman graphs, which have some set ${\mathcal V}$ of vertices and ${\mathcal E}$ of edges. $H_v$ is the ``bulk-boundary'' propagator, associated with a a vertex $v$,  simply given by positive-frequency plane waves. The $G_e(\eta_{v_e},\eta_{v'_e})$'s are the ``bulk-bulk'' propagators associated  with an edge $e$  going from one vertex at time $\eta_{v_e}$ to another one at time $\eta_{v'_e}$. In our scalar case with non-derivative polynomial interactions, the functional form $V_v$ of the vertices is just given by the coupling constant and does not depend on the spatial momenta of the external states.  Explicitly we have
\begin{equation}\label{eq:wvn2}
 \begin{split}
  &H_v(E_v;\,\eta_v)\:=\:e^{iE_v\eta_v},\\
  &G_e(E_e;\,\eta_{v_e},\,\eta_{v'_e})\,=\,\frac{1}{E_e}
   \left[
    e^{-iE_e(\eta_{v_e}-\eta_{v'_e})}\vartheta(\eta_{v_e}-\eta_{v'_e})+e^{+iE_e(\eta_{v_e}-\eta_{v'_e})}\vartheta(\eta_{v'_e}-\eta_{v_e})-e^{E_e(\eta_{v'_e}+\eta_{v_e})}
   \right],
 \end{split}
\end{equation}
where  $E_v = \sum_a |\vec{p}_a|$  is the sum of the energies of the external lines $a$ attached the vertex $v$, and $E_e = |\vec{p}_e|$ is the energy of an internal line. The first two terms in the propagator $G_e$ provide the standard time-ordered Feynman propagator decomposed into its advanced and retarded parts, while the third term is due to the requirement that $\delta \phi(z, \eta \to 0) = 0$ at the boundary. Finally the vertices for our non-polynomial interactions are just time-dependent constants which is convenient to treat in (time) Fourier space;
\begin{equation}\label{eq:endepc}
 \lambda_k(\eta)\:=\:\int_{0}^{\infty} d\varepsilon\,e^{i\varepsilon\eta} \tilde{\lambda}_k(\varepsilon)
\end{equation}
and 
 $\tilde{\psi}_n$ in \eqref{eq:wvn}  is an integral over all the $\varepsilon$'s:
\begin{equation}\label{wvn2}
 \tilde{\psi}\:=\:\int_{0}^{\infty} \prod_v d\varepsilon_v \,\left[\prod_{v\in\mathcal{V}}\tilde{\lambda}_{k_v}(\varepsilon)\right]\,\psi(E_v + \varepsilon_v,E_e),
\end{equation}
where 
\begin{equation}
 \psi(x_v,y_e) = \int_{-\infty}^0 \prod_{v\in\mathcal{V}} d \eta_v \prod_{e\in\mathcal{E}} e^{i x_v \eta_v} G_e(\eta_{v_e},\eta_{v^\prime_e})
\end{equation}
where for future convenience we are introducing a notation ``$x_v$" associated with the (sum of) vertex energies, and $y_e$ associated with internal line energies. We will also find it convenient to draw a slightly simplified version of the Feynman diagrams associated with this integral, where we keep only vertices and internal edges but truncate the propagators going to the boundary. The Feynman diagram associated with a contact interaction is then truncated to a single point, and we find
\begin{equation}
 i \psi(x) = i  \int_{- \infty}^0 e^{i x \eta} = \frac{1}{x}
\end{equation}
In what follows to avoid clutter we will always suppress the factors of $i$ (and coupling constants) that can be trivially re-instated when needed. We will frequently discuss three examples of graphs to illustrate our ideas and methods in this paper: the two-site chain, the three-site chain and the two-site loop\footnote{We use the notation $\psi_{n}^{\mbox{\tiny $(L)$}}$ to denote the $n$-site wavefunction at $L$-loops. We might drop the loop index when the perturbative order we are in is clear.}. The graphs are  shown in Fig. \ref{fig:Graphs}. As we will see the intermediate steps in actually computing the time integrals can be quite complicated but the final answers are much simpler than expected: 
\begin{equation}\label{eq:23sum}
 \begin{split}
  &\psi_{2}^{\mbox{\tiny $(0)$}} \:=\: \frac{1}{(x_1 + x_2)(x_1 + y)(x_2 + y)} \\ 
  &\psi_{3}^{\mbox{\tiny $(0)$}} \:=\: \frac{x_1 + x_3 + 2 x_2  + y_{12} + y_{23}}{(x_1 + y_{12})(x_3 + y_{23})(x_2 + y_{12} + y_{23})(x_1 + x_2 + y_{23})(x_3 + x_2 + y_{12})}\\ 
  &\psi_{2}^{\mbox{\tiny $(1)$}} \:=\: \frac{2(x_1 + x_2 + y_{a} + y_{b})}{(x_1 + x_2 + 2 y_a) (x_1 + x_2 + 2 y_a)(x_1 + y_a + y_b)(x_2 + y_a + y_b)}\\
 \end{split}
\end{equation}

While the simplicity of these expressions is interesting, at first sight nothing in these expressions is suggestive of any connection with an underlying geometry. However experience with scattering amplitudes tells us where to look for such a connection.  Consider the differential form 
\begin{equation}
 \Omega = \prod_{v\in\mathcal{V}}\prod_{e\in\mathcal{E}} dx_v dy_e\,\psi(x_v,y_e).
\end{equation}
We first notice that quite nicely,  if we combine all the $(x,y)$ into a vector ${\cal Y}=(x,y)$ in $\mathbb{R}^{V+E}$, then $\psi$ has precisely the correct weight to correspond to a differential form on the projective space $\mathbb{P}^{V+E-1}$. Furthermore, and quite non-trivially, the reader can verify that in all these examples, $\Omega$ has unit residues! It is this magical property that strongly suggests the identification of $\Omega$ with the ``canonical form'' associated with ``a positive geometry''. In the case at hand where $\Omega$ is naturally defined on a projective space, $\Omega$ should be the form with logarithmic singularities on some polytope, or as is now well-known, the volume of the dual of this polytope. We will describe this polytope in Section \ref{sec:PP}. 

However for the remainder of this section we will focus on understanding the wavefunction $\psi(x,y)$ from a number of different perspectives, each of which makes different aspects of the physics manifest. 

Before engaging in the detailed analysis of the structure of the wavefunctions and their computation, there is a general feature that it is worth to stress and we can already examine. As we already pointed out, the wavefunctions
have support on the spatial momentum conservation sheet, while the presence of space-like boundary at conformal time $\eta\,=\,0$ breaks time-translation symmetry and therefore the total energy is not conserved. However, if we look
carefully at the structure of the {\it integrand} of the wavefunction -- {\it i.e.} before the integration over the energies related to time-dependent coupling constants is performed --, we discover that the sum of the energies
appear in the denominator as an overall factor. It is easy to see how this arises. We can split the integral over all the times $\eta_v$ into one over the ``center of mass co-ordinate" $\bar{\eta}$, and all the differences. Now, as $\bar{\eta} \to -\infty$, there is an oscillatory dependence on $\bar{\eta}$ as $e^{i (\sum_i E_i) \bar{\eta}}$. Thus there is a divergence in the integral as $\bar{\eta} \to - \infty$, when $\sum_i E_i \to 0$. 
Thus, instead of a an energy-conserving $\delta$ function for energy conservation familiar from scattering amplitudes, the integrand of the wavefunction has a pole instead: $\delta(\sum_i  E_i) \to \frac{1}{\sum_i E_i}$.

This argument also tells us a beautiful fact about the residue of the (integrand of the) wavefunction on the pole where $\sum_i E_i \to 0$: the residue is precisely the flat-space scattering amplitude! This is simply because the residue is dominated by the integration where all the vertices go off to the distant past; then the presence of the boundary at $\eta =0$ is immaterial, time-translation invariance is restored and the computation of the coefficient of the divergence is precisely the same as that of the scattering amplitude. Of course in the ``physical region", all these energies $E_i = |\vec{p}_i|$ are positive and we can't reach this pole, but it is remarkable that an analytic continuation of the (integrand of the) completely ``static'' wavefunction contains all the ``dynamical" information of particle scattering! As an illustration, we can see how this works for the a four-point correlator coming from the $s-$channel exchange in a $\phi^3$ theory. The integrand is
\begin{equation}\label{eq:psi4phi3a}
 \psi_{4}^{\mbox{\tiny $(\phi^3)$}}\:=\:\frac{1}{E_1 + E_2 + E_3 + E_4}\,\frac{1}{E_1 + E_2 + E_I}\,\frac{1}{E_3 + E_4 + E_I},
\end{equation}
with $E_i$ ($i\,=\,1,\,\ldots,\,4$) are the energies of the external states while $E_I$ is the energy of the internal one. The residue of this expression as $E_1 + E_2 + E_3 + E_4 \to 0$ is 
\begin{equation}\label{eq:psi4phi3b}
 \begin{split}
  \left.{\rm Res}\left\{\psi_4^{\mbox{\tiny $(\phi^3)$}}\right\}\right|_{\sum_i\,E_i\to 0}\:&=\:\frac{1}{E_1 + E_2 + E_I}\,\frac{1}{E_3 + E_4 + E_I}\:=\:\frac{1}{E_1 + E_2 + E_I}\,\frac{1}{-E_1 - E_2 + E_I}\\ 
                                                                             &=\:\frac{1}{E_I^2-(E_1 + E_2)^2}\:=\:\frac{1}{s}.
 \end{split}
\end{equation}

\begin{figure}
 \centering
 \begin{tikzpicture}[ball/.style = {circle, draw, align=center, anchor=north, inner sep=0}]
    \node[ball,text width=.1cm,fill,color=white] at (0,0) (d1) {};
    \node[ball,text width=.1cm,fill,color=white] at (4,0) (d2) {};
    \node[right=.1cm of d2.east] (tl) {\tiny $\eta\,=\,0$};
    \draw[thick,color=black] (d1) -- (d2);    
    \node[ball,text width=.1cm,fill,color=white] at (0,-4) (d3) {};
    \node[ball,text width=.1cm,fill,color=white] at (4,-4) (d4) {};
    \node[right=.1cm of d4.east] (te) {\tiny $\eta\,=\,-\infty$};    
    \draw[thick,color=black] (d3) -- (d4);  
    \node[ball,text width=.1cm,fill,color=white,above=.2cm of d3.north] (tinf) {};
    \node[ball,text width=.1cm,fill,color=white,below=.2cm of d1.south] (t0) {}; 
    \draw[->,color=black] (tinf) edge node [text width=.18cm,left=-0.05cm,midway] {\tiny $\eta$} (t0);
    \node[ball,text width=.1cm,fill,color=black,right=.5cm of d1, label=above:{\footnotesize $\overrightarrow{p}^{(1)}$}] (p1) {};
    \node[ball,text width=.1cm,fill,color=black,right=.5cm of p1, label=above:{\footnotesize $\:\overrightarrow{p}^{(2)}$}] (p2) {};    
    \node[ball,text width=.0cm,fill,color=black,right=1cm of p2, label=above:{\footnotesize $\ldots$}] (d5) {};        
    \node[ball,text width=.1cm,fill,color=black,left=.5cm of d2, label=above:{\footnotesize $\:\overrightarrow{p}^{(n)}$}] (pn) {};        
    \node[ball,text width=1cm,shade] (S1) at (2,-1) {};  
    \draw[color=black] (S1.west) edge[bend left] (p1.south);
    \draw[color=black] (S1.north west) edge[bend left] (p2.south);    
    \draw[color=black] (S1.east) edge[bend right] (pn.south);   
    \draw[color=red,thick] (S1) circle (.6cm);
    \node[color=red,thick,right=.5cm of S1.south east, label=right:{\tiny Early times}] (et) {$\displaystyle{\bf \xdownarrow{.8cm}}$};
 \end{tikzpicture}
 \caption{Total energy pole. Upon performing analytic continuation in the energy space, it is possible to reach the point where the sum of the all energies vanishes. From the time perspective, this is equivalent to carry
          the interaction at early time, so that it is infinitely far away from the boundary, which appears to be effectively absent from the perspective of the early time process. The energy conservation is restored
          and the process is related to scattering in flat space-time.}\label{fig:Et}
\end{figure}
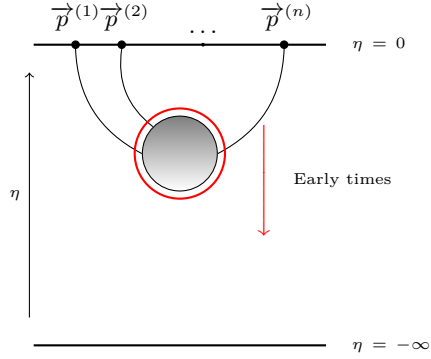


\subsection{``Bulk" Representation as Time Integral}\label{subsec:TimeRep}

In this subsection, we consider the perturbative wavefunction as defined in \eqref{eq:wvn}, with the propagator decomposed into its advanced and retarded parts and a further term due to the condition that it has to vanish at
the boundary $\eta\,=\,0$ \eqref{eq:wvn2}. In such a time-integral representation, a generic contribution to the wavefunction having $v$ sites end $e$ edges will be generated as a sum of $3^{E}$ terms. This is the direct consequence 
of the specific decomposition of the propagator, which makes the flow of time among events manifest. As we will show in what follows, if the sum is explicitly performed, the final answer turns out simplify very significantly.  

Let us analyze the time-integral representation, starting with the simplest example given by the two-site graph:
\begin{equation}\label{eq:2site}
 \begin{tikzpicture}[ball/.style = {circle, draw, align=center, anchor=north, inner sep=0}]
  \node[ball,text width=.18cm,fill,color=black,label=below:$x_1$] at (0,0) (x1) {};
  \node[ball,text width=.18cm,fill,color=black,right=1.5cm of x1.east,label=below:$x_2$] (x2) {};
  \draw[-,thick,color=black] (x1.east) edge node [text width=.18cm,above=-0.05cm,midway] {{$y$}} (x2.west);
  \node[right=.2cm of x2.east] (G2) {$\displaystyle\:=\:\int_{-\infty}^0 d\eta_1 \int_{-\infty}^0d\eta_2\,e^{ix_1\eta_1}e^{ix_2\eta_2}\,G(y;\,\eta_1,\,\eta_2).$};
 \end{tikzpicture}
\end{equation}

Because of the three-term structure of the propagator $G(y;\,\eta_1,\,\eta_2)$, the two-site graph \eqref{eq:2site} is returned as a sum of three contributions:
\begin{equation}\label{eq:2siteDec}
 \begin{tikzpicture}[ball/.style = {circle, draw, align=center, anchor=north, inner sep=0}]
  \node[ball,text width=.18cm,fill,color=black,label=below:$x_1$] at (0,0) (x1) {};
  \node[ball,text width=.18cm,fill,color=black,right=1.5cm of x1.east,label=below:$x_2$] (x2) {};
  \draw[-,thick,color=black] (x1.east) edge node [text width=.18cm,above=-0.05cm,midway] {{$y$}} (x2.west);
  \node[right=.2cm of x2.east] (t1) {$\:=\:\displaystyle\left.\frac{1}{y}\right[$};
  \node[ball,text width=.18cm,fill,color=black,right=.2cm of t1.east,label=below:$x_1$] (x1a) {};
  \node[ball,text width=.18cm,fill,color=black,right=1.5cm of x1a.east,label=below:$x_2$] (x2a) {};
  \draw[-,reverse directed,thick,color=gray] (x1a.east) -- (x2a.west); 
  \node[right=.2cm of x2a.east] (t2) {$\displaystyle\:+\:$};
  \node[ball,text width=.18cm,fill,color=black,right=.2cm of t2.east,label=below:$x_1$] (x1b) {};
  \node[ball,text width=.18cm,fill,color=black,right=1.5cm of x1b.east,label=below:$x_2$] (x2b) {};
  \draw[-,directed,thick,color=gray] (x1b.east) -- (x2b.west);
  \node[right=.2cm of x2b.east] (t3) {$\displaystyle\:-\:$};
  \node[ball,text width=.18cm,fill,color=black,right=.2cm of t3.east,label=below:$x_1$] (x1c) {};
  \node[ball,text width=.18cm,fill,color=black,right=1.5cm of x1c.east,label=below:$x_2$] (x2c) {};
  \draw[-,dashed,thick,color=gray] (x1c.east) -- (x2c.west);
  \node[right=.2cm of x2c.east] (t4) {$\displaystyle\left.\hspace{-.4cm}\phantom{\frac{1}{y}}\right],$};
 \end{tikzpicture}
\end{equation}
where the directed arrows represent the flow of time\footnote{The arrow directed from a site $i$ to another one $j$ is equivalent to $\eta_j\,>\,\eta_i$, {\it i.e.}  it represents the presence of the Heaviside step 
function $\vartheta{(\eta_j-\eta_i)}$.}, while the dashed edge in the third term indicates the absence of time ordering\footnote{Notice that the $1/y$ in the propagator has been stripped off in the right-hand-side of
\eqref{eq:2siteDec}. This is graphically represented by the gray edge. In general, any graph with gray edges has to be understood as having the factors $1/y$'s stripped off.}
\begin{equation}\label{eq:2siteTerm}
 \begin{split}
  &
   \begin{tikzpicture}[ball/.style = {circle, draw, align=center, anchor=north, inner sep=0}]
    \node[ball,text width=.18cm,fill,color=black,right=.2cm of t1.east,label=below:$x_1$] (x1a) {};
    \node[ball,text width=.18cm,fill,color=black,right=1.5cm of x1a.east,label=below:$x_2$] (x2a) {};
    \draw[-,reverse directed,thick,color=gray] (x1a.east) -- (x2a.west);
    \node[right=.2cm of x2a.east] (rhs1) {$\displaystyle\:\equiv\:\int_{-\infty}^0d\eta_1\int_{-\infty}^{0}d\eta_2\,e^{ix_1\eta_1}e^{ix_2\eta_2}e^{-iy(\eta_1-\eta_2)}\vartheta(\eta_1-\eta_2),$};
   \end{tikzpicture}
  \\
  &
   \begin{tikzpicture}[ball/.style = {circle, draw, align=center, anchor=north, inner sep=0}]
    \node[ball,text width=.18cm,fill,color=black,right=.2cm of t1.east,label=below:$x_1$] (x1b) {};
    \node[ball,text width=.18cm,fill,color=black,right=1.5cm of x1a.east,label=below:$x_2$] (x2b) {};
    \draw[-,directed,thick,color=gray] (x1b.east) -- (x2b.west);
    \node[right=.2cm of x2b.east] (rhs1) {$\displaystyle\:\equiv\:\int_{-\infty}^0d\eta_1\int_{-\infty}^{0}d\eta_2\,e^{ix_1\eta_1}e^{ix_2\eta_2}e^{+iy(\eta_1-\eta_2)}\vartheta(\eta_2-\eta_1),$};
   \end{tikzpicture}
  \\
  &
   \begin{tikzpicture}[ball/.style = {circle, draw, align=center, anchor=north, inner sep=0}]
    \node[ball,text width=.18cm,fill,color=black,right=.2cm of t1.east,label=below:$x_1$] (x1c) {};
    \node[ball,text width=.18cm,fill,color=black,right=1.5cm of x1a.east,label=below:$x_2$] (x2c) {};
    \draw[-,dashed,thick,color=gray] (x1c.east) -- (x2c.west);
    \node[right=.2cm of x2c.east] (rhs1) {$\displaystyle\:\equiv\:\int_{-\infty}^0d\eta_1\int_{-\infty}^{0}d\eta_2\,e^{ix_1\eta_1}e^{ix_2\eta_2}e^{+iy(\eta_1+\eta_2)}.$};
   \end{tikzpicture}
 \end{split}
\end{equation}
In the last term, the absence of time-ordering makes the two integration decouple so that it trivially integrates to
\begin{equation}\label{eq:2siteNoTO}
  \begin{tikzpicture}[ball/.style = {circle, draw, align=center, anchor=north, inner sep=0}]
    \node[ball,text width=.18cm,fill,color=black,right=.2cm of t1.east,label=below:$x_1$] (x1c) {};
    \node[ball,text width=.18cm,fill,color=black,right=1.5cm of x1a.east,label=below:$x_2$] (x2c) {};
    \draw[-,dashed,thick,color=gray] (x1c.east) -- (x2c.west);
    \node[right=.2cm of x2c.east] (rhs3) {$\displaystyle\:=\:\frac{1}{(x_1+y)(y+x_2)}$};
   \end{tikzpicture}
\end{equation}
while the two time-ordered terms give us 
\begin{equation}\label{eq:2siteTO}
  \begin{tikzpicture}[ball/.style = {circle, draw, align=center, anchor=north, inner sep=0}]
    \node[ball,text width=.18cm,fill,color=black,right=.2cm of t1.east,label=below:$x_1$] (x1a) {};
    \node[ball,text width=.18cm,fill,color=black,right=1.5cm of x1a.east,label=below:$x_2$] (x2a) {};
    \draw[-,reverse directed,thick,color=gray] (x1a.east) -- (x2a.west);
    \node[right=.2cm of x2a.east] (rhs1) {$\displaystyle\:=\:\frac{1}{(y+x_2)(x_1+x_2)},$};
   \end{tikzpicture}
  \qquad
  \begin{tikzpicture}[ball/.style = {circle, draw, align=center, anchor=north, inner sep=0}]
    \node[ball,text width=.18cm,fill,color=black,right=.2cm of t1.east,label=below:$x_1$] (x1c) {};
    \node[ball,text width=.18cm,fill,color=black,right=1.5cm of x1a.east,label=below:$x_2$] (x2c) {};
    \draw[-,directed,thick,color=gray] (x1c.east) -- (x2c.west);
    \node[right=.2cm of x2c.east] (rhs2) {$\displaystyle\:=\:\frac{1}{(x_1+x_2)(x_1+y)}$.};
   \end{tikzpicture}
\end{equation}
Summing up the three terms \eqref{eq:2siteTO} and \eqref{eq:2siteNoTO} as in \eqref{eq:2siteTerm}, the full expression for the two-site graph is
\begin{equation}\label{eq:2siteFin}
 \begin{split}
  &
   \begin{tikzpicture}[ball/.style = {circle, draw, align=center, anchor=north, inner sep=0}]
    \node[ball,text width=.18cm,fill,color=black,label=below:$x_1$] at (0,0) (x1) {};
    \node[ball,text width=.18cm,fill,color=black,right=1.5cm of x1.east,label=below:$x_2$] (x2) {};
    \draw[-,thick,color=black] (x1.east) edge node [text width=.18cm,above=-0.05cm,midway] {{$y$}} (x2.west);
    \node[right=.2cm of x2.east] (2sf) {$\displaystyle\:=\:\frac{1}{y (x_1 + y)(x_1 + x_2)} +  \frac{1}{y (x_2 + y)(x_1 + x_2)} - \frac{1}{y (x_1 + y)(x_2 + y)}$};
   \end{tikzpicture}
  \\
  &\hspace{2.4cm}=\:\frac{1}{(x_1 + x_2)(x_1 + y)(x_2 + y)}
 \end{split}
\end{equation}
Notice that the final answer simplifies significantly upon summation of the three terms and that the pole in $y\,=\,0$ which appears in each single term coming from the time integration, is actually spurious. The simplification we obtain upon summation is even more striking when we consider more complicated examples. Consider the three-site graph
\begin{equation}\label{eq:3site}
 \begin{tikzpicture}[ball/.style = {circle, draw, align=center, anchor=north, inner sep=0}]
    \node[ball,text width=.18cm,fill,color=black,label=below:$x_1$] at (0,0) (x1) {};
    \node[ball,text width=.18cm,fill,color=black,right=.8cm of x1.east,label=below:$x_2$] (x2) {};
    \node[ball,text width=.18cm,fill,color=black,right=.8cm of x2.east,label=below:$x_3$] (x3) {};
    \draw[-,thick,color=black] (x1.east) edge node [text width=.18cm,above=-0.05cm,midway] {{$y_{\mbox{\tiny $12$}}$}} (x2.west);
    \draw[-,thick,color=black] (x2.east) edge node [text width=.18cm,above=-0.05cm,midway] {{$y_{\mbox{\tiny $23$}}$}} (x3.west);
    \node[right=.2cm of x3.east] (G3) {$\displaystyle\:=\:\int_{-\infty}^0 \prod_{s=1}^3\left[d\eta_s\,e^{ix_i\eta_s}\right]
        G(y_{\mbox{\tiny $12$}};\eta_1,\eta_2)G(y_{\mbox{\tiny $23$}};\eta_2,\eta_3),$};
  \end{tikzpicture}
\end{equation}
whose time-integral representation \eqref{eq:3site} returns a sum of nine terms, given by all the possible combinations among the different time prescriptions for the two propagators:
\begin{equation}\label{eq:3siteTerm}
 \begin{split}
  &
   \begin{tikzpicture}[ball/.style = {circle, draw, align=center, anchor=north, inner sep=0}]
    \node[ball,text width=.18cm,fill,color=black,label=below:$x_1$] at (0,0) (x1) {};
    \node[ball,text width=.18cm,fill,color=black,right=.8cm of x1.east,label=below:$x_2$] (x2) {};
    \node[ball,text width=.18cm,fill,color=black,right=.8cm of x2.east,label=below:$x_3$] (x3) {};
    \draw[-,thick,color=black] (x1.east) edge node [text width=.18cm,above=-0.05cm,midway] {{$y_{\mbox{\tiny $12$}}$}} (x2.west);
    \draw[-,thick,color=black] (x2.east) edge node [text width=.18cm,above=-0.05cm,midway] {{$y_{\mbox{\tiny $23$}}$}} (x3.west);
    \node[right=.2cm of x3.east] (rhs1) {$\displaystyle\:=\:\left.\frac{1}{y_{12}y_{23}}\right[$};
    \node[ball,text width=.18cm,fill,color=black,right=.2cm of rhs1,label=below:$x_1$] (x1a) {};
    \node[ball,text width=.18cm,fill,color=black,right=.8cm of x1a.east,label=below:$x_2$] (x2a) {};
    \node[ball,text width=.18cm,fill,color=black,right=.8cm of x2a.east,label=below:$x_3$] (x3a) {};
    \draw[-,reverse directed, thick,color=gray] (x1a.east) -- (x2a.west);
    \draw[-,reverse directed, thick,color=gray] (x2a.east) -- (x3a.west);
    \node[right=.2cm of x3a.east] (rhs2) {$\displaystyle\:+$};
    \node[ball,text width=.18cm,fill,color=black,right=.2cm of rhs2,label=below:$x_1$] (x1b) {};
    \node[ball,text width=.18cm,fill,color=black,right=.8cm of x1b.east,label=below:$x_2$] (x2b) {};
    \node[ball,text width=.18cm,fill,color=black,right=.8cm of x2b.east,label=below:$x_3$] (x3b) {};
    \draw[-,reverse directed, thick,color=gray] (x1b.east) -- (x2b.west);
    \draw[-,directed, thick,color=gray] (x2b.east) -- (x3b.west);
    \node[right=.2cm of x3b.east] (rhs3) {$\displaystyle\:-$};
    \node[ball,text width=.18cm,fill,color=black,right=.2cm of rhs3,label=below:$x_1$] (x1c) {};
    \node[ball,text width=.18cm,fill,color=black,right=.8cm of x1c.east,label=below:$x_2$] (x2c) {};
    \node[ball,text width=.18cm,fill,color=black,right=.8cm of x2c.east,label=below:$x_3$] (x3c) {};
    \draw[-,reverse directed, thick,color=gray] (x1c.east) -- (x2c.west);
    \draw[-,dashed, thick,color=gray] (x2c.east) -- (x3c.west);
   \end{tikzpicture}
  \\
  &
   \begin{tikzpicture}[ball/.style = {circle, draw, align=center, anchor=north, inner sep=0}]
    \node[ball,text width=.18cm,fill,color=white,label=below:$\phantom{x_1}$] at (0,0) (x1) {};
    \node[ball,text width=.18cm,fill,color=white,right=.8cm of x1.east,label=below:$\phantom{x_2}$] (x2) {};
    \node[ball,text width=.18cm,fill,color=white,right=.8cm of x2.east,label=below:$\phantom{x_3}$] (x3) {};
    \draw[-,thick,color=white] (x1.east) edge node [text width=.18cm,above=-0.05cm,midway] {{$y_{\mbox{\tiny $12$}}$}} (x2.west);
    \draw[-,thick,color=white] (x2.east) edge node [text width=.18cm,above=-0.05cm,midway] {{$y_{\mbox{\tiny $23$}}$}} (x3.west);
    \node[right=.2cm of x3.east] (rhs1) {$\displaystyle\phantom{\:=\:\frac{1}{y_{12}y_{23}}}+$};
    \node[ball,text width=.18cm,fill,color=black,right=.2cm of rhs1,label=below:$x_1$] (x1a) {};
    \node[ball,text width=.18cm,fill,color=black,right=.8cm of x1a.east,label=below:$x_2$] (x2a) {};
    \node[ball,text width=.18cm,fill,color=black,right=.8cm of x2a.east,label=below:$x_3$] (x3a) {};
    \draw[-,directed, thick,color=gray] (x1a.east) -- (x2a.west);
    \draw[-,reverse directed, thick,color=gray] (x2a.east) -- (x3a.west);
    \node[right=.2cm of x3a.east] (rhs2) {$\displaystyle\:+$};
    \node[ball,text width=.18cm,fill,color=black,right=.2cm of rhs2,label=below:$x_1$] (x1b) {};
    \node[ball,text width=.18cm,fill,color=black,right=.8cm of x1b.east,label=below:$x_2$] (x2b) {};
    \node[ball,text width=.18cm,fill,color=black,right=.8cm of x2b.east,label=below:$x_3$] (x3b) {};
    \draw[-,directed, thick,color=gray] (x1b.east) -- (x2b.west);
    \draw[-,directed, thick,color=gray] (x2b.east) -- (x3b.west);
    \node[right=.2cm of x3b.east] (rhs3) {$\displaystyle\:-$};
    \node[ball,text width=.18cm,fill,color=black,right=.2cm of rhs3,label=below:$x_1$] (x1c) {};
    \node[ball,text width=.18cm,fill,color=black,right=.8cm of x1c.east,label=below:$x_2$] (x2c) {};
    \node[ball,text width=.18cm,fill,color=black,right=.8cm of x2c.east,label=below:$x_3$] (x3c) {};
    \draw[-,directed, thick,color=gray] (x1c.east) -- (x2c.west);
    \draw[-,dashed, thick,color=gray] (x2c.east) -- (x3c.west);
   \end{tikzpicture}
  \\
  &
  \begin{tikzpicture}[ball/.style = {circle, draw, align=center, anchor=north, inner sep=0}]
    \node[ball,text width=.18cm,fill,color=white,label=below:$\phantom{x_1}$] at (0,0) (x1) {};
    \node[ball,text width=.18cm,fill,color=white,right=.8cm of x1.east,label=below:$\phantom{x_2}$] (x2) {};
    \node[ball,text width=.18cm,fill,color=white,right=.8cm of x2.east,label=below:$\phantom{x_3}$] (x3) {};
    \draw[-,thick,color=white] (x1.east) edge node [text width=.18cm,above=-0.05cm,midway] {{$y_{\mbox{\tiny $12$}}$}} (x2.west);
    \draw[-,thick,color=white] (x2.east) edge node [text width=.18cm,above=-0.05cm,midway] {{$y_{\mbox{\tiny $23$}}$}} (x3.west);
    \node[right=.2cm of x3.east] (rhs1) {$\displaystyle\phantom{\:=\:\frac{1}{y_{12}y_{23}}}-$};
    \node[ball,text width=.18cm,fill,color=black,right=.2cm of rhs1,label=below:$x_1$] (x1a) {};
    \node[ball,text width=.18cm,fill,color=black,right=.8cm of x1a.east,label=below:$x_2$] (x2a) {};
    \node[ball,text width=.18cm,fill,color=black,right=.8cm of x2a.east,label=below:$x_3$] (x3a) {};
    \draw[-,dashed, thick,color=gray] (x1a.east) -- (x2a.west);
    \draw[-,reverse directed, thick,color=gray] (x2a.east) -- (x3a.west);
    \node[right=.2cm of x3a.east] (rhs2) {$\displaystyle\:-$};
    \node[ball,text width=.18cm,fill,color=black,right=.2cm of rhs2,label=below:$x_1$] (x1b) {};
    \node[ball,text width=.18cm,fill,color=black,right=.8cm of x1b.east,label=below:$x_2$] (x2b) {};
    \node[ball,text width=.18cm,fill,color=black,right=.8cm of x2b.east,label=below:$x_3$] (x3b) {};
    \draw[-,dashed, thick,color=gray] (x1b.east) -- (x2b.west);
    \draw[-,directed, thick,color=gray] (x2b.east) -- (x3b.west);
    \node[right=.2cm of x3b.east] (rhs3) {$\displaystyle\:+$};
    \node[ball,text width=.18cm,fill,color=black,right=.2cm of rhs3,label=below:$x_1$] (x1c) {};
    \node[ball,text width=.18cm,fill,color=black,right=.8cm of x1c.east,label=below:$x_2$] (x2c) {};
    \node[ball,text width=.18cm,fill,color=black,right=.8cm of x2c.east,label=below:$x_3$] (x3c) {};
    \draw[-,dashed, thick,color=gray] (x1c.east) -- (x2c.west);
    \draw[-,dashed, thick,color=gray] (x2c.east) -- (x3c.west);
    \node[right=.2cm of x3c.east] (t4) {$\displaystyle\left.\hspace{-.5cm}\phantom{\frac{1}{y}}\right]$};
  \end{tikzpicture} 
 \end{split}
\end{equation}
As in the two-site case, the term with no time ordering is straightforward to integrate because the three integrations decouple from each other:
\begin{equation}\label{eq:3siteNoTO}
 \begin{tikzpicture}[ball/.style = {circle, draw, align=center, anchor=north, inner sep=0}]
  \node[ball,text width=.18cm,fill,color=black,right=.2cm of rhs3,label=below:$x_1$] (x1c) {};
  \node[ball,text width=.18cm,fill,color=black,right=.8cm of x1c.east,label=below:$x_2$] (x2c) {};
  \node[ball,text width=.18cm,fill,color=black,right=.8cm of x2c.east,label=below:$x_3$] (x3c) {};
  \draw[-,dashed, thick,color=gray] (x1c.east) -- (x2c.west);
  \draw[-,dashed, thick,color=gray] (x2c.east) -- (x3c.west);
  \node[right=.2cm of x3c.east] (t4) {$\displaystyle\:=\:\int_{-\infty}^0\prod_{s=1}^3\left[d\eta_s\,e^{ix_s\eta_s}\right]\,e^{iy_{12}(\eta_1+\eta_2)}e^{iy_{23}(\eta_2+\eta_3)}\:=$};
  \node[below=.2cm of t4.south] (t5) {$\displaystyle=\:\frac{1}{(x_1+y_{12})(y_{12}+x_2+y_{23})(y_{23}+x_3)}.$};
 \end{tikzpicture} 
\end{equation}
In the case of those contributions with where the time ordered is absent along one edge only, just one integration decouples and the remaining integral represents a two-site graph
\begin{equation}\label{eq:3siteMix}
 \begin{tikzpicture}[ball/.style = {circle, draw, align=center, anchor=north, inner sep=0}]
  \node[ball,text width=.18cm,fill,color=black,right=.2cm of rhs2,label=below:$x_1$] (x1b) {};
  \node[ball,text width=.18cm,fill,color=black,right=.8cm of x1b.east,label=below:$x_2$] (x2b) {};
  \node[ball,text width=.18cm,fill,color=black,right=.8cm of x2b.east,label=below:$x_3$] (x3b) {};
    \draw[-,dashed, thick,color=gray] (x1b.east) -- (x2b.west);
    \draw[-,directed, thick,color=gray] (x2b.east) -- (x3b.west);
    \node[right=.2cm of x3b.east] (rhs3) {$\displaystyle\:=\:\int_{-\infty}^0\prod_{s=1}^3\left[d\eta_s\,e^{ix_s\eta_s}\right]e^{iy_{12}(\eta_1+\eta_2)}e^{iy_{23}(\eta_2-\eta_3)}\vartheta(\eta_3-\eta_2)\:=$};
    \node[below=.2cm of rhs3.south] (rhs3b) {$\displaystyle\hspace{1.2cm}=\:\frac{1}{x_1+y_{12}}\underbrace{\int_{-\infty}^0\prod_{s=2}^3d\eta_s\,e^{i(y_{12}+x_2)\eta_2}e^{ix_3\eta_3}e^{iy_{23}(\eta_2-\eta_3)}\vartheta(\eta_3-\eta_2)}_{
         \begin{tikzpicture}[ball/.style = {circle, draw, align=center, anchor=north, inner sep=0}]
          \node[ball,text width=.18cm,fill,color=white] at (-2,0) (reft) {};
          \node[ball,text width=.18cm,fill,color=black,below=.2cm of reft.south,label=below:$\mbox{\footnotesize $y_{12}+x_2$}$] (x1b) {};
          \node[ball,text width=.18cm,fill,color=black,right=1cm of x1b.east,label=below:$\mbox{\footnotesize $x_3$}$] (x2b) {};
          \draw[-,directed,thick,color=gray] (x1b.east) -- (x2b.west);
         \end{tikzpicture}
         }
         \:=$};
    \node[below=.2cm of rhs3b.south] (rhs3c) {$\displaystyle\hspace{-2.3cm}=\:\frac{1}{(x_1+y_{12})(y_{12}+x_2+y_{23})(y_{12}+x_2+x_3)}.$};
 \end{tikzpicture}
\end{equation}
We are finally left with the cases in which both propagators are time-ordered. In two of these situations the time flow has a well-defined direction, {\it i.e.} either we have $\eta_1\,>\,\eta_2\,>\,\eta_3$ or
$\eta_1\,<\,\eta_2\,<\,\eta_3$. They correspond to the two propagators being either both advanced or retarded. The time ordering provides a clear integration ordering, which gives

\begin{equation}\label{eq:3siteAA}
 \begin{tikzpicture}[ball/.style = {circle, draw, align=center, anchor=north, inner sep=0}]
   \node[ball,text width=.18cm,fill,color=white] (reft) {};
   \node[ball,text width=.18cm,fill,color=black,below=.2cm of reft,label=below:$x_1$] (x1a) {};
   \node[ball,text width=.18cm,fill,color=black,right=.8cm of x1a.east,label=below:$x_2$] (x2a) {};
   \node[ball,text width=.18cm,fill,color=black,right=.8cm of x2a.east,label=below:$x_3$] (x3a) {};
   \draw[-,reverse directed, thick,color=gray] (x1a.east) -- (x2a.west);
   \draw[-,reverse directed, thick,color=gray] (x2a.east) -- (x3a.west);
   \node[right=.2cm of x3a.east] (rhs2) {$\displaystyle\:=\:\frac{1}{(x_1+x_2+x_3)(y_{12}+x_2+x_3)(y_{23}+x_3)},$};
   \node[ball,text width=.18cm,fill,color=black,below=1cm of x1a,label=below:$x_1$] (x1b) {};
   \node[ball,text width=.18cm,fill,color=black,right=.8cm of x1b.east,label=below:$x_2$] (x2b) {};
   \node[ball,text width=.18cm,fill,color=black,right=.8cm of x2b.east,label=below:$x_3$] (x3b) {};
   \draw[-,directed, thick,color=gray] (x1b.east) -- (x2b.west);
   \draw[-,directed, thick,color=gray] (x2b.east) -- (x3b.west);
   \node[right=.2cm of x3b.east] (rhs3) {$\displaystyle\:=\:\frac{1}{(x_1+x_2+x_3)(x_1+y_{12})(x_1+x_2+y_{23})}.$};
  \end{tikzpicture}
\end{equation}
In the remaining two cases, $\eta_1$ and $\eta_3$ are both either greater or less than $\eta_2$, but they are not related by any time ordering:
\begin{equation}\label{eq:3siteAR}
 \begin{tikzpicture}[ball/.style = {circle, draw, align=center, anchor=north, inner sep=0}]
   \node[ball,text width=.18cm,fill,color=white] (reft) {};
   \node[ball,text width=.18cm,fill,color=black,below=.2cm of reft,label=below:$x_1$] (x1a) {};
   \node[ball,text width=.18cm,fill,color=black,right=.8cm of x1a.east,label=below:$x_2$] (x2a) {};
   \node[ball,text width=.18cm,fill,color=black,right=.8cm of x2a.east,label=below:$x_3$] (x3a) {};
   \draw[-,reverse directed, thick,color=gray] (x1a.east) -- (x2a.west);
   \draw[-,directed, thick,color=gray] (x2a.east) -- (x3a.west);
   \node[right=.2cm of x3a.east] (rhs2) {$\displaystyle\:=\:\frac{x_1+y_{12}+2x_2+y_{23}+x_3}{(x_1+x_2+x_3)(x_1+x_2+y_{23})(y_{12}+x_2+x_3)(y_{12}+x_2+y_{23})},$};
   \node[ball,text width=.18cm,fill,color=black,below=1cm of x1a,label=below:$x_1$] (x1b) {};
   \node[ball,text width=.18cm,fill,color=black,right=.8cm of x1b.east,label=below:$x_2$] (x2b) {};
   \node[ball,text width=.18cm,fill,color=black,right=.8cm of x2b.east,label=below:$x_3$] (x3b) {};
   \draw[-,directed, thick,color=gray] (x1b.east) -- (x2b.west);
   \draw[-,reverse directed, thick,color=gray] (x2b.east) -- (x3b.west);
   \node[right=.2cm of x3b.east] (rhs3) {$\displaystyle\:=\:\frac{1}{(x_1+x_2+x_3)(x_1+y_{12})(y_{23}+x_3)}.$};
  \end{tikzpicture}
\end{equation}
Recollecting all the terms \eqref{eq:3siteNoTO}, \eqref{eq:3siteMix}, \eqref{eq:3siteAA} and \eqref{eq:3siteAR}, and summing them up according to \eqref{eq:3siteTerm} we get:
\begin{equation}\label{eq:3siteRes}
 \begin{tikzpicture}[ball/.style = {circle, draw, align=center, anchor=north, inner sep=0}]
    \node[ball,text width=.18cm,fill,color=black,label=below:$x_1$] at (0,0) (x1) {};
    \node[ball,text width=.18cm,fill,color=black,right=.8cm of x1.east,label=below:$x_2$] (x2) {};
    \node[ball,text width=.18cm,fill,color=black,right=.8cm of x2.east,label=below:$x_3$] (x3) {};
    \draw[-,thick,color=black] (x1.east) edge node [text width=.18cm,above=-0.05cm,midway] {{$y_{\mbox{\tiny $12$}}$}} (x2.west);
    \draw[-,thick,color=black] (x2.east) edge node [text width=.18cm,above=-0.05cm,midway] {{$y_{\mbox{\tiny $23$}}$}} (x3.west);
    \node[right=.2cm of x3.east] (rhs1) {$\displaystyle\:=\:\frac{x_1+y_{12}+2x_2+y_{23}+x_3}{(x_1+x_2+x_3)(y_{12}+x_2+y_{23})(x_1+y_{12})(y_{23}+x_3)(x_1+x_2+y_{23})(y_{12}+x_2+x_3)},$};
 \end{tikzpicture}
\end{equation}
which is a result enormously simpler than suggested by the nine-term representation \eqref{eq:3siteTerm}! Furthermore, as for the two-site case, the factors of type $1/y$ carried by the propagators disappear so that they actually do not represent poles of the final object. 


\subsubsection{From the time integral to the time diagrammatics}\label{subsubsec:TimeDiag}

Before moving forward and looking for a representation which makes the simplicity just observed manifest, let us analyze in more detail the time integrals. The decomposition of the propagator into its advanced and retarded parts makes the flow of time manifest in the different contributions to the wavefunction. Furthermore, such contributions get expressed {\it almost} in their simplest form: in the three-site example, all but one term (given in the first line of \eqref{eq:3siteAR}) have numerator equal to one. However, notice that the numerator of such a term is equal to the sum of two poles, suggesting the idea that can be further decomposed into two terms, each of which having the very same structure of the others. We would like to obtain a representation for the contributions to the wavefunction as a sum of such a class of terms but still making use of the decomposition \eqref{eq:wvn2} of the propagator.

Time ordering provides a natural integration ordering (from the earliest time to the latest). However, while, as we saw in the two- and three-site cases and we will see more generically later, the terms of the propagators without time ordering factorize the graph without generating any integration ordering issue, the time ordered terms may just fix partially the ordering among the $\eta$'s, as it occurred in \eqref{eq:3siteAR}. Terms of this type may not be in their simplest form. As a first step, we need to understand the effect of a single time integration. 

Let us start with the simplest case and consider a generic graph with the usual decomposition for all the propagators. As for the two simple cases analyzed earlier, the several terms arising, can be represented as the very same graph whose edges are decorated either with an arrow, representing the advanced/retarded prescription, or with a dashed line  for the term related to the vanishing condition at the boundary. In the latter case, we observed in \eqref{eq:2siteNoTO}, \eqref{eq:3siteNoTO} and \eqref{eq:3siteMix} that the time integration over the variable with no time ordering with respect to the others was decoupling. More generally, the structure for these type of terms is given by
\begin{equation}\label{eq:DashInt}
 \begin{tikzpicture}[ball/.style = {circle, draw, align=center, anchor=north, inner sep=0}]
  \node[ball,text width=1cm,shade] (S1) at (0,0) {$\mathcal{L}$};
  \node[ball,text width=1cm,shade,right=1cm of S1.east] (S2) {$\mathcal{R}$};
  \node[ball,text width=.18cm,fill,color=black,right=-.1cm of S1.east, label=below:$\mbox{\tiny $\hspace{.3cm}x_i$}$] (xi) {};
  \node[ball,text width=.18cm,fill,color=black,right=.8cm of xi.east, label=below:$\mbox{\tiny $\hspace{-.3cm}x_j$}$] (xj) {};
  \draw[-,dashed,thick,color=gray] (xi.east) edge node [text width=.18cm,above=-0.05cm,midway] {$\mbox{\tiny $y_{ij}$}$} (xj.west);
  \node[right=.2cm of S2.east] (rhs1) {$\displaystyle\:=\:\int_{-\infty}^0\prod_{v\in\mathcal{L}+\mathcal{R}}d\eta_v\,e^{ix_v\eta_v}\prod_{e\in\mathcal{L}+\mathcal{R}}\tilde{G}(y_{v_e};\eta_{v_e},\eta_{v'_e})e^{iy_{ij}(\eta_i+\eta_j)},$};
\end{tikzpicture}
\end{equation}
where the left ($\mathcal{L}$) and right ($\mathcal{R}$) blobs on the left-hand side indicate a generic graph topology, while $\tilde{G}$ indicates a propagator which can be equivalently advanced/retarded and whose factor $1/y_{v_e}$ has been stripped off. Notice that that the right-hand-side of \eqref{eq:DashInt} can be equivalently written as
\begin{equation}\label{eq:DashInt2}
 \begin{tikzpicture}[ball/.style = {circle, draw, align=center, anchor=north, inner sep=0}]
  \node[ball,text width=1cm,shade] (S1) at (0,0) {$\mathcal{L}$};
  \node[ball,text width=1cm,shade,right=1cm of S1.east] (S2) {$\mathcal{R}$};
  \node[ball,text width=.18cm,fill,color=black,right=-.1cm of S1.east, label=below:$\mbox{\tiny $\hspace{.3cm}x_i$}$] (xi) {};
  \node[ball,text width=.18cm,fill,color=black,right=.8cm of xi.east, label=below:$\mbox{\tiny $\hspace{-.3cm}x_j$}$] (xj) {};
  \draw[-,dashed,thick,color=gray] (xi.east) edge node [text width=.18cm,above=-0.05cm,midway] {$\mbox{\tiny $y_{ij}$}$} (xj.west);
  \node[right=.2cm of S2.east] (for1) {$\displaystyle\:=\:\left[\int_{-\infty}^0d\eta_i\,e^{i(x_i+y_{ij})\eta_i}\prod_{v\in\mathcal{L}\setminus\{i\}}d\eta_v\,e^{ix_v\eta_v}\prod_{e\in\mathcal{L}}
                                        \tilde{G}(y_{v_e};\eta_{v_e},\eta_{v'_e})\right]\otimes$};
  \node[below=.2cm of for1.south] (for2) {$\displaystyle\otimes\left[\int_{-\infty}^0d\eta_j\,e^{i(x_i+y_{ij})\eta_j}\prod_{v\in\mathcal{R}\setminus\{j\}}d\eta_v\,e^{ix_v\eta_v}\prod_{e\in\mathcal{R}}
                                            \tilde{G}(y_{v_e};\eta_{v_e},\eta_{v'_e})\right]\:\equiv$};
  \node[below=2.5cm of S2.south] (for3) {$\:\hspace{2cm}\equiv\:$};
  \node[ball,text width=1cm,shade,right=.2cm of for3.east] (S3) {$\mathcal{L}$};
  \node[ball,text width=.18cm,fill,color=black,right=-.1cm of S3.east, label=below:$\mbox{\tiny $\hspace{.4cm}x_i+y_{ij}$}$] (xi2) {};
  \node[right=.7cm of xi2.east] (op) {$\otimes$};
  \node[ball,text width=1cm,shade,right=.8cm of op.east] (S4) {$\mathcal{R}$};
  \node[ball,text width=.18cm,fill,color=black,right=.7cm of op.east, label=below:$\mbox{\tiny $\hspace{-.4cm}y_{ij}+x_j$}$] (xj2) {};
 \end{tikzpicture}
\end{equation}
Thus, the part of the propagator without time ordering has the effect of mapping the original graph in a direct product of two subgraphs such that the vertices that were originally connected by the propagator now have shifted energies. If the edge without time ordering appears in a loop, then the graph is mapped into a graph with the edge in question and with the shifted energies for the vertices which were its endpoints.

Let us move on to the more interesting case of the contributions from the advanced/retarded parts of the propagators. Interestingly, the integration ordering suggested by the time ordering can be graphically translated in collapsing two vertices together by moving the one at earliest event onto the other: this move produces a graph with one site less times a factor given by (the inverse of the) sum of the energies in the vertex which has been moved. Let us see explicitly how it works in the first no-so-trivial example:
\begin{equation}\label{eq:Tmove1}
 \begin{tikzpicture}[ball/.style = {circle, draw, align=center, anchor=north, inner sep=0}]
  \node[ball,text width=.18cm,fill,color=black, label=below:$\mbox{\tiny $x_1$}$] at (0,0) (x1) {};
  \node[ball,text width=1cm,shade,right=.8cm of x1.east] (S1)  {$\mathcal{B}$};
  \node[ball,text width=.18cm,fill,color=black,right=.7cm of x1.east, label=below:$\mbox{\tiny $\hspace{-.3cm}x_2$}$] (x2) {};
  \draw[-,directed,thick,color=gray] (x1.east) -- (x2.west);
  \node[right=.2cm of S1.east] (for1) {$\displaystyle\:\equiv\:\int_{-\infty}^0\prod_{v\in\mathcal{B}\setminus\{2\}}d\eta_v\,e^{ix_v\eta_v}\int_{-\infty}^0d\eta_2\,e^{ix_2\eta_2}
                                        \prod_{e\in\mathcal{B}}\tilde{G}(y_{v_e};\eta_{v_e},\eta_{v'_e})\int_{-\infty}^{\eta_2}d\eta_1\,e^{ix_1\eta_1}e^{+iy_{12}(\eta_1-\eta_2)}$};
 \end{tikzpicture}
\end{equation}
where $\eta_2\,>\,\eta_1$, which establishes the order of integration between $\eta_1$ and $\eta_2$ (as shown explicitly in \eqref{eq:Tmove1}). The time ordering of $\eta_2$ with respect to the other variables is contained in the Heaviside step functions in the $\tilde{G}$'s. Integrating over $\eta_1$, we obtain:
\begin{equation}\label{eq:Tmove2}
 \begin{tikzpicture}[ball/.style = {circle, draw, align=center, anchor=north, inner sep=0}]
  \begin{scope}
  \node[ball,text width=.18cm,fill,color=black, label=below:$\mbox{\tiny $x_1$}$] at (0,0) (x1) {};
  \node[ball,text width=1cm,shade,right=.8cm of x1.east] (S1)  {$\mathcal{B}$};
  \node[ball,text width=.18cm,fill,color=black,right=.7cm of x1.east, label=below:$\mbox{\tiny $\hspace{-.3cm}x_2$}$] (x2) {};
  \draw[-,directed,thick,color=gray] (x1.east) -- (x2.west);
  \end{scope}
  \node at (8.1,-.85) (for1) {$\displaystyle\:=\:\frac{1}{x_1+y_{12}}\underbrace{\int_{-\infty}^0\prod_{v\in\mathcal{B}\setminus\{2\}}d\eta_v\,e^{ix_v\eta_v}\int_{-\infty}^0d\eta_2\,e^{i(x_1+x_2)\eta_2}
                                        \prod_{e\in\mathcal{B}}\tilde{G}(y_{v_e};\eta_{v_e},\eta_{v'_e})}_{
                                         \begin{tikzpicture}[ball/.style = {circle, draw, align=center, anchor=north, inner sep=0}]
                                           \node[ball,text width=1cm,shade] (S1) at (0,0) {$\mathcal{B}$};
                                          \node[ball,text width=.18cm,fill,color=black,left=-.1cm of S1.west, label=below:$\mbox{\tiny $\hspace{-.6cm}x_1+x_2$}$] (xs) {};
                                         \end{tikzpicture}
                                       }$};
  \end{tikzpicture}
\end{equation}
For the case of the time ordering $\eta_1\,>\,\eta_2$, it works in a similar fashion, with an additional subtlety due to the time-ordering of $\eta_2$ (whose integration, in principle, should be performed first than the one over $\eta_1$). More precisely, if all the propagators of $\mathcal{B}$ having $2$ as an endpoint have time ordering such that $\eta_2$ is always greater than the other $\eta$'s, then the time ordering again suggests a natural integration ordering. If instead, at least one of these propagators has a different time ordering, then it seems that there is no a clear way of ordering the integration. This is exactly the situation we encountered in the first line of \eqref{eq:3siteAR}. If $\eta_3,\,\ldots\,\eta_k$ are the times involved by the propagators having $2$ as an endpoints, with ordering $\eta_j\,>\,\eta_2$ for each $j\,=\,3,\ldots,k$, a canonical choice is to divide the integration space into patches with different ordering among the $\eta_j$'s ($j\,=\,3,\,\ldots,k$). This allows to perform first the integration over $\eta_2$ and then the integration order of the other variables depends on the patch. Diagrammatically, this is equivalent to collapse the site $2$, according to the time arrow, onto every other site $j$ ($j\,=\,3\ldots\,k$), and sum over these contributions. Let us illustrate it in the concrete example \eqref{eq:3siteAR}:
\begin{equation}\label{eq:3siteARrev}
 \begin{tikzpicture}[ball/.style = {circle, draw, align=center, anchor=north, inner sep=0}]
   \node[ball,text width=.18cm,fill,color=white] (reft) {};
   \node[ball,text width=.18cm,fill,color=black,below=.2cm of reft,label=below:$x_1$] (x1a) {};
   \node[ball,text width=.18cm,fill,color=black,right=.8cm of x1a.east,label=below:$x_2$] (x2a) {};
   \node[ball,text width=.18cm,fill,color=black,right=.8cm of x2a.east,label=below:$x_3$] (x3a) {};
   \draw[-,reverse directed, thick,color=gray] (x1a.east) -- (x2a.west);
   \draw[-,directed, thick,color=gray] (x2a.east) -- (x3a.west);
   \node[right=.2cm of x3a.east] (rhs2) {$\displaystyle\:=\:\frac{1}{y_{12}+x_2+y_{23}}$};
    \node[ball,text width=.18cm,fill,color=black,right=.2cm of rhs2.east,label=below:$\mbox{\tiny $x_1+x_2$}$] (x1b) {};
    \node[ball,text width=.18cm,fill,color=black,right=.8cm of x1b.east,label=below:$\mbox{\tiny $x_3$}$] (x2b) {};
    \draw[-,directed,thick,color=gray] (x1b.east) -- (x2b.west);
    \node[right=.2cm of x2b.east] (rhs3) {$\displaystyle\:+\:\frac{1}{y_{12}+x_2+y_{23}}$};
    \node[ball,text width=.18cm,fill,color=black,right=.2cm of rhs3.east,label=below:$\mbox{\tiny $x_1$}$] (x1c) {};
    \node[ball,text width=.18cm,fill,color=black,right=.8cm of x1c.east,label=below:$\mbox{\tiny $x_2+x_3$}$] (x2c) {};
    \draw[-,reverse directed,thick,color=gray] (x1c.east) -- (x2c.west);
    \node[right=.2cm of x2c.east] (rhs4) {$\displaystyle\:=$};
   \node[below=1cm of rhs2.east] (rhs5) {$\displaystyle\:=\:\frac{1}{y_{12}+x_2+y_{23}}\times\frac{1}{(x_1+x_2+y_{23})(x_1+x_2+x_3)}\:+$};
   \node[below=.5cm of rhs5.south] (rhs6) {$\displaystyle\:+\:\frac{1}{y_{12}+x_2+y_{23}}\times\frac{1}{(y_{12}+x_2+x_{3})(x_1+x_2+x_3)}$};
 \end{tikzpicture}
\end{equation}
which is exactly the same result in the first line of \eqref{eq:3siteAR}.

Summarizing, it is possible to canonically define an integration order for the time integrals which returns the time-representation as a sum of products of poles. Given a contribution to the perturbative wavefunction represented by a graph $\mathcal{G}$, the decomposition of the propagator in an advanced, retarded and no time-ordering term, induces a decomposition of the graph 
\begin{equation}\label{eq:GraphDec}
 \mathcal{G}\left(\mathcal{E},\,\mathcal{V}\right)\:=\:\sum_{\sigma\in\mathcal{D}}(-1)^{\xi_{\sigma}}\mathcal{G}\left(\mathcal{E}_{\sigma},\,\mathcal{V}\right)\,\equiv\,
        \left(\prod_{e\in\mathcal{E}}\frac{1}{y_e}\right)\sum_{\sigma\in\mathcal{D}}(-1)^{\xi_{\sigma}}\tilde{\mathcal{G}}\left(\mathcal{E}_{\sigma},\,\mathcal{V}\right),
\end{equation}
where $\mathcal{E}\,\equiv\,(\mathcal{E}_1,\,\ldots,\,\mathcal{E}_{E})$ is the set of the edges of the graph, $\mathcal{V}\,\equiv\,(\mathcal{V}_1,\,\ldots,\,\mathcal{V}_{V})$ the set of vertices, $\mathcal{D}$ is the set of conditions on the propagator (advanced, retarded, no time ordering) which are graphically identified with a decoration (if an edge connects two vertices $i$ and $j$ at times $\eta_i\,<\eta_j$, the edge is decorated via an arrow directed from $i$ to $j$, while if the propagator is not time ordered, the edge is represented by an undirected dashed line); finally, $\mathcal{E}_{\sigma}\,\equiv\,(\mathcal{E}_{\sigma_1},\,\ldots,\,\mathcal{E}_{\sigma_{E}})$ represents the set of edges with decoration $\sigma\,\equiv\,(\sigma_1,\,\ldots,\sigma_{E})$, while $\xi_{\sigma}\,\equiv\,\sum_{r=1}^{E}\xi_{\sigma_r}$ with $\xi_{\sigma_r}\,=\,0,\,1$ depending on whether $\sigma_r$ represents a condition with/without time ordering respectively. In the very right-hand side of \eqref{eq:GraphDec} we have stripped off the factors $1/y$'s from the propagators, and $\tilde{\mathcal{G}}(\mathcal{E}_{\sigma},\,\mathcal{V})$ represented the stripped-off decorated graphs. If a decorated graph has $E$ edges, $n_d$ of which have no time ordering, it is then equivalent to a graph with $E-n_d$ edges and with the energies of the vertices which were connected by these edges shifted (and depending on the location of the edges in the original graph, the new graph can become a direct product of subgraphs). As far as the time-ordered edges are concerned the time integration can be performed as diagrammatic moves: two vertices are contracted according to the time direction along the edge which connected, generating a graph with one vertex less with energy given by the sum of the energies of the two vertices which have been collapsed, divided by the sum of the energies at the vertex which have been moved. If a vertex can be moved in several different directions (getting then collapsed with different vertices), one needs to sum over all the possible moves.


\subsection{``Boundary" Representation as Old-Fashioned Perturbation Theory}\label{subsec:OFPT}

An important lesson learnt so far is that if we want to keep the time-evolution manifest in a representation for the perturbative wavefunction, the price we pay is to have a proliferation of terms with spurious singularities, with the final answer being much simpler than what such a representation shows. In other words, in order to have the flow of time manifest, an intrinsic simplicity of the wavefunction gets hidden. It is important thus, to understand how such a simplicity can be made manifest.

Let us consider the general representation for the wavefunction \eqref{eq:wvn}, but inserting the time translation operator $\Delta$ in such a way that it acts on the external states only:
\begin{equation}\label{eq:GenOFPT}
 \mathcal{O}\:\equiv\:\int_{-\infty}^0\prod_{v\in\mathcal{V}}d\eta_v\,\Delta\left[\prod_{v\in\mathcal{V}}e^{ix_v\eta_v}\right]\prod_{e\in\mathcal{E}}G(y_e;\eta_e,\eta'_e),\qquad
 \Delta\,\equiv\:-i\sum_{v\in\mathcal{V}}\partial_{\eta_v}.
\end{equation}
We can obtain two different expression for \eqref{eq:GenOFPT} on one hand by writing explicitly the action of the time-translation operator $\Delta$, and on the other, by integrating by part so that $\Delta$ acts on the product of the propagators:
\begin{equation}\label{eq:GenOFPT2}
 \mathcal{O}\:\equiv\:\left[\sum_{v\in\mathcal{V}}x_v\right]\psi(x,y)\:=\:-\int_{-\infty}^0\left[\prod_{v\in\mathcal{V}}d\eta_v\,e^{ix_v\eta_v}\right]\Delta\left[\prod_{e\in\mathcal{E}}G(y_e;\eta_e,\eta'_e)\right],
\end{equation}
where the left-hand-side is given by the explicit action of the operator $\Delta$ in \eqref{eq:GenOFPT}, which brings down the total energy and leaving the original object we would like to compute, while the right-hand-side is
the result of the integration by parts. More explicitly, we can write the equation \eqref{eq:GenOFPT2} as
\begin{equation}\label{eq:GenOFPT3}
 \left[\sum_{v\in\mathcal{V}}x_v\right]\psi(x,y)\:=\:-\int_{-\infty}^0\left[\prod_{v\in\mathcal{V}}d\eta_v\,e^{ix_v\eta_v}\right]\sum_{e\in\mathcal{E}}
          \left[\left(\Delta\,G_e\right)\prod_{\tilde{e}\in\mathcal{E}\setminus\{e\}}G_{\tilde{e}}\right],
\end{equation}
where for convenience we shortened the notation by writing $G_e\,\equiv\,G(y_e;\eta_e,\,\eta'_e)$. Interestingly, the time-ordered part of the propagator \eqref{eq:wvn2} is time translational invariant and, consequently, the action of the time-translation operator on it is zero. Hence, the only non trivial contribution comes from the term of the propagator {\it without} time-ordering:
\begin{equation}\label{eq:TimeTprop}
 \Delta\,G(y_e,\,\eta_e,\,\eta'_e)\:\equiv\:\Delta\left[-\frac{e^{iy_e(\eta_e+\eta'_e)}}{y_e}\right]\:=\:-2\,e^{iy_e(\eta_e+\eta'_e)}.
\end{equation}
Thus, \eqref{eq:GenOFPT3} becomes
\begin{equation}\label{eq:OFPT}
 \left[\sum_{v\in\mathcal{V}}x_v\right]\psi(x,y)\:=\:2\sum_{e\in\mathcal{E}}\int_{-\infty}^0\left[\prod_{v\in\tilde{\mathcal{V}}}d\eta_v\,e^{ix_v\eta_v}\right]e^{i(x_{v_e}+y_e)\eta_e}e^{i(x_{v'_e}+y_e)\eta'_e}
        \prod_{\tilde{e}\in\mathcal{E}\setminus\{e\}}G_{\tilde{e}},
\end{equation}
where $\tilde{\mathcal{V}}\,\equiv\,\mathcal{V}\setminus\{v_e,\,v'_e\}$, $v_e$ and $v'_e$ being the two vertices connected by the edge $e$. The right-hand-side represents a sum over the very same function $\psi$ with one edge less (and the energy of the vertices which were connected by the erased edge, shifted by $y_e$), which at tree-level translates into a factorization diagram, while at loop level $L$ it represents a $(L-1)$-loop graph:
\begin{equation}\label{eq:RR}
 \begin{split}
  &
  \begin{tikzpicture}[ball/.style = {circle, draw, align=center, anchor=north, inner sep=0}]
   \fill[shade,thick] (0,0) circle (.8);
   \node[text width=.18cm,color=black] at (0,0) (Fn) {$\displaystyle\psi_n$};
   \node[text width=.18cm,color=black] at (-2.8,-0.1) (sum) {$\displaystyle\left(\sum_{v\in\mathcal{V}}x_v\right)$}; 
   \node[ball,text width=.18cm,fill,color=black,label=left:$\mbox{\tiny $x_1$}$] at (-.8,0) (x1) {}; 
   \node[ball,text width=.18cm,fill,color=black,label=left:$\mbox{\tiny $x_2$}$] at ({.8*cos(150)},{.8*sin(150)}) (x2) {}; 
   \node[ball,text width=.18cm,fill,color=black,label={right:$\mbox{\tiny $x_{i-1}$}$}] at ({.8*cos(30)},{.8*sin(30)}) (xi1) {};
   \node[ball,text width=.18cm,fill,color=black,label={right:$\mbox{\tiny $x_i$}$}] at ({.8},{0}) (xi) {};   
   \node[ball,text width=.18cm,fill,color=black,label={right:$\mbox{\tiny $x_{i+1}$}$}] at ({.75*cos(-30)},{.75*sin(-30)}) (xi1b) {};  
   \node[ball,text width=.18cm,fill,color=black,label=left:$\mbox{\tiny $x_n$}$] at ({.8*cos(150)},{-.8*sin(150)}) (xn) {};   
   \node[color=black] at (1.8,-0.25) (eq) {$\displaystyle\quad=\:\sum_{e\in\mathcal{E}}$};
   \fill[shade,thick] (4,0) circle (.8);  
   \node[text width=.18cm,color=black] at (4,0) (Fl) {$\displaystyle\psi_{\mathcal{L}}$};
   \node[ball,text width=.18cm,fill,color=black] at (3.2,0) (x1) {}; 
   \node[ball,text width=.18cm,fill,color=black] at ({.8*cos(150)+4},{.8*sin(150)}) (x2) {}; 
   \node[ball,text width=.18cm,fill,color=red,label={[label distance=.05mm]30:$\hspace{-.1cm}\mbox{\tiny $x_{v_e}+y_e$}$}] at ({.8+4},{0}) (xv) {};   
   \node[ball,text width=.18cm,fill,color=black] at ({.8*cos(150)+4},{-.8*sin(150)}) (xn) {};   
   \node[color=black] at (1.8,-0.25) (eq) {$\displaystyle\quad=\:\sum_{e\in\mathcal{E}}$};
   \fill[shade,thick] (8,0) circle (.8);  
   \node[text width=.18cm,color=black] at (8,0) (Fr) {$\displaystyle\psi_{\mathcal{R}}$};
   \node[ball,text width=.18cm,fill,color=red,label={[left=1cm]-30:$\hspace{-.1cm}\mbox{\tiny $x_{v'_e}+y_e$}$}] at (7.2,0) (xv2) {}; 
   \node[ball,text width=.18cm,fill,color=black] at ({.8*cos(30)+8},{.8*sin(30)}) (x2) {}; 
   \node[ball,text width=.18cm,fill,color=black] at ({.8+8},{0}) (xt) {};   
   \node[ball,text width=.18cm,fill,color=black] at ({.8*cos(-30)+8},{.8*sin(-30)}) (xn) {};   
   \draw[-,dashed,ultra thick,color=red] (xv) -- (xv2);
   \node[color=black] at (9.8,-0.1) (eq2) {$\displaystyle\quad+$};
  \end{tikzpicture}
  \\
  &\vspace{.5cm}\\
  &
  \hspace{4cm}
  \begin{tikzpicture}[ball/.style = {circle, draw, align=center, anchor=north, inner sep=0},every loop/.style={}]
   \node[color=black] at (1.8,-0.25) (eq) {$\displaystyle\quad+\:\sum_{e\in\mathcal{E}}$};
   \fill[shade,thick] (4,0) circle (.8);
   \node[ball,text width=.18cm,fill,color=red,label={[left=1.2cm]-30:$\mbox{\tiny $x_{v_e}+y_e$}$}] at ({.7*cos(-120)+4},{.7*sin(-120)}) (xv1) {};
   \node[ball,text width=.18cm,fill,color=red,label={[right=-.2cm]-30:$\mbox{\tiny $x_{v'_e}+y_e$}$}] at ({.7*cos(-60)+4},{.7*sin(-60)}) (xv2) {};
   \draw[-,dashed,ultra thick,color=red] (xv1.south) edge[bend right=100] (xv2.south);
  \end{tikzpicture}
 \end{split}
\end{equation}
where the dashed lines indicates the edges which get erased. Strikingly, the above recursive relation is reminiscent of the one known in scattering amplitudes\cite{ArkaniHamed:2010kv, ArkaniHamed:2012nw, Benincasa:2015zna}! Thus, the contributions to the perturbative wavefunction can be computed as a sum of terms obtained by erasing an edge from the original graph and shifting the energies of the pair of vertices $(v_e,\,v'_e)$ the edge in question was connecting by $y_e$ ({\it i.e.} the energy associated to the edge itself). One of the main differences between the latter and \eqref{eq:RR}, is the presence of the total energy factor on the left-hand-side, which is thus always a pole of $\psi_n$.

The recursion relation \eqref{eq:RR} can be implemented via a combinatorial operation: given a graph $\mathcal{G}$ representing the wavefunction $\psi_n$, the recursion relation is equivalent in summing up all the possible ways in which the graph $\mathcal{G}$ can be iteratively divided into {\it connected} subgraphs, to which one associate (the inverse of) the sum of the {\it external}\footnote{Here for ``sum of external energies'' of a subgraph, we refer to the sum of the energies of the vertices contained in it plus the energies of the edges which connect the subgraph to other subgraphs.} energies of the subgraphs themselves. Notice that the representation obtained in this way is exactly the {\it old-fashioned perturbation theory} (OFPT) which one would obtain via the Lippmann-Schwinger kernel. Let us illustrate these computations explicitly with some simple examples.  


\subsubsection{The two-site graph at tree level}\label{subsubsec:2stg}

The simplest case is provided by the two-site case which we already computed from the time integral representation \eqref{eq:2site}. Following the recursion relation \eqref{eq:RR}, we find:
\begin{equation}\label{eq:2sRR}
 (x_1+x_2)\,\psi_2(x_1,\,x_2;\,y)\:=\:\psi_1(x_1+y)\otimes\psi_1(y+x_2),
\end{equation}
with $\psi_2$ being the two-site graph and the $\psi_1$'s two factorized single site graphs. Amusingly, we straightforwardly obtained the final {\it simplified} result \eqref{eq:2siteFin}.

However, as described above, the right-hand-side of \eqref{eq:RR} (divided by the sum of all the energies) can be obtained as a sum of all the possible way of iteratively dividing the graph into connected subgraph to which the inverse of the sum of the external energies in the subgraph itself is associated. The first connected subgraph is the graph itself, which provides the total energy pole; there is then a unique way to identify connected subgraphs, by picking the two vertices, whose total energies are given by the sum of the energies $x$ of the external states and the energy $y$ of the intermediate one:

\begin{equation}\label{eq:2sOFPT}
  \begin{tikzpicture}[ball/.style = {circle, draw, align=center, anchor=north, inner sep=0}]
  \node[ball,text width=.18cm,fill,color=black,label=below:$x_1$] at (0,0) (x1) {};
  \node[ball,text width=.18cm,fill,color=black,right=1.5cm of x1.east,label=below:$x_2$] (x2) {};
  \draw[-,thick,color=black] (x1.east) edge node [text width=.18cm,above=-0.05cm,midway] {{$y$}} (x2.west);
  \draw[decorate,decoration={brace,mirror,raise=6pt,amplitude=10pt}, thick] (5,1.9)--(5,-2.1) ;
  \path[draw=black,solid,line width=.1cm,shade,preaction={-triangle 90,thin,draw,color=black,shorten >=-1mm}] (2.5,-0.1) -- (4,-0.1);
  \node[ball,text width=.18cm,fill,color=black,label=below:$x_1$] at (5.5,1.2) (x1b) {};
  \node[ball,text width=.18cm,fill,color=black,right=1.5cm of x1b.east,label=below:$x_2$] (x2b) {};
  \draw[-,thick,color=black] (x1b.east) edge node [text width=.18cm,above=-0.05cm,midway] {{$y$}} (x2b.west);
  \draw[color=blue] (6.4,1) ellipse (1.5cm and .5cm);
  \node[ball,text width=.18cm,fill,color=black,below=2cm of x1b,label=below:$x_1$] (x1c) {};
  \node[ball,text width=.18cm,fill,color=black,right=1.5cm of x1c.east,label=below:$x_2$] (x2c) {};
  \draw[-,thick,color=black] (x1c.east) edge node [text width=.18cm,above=-0.05cm,midway] {{$y$}} (x2c.west);
  \draw[color=blue] (6.4,-1.2) ellipse (1.5cm and .5cm);
  \draw[color=red] (5.6,-1.1) ellipse (.3cm and .2cm);
  \draw[color=red] (7.1,-1.1) ellipse (.3cm and .2cm);  
  \node[right=.8cm of x2b] (sbd1) {$\displaystyle\:=\:\frac{1}{x_1+x_2}$};
  \node[right=.8cm of x2c] (sbd2) {$\displaystyle\:=\:\frac{1}{x_1+x_2}\times\frac{1}{x_1+y}\times\frac{1}{y+x_2}$};
 \end{tikzpicture}
\end{equation} 


\subsubsection{The three-site graph at tree level}\label{subsubsec:3stg}

Let us now consider the three-site graph, which is the first example which shows the simplicity and power of our combinatorial method to implement \eqref{eq:RR}. The first subgraph is always the graph itself which the total
energy pole is associated to. Contrarily to the two-site case, there are two inequivalent ways of dividing the graph in two connected subgraphs, which one needs to sum on: we can take either direct product 
$\psi_2(x_1,x_2+y_{\mbox{\tiny $23$}};y_{\mbox{\tiny $12$}})\,\otimes\,\psi_1(y_{\mbox{\tiny $12$}}+x_3)$ of the two-site subgraph having vertices $(1,\,2)$, with the one-site graph with vertex $3$; or 
$\psi_{1}(x_1+y_{\mbox{\tiny $12$}})\,\otimes\,\psi_2(y_{\mbox{\tiny $12$}}+x_2,x_3;\,y_{\mbox{\tiny $23$}})$. Diagrammatically:

\begin{equation}\label{eq:3sOFPT}
 \begin{tikzpicture}[ball/.style = {circle, draw, align=center, anchor=north, inner sep=0}]
    \node[ball,text width=.18cm,fill,color=black,label=below:$x_1$] at (0,0) (x1) {};
    \node[ball,text width=.18cm,fill,color=black,right=.8cm of x1.east,label=below:$x_2$] (x2) {};
    \node[ball,text width=.18cm,fill,color=black,right=.8cm of x2.east,label=below:$x_3$] (x3) {};
    \draw[-,thick,color=black] (x1.east) edge node [text width=.18cm,above=-0.05cm,midway] {{$y_{\mbox{\tiny $12$}}$}} (x2.west);
    \draw[-,thick,color=black] (x2.east) edge node [text width=.18cm,above=-0.05cm,midway] {{$y_{\mbox{\tiny $23$}}$}} (x3.west);
    \node[below=.5cm of x2.south] (d1) {};
    \node[below=1cm of d1.south] (d2) {};
    \path[draw=black,solid,line width=.1cm,shade,preaction={-triangle 90,thin,draw,color=black,shorten >=-1mm}] (d1) -- (d2);
    \node[ball,text width=.18cm,fill,color=black,label=below:$x_1$] at (-2,-2.5) (x1b) {};
    \node[ball,text width=.18cm,fill,color=black,right=.8cm of x1b.east,label=below:$x_2$] (x2b) {};
    \node[ball,text width=.18cm,fill,color=black,right=.8cm of x2b.east,label=below:$x_3$] (x3b) {};
    \draw[-,thick,color=black] (x1b.east) edge node [text width=.18cm,above=-0.05cm,midway] {{$y_{\mbox{\tiny $12$}}$}} (x2b.west);
    \draw[-,thick,color=black] (x2b.east) edge node [text width=.18cm,above=-0.05cm,midway] {{$y_{\mbox{\tiny $23$}}$}} (x3b.west);
    \draw[color=blue] (-1,-2.7) ellipse (1.5cm and .5cm);
    \node[right=.7cm of x3b.west] (sbd1) {$\displaystyle\:=\:\frac{1}{x_1+x_2+x_3}$};
    \node[below=.5cm of x2b.south] (d3) {};
    \node[below=.1cm of sbd1.south] (d4) {};
    \node[below=.5cm of d3.south] (tmp3) {};
    \node[left=1cm of tmp3.west] (d5) {};
    \node[below=.5cm of d4.south] (tmp4) {};
    \node[right=1cm of tmp4.east] (d6) {};
    \draw[solid,line width=.1cm,shade,preaction={-triangle 90,thin,draw,color=black,shorten >=-1mm}] (d3) -- (d5);    
    \draw[solid,line width=.1cm,shade,preaction={-triangle 90,thin,draw,color=black,shorten >=-1mm}] (d4) -- (d6);
    \node[below=.5cm of d5.south] (tmp5) {};
    \node[ball,text width=.13cm,fill,color=black,left=.7cm of tmp5.west,label=below:{\tiny $x_3$}] (x3c) {};
    \node[ball,text width=.13cm,fill,color=black,left=.6cm of x3c.west,label=below:{\tiny $x_2$}] (x2c) {};
    \node[ball,text width=.13cm,fill,color=black,left=.6cm of x2c.west,label=below:{\tiny $x_1$}] (x1c) {};
    \draw[-,thick,color=black] (x1c.east) edge node [text width=.18cm,above=-0.05cm,midway] {{\tiny $y_{12}$}} (x2c.west);
    \draw[-,thick,color=black] (x2c.east) edge node [text width=.18cm,above=-0.05cm,midway] {{\tiny $y_{23}$}} (x3c.west);
    \node[right=.1cm of x3c.east] (sdb2) {\tiny{$\displaystyle\:=\:\frac{1}{x_1+x_2+x_3}\times\frac{1}{x_1+x_2+y_{23}}\times$}};
    \node[below=.2cm of sdb2.south] (sdb2a) {\tiny{$\displaystyle\:\hspace{-2cm}\times\frac{1}{y_{23}+x_3}$}};
    \draw[color=blue] (x2c) ellipse (1cm and .55cm);
    \draw[color=red] (-4.3,-4.85) ellipse (.5cm and .275cm);
    \draw[color=red] (x3c.west) ellipse (.2cm and .1375cm);
    \node[below=1cm of d4.south] (tmp6) {};
    \node[right=.6cm of tmp6.east] (tmp7) {};
    \node[ball,text width=.13cm,fill,color=black,below=.0cm of tmp7,label=below:{\tiny $x_2$}] (x2d) {};
    \node[ball,text width=.13cm,fill,color=black,right=.6cm of x2d.east,label=below:{\tiny $x_3$}] (x3d) {};
    \node[ball,text width=.13cm,fill,color=black,left=.6cm of x2d.west,label=below:{\tiny $x_1$}] (x1d) {};
    \draw[-,thick,color=black] (x1d.east) edge node [text width=.18cm,above=-0.05cm,midway] {{\tiny $y_{12}$}} (x2d.west);
    \draw[-,thick,color=black] (x2d.east) edge node [text width=.18cm,above=-0.05cm,midway] {{\tiny $y_{23}$}} (x3d.west);
    \node[right=.1cm of x3d.east] (sdb3) {\tiny{$\displaystyle\:=\:\frac{1}{x_1+x_2+x_3}\times\frac{1}{y_{\mbox{\tiny $12$}}+x_2+x_3}$}};
    \node[below=.2cm of sdb3.south] (sdb3a) {\tiny{$\displaystyle\:\hspace{-2cm}\times\frac{1}{x_1+y_{12}}$}};
    \draw[color=blue] (x2d) ellipse (1cm and .55cm);
    \draw[color=red] (x1d.east) ellipse (.2cm and .1375cm);
    \draw[color=red] (+3.35,-4.85) ellipse (.5cm and .275cm);
    \node[below=.1cm of sdb2a.south] (d7) {};
    \node[below=.9cm of d7.south] (d8) {};
    \draw[solid,line width=.1cm,shade,preaction={-triangle 90,thin,draw,color=black,shorten >=-1mm}] (d7) -- (d8);   
    \node[below=.1cm of sdb3a.south] (d9) {};
    \node[below=.9cm of d9.south] (d10) {};
    \draw[solid,line width=.1cm,shade,preaction={-triangle 90,thin,draw,color=black,shorten >=-1mm}] (d9) -- (d10);
    \node[below=.5cm of d8.south] (tmp8) {};
    \node[ball,text width=.13cm,fill,color=black,left=2.2cm of tmp8.west,label=below:{\tiny $x_3$}] (x3e) {};
    \node[ball,text width=.13cm,fill,color=black,left=.6cm of x3e.west,label=below:{\tiny $x_2$}] (x2e) {};
    \node[ball,text width=.13cm,fill,color=black,left=.6cm of x2e.west,label=below:{\tiny $x_1$}] (x1e) {};
    \draw[-,thick,color=black] (x1e.east) edge node [text width=.18cm,above=-0.05cm,midway] {{\tiny $y_{12}$}} (x2e.west);
    \draw[-,thick,color=black] (x2e.east) edge node [text width=.18cm,above=-0.05cm,midway] {{\tiny $y_{23}$}} (x3e.west);
    \node[right=.1cm of x3e.east] (sdb4) {\tiny{$\displaystyle\:=\:\frac{1}{x_1+x_2+x_3}\times\frac{1}{x_1+x_2+y_{23}}\times$}};
    \node[below=.2cm of sdb4.south] (sdb4a) {\tiny{$\displaystyle\hspace{-1.1cm}\times\frac{1}{y_{23}+x_3}\times\frac{1}{(x_1+y_{12})(y_{12}+x_2+y_{23})}$}};
    \draw[color=blue] (x2e) ellipse (1cm and .55cm);
    \draw[color=red] (-4.3,-8.45) ellipse (.5cm and .275cm);
    \draw[color=red] (x3e.west) ellipse (.2cm and .1375cm);
    \draw[color=green] (x1e.east) ellipse (.2cm and .1375cm);    
    \draw[color=green] (x2e.west) ellipse (.2cm and .1375cm);    
    \node[below=.5cm of d10.south] (tmp9) {};
    \node[ball,text width=.13cm,fill,color=black,left=2.8cm of tmp9.west,label=below:{\tiny $x_2$}] (x2f) {};
    \node[ball,text width=.13cm,fill,color=black,right=.6cm of x2f.east,label=below:{\tiny $x_3$}] (x3f) {};
    \node[ball,text width=.13cm,fill,color=black,left=.6cm of x2f.west,label=below:{\tiny $x_1$}] (x1f) {};
    \draw[-,thick,color=black] (x1f.east) edge node [text width=.18cm,above=-0.05cm,midway] {{\tiny $y_{12}$}} (x2f.west);
    \draw[-,thick,color=black] (x2f.east) edge node [text width=.18cm,above=-0.05cm,midway] {{\tiny $y_{23}$}} (x3f.west);
    \node[right=.1cm of x3f.east] (sdb5) {\tiny{$\displaystyle\:=\:\frac{1}{x_1+x_2+x_3}\times\frac{1}{y_{\mbox{\tiny $12$}}+x_2+x_3}\times$}};
    \node[below=.2cm of sdb5.south] (sdb5a) {\tiny{$\displaystyle\hspace{-1.1cm}\times\frac{1}{x_1+y_{12}}\times\frac{1}{(y_{12}+x_2+y_{23})(y_{23}+x_3)}$}};    
    \draw[color=blue] (x2f) ellipse (1cm and .55cm);
    \draw[color=red] (+3.35,-8.45) ellipse (.5cm and .275cm);
    \draw[color=red] (x3f.west) ellipse (.2cm and .1375cm);
    \draw[color=green] (x1f.east) ellipse (.2cm and .1375cm);    
    \draw[color=green] (x2f.east) ellipse (.2cm and .1375cm);    
   \end{tikzpicture}
\end{equation}
The final answer for $\psi_3(x,y)$ is given by the sum of the two terms in the last line of \eqref{eq:3sOFPT}, which again is strikingly simpler than the nine term representation provided by the time integral.


\subsubsection{The four-site star graph}\label{subsubsec:4s0L}

In order to further clarify the combinatorial rules which allow us to straightforwardly derive the OFPT representation of the wavefunction, let us consider a slightly more complicated example, provided by the star graph appearing in Fig \ref{fig:Graphs}.

\begin{equation}\label{eq:4s0L}
 \begin{tikzpicture}[ball/.style = {circle, draw, align=center, anchor=north, inner sep=0}] 
  \node[ball,text width=.18cm,fill,color=black,label=left:{\tiny $x_4$}] at (0,0) (w4) {};    
   \node[ball,text width=.18cm,fill,color=black,label=below:{\tiny $x_3$}] at ({1.5*cos(0)},{1.5*sin(0)}) (w3) {};    
   \node[ball,text width=.18cm,fill,color=black,label=left:{\tiny $x_1$}] at ({1.5*cos(120)},{1.5*sin(120)}) (w1) {};
   \node[ball,text width=.18cm,fill,color=black,label=left:{\tiny $x_2$}] at ({1.5*cos(240)},{1.5*sin(240)}) (w2) {};    
   \draw[-,thick,color=black] (w1) edge node [text width=.18cm,left=.3cm,midway] {\tiny $y_{14}$} (w4);    
   \draw[-,thick,color=black] (w2) edge node [text width=.18cm,right=-.1cm,midway] {\tiny $y_{24}$} (w4);        
   \draw[-,thick,color=black] (w3) edge node [text width=.18cm,above=-.05cm,midway] {\tiny $y_{34}$} (w4);

   \node[right=.3cm of w3.east] (f1) {$\displaystyle\:=\:\frac{1}{\sum_{i=1}^4x_i}\left(\prod_{j=1}^3\frac{1}{x_j+y_{j4}}\right)\frac{1}{x_4+y_{14}+y_{24}+y_{34}}\times$};
   \node[below=.5cm of f1.south] (f2) {$\displaystyle\hspace{1cm}\times\left[\frac{1}{x_4+x_1+x_2+y_{34}}\left(\frac{1}{x_4+x_1+y_{24}+y_{34}}+\frac{1}{x_4+x_2+y_{14}+y_{34}}\right)\right.+$};
   \node[below=2cm of f1.south] (f3) {$\displaystyle\hspace{1.25cm}+\frac{1}{x_4+x_2+x_3+y_{14}}\left(\frac{1}{x_4+x_2+y_{34}+y_{14}}+\frac{1}{x_4+x_3+y_{24}+y_{14}}\right)+$};   
   \node[below=3.5cm of f1.south] (f4) {$\displaystyle\hspace{1.25cm}\left.+\frac{1}{x_4+x_3+x_1+y_{24}}\left(\frac{1}{x_4+x_3+y_{14}+y_{24}}+\frac{1}{x_4+x_1+y_{34}+y_{24}}\right)\right].$};      
 \end{tikzpicture}
\end{equation}

As for the previous two case, we are going to illustrate the rules to obtain the right-hand-side of \eqref{eq:4s0L} in some detail. As usual, the first subgraph is the graph itself which provides the total energy pole:
\begin{equation}\label{eq:4s0Lr1}
 \begin{tikzpicture}[ball/.style = {circle, draw, align=center, anchor=north, inner sep=0}] 
  \node[ball,text width=.18cm,fill,color=black,label=left:{\tiny $x_4$}] at (0,0) (w4) {};    
   \node[ball,text width=.18cm,fill,color=black,label=below:{\tiny $x_3$}] at ({1.5*cos(0)},{1.5*sin(0)}) (w3) {};    
   \node[ball,text width=.18cm,fill,color=black,label=left:{\tiny $x_1$}] at ({1.5*cos(120)},{1.5*sin(120)}) (w1) {};
   \node[ball,text width=.18cm,fill,color=black,label=left:{\tiny $x_2$}] at ({1.5*cos(240)},{1.5*sin(240)}) (w2) {};    
   \draw[-,thick,color=black] (w1) edge node [text width=.18cm,left=.3cm,midway] {\tiny $y_{14}$} (w4);    
   \draw[-,thick,color=black] (w2) edge node [text width=.18cm,right=-.1cm,midway] {\tiny $y_{24}$} (w4);        
   \draw[-,thick,color=black] (w3) edge node [text width=.18cm,above=-.05cm,midway] {\tiny $y_{34}$} (w4);
   \draw[color=blue] (w4) circle (1.75cm);
   \node[right=.3cm of w3.east] (f1) {$\displaystyle\:=\:\frac{1}{\sum_{i=1}^4x_i}.$};    
 \end{tikzpicture}
\end{equation}
Now there are three inequivalent ways to divide the above graph in two subgraphs, by grouping two external nodes and the internal one, as well as the edges connecting them:
\begin{equation}\label{eq:4s0Lr2}
 \begin{tikzpicture}[ball/.style = {circle, draw, align=center, anchor=north, inner sep=0}] 
  \node[ball,text width=.18cm,fill,color=black,label=left:{\tiny $x_4$}] at (0,0) (w4) {};    
   \node[ball,text width=.18cm,fill,color=black,label=below:{\tiny $x_{j+2}$}] at ({1.5*cos(0)},{1.5*sin(0)}) (w3) {};    
   \node[ball,text width=.18cm,fill,color=black,label=left:{\tiny $x_j$}] at ({1.5*cos(120)},{1.5*sin(120)}) (w1) {};
   \node[ball,text width=.18cm,fill,color=black,label=left:{\tiny $x_{j+1}$}] at ({1.5*cos(240)},{1.5*sin(240)}) (w2) {};    
   \draw[-,thick,color=black] (w1) edge node [text width=.18cm,left=.3cm,midway] {\tiny $y_{j4}$} (w4);    
   \draw[-,thick,color=black] (w2) edge node [text width=.18cm,right=-.1cm,midway] {\tiny $y_{(j+1)4}$} (w4);        
   \draw[-,thick,color=black] (w3) edge node [text width=.18cm,above=-.05cm,midway] {\tiny $y_{(j+2)4}$} (w4);
   \draw[color=blue] (w4) circle (1.75cm);
   \coordinate (a1) at (w1.north);
   \coordinate (a3) at (w2.south);
   \coordinate (a2) at (-1.5,0);
   \coordinate (a4) at ($(w4)!.25!(w3)$);
   \draw[red] plot [smooth cycle] coordinates {(a1) (a2) (a3) (a4)};
   \draw[red] (w3) circle (.2cm);
   \node[right=.3cm of w3.east] (f1) {$\displaystyle\:=\:\frac{1}{\sum_{i=1}^4x_i}\times\frac{1}{x_4+x_j+x_{j+1}+y_{(j+2)4}}\times\frac{1}{x_{j+2}+y_{(j+2)4}},$};    
 \end{tikzpicture}
\end{equation}
with $j\,=\,1,\,2,\,3$ (mod$\{3\}$). Notice that the second factor on the right-hand-side corresponds to the subgraph containing $x_j$, $x_{j+1}$ and $x_4$, and it is obtained as the inverse of the sum of all the related external energies ({\it i.e.} $x_j$, $x_{j+1}$, $x_4$ and $y_{(j+2)4}$). Notice that just one of the two subgraphs encircled in red in \eqref{eq:4s0Lr2} can be further decomposed into connected subgraphs. Such a subgraph is exactly the three node graph studied in the previous section, where we have seen that there exist two inequivalent ways to decompose it:
\begin{equation}\label{eq:4s0Lr3}
 \begin{tikzpicture}[ball/.style = {circle, draw, align=center, anchor=north, inner sep=0}] 
  \node[ball,text width=.18cm,fill,color=black,label=left:{\tiny $x_4$}] at (0,0) (w4) {};    
   \node[ball,text width=.18cm,fill,color=black,label=below:{\tiny $x_{j+2}$}] at ({1.5*cos(0)},{1.5*sin(0)}) (w3) {};    
   \node[ball,text width=.18cm,fill,color=black,label=left:{\tiny $x_j$}] at ({1.5*cos(120)},{1.5*sin(120)}) (w1) {};
   \node[ball,text width=.18cm,fill,color=black,label=left:{\tiny $x_{j+1}$}] at ({1.5*cos(240)},{1.5*sin(240)}) (w2) {};    
   \draw[-,thick,color=black] (w1) edge node [text width=.18cm,left=.3cm,midway] {\tiny $y_{j4}$} (w4);    
   \draw[-,thick,color=black] (w2) edge node [text width=.18cm,right=-.1cm,midway] {\tiny $y_{(j+1)4}$} (w4);        
   \draw[-,thick,color=black] (w3) edge node [text width=.18cm,above=-.05cm,midway] {\tiny $y_{(j+2)4}$} (w4);
   \draw[color=blue] (w4) circle (1.75cm);
   \coordinate (a1) at (w1.north);
   \coordinate (a3) at (w2.south);
   \coordinate (a2) at (-1.5,0);
   \coordinate (a4) at ($(w4)!.25!(w3)$);
   \draw[red] plot [smooth cycle] coordinates {(a1) (a2) (a3) (a4)};
   \draw[red] (w3) circle (.2cm);
   \draw[rotate=120,color=green!70!black] ($(w1)!.5!(w4)$) ellipse (.9cm and .3cm);
   \draw[green!70!black] (w2) circle (.2cm);
   \draw[cyan] (w1) circle (.2cm);
   \draw[cyan] (w4) circle (.2cm);
   \node[right=.3cm of w3.east] (f1) {$\displaystyle\:=\:\frac{1}{\sum_{i=1}^4x_i}\times\frac{1}{x_4+x_j+x_{j+1}+y_{(j+2)4}}\,\frac{1}{x_{j+2}+y_{(j+2)4}}\times$};
   \node[below=.5cm of f1.south] (f2) {$\displaystyle\hspace{.45cm}\times\frac{1}{x_4+x_j+y_{(j+1)4}+y_{(j+2)4}}\,\frac{1}{x_{j+1}+y_{(j+1)4}}\times\frac{1}{x_j+y_{j4}}\,\frac{1}{x_4+y_{14}+y_{24}+y_{34}}$};   
  \begin{scope}[shift={(0,-4.5)}]
  \node[ball,text width=.18cm,fill,color=black,label=left:{\tiny $x_4$}] at (0,0) (w4) {};    
   \node[ball,text width=.18cm,fill,color=black,label=below:{\tiny $x_{j+2}$}] at ({1.5*cos(0)},{1.5*sin(0)}) (w3) {};    
   \node[ball,text width=.18cm,fill,color=black,label=left:{\tiny $x_j$}] at ({1.5*cos(120)},{1.5*sin(120)}) (w1) {};
   \node[ball,text width=.18cm,fill,color=black,label=left:{\tiny $x_{j+1}$}] at ({1.5*cos(240)},{1.5*sin(240)}) (w2) {};    
   \draw[-,thick,color=black] (w1) edge node [text width=.18cm,left=.3cm,midway] {\tiny $y_{j4}$} (w4);    
   \draw[-,thick,color=black] (w2) edge node [text width=.18cm,right=-.1cm,midway] {\tiny $y_{(j+1)4}$} (w4);        
   \draw[-,thick,color=black] (w3) edge node [text width=.18cm,above=-.05cm,midway] {\tiny $y_{(j+2)4}$} (w4);
   \draw[color=blue] (w4) circle (1.75cm);
   \coordinate (a1) at (w1.north);
   \coordinate (a3) at (w2.south);
   \coordinate (a2) at (-1.5,0);
   \coordinate (a4) at ($(w4)!.25!(w3)$);
   \draw[red] plot [smooth cycle] coordinates {(a1) (a2) (a3) (a4)};
   \draw[red] (w3) circle (.2cm);
   \draw[rotate=240,color=green!70!black] ($(w2)!.5!(w4)$) ellipse (.9cm and .3cm);
   \draw[green!70!black] (w1) circle (.2cm);
   \draw[cyan] (w2) circle (.2cm);
   \draw[cyan] (w4) circle (.2cm);
   \node[right=.3cm of w3.east] (f1) {$\displaystyle\:=\:\frac{1}{\sum_{i=1}^4x_i}\times\frac{1}{x_4+x_j+x_{j+1}+y_{(j+2)4}}\,\frac{1}{x_{j+2}+y_{(j+2)4}}\times$};
   \node[below=.5cm of f1.south] (f2) {$\displaystyle\hspace{.45cm}\times\frac{1}{x_4+x_{j+1}+y_{j4}+y_{(j+2)4}}\,\frac{1}{x_{j}+y_{j4}}\times\frac{1}{x_{j+1}+y_{(j+1)4}}\,\frac{1}{x_4+y_{14}+y_{24}+y_{34}},$}; 
  \end{scope}
 \end{tikzpicture}
\end{equation}
where the first two factors in the second line of the right-hand-side of both contributions correspond to the subgraphs encircled in green, while the last two correspond to the atomic subgraphs encircled in cyan.

It is easy to check that summing the two expressions in \eqref{eq:4s0Lr3} and then over $j$ ($j\,=\,1,\,2,\,3$), we obtain the OFPT expression in \eqref{eq:4s0L}.


\subsubsection{The two-site graph at one loop}\label{subsubsec:2s1L}

The three examples discussed above concerned tree level graphs. As from our general discussion, our combinatorial way of writing down the wavefunction $\psi_n(x,y)$ does not depend on the perturbative order. It is instructive
to illustrate the method for the simplest example at loop level, which is given by the following two-site graph:

\begin{equation}\label{eq:2s1l}
 \begin{tikzpicture}[ball/.style = {circle, draw, align=center, anchor=north, inner sep=0}]
   \node[ball,text width=.18cm,fill,color=black] at (0,0) (x1) {};
   \node[left=.0cm of x1.west,label=below:$x_1$] (x1l) {};
   \node[ball,text width=.18cm,fill,color=black,right=1.5cm of x1.east] (x2) {};
   \node[right=.0cm of x2.east,label=below:$x_2$] (x2l) {};
   \draw[thick] (.85,0) circle (.85);
   \node at (.85,1) {$\displaystyle y_a$};
   \node at (.85,-1) {$\displaystyle y_b$};   
   \node[right=.2cm of x2.east] (f1) {$\displaystyle\:=\:\frac{1}{(x_1+x_2)(x_1+y_a+y_b)(x_2+y_a+y_b)}\left[\frac{1}{x_1+x_2+2y_a}+\frac{1}{x_1+x_2+2y_b}\right],$};
 \end{tikzpicture} 
\end{equation}
where the right-hand-side is already its OFPT representation (to be contrasted with the time integral one which is a sum of nine terms). As in the three-site graph at tree level, the two terms of \eqref{eq:2s1l} are due to the fact that the one-loop graph in question can be divided in two subgraphs in two inequivalent ways and the factors $(x_1+x_2+2y_a)$ and $(x_1+x_2+2y_b)$ are their total energies:
\begin{equation}\label{eq:2s1lb}
 \begin{split}
  &
  \begin{tikzpicture}[ball/.style = {circle, draw, align=center, anchor=north, inner sep=0}]
   \node[ball,text width=.18cm,fill,color=black] at (0,0) (x1) {};
   \node[left=.0cm of x1.west,label=below:$x_1$] (x1l) {};
   \node[ball,text width=.18cm,fill,color=black,right=1.5cm of x1.east] (x2) {};
   \node[right=.0cm of x2.east,label=below:$x_2$] (x2l) {};
   \draw[thick] (.85,0) circle (.85);
   \node at (.85,1.3) (yal) {$\displaystyle y_a$};
   \node at (.85,-1.3) (ybl) {$\displaystyle y_b$}; 
   \draw (.85,0) circle (1.05);
   \coordinate (a1) at (x1.north);
   \coordinate (a2) at (x1.west);
   \coordinate (b2) at (.25,-.7);
   \coordinate (a3) at (.9,-.9);
   \coordinate (b3) at (1.5,-.7);
   \coordinate (a4) at (x2.east);
   \coordinate (a5) at (x2.north);
   \coordinate (a6) at (x2.west);
   \coordinate (b6) at (1.3,-.6);
   \coordinate (a7) at (.9,-.75);
   \coordinate (b7) at (.4,-.6);
   \coordinate (a8) at (x1.east);
   \draw[black] plot [smooth cycle] coordinates {(a1) (a2) (b2) (a3) (b3) (a4) (a5) (a6) (b6) (a7) (b7) (a8)};
   \node[right=.5cm of x2.east] (res1) {$\displaystyle\:=\:\frac{1}{(x_1+x_2)(x_1+y_a+y_b)(x_2+y_a+y_b)(x_1+x_2+2y_a)}$};
  \end{tikzpicture}
  \\
  &
  \begin{tikzpicture}[ball/.style = {circle, draw, align=center, anchor=north, inner sep=0}]
   \node[ball,text width=.18cm,fill,color=black] at (0,0) (x1) {};
   \node[left=.0cm of x1.west,label=below:$x_1$] (x1l) {};
   \node[ball,text width=.18cm,fill,color=black,right=1.5cm of x1.east] (x2) {};
   \node[right=.0cm of x2.east,label=below:$x_2$] (x2l) {};
   \draw[thick] (.85,0) circle (.85);
   \node at (.85,1.3) (yal) {$\displaystyle y_a$};
   \node at (.85,-1.3) (ybl) {$\displaystyle y_b$}; 
   \draw (.85,0) circle (1.05);
   \coordinate (a1) at (x1.south);
   \coordinate (a2) at (x1.west);
   \coordinate (b2) at (.15,.65);
   \coordinate (a3) at (.9,.9);
   \coordinate (b3) at (1.55,.7);
   \coordinate (a4) at (x2.east);
   \coordinate (a5) at (x2.south);
   \coordinate (a6) at (x2.west);
   \coordinate (b6) at (1.35,.6);
   \coordinate (a7) at (.9,.75);
   \coordinate (b7) at (.4,.6);
   \coordinate (a8) at (x1.east);
   \draw[black] plot [smooth cycle] coordinates {(a1) (a2) (b2) (a3) (b3) (a4) (a5) (a6) (b6) (a7) (b7) (a8)};
   \node[right=.5cm of x2.east] (res1) {$\displaystyle\:=\:\frac{1}{(x_1+x_2)(x_1+y_a+y_b)(x_2+y_a+y_b)(x_1+x_2+2y_b)}$};
  \end{tikzpicture}
 \end{split}
\end{equation}


\subsection{Recursion Relations At Tree-Level}\label{subsec:TreeRR}

The graphs which represent tree-level processes turn out to admit a further representation. In order to show it, let us consider a tree-level graph with a generic topology as a time integral and let us focus on one of its external edges:
\begin{equation}\label{eq:TreeInt}
  \begin{tikzpicture}[ball/.style = {circle, draw, align=center, anchor=north, inner sep=0}]
  \node[ball,text width=.18cm,fill,color=black, label=below:$\mbox{\tiny $x_1$}$] at (0,0) (x1) {};
  \node[ball,text width=1cm,shade,right=.8cm of x1.east] (S1)  {$\mathcal{B}$};
  \node[ball,text width=.18cm,fill,color=black,right=.7cm of x1.east, label=below:$\mbox{\tiny $\hspace{-.3cm}x_2$}$] (x2) {};
  \draw[-,thick,color=black] (x1.east) -- (x2.west);
  \node[right=.2cm of S1.east] (for1) {$\displaystyle\:\equiv\:\int_{-\infty}^0\prod_{v\in\mathcal{B}\setminus\{2\}}d\eta_v\,e^{ix_v\eta_v}\int_{-\infty}^0d\eta_2\,e^{ix_2\eta_2}
                                       \prod_{e\in\mathcal{B}}\tilde{G}(y_{v_e};\eta_{v_e},\eta_{v'_e})I_1(y_{12},\eta_2)$};
 \end{tikzpicture}
\end{equation}
where $I_1(y_{12},\eta_2)$ is the integral over $\eta_1$, whose explicit expression is
\begin{equation}\label{eq:I1}
 I_1(y_{12},\eta_2)\:\equiv\:\int_{-\infty}^{0}d\eta_1\,e^{ix_1\eta_1}G(y_{12},\eta_1,\eta_2).
\end{equation}
The propagator $G(y_{12},\eta_1,\eta_2)$ can be written as an integral over frequencies and $I_1$ becomes
\begin{equation}\label{eq:I1b}
  I_1(y_{12},\eta_2)\:=\:\frac{2}{2\pi i}\int_{-\infty}^{+\infty}d\omega\int_{-\infty}^{0}d\eta_1\,e^{ix_1\eta_1}
   \left[
    \frac{e^{i\omega(\eta_1-\eta_2)}}{\omega^2-y^{2}_{12}+i\varepsilon}-\frac{e^{i\omega(\eta_1+\eta_2)}}{\omega^2-y^2_{12}+i\varepsilon}
   \right].
\end{equation}
Performing the integration over $\eta_1$  first, we obtain
\begin{equation}\label{eq:I1c}
 I_1(y_{12},\eta_2)\:=\:\frac{2}{2\pi i}\int_{-\infty}^{+\infty}d\omega\,\frac{1}{\omega-(-x_1+i\varepsilon)}
   \left[
    \frac{e^{-i\omega\eta_2}}{\omega^2-y^{2}_{12}+i\varepsilon}-\frac{e^{i\omega\eta_2}_{12}}{\omega^2-y^2_{12}+i\varepsilon}
   \right].
\end{equation}
Being $\eta_2$ always negative, for the first term in \eqref{eq:I1c} we can close the contour of integration in the upper-half $\omega$-plane, and thus the contribution comes from the poles $\omega\,=\,-x_1+i\varepsilon$ and $\omega\,=\,-y_{12}+i\varepsilon$. For the second term, we need to close in the lower-half $\omega$-plane, so that the integral is given by the residue of the the pole $\omega\,=\,y_{12}-i\varepsilon$ only. Hence:
\begin{equation}\label{eq:I1fin}
 I_{1}\:=\:\frac{1}{y_{12}^2-x_1^2}\left[e^{ix_1\eta_2}-e^{i y_{12}\eta_2}\right].
\end{equation}
Notice that putting this expression into \eqref{eq:TreeInt}, we can view it as the difference of two terms which are still tree-level graphs but with one edge less and shifted energies in the site at time $\eta_2$:
\begin{equation}\label{eq:TreeRRf}
 \begin{tikzpicture}[ball/.style = {circle, draw, align=center, anchor=north, inner sep=0}]
  \node[ball,text width=.18cm,fill,color=black, label=below:$\mbox{\tiny $x_1$}$] at (0,0) (x1) {};
  \node[ball,text width=1cm,shade,right=.8cm of x1.east] (S1)  {$\mathcal{B}$};
  \node[ball,text width=.18cm,fill,color=black,right=.7cm of x1.east, label=below:$\mbox{\tiny $\hspace{-.3cm}x_2$}$] (x2) {};
  \draw[-,thick,color=black] (x1.east) -- (x2.west);
  \node[right=.2cm of S1.east] (for1) {$\displaystyle\:=\:\frac{1}{y_{12}^2-x_1^2}\left[\phantom{\frac{1}{y}}\right.$};
  \node[ball,text width=1cm,shade,right=.2cm of for1.east] (S2)  {$\mathcal{B}$};
  \node[ball,text width=.18cm,fill,color=black,right=.1cm of for1.east, label=below:$\mbox{\tiny $\hspace{-.4cm}x_1+x_2$}$] (x2a) {};
  \node[right=.2cm of S2.east] (for2) {$\displaystyle\:-\:$};  
  \node[ball,text width=1cm,shade,right=.2cm of for2.east] (S3)  {$\mathcal{B}$};
  \node[ball,text width=.18cm,fill,color=black,right=.1cm of for2.east, label=below:$\mbox{\tiny $\hspace{-.4cm}y_{12}+x_2$}$] (x2b) {};
  \node[right=.2cm of S3.east] (for3) {$\displaystyle\hspace{-.3cm}{\left.\phantom{\frac{1}{y}}\right].}$};    
 \end{tikzpicture}
\end{equation}

The equation \eqref{eq:TreeRRf} can be seen as a diagrammatic operation of collapsing two vertices (one being external), and it constitutes a tree-level recursive relation which connects the $n$-point wavefunction to the $(n-1)$-point one. Iterating the recursive relation, the $n$-point wavefunction at tree-level gets expressed as a sum of $2^{E}$ terms (given by isolated vertices), weighted by factors of the form $\prod(y_I^2-x_I^2)^{-1}$. 

Notice that \eqref{eq:TreeRRf} introduces spurious poles which reorganize themselves with physical poles to form Mandelstam-like quantities. As a last remark, notice that the \eqref{eq:TreeRRf} is valid whenever a graph has a tree structure with some external edge.

It is instructive to actually compute some simple explicit example using this recursion relation. As usual, the simplest example is provided by the two-site graph:
\begin{equation}\label{eq:TreeRRex1}
 \psi_2(x_1,x_2;y)\:=\:\frac{1}{y^2-x_1^2}\left(\frac{1}{x_1+x_2}-\frac{1}{y+x_2}\right)\:\equiv\:\frac{1}{(x_1+x_2)(x_1+y)(y+x_2)}.
\end{equation}
From the time integral perspective \eqref{eq:TreeInt}, this expression implies that the integration over the time $\eta_1$ has been performed. However, in this particular case, we can decide to integrate $\eta_2$ first, obtaining a representation of the very same form \eqref{eq:TreeRRex1} but with $x_1$ and $x_2$ exchanged.

Let us move to a bit less trivial example given by the three-site graph:
\begin{equation}\label{eq:TreeRRex2}
 \begin{split}
  \psi_3(x_1,x_2,x_3;y_{12},y_{23})\:&=\:\frac{1}{y_{12}^2-x_1^2}\left[\psi_{2}(x_1+x_2,x_3;y_{23})-\psi_{2}(y_{12}+x_2,x_3;y_{23})\right]\:=\\
                                     &=\:\frac{1}{y_{12}^2-x_1^2}\left[\frac{1}{(x_1+x_2+x_3)(x_1+x_2+y_{23})(y_{23}+x_3)}-\right.\\
                                     &\:-\left.\frac{1}{(y_{12}+x_2+x_3)(y_{12}+x_2+y_{23})(y_{23}+x_3)}\right]
 \end{split}.
\end{equation}
It is straightforward to see that, explicitly summing the two terms in the last line of \eqref{eq:TreeRRex2}, the $y_{12}-x_1$ pole disappears leaving the result for this graph previously computed via OFPT. Interestingly, the representation splits the wavefunction into a term which contains the total energy pole and one which does not, so that the information about the flat-space amplitude is encoded by one term only. Notice that \eqref{eq:TreeRRex2} boils down to collapsing the vertex $1$ onto $2$:
\begin{equation}\label{eq:TreeRRex2b}
 \begin{tikzpicture}[ball/.style = {circle, draw, align=center, anchor=north, inner sep=0}]
    \node[ball,text width=.18cm,fill,color=black,label=below:$x_1$] at (0,0) (x1) {};
    \node[ball,text width=.18cm,fill,color=black,right=.8cm of x1.east,label=below:$x_2$] (x2) {};
    \node[ball,text width=.18cm,fill,color=black,right=.8cm of x2.east,label=below:$x_3$] (x3) {};
    \draw[-,thick,color=black] (x1.east) edge node [text width=.18cm,above=-0.05cm,midway] {{$y_{\mbox{\tiny $12$}}$}} (x2.west);
    \draw[-,thick,color=black] (x2.east) edge node [text width=.18cm,above=-0.05cm,midway] {{$y_{\mbox{\tiny $23$}}$}} (x3.west);
    \node[right=.2cm of x3.east] (rhs1) {$\displaystyle\:=\:\frac{1}{y_{12}^2-x_1^2}\left[\phantom{\frac{1}{y_{12}}}\right.$};
    \node[ball,text width=.18cm,fill,color=black,right=-.1cm of rhs1.east,label=below:{\footnotesize $x_1+x_2$}] (x1a) {};
  \node[ball,text width=.18cm,fill,color=black,right=1.5cm of x1a.east,label=below:{\footnotesize $x_3$}] (x3a) {};
  \draw[-,thick,color=black] (x1a.east) edge node [text width=.18cm,above=-0.05cm,midway] {{\footnotesize $y_{23}$}} (x3a.west);
  \node[right=.2cm of x3a.east] (t1) {$\displaystyle\:-\:$};
   \node[ball,text width=.18cm,fill,color=black,right=.1cm of t1.east,label=below:{\footnotesize $y_{12}+x_2$}] (x1b) {};
  \node[ball,text width=.18cm,fill,color=black,right=1.5cm of x1b.east,label=below:{\footnotesize $x_3$}] (x3b) {};
  \draw[-,thick,color=black] (x1b.east) edge node [text width=.18cm,above=-0.05cm,midway] {{\footnotesize $y_{23}$}} (x3b.west);
  \node[right=.2cm of x3b.east] (t2) {$\displaystyle\hspace{-.2cm}{\left.\phantom{\frac{1}{y}}\right].}$};
 \end{tikzpicture}
\end{equation}

There is however a second representation of this type which diagrammatically is obtained by collapsing vertex $3$ onto vertex $2$, {\it i.e.} performing in \eqref{eq:TreeInt} the integration over $\eta_3$ first. Its explicit expression can be obtained from \eqref{eq:TreeRRex2b} through the label exchanges $(x_1,\,y_{12})\,\longleftrightarrow\,(x_3,\,y_{23})$.

The recursion relation \eqref{eq:TreeRRex2b} can be iterated to express $\psi_3$ in terms of a sum of vertices $\psi_1$, whose coefficients are products of Mandelstam-like combination of the energies:
\begin{equation}\label{eq:TreeRRex2c}
 \begin{tikzpicture}[ball/.style = {circle, draw, align=center, anchor=north, inner sep=0}]
    \node[ball,text width=.18cm,fill,color=black,label=below:$x_1$] at (0,0) (x1) {};
    \node[ball,text width=.18cm,fill,color=black,right=.8cm of x1.east,label=below:$x_2$] (x2) {};
    \node[ball,text width=.18cm,fill,color=black,right=.8cm of x2.east,label=below:$x_3$] (x3) {};
    \draw[-,thick,color=black] (x1.east) edge node [text width=.18cm,above=-0.05cm,midway] {{$y_{\mbox{\tiny $12$}}$}} (x2.west);
    \draw[-,thick,color=black] (x2.east) edge node [text width=.18cm,above=-0.05cm,midway] {{$y_{\mbox{\tiny $23$}}$}} (x3.west);
    \node[right=.2cm of x3.east] (rhs1) {$\displaystyle\:=\:\frac{1}{(y_{12}^2-x_1^2)(y_{23}^2-x_3^2)}\left[\phantom{\frac{1}{y_{12}}}\right.$};
    \node[ball,text width=.18cm,fill,color=black,right=-.1cm of rhs1.east,label=below:{\tiny $x_1+x_2+x_3$}] (xtot) {};    
    \node[right=.6cm of xtot.east] (rhs2) {$\displaystyle\:-\:$};
    \node[ball,text width=.18cm,fill,color=black,right=.6cm of rhs2.east,label=below:{\tiny $x_1+x_2+y_{23}$}] (xint1) {};        
    \node[right=.6cm of xint1.east] (rhs3) {$\displaystyle\:-\:$};
    \node[ball,text width=.18cm,fill,color=black,right=.6cm of rhs3.east,label=below:{\tiny $y_{12}+x_2+x_3$}] (xint2) {};        
    \node[right=.6cm of xint2.east] (rhs4) {$\displaystyle\:+\:$};
    \node[ball,text width=.18cm,fill,color=black,right=.6cm of rhs4.east,label=below:{\tiny $y_{12}+x_2+y_{23}$}] (xint3) {};            
    \node[right=.3cm of xint3.east] (rhs5) {$\displaystyle{\left.\phantom{\frac{1}{y}}\right]}$};    
 \end{tikzpicture}
\end{equation}

For these graphs with line topology, one can iteratively collapse one among the most external nodes. Actually, this procedure does not change for different topologies. As a concrete example, let us take the star graph and let us compute it by using the recursion relation \eqref{eq:TreeRRf}:

\begin{equation}\label{eq:TreeRRex3}
 \begin{tikzpicture}[ball/.style = {circle, draw, align=center, anchor=north, inner sep=0}]
    \node[ball,text width=.18cm,fill,color=black,label=left:$x_4$] at (0,0) (x4) {};  
    \node[ball,text width=.18cm,fill,color=black,label=right:$x_3$] at ({1.5*cos(0)},{1.5*sin(0)}) (x3) {};    
    \node[ball,text width=.18cm,fill,color=black,label=left:$x_1$] at ({1.5*cos(120)},{1.5*sin(120)}) (x1) {};
    \node[ball,text width=.18cm,fill,color=black,label=left:$x_2$] at ({1.5*cos(240)},{1.5*sin(240)}) (x2) {};    
    \draw[-,thick,color=black] (x1) edge node [text width=.18cm,left=.3cm,midway] {\footnotesize $y_{14}$} (x4);    
    \draw[-,thick,color=black] (x2) edge node [text width=.18cm,right=-.1cm,midway] {\footnotesize $y_{24}$} (x4);        
    \draw[-,thick,color=black] (x3) edge node [text width=.18cm,above=-.05cm,midway] {\footnotesize $y_{34}$} (x4);
    \node[right=.4cm of x3.east] (rhs1) {$\displaystyle\:=\:\frac{1}{y_{14}^2-x_1^2}\left[\phantom{\frac{1}{y_{12}}}\right.$};    
    \node[ball,text width=.18cm,fill,color=black,right=-.5cm of rhs1.east, label=below:{\footnotesize $x_2$}] (x2a) {};
    \node[ball,text width=.18cm,fill,color=black,right=.8cm of x2a.east,label=below:{\footnotesize $x_1+x_4$}] (x4a) {};
    \node[ball,text width=.18cm,fill,color=black,right=.8cm of x4a.east,label=below:{\footnotesize $x_3$}] (x3a) {};
    \draw[-,thick,color=black] (x2a.east) edge node [text width=.18cm,above=-0.05cm,midway] {{$y_{\mbox{\tiny $24$}}$}} (x4a.west);
    \draw[-,thick,color=black] (x4a.east) edge node [text width=.18cm,above=-0.05cm,midway] {{$y_{\mbox{\tiny $34$}}$}} (x3a.west);
    \node[right=.4cm of x3a.east] (rhs2) {$\displaystyle\:-\:$};
    \node[ball,text width=.18cm,fill,color=black,right=.3cm of rhs2.east, label=below:{\footnotesize $x_2$}] (x2b) {};
    \node[ball,text width=.18cm,fill,color=black,right=.8cm of x2b.east,label=below:{\footnotesize $y_{14}+x_4$}] (x4b) {};
    \node[ball,text width=.18cm,fill,color=black,right=.8cm of x4b.east,label=below:{\footnotesize $x_3$}] (x3b) {};
    \draw[-,thick,color=black] (x2b.east) edge node [text width=.18cm,above=-0.05cm,midway] {{$y_{\mbox{\tiny $24$}}$}} (x4b.west);
    \draw[-,thick,color=black] (x4b.east) edge node [text width=.18cm,above=-0.05cm,midway] {{$y_{\mbox{\tiny $34$}}$}} (x3b.west);
    \node[right=-.1cm of x3b.east] (rhs3) {$\displaystyle{\left.\phantom{\frac{1}{y}}\right].}$};        
 \end{tikzpicture}
\end{equation}
At this point, we could either substitute the explicit expression for the three-site graph previously calculated, or we could pretend not to know the functional expression for such a graph, and then iterate the recursion relation until we map the star graph in a linear combination of isolated nodes with coefficients given by Mandelstam-like combination of the energies. The main difference between these two representations is the number of spurious poles that the second one generates. However, if we look at the total energy pole, in the isolated-node representation there is a single term containing such a pole, making it easy to identify the flat-space scattering amplitude. 


\subsection{Singularities and Factorization}\label{subsub:Brr}

The representations discussed so far have all been obtained by suitably manipulating the time integral, despite the fact that we were able to provide diagrammatic rules which make no reference to the time bulk, except for the time-representation ({\it nomen omen}) where the time flow plays an important role in the diagrammatics itself. Is there a way to find representations for the wavefunction without making reference to any time integration at all?

Given the $\psi_n$, if we assume to know the location of its poles, then we can introduce a one parameter deformation of the energy space, in a similar fashion (but not quite) to the BCFW deformation for amplitudes  \cite{Britto:2005fq}\footnote{See Section 4.1 and its Subsections in \cite{Benincasa:2013faa} for a general discussion of these deformation in the context of scattering amplitudes.}. In this case the space of deformations we can define is much larger because there is no further constraint that a deformation needs to satisfy: while in the amplitude case it had to be chosen to preserve both momentum conservation and on-shell condition, for the wavefunction there is no energy conservation to be respected. The minimal deformation one can define is obtained by just shifting one energy variable $x_i$ only:
\begin{equation}\label{eq:EnDef}
 x_i\:\longrightarrow\:x_i+\zeta.
\end{equation}
More generally, one can deform the energy space in the following way
\begin{equation}\label{eq:EnDef2}
 \left(\{x_i\},\,\{y_j\}\right)\:\longrightarrow\:\left(\{x_i\},\,\{y_j\}\right)+\left(\{\alpha_i\},\,\{\beta_j\}\right)\zeta,
\end{equation}
where $\alpha_i$ and $\beta_j$ are the coefficients for the deformation of $x_i$ and $y_j$ respectively, and they can be tuned simply on the basis of how many and which poles in the deformation parameter $\zeta$ one would like to have. In any case, for a given deformation, if the one-parameter family of wavefunctions $\psi_n(\zeta)$ vanishes as $\zeta$ goes to infinity, then $\psi_n$ can be expressed as a sum of its residues:
\begin{equation}\label{eq:EnDefInt}
 0\:=\:\frac{1}{2\pi i}\oint_{\hat{\mathbb{C}}}\frac{d\zeta}{\zeta}\,\psi_n(\zeta)\qquad\Longrightarrow\qquad\psi_n\:=\:\sum_{i\in\mathcal{P}}\mbox{Res}\left\{-\frac{\psi_n(\zeta)}{\zeta},\zeta\,=\,\zeta_i\right\},
\end{equation}
where $\mathcal{P}$ is the set of poles in $\zeta$. Thinking about $\psi_n$ as decomposed into meromorphic functions with numerator equal to one and physical poles only, then the relations \eqref{eq:EnDefInt} holds if a deformation induces at least one pole in each term.

A crucial point is now the physical interpretation of the residues in \eqref{eq:EnDefInt}. First, notice that physical poles appear in correspondence of sums of energies of a subset of {\it consecutive} vertices. This means that going to the point of energy space where one of these sums vanishes imposes energy conservation on a subgraph: at this point in energy space a subgraph is mapped to an amplitude. This then translates to a factorization into a product of such an amplitude with a lower-point wavefunction (times the total energy computed at this point) if and only if the energy conservation is imposed on a codimension-$1$ subgraph, {\it i.e.} in $\psi_n$, this exact factorization occurs  if the energy conservation involves $n-1$ vertices: $\psi_n\,\longrightarrow\,A_{n-1}\otimes\psi_3$. For higher codimension subgraphs, the residues can be seen as sum of products of lower point wavefunctions. 

As a first example let us consider the two-site graph and its representation under the energy deformation $x_2\,\longrightarrow\,x_2+\zeta$. Two out of its three poles acquire a $\zeta$-dependence, {\it i.e.} the total energy pole ($x_1+x_2+\zeta$) and the pole related to the vertex $2$ ($y+x_2+\zeta$). The residue of the total energy pole is the amplitude in the representation which {\it does not} depend on $x_2$. On the location of the other pole, the vertex $2$ is mapped to a three-point amplitude (which in this case it is a constant). The resulting representation for the two-site graph is therefore 
\begin{equation}\label{eq:2sEnRep}
 \psi_2(x_1,y,x_2)\:=\:\frac{A_2(x_1,y)}{x_1+x_2}+\frac{\psi_1(x_1+y)\tilde{\otimes}A_1}{y+x_2},\qquad A_2(x_1,y)\:\equiv\:\frac{1}{y^2-x_1^2},
\end{equation}
where $A_i$ are the amplitudes with $i$ nodes, while the operator $\tilde{\otimes}$ indicates that a factor with the total energy ($x_1+x_2+\hat{\zeta}\,\equiv\,x_1-y$) of $\psi_1$ and $A_1$ need to be included. Notice that the explicit functional expression for \eqref{eq:2sEnRep} returns
\begin{equation}\label{eq:2sEnRep2}
 \psi_2(x_1,y,x_2)\:=\:\frac{1}{y^2-x_1^2}\left(\frac{1}{x_1+x_2}-\frac{1}{y+x_2}\right),
\end{equation}
which is exactly the representation \eqref{eq:TreeRRex1} obtained by writing the propagator as an integral in the frequency space.

For the three-site case, if we deform the energy of one of the outer sites, namely we consider $x_3\,\longrightarrow\,x_3+\zeta$, then the poles in $\zeta$ are given by $x_1+x_2+x_3+\zeta$, $y_{12}+x_2+x_3+\zeta$ and $y_{23}+x_3+\zeta$. This recursive representation can then be written as
\begin{equation}\label{eq:3sEnRep}
 \psi_3(x,y)\:=\:\frac{A_3}{x_1+x_2+x_3}+\frac{\psi_1\tilde{\otimes}A_{2}}{y_{12}+x_2+x_3}+\frac{\psi_2\tilde{\otimes}A_1+\psi_1\tilde{\otimes}(\psi_1\tilde{\otimes}A_1)}{y_{34}+x_3},
\end{equation}
where
\begin{equation}\label{eq:3sRes}
 A_3\:=\:\frac{1}{(y_{12}^2-x_1^2)(y_{23}^2-(x_1+x_2)^2)},\qquad A_2\:=\:\frac{1}{y_{23}^2-(x_1+x_2)^2}.
\end{equation}

Rather than providing more examples of representations for $\psi_n$ generated with a single energy shift, let us consider the following deformation
\begin{equation}\label{eq:EnDefM}
 x_2\,\longrightarrow\,x_2+\zeta,\qquad x_3\,\longrightarrow\,x_3-\zeta,\qquad y_{23}\,\longrightarrow\,y_{23}+\zeta,
\end{equation}
such that at least one between $x_2$ and $x_3$ is related to an internal site. If we apply this deformation to the three-site graph, then the only $\zeta$-dependent poles are the ones containing the sum $x_2+y_{23}$, {\it i.e.}  $x_1+x_2+y_{23}+2\zeta$ and $y_{12}+x_2+y_{23}+2\zeta$. Following the same rules as before, we get
\begin{equation}\label{eq:3sEnM2}
 \begin{split}
  \psi_3(x,y)\:&=\:\frac{A_2\tilde{\otimes}\psi_1}{x_1+x_2+y_{23}}+\frac{(\psi_1\tilde{\otimes}A_1)\tilde{\otimes}\psi_1+\psi_1\tilde{\otimes}(A_1\tilde{\otimes}\psi_1)}{y_{12}+x_2+y_{23}}\:=\\
               &=\:\frac{1}{y_{12}^2-x_1^2}\left[\frac{1}{(x_1+x_2+x_3)(x_1+x_2+y_{23})(y_{23}+x_3)}-\right.\\
               &\hspace{3cm}-\left.\frac{1}{(y_{12}+x_2+x_3)(y_{12}+x_2+y_{23})(y_{23}+x_3)}\right],
 \end{split}
\end{equation}
which is the two term representation \eqref{eq:TreeRRex2b}. More generally, for any graph of the following topology, the deformation \eqref{eq:EnDefM} return the representation \eqref{eq:TreeRRex2b}:
\begin{equation}\label{eq:nsEnMf}
 \begin{tikzpicture}[ball/.style = {circle, draw, align=center, anchor=north, inner sep=0}]
  \node[ball,text width=.18cm,fill,color=black, label=below:$\mbox{\tiny $x_1$}$] at (0,0) (x1) {};
  \node[ball,text width=1cm,shade,right=.175cm of x3.east] (S1)  {$\mathcal{B}$};
  \node[ball,text width=.18cm,fill,color=black,right=.7cm of x1.east, label=below:$\mbox{\tiny $\hspace{-.3cm}x_2$}$] (x2) {};
  \node[ball,text width=.18cm,fill,color=black,right=.7cm of x2.east, label=below:$\mbox{\tiny $\hspace{-.3cm}x_3$}$] (x3) {};  
  \draw[-,thick,color=black] (x1.east) -- (x2.west);
  \draw[-,thick,color=black] (x2.east) -- (x3.west);  
  \node[right=.2cm of S1.east] (for1) {$\displaystyle\:=\:\frac{1}{y_{12}^2-x_1^2}\left[\phantom{\frac{1}{y}}\right.$};
  \node[ball,text width=.18cm,fill,color=black,right=.1cm of for1.east, label=below:$\mbox{\tiny $\hspace{-.4cm}x_1+x_2$}$] (x1a) {}; 
  \node[ball,text width=1cm,shade,right=.8cm of x1a.east] (S2)  {$\mathcal{B}$};    
  \node[ball,text width=.18cm,fill,color=black,right=.7cm of x1a.east, label=below:$\mbox{\tiny $\hspace{-.4cm}x_3$}$] (x3a) {};
  \draw[-,thick,color=black] (x1a.east) -- (x3a.west);  
  \node[right=.2cm of S2.east] (for2) {$\displaystyle\:-\:$};  
  \node[ball,text width=.18cm,fill,color=black,right=.1cm of for2.east, label=below:$\mbox{\tiny $\hspace{-.4cm}y_{12}+x_2$}$] (x1b) {}; 
  \node[ball,text width=1cm,shade,right=.8cm of x1b.east] (S3)  {$\mathcal{B}$};    
  \node[ball,text width=.18cm,fill,color=black,right=.7cm of x1b.east, label=below:$\mbox{\tiny $\hspace{-.4cm}x_3$}$] (x3b) {};
  \draw[-,thick,color=black] (x1b.east) -- (x3b.west);  
  \node[right=.2cm of S3.east] (for3) {$\displaystyle\hspace{-.3cm}{\left.\phantom{\frac{1}{y}}\right].}$};   
 \end{tikzpicture}
\end{equation}
for those graph topologies such that $m$ edges ($m\,\ge\,3$) are connected to the node $2$, then the deformation \eqref{eq:EnDefM} induces a representation with a higher number of terms.

Summarizing, the energy space deformation provides a general method to derive directly from a boundary perspective a plethora of representations for the wavefunction. In this analysis we focused on a single graph, however this method allows to derive new representations for the full $\psi_n$ ({\it i.e.} considering all the graphs contributing to it). A discussion on the full wavefunction goes beyond the purpose of this paper and we leave it for future work.


\section{Cosmological Polytopes}\label{sec:CosmP}

We will now switch gears and describe a class of polytopes -- which we will refer to as ``{\it Cosmological Polytopes}'' -- that provide a deeper combinatorial and geometrical understanding for the ``wavefunction of the universe'' of the toy scalar model we have been studying. We will begin with giving the most intrinsic and fundamental definition of these polytope, and see how its facet structure, and ultimately the connection to physics via its canonical form, emerge from this definition.

We begin with considering a collection of $E$ triangles $\triangle_i$, whose vertices are given by the vectors $({\bf a}_i,\,{\bf b}_i,\,{\bf c}_i)$. If the vectors are all linearly independent, then we simply have $3E$ independent vectors in a $3E$-dimensional vector space. However, we will now allow these triangles to intersect each other in a particular way. Each triangle is characterized by having two sides $S_j^{\mbox{\tiny $(f)$}}$ $(j\,=\,1,\,2)$, on which it can intersect another triangle on the midpoint of the side itself, and one edge $S^{\mbox{\tiny (u)}}$ on which it is not allowed to intersect any other triangle. Without loss of generality we can take the edges $S_j^{\mbox{\tiny $(f)$}}$ $(j\,=\,1,\,2)$, for any triangle $\triangle_i$ to be $({\bf a}_i,\,{\bf b}_i)$ and $({\bf a}_i,\,{\bf c}_i)$, while the edge $S^{\mbox{\tiny $(u)$}}$ to be $({\bf b}_i,\,{\bf c}_i)$.

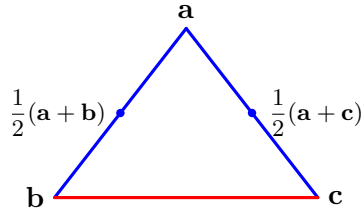
\begin{equation*}
 \begin{tikzpicture}[shift={(1,0)}, line join = round, line cap = round, ball/.style = {circle, draw, align=center, anchor=north, inner sep=0}]
  \coordinate [label=above:${\bf a}$] (A) at (0,0);
  \coordinate [label=left:${\bf b}$] (B) at (-1.75,-2.25);
  \coordinate [label=right:${\bf c}$] (C) at (+1.75,-2.25);
  \coordinate [label=left:{\footnotesize $\displaystyle \frac{1}{2}({\bf a}+{\bf b})\,$}] (m1) at ($(A)!0.5!(B)$);
  \coordinate [label=right:{\footnotesize $\displaystyle \;\frac{1}{2}({\bf a}+{\bf c})$}] (m2) at ($(A)!0.5!(C)$);
  \tikzset{point/.style={insert path={ node[scale=2.5*sqrt(\pgflinewidth)]{.} }}} 

  \draw[-, very thick, color=blue] (B) -- node[point,text width=.08cm, color=blue] {} (A);
  \draw[-, very thick, color=blue] (A) -- node[point, text width=.08cm,color=blue] {} (C);  
  \draw[-, very thick, color=red] (B) -- (C);    


\end{tikzpicture} 
\end{equation*}

We pause here very early in our exposition to motivate why this seemingly odd construction should have anything to do with the physics of ``time'' -- why ``triangles'', and why the asymmetry between the two $S_j^{\mbox{\tiny $(f)$}}$'s and the single $S^{\mbox{\tiny $(u)$}}$ sides of the triangle?

The obvious connection is seen in the familiar space-time picture of the causal relationship between the events:

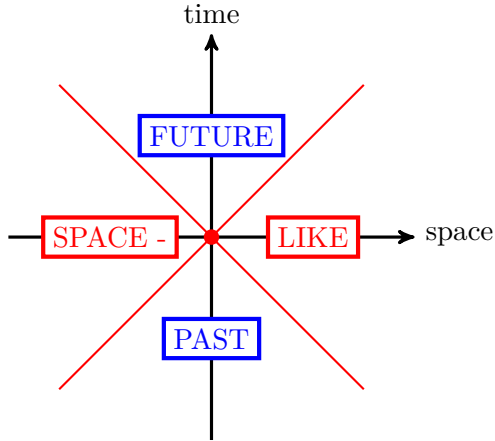
\begin{equation*}
 \begin{tikzpicture}[
    scale=4.5,
    axis/.style={very thick, ->, >=stealth'},
    pile/.style={thick, ->, >=stealth', shorten <=2pt, shorten
    >=2pt},
    every node/.style={color=black}
    ]
  \draw[axis] (-0.6,0) -- (+0.6,0) node(xline)[right]{space};
  \draw[axis] (0,-0.6) -- (0,+0.6) node(yline)[above]{time};
  \fill[red] (0,0) circle (.65pt);    
  \draw[-, thick, color=red] (-0.45,+0.45) -- (+0.45,-0.45);
  \draw[-, thick, color=red] (-0.45,-0.45) -- (+0.45,+0.45);
  \node[draw, ultra thick, align=center, color=blue, fill=white] at (0,+0.3) {FUTURE};
  \node[draw, ultra thick, align=center, color=blue, fill=white] at (0,-0.3) {PAST};
  \node[draw, ultra thick, align=center, color=red, fill=white] at (-0.3,0) {SPACE -};
  \node[draw, ultra thick, align=center, color=red, fill=white] at (+0.3,0) {LIKE}; 
 \end{tikzpicture} 
\end{equation*}

The space-time diagram is broken in three regions --  two of one type (with a well-defined causal relationship of either being in the ``{\it past}'' or ``{\it future}''), and one of another type (with no fixed causal relationship). This is also reflected into the structure of the propagator $G(\eta,\,\eta')\:=\:\vartheta(\eta-\eta')\mbox{exp}\{i(\eta-\eta')\}+\vartheta(\eta'-\eta)\mbox{exp}\{i(\eta'-\eta)\}-\mbox{exp}\{i(\eta+\eta')\}$, with the two terms with the positive sign being associated with the 2 time orderings, and the term with the negative sign with no time ordering. 

When two triangles $\triangle$, $\triangle'$ intersect, say on the $({\bf a}, {\bf b})$, $({\bf a'},{\bf b'})$ sides, then the relation ${\bf a}+{\bf b}\,=\,{\bf a'}+{\bf b'}$ is implied. Thus, we can also characterize the geometry as being given by a collection of $3E$ vectors $({\bf a}_i,\,{\bf b}_i,\,{\bf c}_i)$, where for pairs $(i,j)$ we have either the relations between $({\bf a}_i,\,{\bf b}_i,\,{\bf c}_i)$ and $({\bf a}_j,\,{\bf b}_j,\,{\bf c}_j)$, or relations of the form ${\bf a}_i+{\bf b}_i\,=\,{\bf a}_j+{\bf b}_j$ and/or ${\bf a}_i+{\bf c}_i\,=\,{\bf a}_j+{c\bf }_j$, but never ${\bf b}_i+{\bf c}_i\,=\,{\bf b}_j+{\bf c}_j$. We require enough intersections for the collection of triangles to be connected. 

We then consider, projectively, the convex hull of all the vertices, {\it i.e.} the space of all $\mathcal{Y}$'s of the form
\begin{equation}\label{eq:Es}
 \mathcal{Y}\:=\:\alpha_a {\bf a}_a+\beta_a {\bf b}_a + \gamma_a {\bf c}_a
\end{equation}
with $\alpha_a,\,\beta_a,\,\gamma_a\,\ge\,0$. This defines a ``{\it Cosmological Polytope}''.

Let us illustrate some of the simplest examples. The most trivial one is a single triangle with no intersections. Let us consider the case of two triangles. We need at least a single relation to make them connected, which we can take to be ${\bf a}_1+{\bf b}_1\,=\,{\bf a}_2+{\bf b}_2$. We start from a $2\times3\,=\,6$-dimensional space. With this relation we have a five-dimensional space, so the projective polytope is $(5-1)$-dimensional and lives in $\mathbb{P}^4$. We can nearly visualize this polytope:
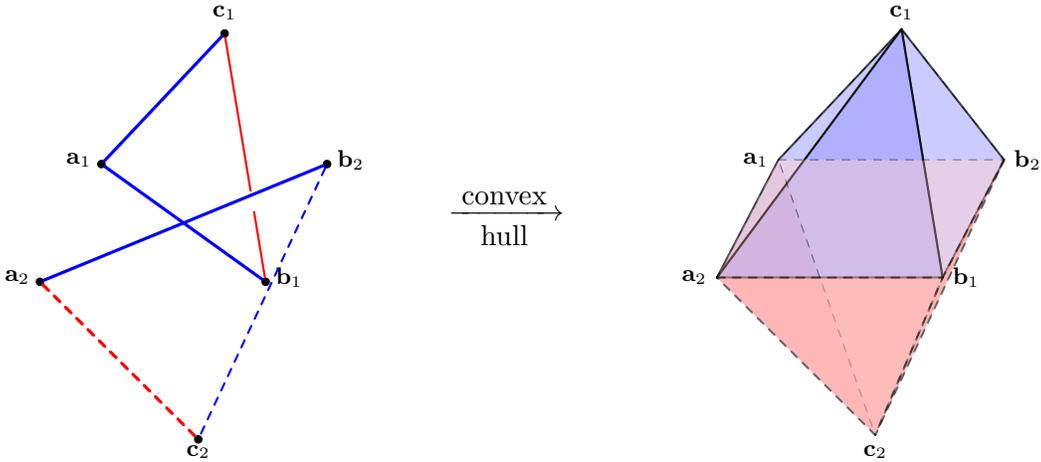
\begin{equation*}
 \begin{tikzpicture}[line join = round, line cap = round, ball/.style = {circle, draw, align=center, anchor=north, inner sep=0}]
  \begin{scope}
   \pgfmathsetmacro{\factor}{1/sqrt(2)};  
   \coordinate [label=right:{\footnotesize ${\bf b}_2$}] (B2) at (1.5,-3,-1.5*\factor);
   \coordinate [label=left:{\footnotesize ${\bf a}_1$}] (A1) at (-1.5,-3,-1.5*\factor);
   \coordinate [label=right:{\footnotesize ${\bf b}_1$}] (B1) at (1.5,-3.75,1.5*\factor);
   \coordinate [label=left:{\footnotesize ${\bf a}_2$}] (A2) at (-1.5,-3.75,1.5*\factor);  
   \coordinate [label=above:{\footnotesize ${\bf c}_1$}] (C1) at (0.75,-.65,.75*\factor);
   \coordinate [label=below:{\footnotesize ${\bf c}_2$}] (C2) at (0.4,-6.05,.75*\factor);
   \coordinate (Int) at (intersection of A2--B2 and B1--C1);
   \coordinate (Int2) at (intersection of A1--B1 and A2--B2);

   \coordinate [label=right:{\footnotesize ${\bf b}_2$}] (B2c) at (10.5,-3,-1.5*\factor);
   \coordinate [label=left:{\footnotesize ${\bf a}_1$}] (A1c) at (7.5,-3,-1.5*\factor);
   \coordinate [label=right:{\footnotesize ${\bf b}_1$}] (B1c) at (10.5,-3.75,1.5*\factor);
   \coordinate [label=left:{\footnotesize ${\bf a}_2$}] (A2c) at (7.5,-3.75,1.5*\factor);  
   \coordinate [label=above:{\footnotesize ${\bf c}_1$}] (C1c) at (9.75,-.65,.75*\factor);
   \coordinate [label=below:{\footnotesize ${\bf c}_2$}] (C2c) at (9.4,-6.05,.75*\factor);
   \coordinate (Int3) at (intersection of A2c--B2c and B1c--C1c);

   \tikzstyle{interrupt}=[
    postaction={
        decorate,
        decoration={markings,
                    mark= at position 0.5 
                          with
                          {
                            \node[rectangle, color=white, fill=white, below=.0006 of Int] {};
                          }}}
   ]

   \node[ball,text width=.1cm,fill,color=black] at (A1) (A1b) {};
   \node[ball,text width=.1cm,fill,color=black] at (B2) (B2b) {};
   \node[ball,text width=.1cm,fill,color=black] at (B1) (B1b) {};
   \node[ball,text width=.1cm,fill,color=black] at (A2) (A2b) {};
   \node[ball,text width=.1cm,fill,color=black] at (C1) (C1b) {};
   \node[ball,text width=.1cm,fill,color=black] at (C2) (C2b) {};

   \draw[interrupt,thick,color=red] (B1b) -- (C1b);
   \draw[-,very thick,color=blue] (A1b) -- (B1b);
   \draw[-,very thick,color=blue] (A2b) -- (B2b);
   \draw[-,very thick,color=blue] (A1b) -- (C1b);
   \draw[-, dashed, very thick, color=red] (A2b) -- (C2b);
   \draw[-, dashed, thick, color=blue] (B2b) -- (C2b);

   \node[thick, right=3cm of Int2] (arr) {$\displaystyle\quad\xrightarrow[\mbox{hull}]{\;\mbox{convex}\;}\quad$}; 

   \node at (A1c) (A1d) {};
   \node at (B2c) (B2d) {};
   \node at (B1c) (B1d) {};
   \node at (A2c) (A2d) {};
   \node at (C1c) (C1d) {};
   \node at (C2c) (C2d) {};

   \draw[-,dashed,fill=blue!30, opacity=.7] (A1c) -- (B2c) -- (C1c) -- cycle;
   \draw[-,thick,fill=blue!20, opacity=.7] (A1c) -- (A2c) -- (C1c) -- cycle;
   \draw[-,thick,fill=blue!20, opacity=.7] (B1c) -- (B2c) -- (C1c) -- cycle;
   \draw[-,thick,fill=blue!35, opacity=.7] (A2c) -- (B1c) -- (C1c) -- cycle;

   \draw[-,dashed,fill=red!30, opacity=.3] (A1c) -- (B2c) -- (C2c) -- cycle;
   \draw[-,dashed, thick, fill=red!50, opacity=.5] (B2c) -- (B1c) -- (C2c) -- cycle;
   \draw[-,dashed,fill=red!40, opacity=.3] (A1c) -- (A2c) -- (C2c) -- cycle;
   \draw[-,dashed, thick, fill=red!45, opacity=.5] (A2c) -- (B1c) -- (C2c) -- cycle;

  \end{scope}
 \end{tikzpicture}
\end{equation*}
This is a {\it double square pyramid}, with $({\bf a}_1,\,{\bf a}_2,\,{\bf b}_1,\,{\bf b}_2)$ the vertices of a plane quadrilateral, and two other vertices ${\bf c}_1$, ${\bf c}_2$ in a different third and fourth dimension respectively. Staying with two triangles, we can consider having both relations ${\bf a}_1+{\bf b}_1\,=\,{\bf a}_2+{\bf b}_2$ and ${\bf a}_1+{\bf c}_1\,=\,{\bf a}_2+{\bf c}_2$. The resulting polytope lives in $\mathbb{P}^3$ and can be directly visualized:
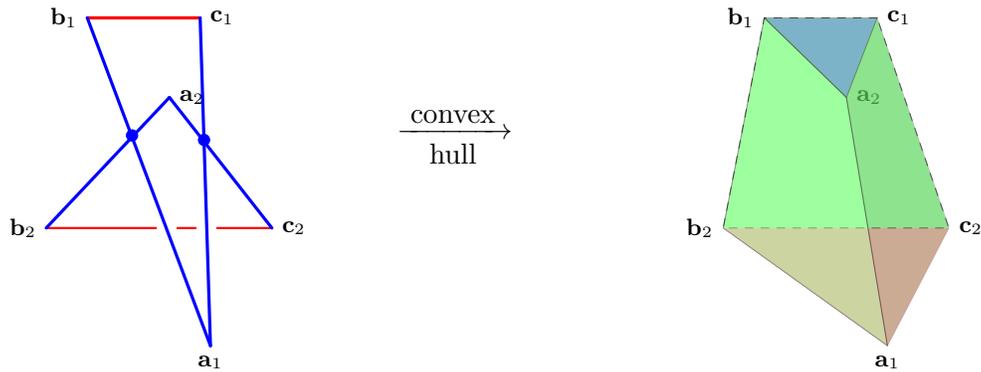
\begin{equation*}
\begin{tikzpicture}[line join = round, line cap = round, ball/.style = {circle, draw, align=center, anchor=north, inner sep=0}]
 \begin{scope}
  \pgfmathsetmacro{\factor}{1/sqrt(2)};
  \coordinate [label=right:{\footnotesize ${\bf c}_1$}] (c1b) at (0.75,0,-.75*\factor);
  \coordinate [label=left:{\footnotesize ${\bf b}_1$}] (b1b) at (-.75,0,-.75*\factor);
  \coordinate [label=right:{\footnotesize ${\bf a}_2$}] (a2b) at (0.75,-.65,.75*\factor);
  \coordinate [label=right:{\footnotesize ${\bf c}_1$}] (c1a) at (9.75,0,-.75*\factor);
  \coordinate [label=left:{\footnotesize ${\bf b}_1$}] (b1a) at (8.25,0,-.75*\factor);
  \coordinate [label=right:{\footnotesize ${\bf a}_2$}] (a2a) at (9.75,-.65,.75*\factor);
 
  \coordinate (Ab) at ($(B)!.5!(C)$); 
  \coordinate (Bb) at ($(A)!.5!(C)$);
  \coordinate (G) at (intersection of A--Ab and B--Bb);

  \coordinate [label=right:{\footnotesize ${\bf c}_2$}] (c2b) at (1.5,-3,-1.5*\factor);
  \coordinate [label=left:{\footnotesize ${\bf b}_2$}] (b2b) at (-1.5,-3,-1.5*\factor);
  \coordinate [label=below:{\footnotesize ${\bf a}_1$}] (a1b) at (1.5,-3.75,1.5*\factor);
  \coordinate [label=right:{\footnotesize ${\bf c}_2$}] (c2a) at (10.5,-3,-1.5*\factor);
  \coordinate [label=left:{\footnotesize ${\bf b}_2$}] (b2a) at (7.5,-3,-1.5*\factor);
  \coordinate [label=below:{\footnotesize ${\bf a}_1$}] (a1a) at (10.5,-3.75,1.5*\factor);

  \coordinate (Int1) at (intersection of b2b--c2b and b1b--a1b);
  \coordinate (Int2) at (intersection of b2b--c2b and c1b--a1b);
  \coordinate (Int3) at (intersection of b2b--a2b and b1b--a1b);
  \coordinate (Int4) at (intersection of a2b--c2b and c1b--a1b);
   \tikzstyle{interrupt}=[
    postaction={
        decorate,
        decoration={markings,
                    mark= at position 0.5 
                          with
                          {
                            \node[rectangle, color=white, fill=white] at (Int1) {};
                            \node[rectangle, color=white, fill=white] at (Int2) {};                            
                          }}}
   ]

   \node at (c1b) (c1c) {};
   \node at (b1b) (b1c) {};
   \node at (a2b) (a2c) {};
   \node at (c2b) (c2c) {};
   \node at (b2b) (b2c) {};
   \node at (a1b) (a1c) {};

   \draw[interrupt,thick,color=red] (b2b) -- (c2b);
   \draw[-,very thick,color=red] (b1b) -- (c1b);
   \draw[-,very thick,color=blue] (b1b) -- (a1b);
   \draw[-,very thick,color=blue] (a1b) -- (c1b);   
   \draw[-,very thick,color=blue] (b2b) -- (a2b);
   \draw[-,very thick,color=blue] (a2b) -- (c2b);

   \node[ball,text width=.15cm,fill,color=blue, above=-.06cm of Int3] (Inta) {};
   \node[ball,text width=.15cm,fill,color=blue, above=-.06cm of Int4] (Intb) {};

   \node[thick, right=3cm of Int3] (arr) {$\displaystyle\quad\xrightarrow[\mbox{hull}]{\;\mbox{convex}\;}\quad$};       


  \draw[-,dashed,fill=green!50,opacity=.6] (c1a) -- (b1a) -- (b2a) -- (c2a) -- cycle;
  \draw[draw=none,fill=red!60, opacity=.45] (c2a) -- (b2a) -- (a1a) -- cycle;
  \draw[-,fill=blue!,opacity=.3] (c1a) -- (b1a) -- (a2a) -- cycle; 
  \draw[-,fill=green!50,opacity=.4] (b1a) -- (a2a) -- (a1a) -- (b2a) -- cycle;
  \draw[-,fill=green!45!black,opacity=.2] (c1a) -- (a2a) -- (a1a) -- (c2a) -- cycle; 
 \end{scope}
 \end{tikzpicture}
\end{equation*}
The cosmological polytope is a truncated tetrahedron in this case.

We can characterize all the cosmological polytopes more systematically by introducing some graphical notation. We will associate each triangle $\triangle$ with a line segment and its two endpoints
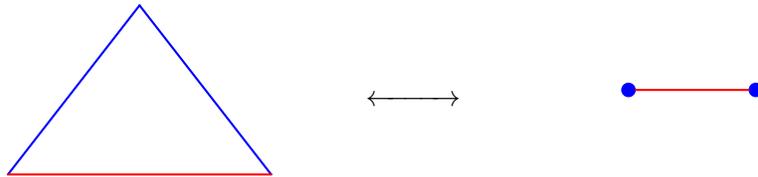
\begin{equation*}
 \begin{tikzpicture}[shift={(1,0)}, line join = round, line cap = round, ball/.style = {circle, draw, align=center, anchor=north, inner sep=0}]
  \coordinate (A) at (0,0);
  \coordinate (B) at (-1.75,-2.25);
  \coordinate (C) at (+1.75,-2.25);
  \coordinate  (m2) at ($(A)!0.5!(C)$);
  
  \draw[-, thick, color=blue] (B) -- (A); 
  \draw[-, thick, color=blue] (A) -- (C); 
  \draw[-, thick, color=red] (B) -- (C);    


   \node[right=2cm of m2.east] (lra) {$\xleftrightarrow{\phantom{convext}}$};
   \node[ball,text width=.18cm,fill,color=blue,right=2cm of lra.east] (x1) {};
    \node[ball,text width=.18cm,fill,color=blue,right=1.5cm of x1.east] (x2) {};
    \draw[-,thick,color=red] (x1.east) -- (x2.west);
   \end{tikzpicture}
\end{equation*}

The two endpoints represent the two $S_j^{\mbox{\tiny $(f)$}}$ edges of the triangle. Thus, if two triangles intersect on the midpoint of some edge, the line segments share a common endpoint. Thus, our two examples with two triangles are associated with the graphs
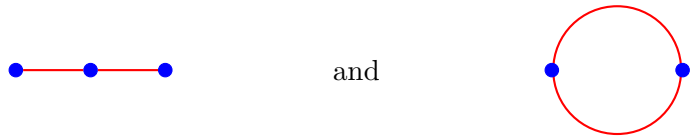
\begin{equation*}
 \begin{tikzpicture}[ball/.style = {circle, draw, align=center, anchor=north, inner sep=0}]
    \node[ball,text width=.18cm,fill,color=blue] at (0,0) (x1) {};
    \node[ball,text width=.18cm,fill,color=blue,right=.8cm of x1.east] (x2) {};
    \node[ball,text width=.18cm,fill,color=blue,right=.8cm of x2.east] (x3) {};
    \draw[-,thick,color=red] (x1.east) -- (x2.west);
    \draw[-,thick,color=red] (x2.east) -- (x3.west);
  
    \node[right=2cm of x3.east] (c) {and};
    \node[right=2cm of c.east] (n1) {};
   \node[right=1.5cm of n1.east] (n2)  {};
   \draw[thick, color=red] ($(n1)!0.5!(n2)$) circle (.85);
   \node[ball,text width=.18cm,fill,color=blue, right=2.055cm of c.east] (x1) {};
   \node[ball,text width=.18cm,fill,color=blue,right=1.545cm of x1.east] (x2) {};
 \end{tikzpicture}
\end{equation*}

More generally, it is clear that every cosmological polytope is associated with a connected graph $\mathcal{G}$, and vice-versa. We simply begin with a collection of line segments and glue them together in any way we like to get a connected graph.  Some examples starting with four triangles are
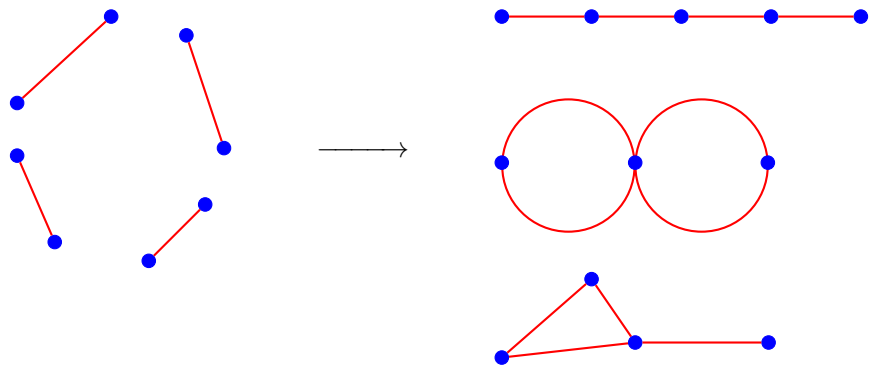
\begin{equation*}
 \begin{tikzpicture}[ball/.style = {circle, draw, align=center, anchor=north, inner sep=0}]
  \node[ball,text width=.18cm,fill,color=blue] at (-.5,0) (x1) {};
  \node[ball,text width=.18cm,fill,color=blue] at (-1.75,-1.15) (x2) {};
  \node[ball,text width=.18cm,fill,color=blue] at (-1.75,-1.85) (x3) {};
  \node[ball,text width=.18cm,fill,color=blue] at (.5,-.25) (x4) {};
  \node[ball,text width=.18cm,fill,color=blue] at (1,-1.75) (x5) {};
  \node[ball,text width=.18cm,fill,color=blue] at (-1.25,-3) (x6) {};
  \node[ball,text width=.18cm,fill,color=blue] at (.75,-2.5) (x7) {};
  \node[ball,text width=.18cm,fill,color=blue] at (0,-3.25) (x8) {};

  \draw[-,thick,color=red] (x1) -- (x2);
  \draw[-,thick,color=red] (x4) -- (x5);
  \draw[-,thick,color=red] (x3) -- (x6);
  \draw[-,thick,color=red] (x7) -- (x8);

  \node[right=1cm of x5.east] (arr) {$\displaystyle \xrightarrow{\hspace{1cm}}$};

  \node[ball,text width=.18cm,fill,color=blue,right=5cm of x1.east] (xa) {};
  \node[ball,text width=.18cm,fill,color=blue,right=1cm of xa.east] (xb) {};
  \node[ball,text width=.18cm,fill,color=blue,right=1cm of xb.east] (xc) {};
  \node[ball,text width=.18cm,fill,color=blue,right=1cm of xc.east] (xd) {};
  \node[ball,text width=.18cm,fill,color=blue,right=1cm of xd.east] (xe) {};  
  \draw[-,thick,color=red] (xa.east) -- (xb.west);
  \draw[-,thick,color=red] (xb.east) -- (xc.west);
  \draw[-,thick,color=red] (xc.east) -- (xd.west);
  \draw[-,thick,color=red] (xd.east) -- (xe.west);
 
  \node[below=1.75cm of xa] (t1) {};
  \node[right=1.5cm of t1] (t2) {};
  \node[right=1.5cm of t2] (t3) {};

   \draw[thick, color=red] ($(t1)!0.5!(t2)$) circle (.88);  
   \draw[thick, color=red] ($(t2)!0.5!(t3)$) circle (.88);  

  \node[ball,text width=.18cm,fill,color=blue,below=1.75cm of xa] (xf) {};
  \node[ball,text width=.18cm,fill,color=blue,right=1.58cm of xf.east] (xg) {};
  \node[ball,text width=.18cm,fill,color=blue,right=1.57cm of xg.east] (xh) {};

  \node[ball,text width=.18cm,fill,color=blue,below=3.3cm of xb] (xi) {};  
  \node[ball,text width=.18cm,fill,color=blue,below=2.4cm of xf] (xj) {};
  \node[ball,text width=.18cm,fill,color=blue,below=2.2cm of xg] (xk) {};
  \node[ball,text width=.18cm,fill,color=blue,right=1.58cm of xk.east] (xl) {};

  \draw[-,thick,color=red] (xi) -- (xj) -- (xk) -- (xi);
  \draw[-,thick,color=red] (xk) -- (xl);

 \end{tikzpicture}  
\end{equation*}

In our description so far we are still imposing relations on the vertices $(a_i\,b_i,\,c_i)$ of the triangles $\triangle_i$; It is also convenient to label the triangle non-redundantly by the midpoints of their sides. This amounts to labelling the graph $\mathcal{G}$ with vectors $x_v$ for each vertex and $y_e$ for each edge. Each edge of the graph is thus associated with three vertices of the polytope as:
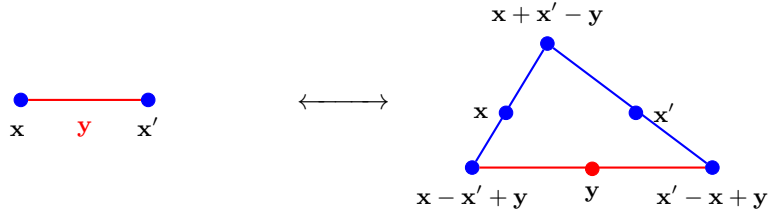
\begin{equation*}
 \begin{tikzpicture}[ball/.style = {circle, draw, align=center, anchor=north, inner sep=0}]
  \node[ball,text width=.18cm,fill,color=blue,label=below:{\footnotesize ${\bf x\phantom{'}}$}] at (0,0) (x1) {};
  \node[ball,text width=.18cm,fill,color=blue,right=1.5cm of x1.east,label=below:{\footnotesize ${\bf x'}$}] (x2) {};
  \draw[-,thick,color=red] (x1) edge node[midway, color=red,label=below:{\footnotesize ${\bf y}$}] {} (x2);

  \node[right=1.75cm of x2.east] (t1) {$\displaystyle \xleftrightarrow{\hspace{1cm}}$};

  \node[ball,text width=.18cm,fill,color=blue,label=above:{\footnotesize${\bf x}+{\bf x'}-{\bf y}$}] at (7,.75) (x3) {};
  \node[ball,text width=.18cm,fill,color=blue,label=below:{\footnotesize${\bf x}-{\bf x'}+{\bf y}$}] at (6,-.9) (x4) {};

  \node[ball,text width=.18cm,fill,color=blue,label=below:{\footnotesize${\bf x'}-{\bf x}+{\bf y}$},right=3cm of x4.east] (x5) {};

  \draw[-,thick,color=blue] (x3) -- (x4);
  \draw[-,thick,color=blue] (x3) -- (x5);
  \draw[-,thick,color=red] (x4) -- (x5);

  \node[ball,text width=.18cm,fill,color=blue,label=left:{\footnotesize${\bf x}$}] at ($(x3)!0.5!(x4)-(.05,0)$) {};
  \node[ball,text width=.18cm,fill,color=blue,label=right:{\footnotesize${\bf x'}$}] at ($(x3)!0.5!(x5)+(.08,0)$) {};
  \node[ball,text width=.18cm,fill,color=red,label=below:{\footnotesize${\bf y}$}] at ($(x4)!0.5!(x5)+(0,.08)$) {};
 \end{tikzpicture} 
\end{equation*}
In this way, with every graph $\mathcal{G}$ we have an associated cosmological polytope $\mathcal{P}_{\mbox{\tiny $\mathcal{G}$}}$. Thus, for instance, the polytope associated with
\begin{equation*}
 \begin{tikzpicture}[ball/.style = {circle, draw, align=center, anchor=north, inner sep=0}]
   \draw[thick,color=red] (.85,0) circle (.85);
   \node[ball,text width=.18cm,fill,color=blue] at (0,0) (x1) {};
   \node[left=.0cm of x1.west,label=below:${\bf x_1}$] (x1l) {};
   \node[ball,text width=.18cm,fill,color=blue,right=1.5cm of x1.east] (x2) {};
   \node[right=.0cm of x2.east,label=below:${\bf x_2}$] (x2l) {};
   \node at (.85,1) {$\displaystyle {\bf y_a}$};
   \node at (.85,-1) {$\displaystyle {\bf y_b}$};   
 \end{tikzpicture}
\end{equation*}
is the convex hull of the six points
\begin{equation}\label{6ptG}
 \left\{
  {\bf x}_1 + {\bf x}_2 - {\bf y}_a,\, {\bf x}_1 + {\bf y}_a - {\bf x}_2,\, {\bf x}_2 + {\bf y}_a - {\bf x}_1;  {\bf x}_1 + {\bf x}_2 - {\bf y}_b,\, {\bf x}_1 + {\bf y}_b - {\bf x}_2,\, {\bf x}_2 + {\bf y}_b-{\bf x}_1
 \right\},
\end{equation}
as seen in 
\begin{equation*}
\begin{tikzpicture}[line join = round, line cap = round, ball/.style = {circle, draw, align=center, anchor=north, inner sep=0}]
 \begin{scope}
  \pgfmathsetmacro{\factor}{1/sqrt(2)};
  \coordinate [label=right:{\footnotesize ${\bf x}_2 + {\bf y}_b - {\bf x}_1$}] (c1a) at (9.75,0,-.75*\factor);
  \coordinate [label=left:{\footnotesize ${\bf x}_1 + {\bf y}_b - {\bf x}_2$}] (b1a) at (8.25,0,-.75*\factor);
  \coordinate [label=right:{\footnotesize ${\bf x}_1 + {\bf x}_2 - {\bf y}_a$}] (a2a) at (9.75,-.65,.75*\factor);
 
  \coordinate (Ab) at ($(B)!.5!(C)$); 
  \coordinate (Bb) at ($(A)!.5!(C)$);
  \coordinate (G) at (intersection of A--Ab and B--Bb);

  \coordinate [label=right:{\footnotesize ${\bf x}_2 + {\bf y}_a - {\bf x}_1$}] (c2a) at (10.5,-3,-1.5*\factor);
  \coordinate [label=left:{\footnotesize ${\bf x}_1 + {\bf y}_a - {\bf x}_2$}] (b2a) at (7.5,-3,-1.5*\factor);
  \coordinate [label=below:{\footnotesize ${\bf x}_1 + {\bf x}_2 - {\bf y}_b$}] (a1a) at (10.5,-3.75,1.5*\factor);

  \draw[-,dashed,fill=green!50,opacity=.6] (c1a) -- (b1a) -- (b2a) -- (c2a) -- cycle;
  \draw[draw=none,fill=red!60, opacity=.45] (c2a) -- (b2a) -- (a1a) -- cycle;
  \draw[-,fill=blue!,opacity=.3] (c1a) -- (b1a) -- (a2a) -- cycle; 
  \draw[-,fill=green!50,opacity=.4] (b1a) -- (a2a) -- (a1a) -- (b2a) -- cycle;
  \draw[-,fill=green!45!black,opacity=.2] (c1a) -- (a2a) -- (a1a) -- (c2a) -- cycle; 
 \end{scope}                                              
 \end{tikzpicture}
\end{equation*}
while
$
 \begin{tikzpicture}[ball/.style = {circle, draw, align=center, anchor=north, inner sep=0}]
  \node[ball,text width=.18cm,fill,color=blue,label=above:{\tiny ${\bf x_1}$}] at (0,0) (x1) {};
  \node[ball,text width=.18cm,fill,color=blue,right=1.2cm of x1.east,label=above:{\tiny ${\bf x_2}$}] (x2) {};
  \node[ball,text width=.18cm,fill,color=blue,right=1.2cm of x2.east,label=above:{\tiny ${\bf x_3}$}] (x3) {};  
  \draw[-,red,thick] (x1) edge node[midway, color=red,label=above:{\tiny ${\bf y_1}$}] {} (x2);
  \draw[-,red,thick] (x2) edge node[midway, color=red,label=above:{\tiny ${\bf y_2}$}] {} (x3);
 \end{tikzpicture} 
$
is the convex hull of the six points 
\begin{equation}\label{6ptG2}
 \left\{
  {\bf x}_1 + {\bf x}_2 - {\bf y}_1,\, {\bf x}_1 + {\bf y}_1 - {\bf x}_2,\, {\bf x}_2 + {\bf y}_1 - {\bf x}_1;  {\bf x}_2 + {\bf x}_3 - {\bf y}_2,\, {\bf x}_2 + {\bf y}_2 - {\bf x}_3,\, {\bf x}_3 + {\bf y}_2 - {\bf x}_2
 \right\},
\end{equation}
as seen in 
\begin{equation*}
 \begin{tikzpicture}[line join = round, line cap = round, ball/.style = {circle, draw, align=center, anchor=north, inner sep=0}]
  \begin{scope}
   \pgfmathsetmacro{\factor}{1/sqrt(2)};  
   \coordinate [label=right:{\footnotesize ${\bf x}_2 + {\bf x}_3 - {\bf y}_2$}] (B2c) at (10.5,-3,-1.5*\factor);
   \coordinate [label=left:{\footnotesize ${\bf x}_2 + {\bf y}_1 - {\bf x}_1$}] (A1c) at (7.5,-3,-1.5*\factor);
   \coordinate [label=right:{\footnotesize ${\bf x}_1 + {\bf x}_2 - {\bf y}_1$}] (B1c) at (10.5,-3.75,1.5*\factor);
   \coordinate [label=left:{\footnotesize ${\bf x}_2 + {\bf y}_2 - {\bf x}_3$}] (A2c) at (7.5,-3.75,1.5*\factor);  
   \coordinate [label=above:{\footnotesize ${\bf x}_1 + {\bf y}_1 - {\bf x}_2$}] (C1c) at (9.75,-.65,.75*\factor);
   \coordinate [label=below:{\footnotesize ${\bf x}_3 + {\bf y}_2 - {\bf x}_2$}] (C2c) at (9.4,-6.05,.75*\factor);

   \node at (A1c) (A1d) {};
   \node at (B2c) (B2d) {};
   \node at (B1c) (B1d) {};
   \node at (A2c) (A2d) {};
   \node at (C1c) (C1d) {};
   \node at (C2c) (C2d) {};

   \draw[-,dashed,fill=blue!30, opacity=.7] (A1c) -- (B2c) -- (C1c) -- cycle;
   \draw[-,thick,fill=blue!20, opacity=.7] (A1c) -- (A2c) -- (C1c) -- cycle;
   \draw[-,thick,fill=blue!20, opacity=.7] (B1c) -- (B2c) -- (C1c) -- cycle;
   \draw[-,thick,fill=blue!35, opacity=.7] (A2c) -- (B1c) -- (C1c) -- cycle;

   \draw[-,dashed,fill=red!30, opacity=.3] (A1c) -- (B2c) -- (C2c) -- cycle;
   \draw[-,dashed,thick,fill=red!50, opacity=.5] (B2c) -- (B1c) -- (C2c) -- cycle;
   \draw[-,dashed,fill=red!40, opacity=.3] (A1c) -- (A2c) -- (C2c) -- cycle;
   \draw[-,dashed,thick,fill=red!45, opacity=.5] (A2c) -- (B1c) -- (C2c) -- cycle;
  \end{scope}
 \end{tikzpicture}
\end{equation*}
Notice that this is a four-dimensional object, a ``double square pyramid'', which we have represented in three dimensions as a square in two dimensions, with the addition of two points in the third and fourth dimensions.

For a general graph with $E$ edges and $V$ vertices, we always have a projective polytope living in $\mathbb{P}^{E+V-1}$, with each edge of the graph associated with three vertices of the polytope,  for of a total of $3E$ vertices. It is interesting that the association of the cosmological polytope with a graph arises in such a simple way. Note however that we are let to an ``edge-centered'' description of the graph -- where the edges are glued together at their endpoints to yield a graph -- rather than the ``vertex-centered'' description with vertices are given and connected to each other by edge to yield the graph. 

We stress that none of the vertices $(x_v,\,y_e)$ are vertices of the cosmological polytope. Instead the $(x_v,\,y_e)$ form a natural basis of $(E,\,V)$ vertices of the $\mathbb{P}^{E+V-1}$ space the polytope lives in. Any point in this space can be written as 
\begin{equation}\label{PPP}
 \mathcal{Y}\:=\:\sum_{v}x_v{\bf X}_v + y_e{\bf Y}_e.
\end{equation}
Thus, given a graph $\mathcal{G}$, there is an associated cosmological polytope $\mathcal{P}_{\mathcal{G}}$, which is further associated with a canonical form $\Omega(\mathcal{Y},\,\mathcal{P}_{\mathcal{G}})$, which has logarithmic singularities on (and only on) all the faces of $\mathcal{P}_{\mathcal{G}}$. This gives us the connection between the cosmological polytope and the wavefunction of the universe for a toy scalar model; The part of the wavefunction associated with the graph $\mathcal{G}$, $\Psi_{\mathcal{G}}(x_v,\,y_e)$ is identified with the canonical form
\begin{equation}\label{CanForm}
 \Omega(\mathcal{Y};\,\mathcal{P}_{\mathcal{G}})\:=\:
  \left(
   \prod_{v,e}dx_v dy_e
  \right)
  \Psi_{\mathcal{G}}(x_v,\,y_e).
\end{equation}


\subsection{Faces of Cosmological Polytopes}\label{subsec:FCP}

We have defined the cosmological polytope as the convex hull of a collection of vertices. We would now like to go to the opposite extreme and find all the co-dimension one faces of the polytope -- or, what is the same, determine the vertices of all the dual cosmological polytope $\tilde{\mathcal{P}}$. For a general polytope, there is a systematic but (NP) hard problem to determine the faces, given the vertices. Happily, we will see that this is a completely solvable problem for cosmological polytopes.

Given a collection of vertices ${\bf V}_a^{I}$ (for $a\,=\,1,\ldots,\,3E$ in the case of the cosmological polytope), which are vectors in $\mathbb{R}^{E+V}$ (or points in $\mathbb{P}^{E+V-1}$), we would like to find faces $\mathcal{W}_I$, which are co-vectors in $\mathbb{R}^{E+V}$ (and correspond to hyperplanes in $\mathbb{P}^{E+V-1}$). We will do this by finding all the vertices on the faces. In other words, we will solve the linear equations $\mathcal{W}_I{\bf V}^I_a\,\ge\,0$, and find solutions where as many of the $\mathcal{W}_I{\bf V}^I_a$ can be set to zero as possible keeping all the rest positive. 

Now we have a basis for $\mathbb{R}^{E+V}$ given by the vector $x_v$ and $y_e$ associated with the vertices and edges of the graph; For convenience we will also introduce a basis of co-vectors ${\bf \tilde{X}}_{v_I}$, ${\bf \tilde{Y}}_{e_I}$ with the defining properties that $(\tilde{x}_v\cdot x_{v'})\,=\,\delta_{v\,v'}$, $(\tilde{x}_e\cdot x_{e'})\,=\,\delta_{e\,e'}$, while $\tilde{x}_v\cdot y_e\,=\,0$, $\tilde{y}_v\cdot x_e\,=\,0$. We can thus expand
\begin{equation}\label{eq:WIexp}
 \mathcal{W}_I\:=\:\tilde{x}_v{\bf \tilde{X}}_{v_I}+\tilde{y}_e{\bf \tilde{Y}}_{e_I}.
\end{equation}
Now, for every edge $e$ connecting $v$ and $v'$ in the graph, we have $3$ vertices of the cosmological polytope $\{{\bf x}_v+{\bf x}_{v'}-{\bf y}_e,\,{\bf x}_v+{\bf y}_e-{\bf x}_{v'},\,{\bf x}_{v'}+{\bf y}_e-{\bf x}_v\}$ and thus three equations for $\mathcal{W}_I{\bf V}^I_a\,\ge\,0$:
\begin{equation}\label{eq:eqsWV}
 \tilde{x}_v+\tilde{x}_{v'}-\tilde{y}_e\:\ge\:0,\qquad
 \tilde{x}_v+\tilde{y}_e-\tilde{x}_{v'}\:\ge\:0,\qquad
 \tilde{x}_{v'}+\tilde{y}_e-\tilde{x}_v\:\ge\:0.
\end{equation}
Now, mirroring the definition of the polytope itself, let us define for each edge $e$ connection $v$ and $v'$, the triple of variables
\begin{equation}\label{eq:varA}
 \alpha_{\mbox{\tiny $(e,e)$}}\:=\: \tilde{x}_v+\tilde{x}_{v'}-\tilde{y}_e,\qquad
 \alpha_{\mbox{\tiny $(e,v)$}}\:=\: \tilde{x}_v+\tilde{y}_e-\tilde{x}_{v'},\qquad
 \alpha_{\mbox{\tiny $(e,v')$}}\:=\: \tilde{x}_{v'}+\tilde{y}_e-\tilde{x}_v.
\end{equation}
So we are setting for all 
\begin{equation}\label{eq:Ain}
 \alpha_{\mbox{\tiny $(e,e)$}},\;\alpha_{\mbox{\tiny $(e,v)$}},\;\alpha_{\mbox{\tiny $(e,v')$}}\:\ge\:0
\end{equation}
while these $\alpha$ variables satisfy the same linear constraints that occurred in the definition of the polytope, {\it i.e.} for any two edges $e$, $e'$ meeting at a common vertex $v$ we have
\begin{equation}\label{eq:Aconst}
 \alpha_{\mbox{\tiny $(e,e)$}}\:+\:\alpha_{\mbox{\tiny $(e,v)$}}\:=\:\alpha_{\mbox{\tiny $(e',e')$}}\:+\:\alpha_{\mbox{\tiny $(e',v)$}}.
\end{equation}
Now, if we send any of the $\alpha$'s to zero, the corresponding vertex of the polytope is on the hyperplane given by $\mathcal{W}_I$. Obviously, we can't send all the $\alpha$'s to zero. The task of finding a face then reduces to the following problem. We have to set as many of the $\alpha$'s to zero as we can -- but not all of them -- compatible with the constraint \eqref{eq:Aconst}. In other words, we have to find a pattern of zeros for the $\alpha$'s, compatible with \eqref{eq:Aconst} with at least one non-zero $\alpha$, such that setting any further $\alpha$ to zero would, by compatibility with \eqref{eq:Aconst}, force them all to vanish. This will tell us which vertices lie on a face, and armed with this information we will easily be able to exhibit the $\mathcal{W}_I$ itself. 

Since we want to keep track of the $\alpha$'s that are non-zero, it is convenient to denote these graphically in an obvious way, by marking and edge $e$ as:
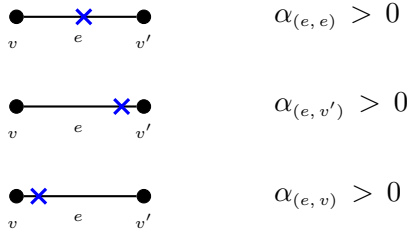
\begin{equation*}
 \begin{tikzpicture}[ball/.style = {circle, draw, align=center, anchor=north, inner sep=0}, cross/.style={cross out, draw, minimum size=2*(#1-\pgflinewidth), inner sep=0pt, outer sep=0pt}]
  \node[ball,text width=.18cm,fill,color=black,label=below:{\tiny $v\phantom{'}$}] at (0,0) (v1) {};
  \node[ball,text width=.18cm,fill,color=black,label=below:{\tiny $v'$},right=1.5cm of v1.east] (v2) {};  
  \draw[-,thick,color=black] (v1.east) edge node [text width=.18cm,below=.1cm,midway] {\tiny $e$} (v2.west);
  \node[very thick, cross=4pt, rotate=0, color=blue, right=.7cm of v1.east]{};
  \node[right=1.5cm of v2.east] (lb1) {$\alpha_{\mbox{\tiny $(e,e)$}}\,>\,0$};

  \node[ball,text width=.18cm,fill,color=black,label=below:{\tiny $v\phantom{'}$}, below=1cm of v1.south] (v3) {};
  \node[ball,text width=.18cm,fill,color=black,label=below:{\tiny $v'$},right=1.5cm of v3.east] (v4) {};  
  \draw[-,thick,color=black] (v3.east) edge node [text width=.18cm,below=.1cm,midway] {\tiny $e$} (v4.west);
  \node[very thick, cross=4pt, rotate=0, color=blue, left=.1cm of v4.west]{};
  \node[right=1.5cm of v4.east] (lb2) {$\alpha_{\mbox{\tiny $(e,v')$}}\,>\,0$};

  \node[ball,text width=.18cm,fill,color=black,label=below:{\tiny $v\phantom{'}$}, below=1cm of v3.south] (v5) {};
  \node[ball,text width=.18cm,fill,color=black,label=below:{\tiny $v'$},right=1.5cm of v5.east] (v6) {};  

  \draw[-,thick,color=black] (v5.east) edge node [text width=.18cm,below=.1cm,midway] {\tiny $e$} (v6.west);
  \node[very thick, cross=4pt, rotate=0, color=blue, right=.1cm of v5.east]{};

  \node[right=1.5cm of v6.east] (lb2) {$\alpha_{\mbox{\tiny $(e,v)$}}\,>\,0$};

 \end{tikzpicture}
\end{equation*}
while any unmarked edge has all the associated $\alpha$'s set to zero.

This notation also doubles to denote the vertices of the polytope that {\it do not} lie on $\mathcal{W}_I$; {\it e.g.} 
$
 \begin{tikzpicture}[ball/.style = {circle, draw, align=center, anchor=north, inner sep=0}, cross/.style={cross out, draw, minimum size=2*(#1-\pgflinewidth), inner sep=0pt, outer sep=0pt}] 
  \node[ball,text width=.18cm,fill,color=black,label=above:{\tiny $v\phantom{'}$}] at (0,0) (v3) {};
  \node[ball,text width=.18cm,fill,color=black,label=above:{\tiny $v'$},right=1.5cm of v3.east] (v4) {};  
  \draw[-,thick,color=black] (v3.east) edge node [text width=.18cm,above=-0.05cm,midway] {\tiny $e$} (v4.west);
  \node[very thick, cross=4pt, rotate=0, color=blue, left=.1cm of v4.west]{};
 \end{tikzpicture} 
$ 
means that $\alpha_{\mbox{\tiny $(e,v')$}}\,=\,\mathcal{W}\cdot({\bf x}_{v'}+{\bf y}_e-{\bf x}_v)\,>\,0$ and so $({\bf x}_{v'}+{\bf y}_e-{\bf x}_v)$ does {\it not} lie on $\mathcal{W}_I$.

Now, we want to put as many $\alpha$'s as we can (but not all) to zero, compatible with \eqref{eq:Aconst}. Given that the left-hand-side and right-hand-side of \eqref{eq:Aconst} are a sum of non-negative terms, if {\it e.g.} both of the terms on the left-hand-side are set to zero, then both terms on the right-hand-side must also vanish. This tells us that the following marking of the graph, zooming into a vertex, are not-allowed:
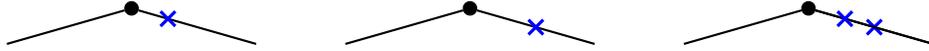
\begin{equation}\label{eq:AlConfs}
 \begin{tikzpicture}[ball/.style = {circle, draw, align=center, anchor=north, inner sep=0}, cross/.style={cross out, draw, minimum size=2*(#1-\pgflinewidth), inner sep=0pt, outer sep=0pt}]
  \node[ball,text width=.18cm,fill,color=black] at (0,0) (v1) {};  
  \node[ball,text width=.18cm,fill=white,color=white] at (-1.75,-.5) (v2) {};  
  \node[ball,text width=.18cm,fill=white,color=white] at (1.75,-.5) (v3) {};  
  \draw[-,color=black,thick] (v1) -- (v2);
  \draw[-,color=black,thick] (v1) edge node [very thick,cross=4pt,rotate=0,color=blue,near start]{} (v3);  
  \node[ball,text width=.18cm,fill,color=black] at (4.5,0) (v4) {};  
  \node[ball,text width=.18cm,fill=white,color=white] at (2.75,-.5) (v5) {};  
  \node[ball,text width=.18cm,fill=white,color=white] at (6.25,-.5) (v6) {};  
  \draw[-,color=black,thick] (v4) -- (v5);
  \draw[-,color=black,thick] (v4) edge node [very thick,cross=4pt,rotate=0,color=blue,midway]{} (v6);

  \node[ball,text width=.18cm,fill,color=black] at (9,0) (v7) {};  
  \node[ball,text width=.18cm,fill=white,color=white] at (7.25,-.5) (v8) {};  
  \node[ball,text width=.18cm,fill=white,color=white] at (10.75,-.5) (v9) {};  
  \draw[-,color=black,thick] (v7) -- (v8);
  \draw[-,color=black,thick] (v7) edge node [very thick,cross=4pt,rotate=0,color=blue,midway]{} (v9);
  \draw[-,color=black,thick] (v7) edge node [very thick,cross=4pt,rotate=0,color=blue,near start]{} (v9);  

  \node[draw, ultra thick, align=center, color=blue, fill=white, below=1.25cm of v4.south] {N $\:$ O $\:$ T $\qquad$ A $\:$ L $\:$ L $\:$ O $\:$ W $\:$ E $\:$ D};  
 \end{tikzpicture}  
\end{equation}
Since we want to find the $\mathcal{W}_I$ with as many vertices on it as possible, we are looking for  a marking of the graph which is allowed, {\it i.e.} not including any configuration that looks like \eqref{eq:AlConfs}, but with the further property that deleting any number of further markings would either remove all of them or give a marking which is not allowed. This rule allows us to easily determine all the faces of a cosmological polytope, but, before giving the general answer, let us give some illustrative examples. For the sake of concreteness, we will choose a graph $\mathcal{G}$ which is complicated enough to exhibit all the interesting features: 
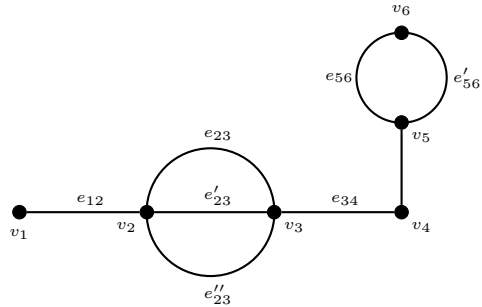
\begin{equation*}
 \begin{tikzpicture}[ball/.style = {circle, draw, align=center, anchor=north, inner sep=0}, cross/.style={cross out, draw, minimum size=2*(#1-\pgflinewidth), inner sep=0pt, outer sep=0pt}]
  \node[ball,text width=.18cm,fill,color=black,label=below:{\tiny $v_1$}] at (0,0) (v1) {};  
  \node[ball,text width=.18cm,fill,color=black,right=1.5cm of v1.east,label={[label distance=-.1cm]250:{\tiny $v_2$}}] (v2) {};      
  \node[ball,text width=.18cm,fill,color=black,right=1.5cm of v2.east,label={[label distance=-.1cm]320:{\tiny $v_3$}}] (v3) {};        
  \node[ball,text width=.18cm,fill,color=black,right=1.5cm of v3.east,label={[label distance=-.1cm]320:{\tiny $v_4$}}] (v4) {};        
  \node[ball,text width=.18cm,fill,color=black,above=1cm of v4.north,label={[label distance=-.1cm]320:{\tiny $v_5$}}] (v5) {};        
  \node[ball,text width=.18cm,fill,color=black,above=1cm of v5.north,label=above:{\tiny $v_6$}] (v6) {};          
  \draw[-,thick,color=black] (v1) edge node[text width=.18cm,above=-0.05cm,midway] {\tiny $e_{12}$} (v2);
  \draw[-,thick,color=black] (v2) edge node[text width=.18cm,above=-0.05cm,midway] {\tiny $e'_{23}$} (v3);  
  \draw[-,thick,color=black] (v3) edge node[text width=.18cm,above=-0.05cm,midway] {\tiny $e_{34}$} (v4);    
  \draw[-,thick,color=black] (v4) -- (v5);
  \draw[thick,color=black] ($(v2)!0.5!(v3)$) circle (.85cm);
  \draw[thick,color=black] ($(v5)!0.5!(v6)$) circle (.6cm);  

  \coordinate (t1) at ($(v2)!0.5!(v3)$);
  \node[color=black,text width=.18cm,above=.825cm of t1.north] {\tiny $e_{23}$}; 
  \node[color=black,text width=.18cm,below=.825cm of t1.south] {\tiny $e''_{23}$};  

  \coordinate (t2) at ($(v5)!0.5!(v6)$);
  \node[color=black,text width=.18cm,left=.7cm of t2.west] {\tiny $e_{56}$}; 
  \node[color=black,text width=.18cm,right=.55cm of t2.east] {\tiny $e'_{56}$};
 \end{tikzpicture}   
\end{equation*}
First, for any graph $\mathcal{G}$, we have a marking where the middle of each edge is marked, corresponding to $ \alpha_{\mbox{\tiny $(e,e)$}}\,>\,0$, with all the other $\alpha$'s set to zero.
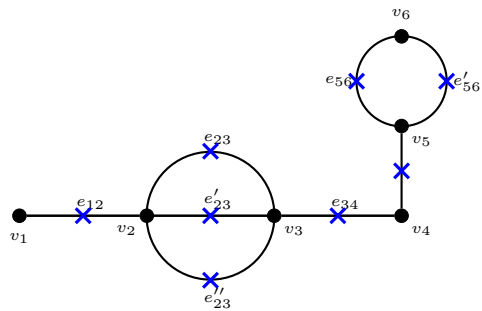
\begin{equation*}
 \begin{tikzpicture}[ball/.style = {circle, draw, align=center, anchor=north, inner sep=0}, cross/.style={cross out, draw, minimum size=2*(#1-\pgflinewidth), inner sep=0pt, outer sep=0pt}]
  \node[ball,text width=.18cm,fill,color=black,label=below:{\tiny $v_1$}] at (0,0) (v1) {};  
  \node[ball,text width=.18cm,fill,color=black,right=1.5cm of v1.east,label={[label distance=-.1cm]250:{\tiny $v_2$}}] (v2) {};      
  \node[ball,text width=.18cm,fill,color=black,right=1.5cm of v2.east,label={[label distance=-.1cm]320:{\tiny $v_3$}}] (v3) {};        
  \node[ball,text width=.18cm,fill,color=black,right=1.5cm of v3.east,label={[label distance=-.1cm]320:{\tiny $v_4$}}] (v4) {};        
  \node[ball,text width=.18cm,fill,color=black,above=1cm of v4.north,label={[label distance=-.1cm]320:{\tiny $v_5$}}] (v5) {};        
  \node[ball,text width=.18cm,fill,color=black,above=1cm of v5.north,label=above:{\tiny $v_6$}] (v6) {};          
  \draw[-,thick,color=black] (v1) edge node[text width=.18cm,above=-0.05cm,midway] {\tiny $e_{12}$} (v2);
  \draw[-,thick,color=black] (v1) edge node[very thick,cross=4pt,rotate=0,color=blue,midway] {} (v2);  
  \draw[-,thick,color=black] (v2) edge node[text width=.18cm,above=-0.05cm,midway] {\tiny $e'_{23}$} (v3);  
  \draw[-,thick,color=black] (v2) edge node[very thick,cross=4pt,rotate=0,color=blue,midway] {} (v3);    
  \draw[-,thick,color=black] (v3) edge node[text width=.18cm,above=-0.05cm,midway] {\tiny $e_{34}$} (v4);    
  \draw[-,thick,color=black] (v3) edge node[very thick,cross=4pt,rotate=0,color=blue,midway] {} (v4);      
  \draw[-,thick,color=black] (v4) edge node[very thick,cross=4pt,rotate=0,color=blue,midway] {} (v5);
  \draw[thick,color=black] ($(v2)!0.5!(v3)$) circle (.85cm);
  \draw[thick,color=black] ($(v5)!0.5!(v6)$) circle (.6cm);  

  \coordinate (t1) at ($(v2)!0.5!(v3)$);
  \node[color=black,text width=.18cm,above=.825cm of t1.north] {\tiny $e_{23}$}; 
  \node[very thick,cross=4pt,rotate=0,color=blue,above=.76cm of t1.north] {};
  \node[color=black,text width=.18cm,below=.825cm of t1.south] {\tiny $e''_{23}$};  
  \node[very thick,cross=4pt,rotate=0,color=blue,below=.76cm of t1.south] {};

  \coordinate (t2) at ($(v5)!0.5!(v6)$);
  \node[color=black,text width=.18cm,left=.7cm of t2.west] {\tiny $e_{56}$}; 
  \node[very thick,cross=4pt,rotate=0,color=blue,left=.5cm of t2.west] {};  
  \node[color=black,text width=.18cm,right=.55cm of t2.east] {\tiny $e'_{56}$};
  \node[very thick,cross=4pt,rotate=0,color=blue,right=.5cm of t2.east] {};    
 \end{tikzpicture}  
\end{equation*}
It is obvious that setting any further $\alpha$ to zero would give a not-allowed marking or, what is the same, would by \eqref{eq:Aconst} force all the $\alpha$'s to vanish. This tells us for every edge $e$ connecting $v$ and $v'$ in the graph, that the vertices $\{{\bf x}_v+{\bf y}_e-{\bf x}_{v'},\,{\bf x}_{v'}+{\bf y}_e-{\bf x}_v\}$ lie on this face. Since $2E\,\ge\,E+V$, there are enough vertices to form a co-dimension one face. Thus, in order to explicitly determine the $\mathcal{W}_I$ for this face, it suffices to exhibit a $\mathcal{W}_I$ such that $\mathcal{W}\,\cdot\,{\bf V}\,=\,0$ for all these $2E$ vertices ${\bf V}_I$. This is very easy to do, we simply have 
\begin{equation}\label{eq:WiXv}
 \mathcal{W}_I\:=\:\sum_v{\bf\tilde{X}}_{v_I},
\end{equation}
which obviously satisfies $\mathcal{W}\,\cdot\,\left({\bf x}_v+{\bf y}_e-{\bf x}_{v'}\right)\,=\,0$ .

Let us now consider an opposite extreme, where none of the middles of edges are marked. Here are some examples:
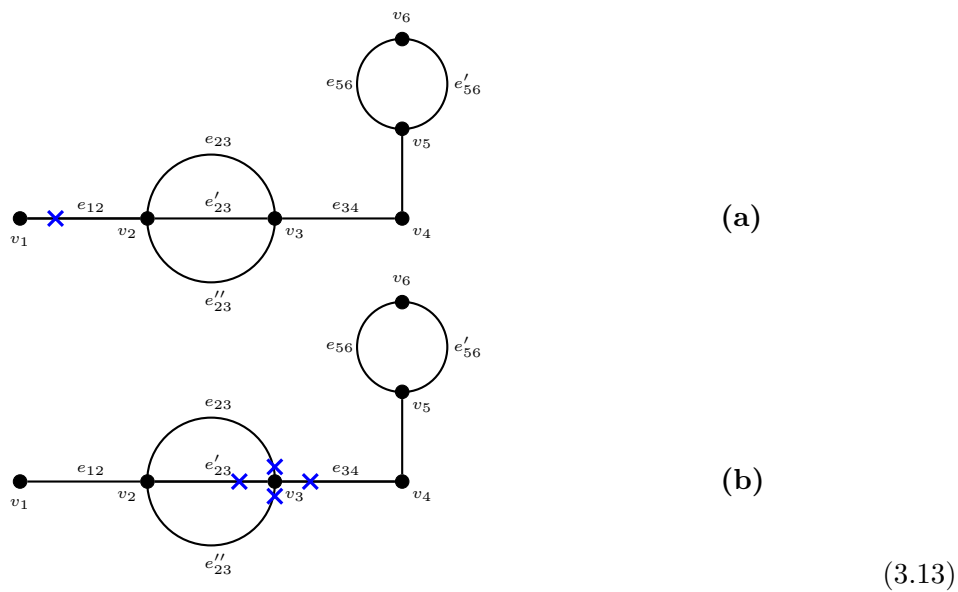
\begin{equation}
 \begin{tikzpicture}[ball/.style = {circle, draw, align=center, anchor=north, inner sep=0}, cross/.style={cross out, draw, minimum size=2*(#1-\pgflinewidth), inner sep=0pt, outer sep=0pt}]
  \node[ball,text width=.18cm,fill,color=black,label=below:{\tiny $v_1$}] at (0,0) (v1) {};  
  \node[ball,text width=.18cm,fill,color=black,right=1.5cm of v1.east,label={[label distance=-.1cm]250:{\tiny $v_2$}}] (v2) {};      
  \node[ball,text width=.18cm,fill,color=black,right=1.5cm of v2.east,label={[label distance=-.1cm]320:{\tiny $v_3$}}] (v3) {};        
  \node[ball,text width=.18cm,fill,color=black,right=1.5cm of v3.east,label={[label distance=-.1cm]320:{\tiny $v_4$}}] (v4) {};        
  \node[ball,text width=.18cm,fill,color=black,above=1cm of v4.north,label={[label distance=-.1cm]320:{\tiny $v_5$}}] (v5) {};        
  \node[ball,text width=.18cm,fill,color=black,above=1cm of v5.north,label=above:{\tiny $v_6$}] (v6) {};          
  \draw[-,thick,color=black] (v1) edge node[text width=.18cm,above=-0.05cm,midway] {\tiny $e_{12}$} (v2);
  \draw[-,thick,color=black] (v1) edge node[very thick,cross=4pt,rotate=0,color=blue,near start] {} (v2);    
  \draw[-,thick,color=black] (v2) edge node[text width=.18cm,above=-0.05cm,midway] {\tiny $e'_{23}$} (v3);  
  \draw[-,thick,color=black] (v3) edge node[text width=.18cm,above=-0.05cm,midway] {\tiny $e_{34}$} (v4);    
  \draw[-,thick,color=black] (v4) -- (v5);
  \draw[thick,color=black] ($(v2)!0.5!(v3)$) circle (.85cm);
  \draw[thick,color=black] ($(v5)!0.5!(v6)$) circle (.6cm);  

  \coordinate (t1) at ($(v2)!0.5!(v3)$);
  \node[color=black,text width=.18cm,above=.825cm of t1.north] {\tiny $e_{23}$}; 
  \node[color=black,text width=.18cm,below=.825cm of t1.south] {\tiny $e''_{23}$};  

  \coordinate (t2) at ($(v5)!0.5!(v6)$);
  \node[color=black,text width=.18cm,left=.7cm of t2.west] {\tiny $e_{56}$}; 
  \node[color=black,text width=.18cm,right=.55cm of t2.east] {\tiny $e'_{56}$};

  \node[text width=.18cm, color=black, right=4cm of v4.east] {\bf (a)};

  \node[ball,text width=.18cm,fill,color=black,label=below:{\tiny $v_1$}] at (0,-3.5) (x1) {};  
  \node[ball,text width=.18cm,fill,color=black,right=1.5cm of x1.east,label={[label distance=-.1cm]250:{\tiny $v_2$}}] (x2) {};      
  \node[ball,text width=.18cm,fill,color=black,right=1.5cm of x2.east,label={[label distance=-.1cm]320:{\tiny $v_3$}}] (x3) {};        
  \node[ball,text width=.18cm,fill,color=black,right=1.5cm of x3.east,label={[label distance=-.1cm]320:{\tiny $v_4$}}] (x4) {};        
  \node[ball,text width=.18cm,fill,color=black,above=1cm of x4.north,label={[label distance=-.1cm]320:{\tiny $v_5$}}] (x5) {};        
  \node[ball,text width=.18cm,fill,color=black,above=1cm of x5.north,label=above:{\tiny $v_6$}] (x6) {};          
  \draw[-,thick,color=black] (x1) edge node[text width=.18cm,above=-0.05cm,midway] {\tiny $e_{12}$} (x2);
  \draw[-,thick,color=black] (x2) edge node[text width=.18cm,above=-0.05cm,midway] {\tiny $e'_{23}$} (x3);  
  \draw[-,thick,color=black] (x2) edge node[very thick,cross=4pt,rotate=0,color=blue,near end] {} (x3);      
  \draw[-,thick,color=black] (x3) edge node[text width=.18cm,above=-0.05cm,midway] {\tiny $e_{34}$} (x4);    
  \draw[-,thick,color=black] (x3) edge node[very thick,cross=4pt,rotate=0,color=blue,near start] {} (x4);      
  \draw[-,thick,color=black] (x4) -- (x5);
  \draw[thick,color=black] ($(x2)!0.5!(x3)$) circle (.85cm);
  \node[very thick,cross=4pt,rotate=0,color=blue,above=.0cm of x3.north]{};
  \node[very thick,cross=4pt,rotate=0,color=blue,below=.0cm of x3.south]{};  
  \draw[thick,color=black] ($(x5)!0.5!(x6)$) circle (.6cm);  

  \coordinate (t3) at ($(x2)!0.5!(x3)$);
  \node[color=black,text width=.18cm,above=.825cm of t3.north] {\tiny $e_{23}$}; 
  \node[color=black,text width=.18cm,below=.825cm of t3.south] {\tiny $e''_{23}$};  

  \coordinate (t4) at ($(x5)!0.5!(x6)$);
  \node[color=black,text width=.18cm,left=.7cm of t4.west] {\tiny $e_{56}$}; 
  \node[color=black,text width=.18cm,right=.55cm of t4.east] {\tiny $e'_{56}$};

  \node[text width=.18cm, color=black, right=4cm of x4.east] {\bf (b)};
 \end{tikzpicture} 
\end{equation}

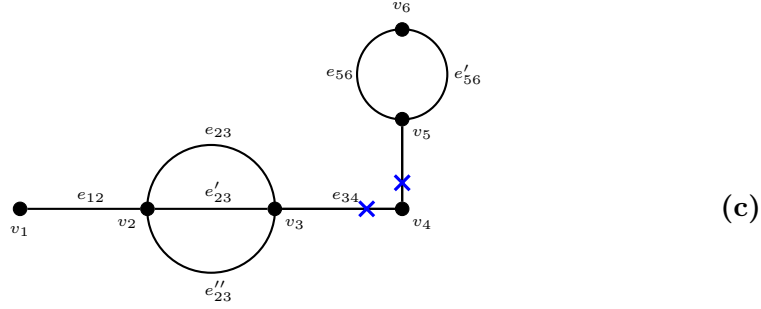
\begin{equation*}
 \begin{tikzpicture}[ball/.style = {circle, draw, align=center, anchor=north, inner sep=0}, cross/.style={cross out, draw, minimum size=2*(#1-\pgflinewidth), inner sep=0pt, outer sep=0pt}]
  \node[ball,text width=.18cm,fill,color=black,label=below:{\tiny $v_1$}] at (0,0) (z1) {};  
  \node[ball,text width=.18cm,fill,color=black,right=1.5cm of z1.east,label={[label distance=-.1cm]250:{\tiny $v_2$}}] (z2) {};      
  \node[ball,text width=.18cm,fill,color=black,right=1.5cm of z2.east,label={[label distance=-.1cm]320:{\tiny $v_3$}}] (z3) {};        
  \node[ball,text width=.18cm,fill,color=black,right=1.5cm of z3.east,label={[label distance=-.1cm]320:{\tiny $v_4$}}] (z4) {};        
  \node[ball,text width=.18cm,fill,color=black,above=1cm of z4.north,label={[label distance=-.1cm]320:{\tiny $v_5$}}] (z5) {};        
  \node[ball,text width=.18cm,fill,color=black,above=1cm of z5.north,label=above:{\tiny $v_6$}] (z6) {};          
  \draw[-,thick,color=black] (z1) edge node[text width=.18cm,above=-0.05cm,midway] {\tiny $e_{12}$} (z2);
  \draw[-,thick,color=black] (z2) edge node[text width=.18cm,above=-0.05cm,midway] {\tiny $e'_{23}$} (z3);  
  \draw[-,thick,color=black] (z3) edge node[text width=.18cm,above=-0.05cm,midway] {\tiny $e_{34}$} (z4);    
  \draw[-,thick,color=black] (z3) edge node[very thick,cross=4pt,rotate=0,color=blue,near end] {} (z4);      
  \draw[-,thick,color=black] (z4) -- (z5);
  \draw[-,thick,color=black] (z4) edge node[very thick,cross=4pt,rotate=0,color=blue,near start] {} (z5);
  \draw[thick,color=black] ($(z2)!0.5!(z3)$) circle (.85cm);
  \draw[thick,color=black] ($(z5)!0.5!(z6)$) circle (.6cm);  
  \coordinate (t5) at ($(z2)!0.5!(z3)$);
  \node[color=black,text width=.18cm,above=.825cm of t5.north] {\tiny $e_{23}$}; 
  \node[color=black,text width=.18cm,below=.825cm of t5.south] {\tiny $e''_{23}$};  

  \coordinate (t6) at ($(z5)!0.5!(z6)$);
  \node[color=black,text width=.18cm,left=.7cm of t6.west] {\tiny $e_{56}$}; 
  \node[color=black,text width=.18cm,right=.55cm of t6.east] {\tiny $e'_{56}$};

  \node[text width=.18cm, color=black, right=4cm of z4.east] {\bf (c)};
 \end{tikzpicture} 
\end{equation*}
Again clearly we cannot erase any marking without getting a not-allowed configuration. It is clear that if there are to be no marked middles, then all the edges $e$ ending on some vertex $v$ must be marked next to $v$.  What is the $\mathcal{W}_I$ corresponding to this face? Let us take the graph (3.13b) as an example. Clearly
\begin{equation}\label{eq:WiMarkB}
 \mathcal{W}_I\:=\:{\bf \tilde{X}}_{v_3}+{\bf\tilde{Y}}_{e_{23}}+{\bf\tilde{Y}}_{e'_{23}}+{\bf\tilde{Y}}_{e''_{23}}+{\bf\tilde{Y}}_{e_{34}}
\end{equation}
satisfies that $(\mathcal{W}\cdot {\bf V})\,=\,0$ for all the unmarked vertices of the polytope. This is trivial for all the polytope vertices associated with edges other than $e_{23}$, $e'_{23}$, $e''_{23}$, $e_{34}$, while {\it e.g.} for the two marked vertices associated with $e_{23}$, $\mathcal{W}\,\cdot\,\left({\bf x}_{v_2}+{\bf y}_{e_{23}}-{\bf x}_{v_3}\right)\,=\,0$ and $\mathcal{W}\,\cdot\,\left({\bf x}_{v_4}-{\bf y}_{e_{23}}+{\bf x}_{v_2}\right)\,=\,0$. Obviously, this also generalizes to any graph: there is a face where all the edges $e$ ending on $v$ are marked next to $v$, and the corresponding $\mathcal{W}_I$ is
\begin{equation}\label{eq:WiF}
 \mathcal{W}_I\:=\:{\bf\tilde{X}}_{v_I}+\sum_{\substack{e\,\mbox{\tiny ending} \\ \mbox{\tiny on } v}}{\bf \tilde{Y}}_{e_I}.
\end{equation}
The general rule for all the faces interpolates between the two cases we have looked at above. A consistent marking is associated with any set of vertices of $\mathcal{G}$ that forms a connected subgraph $\mathfrak{g}$, with any number of external edges attached to the vertices of $\mathfrak{g}$. The edges internal to $\mathfrak{g}$ are marked on their middles, while the external edges ending on vertices $v$ in $\mathfrak{g}$ are marked next to the $v$. Taking our representative graph, the examples we have looked at already correspond to the two extremes where $\mathfrak{g}$ consists of the entire graph $\mathcal{G}$ or a single vertex $v$:
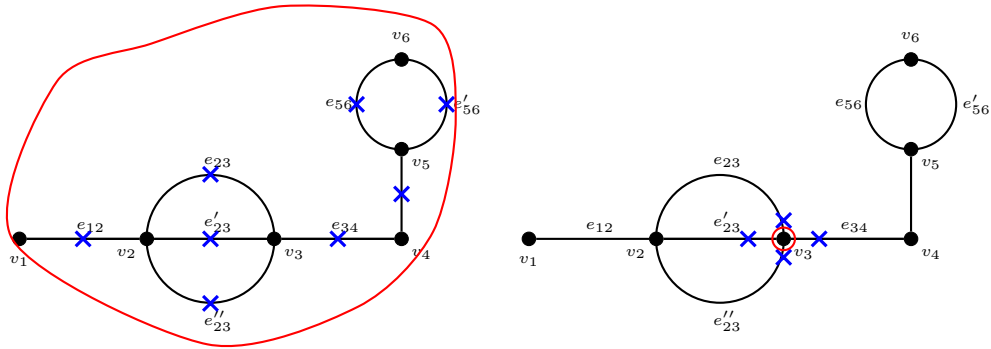
\begin{equation*}
 \begin{tikzpicture}[ball/.style = {circle, draw, align=center, anchor=north, inner sep=0}, cross/.style={cross out, draw, minimum size=2*(#1-\pgflinewidth), inner sep=0pt, outer sep=0pt}]
  \node[ball,text width=.18cm,fill,color=black,label=below:{\tiny $v_1$}] at (0,0) (v1) {};  
  \node[ball,text width=.18cm,fill,color=black,right=1.5cm of v1.east,label={[label distance=-.1cm]250:{\tiny $v_2$}}] (v2) {};      
  \node[ball,text width=.18cm,fill,color=black,right=1.5cm of v2.east,label={[label distance=-.1cm]320:{\tiny $v_3$}}] (v3) {};        
  \node[ball,text width=.18cm,fill,color=black,right=1.5cm of v3.east,label={[label distance=-.1cm]320:{\tiny $v_4$}}] (v4) {};        
  \node[ball,text width=.18cm,fill,color=black,above=1cm of v4.north,label={[label distance=-.1cm]320:{\tiny $v_5$}}] (v5) {};        
  \node[ball,text width=.18cm,fill,color=black,above=1cm of v5.north,label=above:{\tiny $v_6$}] (v6) {};          
  \draw[-,thick,color=black] (v1) edge node[text width=.18cm,above=-0.05cm,midway] {\tiny $e_{12}$} (v2);
  \draw[-,thick,color=black] (v1) edge node[very thick,cross=4pt,rotate=0,color=blue,midway] {} (v2);  
  \draw[-,thick,color=black] (v2) edge node[text width=.18cm,above=-0.05cm,midway] {\tiny $e'_{23}$} (v3);  
  \draw[-,thick,color=black] (v2) edge node[very thick,cross=4pt,rotate=0,color=blue,midway] {} (v3);    
  \draw[-,thick,color=black] (v3) edge node[text width=.18cm,above=-0.05cm,midway] {\tiny $e_{34}$} (v4);    
  \draw[-,thick,color=black] (v3) edge node[very thick,cross=4pt,rotate=0,color=blue,midway] {} (v4);      
  \draw[-,thick,color=black] (v4) edge node[very thick,cross=4pt,rotate=0,color=blue,midway] {} (v5);
  \draw[thick,color=black] ($(v2)!0.5!(v3)$) circle (.85cm);
  \draw[thick,color=black] ($(v5)!0.5!(v6)$) circle (.6cm);  

  \coordinate (t1) at ($(v2)!0.5!(v3)$);
  \node[color=black,text width=.18cm,above=.825cm of t1.north] {\tiny $e_{23}$}; 
  \node[very thick,cross=4pt,rotate=0,color=blue,above=.76cm of t1.north] {};
  \node[color=black,text width=.18cm,below=.825cm of t1.south] {\tiny $e''_{23}$};  
  \node[very thick,cross=4pt,rotate=0,color=blue,below=.76cm of t1.south] {};

  \coordinate (t2) at ($(v5)!0.5!(v6)$);
  \node[color=black,text width=.18cm,left=.7cm of t2.west] {\tiny $e_{56}$}; 
  \node[very thick,cross=4pt,rotate=0,color=blue,left=.5cm of t2.west] {};  
  \node[color=black,text width=.18cm,right=.55cm of t2.east] {\tiny $e'_{56}$};
  \node[very thick,cross=4pt,rotate=0,color=blue,right=.5cm of t2.east] {};  
  
  \node[ball,text width=.18cm,fill,color=black,label=below:{\tiny $v_1$},right=1.5cm of v4.east] (x1) {};  
  \node[ball,text width=.18cm,fill,color=black,right=1.5cm of x1.east,label={[label distance=-.1cm]250:{\tiny $v_2$}}] (x2) {};      
  \node[ball,text width=.18cm,fill,color=black,right=1.5cm of x2.east,label={[label distance=-.1cm]320:{\tiny $v_3$}}] (x3) {};        
  \node[ball,text width=.18cm,fill,color=black,right=1.5cm of x3.east,label={[label distance=-.1cm]320:{\tiny $v_4$}}] (x4) {};        
  \node[ball,text width=.18cm,fill,color=black,above=1cm of x4.north,label={[label distance=-.1cm]320:{\tiny $v_5$}}] (x5) {};        
  \node[ball,text width=.18cm,fill,color=black,above=1cm of x5.north,label=above:{\tiny $v_6$}] (x6) {};          
  \draw[-,thick,color=black] (x1) edge node[text width=.18cm,above=-0.05cm,midway] {\tiny $e_{12}$} (x2);
  \draw[-,thick,color=black] (x2) edge node[text width=.18cm,above=-0.05cm,midway] {\tiny $e'_{23}$} (x3);  
  \draw[-,thick,color=black] (x2) edge node[very thick,cross=4pt,rotate=0,color=blue,near end] {} (x3);      
  \draw[-,thick,color=black] (x3) edge node[text width=.18cm,above=-0.05cm,midway] {\tiny $e_{34}$} (x4);    
  \draw[-,thick,color=black] (x3) edge node[very thick,cross=4pt,rotate=0,color=blue,near start] {} (x4);      
  \draw[-,thick,color=black] (x4) -- (x5);
  \draw[thick,color=black] ($(x2)!0.5!(x3)$) circle (.85cm);
  \node[very thick,cross=4pt,rotate=0,color=blue,above=.05cm of x3.north]{};
  \node[very thick,cross=4pt,rotate=0,color=blue,below=.05cm of x3.south]{};  
  \draw[thick,color=black] ($(x5)!0.5!(x6)$) circle (.6cm);  

  \coordinate (t3) at ($(x2)!0.5!(x3)$);
  \node[color=black,text width=.18cm,above=.825cm of t3.north] {\tiny $e_{23}$}; 
  \node[color=black,text width=.18cm,below=.825cm of t3.south] {\tiny $e''_{23}$};  

  \coordinate (t4) at ($(x5)!0.5!(x6)$);
  \node[color=black,text width=.18cm,left=.7cm of t4.west] {\tiny $e_{56}$}; 
  \node[color=black,text width=.18cm,right=.55cm of t4.east] {\tiny $e'_{56}$};

  \coordinate (a1) at (v1.west);
  \coordinate (a2) at (0.5,2);
  \coordinate (b2) at (1.75,2.5);
  \coordinate (a3) at (3.5,3);
  \coordinate (a4) at (5.5,2.75);
  \coordinate (a5) at (5.8,1.5);
  \coordinate (a6) at (5.5,0);
  \coordinate (a7) at (4.5,-1);
  \coordinate (a8) at (2.5,-1.5);

  \draw[thick,red] plot [smooth cycle] coordinates {(a1) (a2) (b2) (a3) (a4) (a5) (a6) (a7) (a8)};  
  \draw[thick,red] (x3) ellipse (.15cm and .15cm);
 \end{tikzpicture} 
\end{equation*}
However, we also have other possibilities, {\it e.g.}
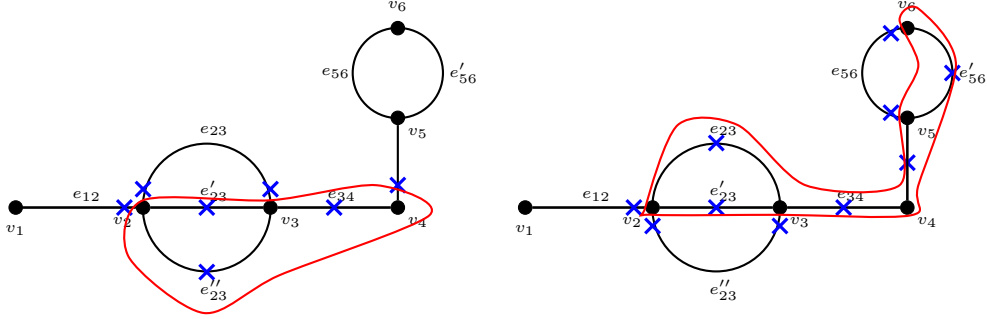
\begin{equation*}
 \begin{tikzpicture}[ball/.style = {circle, draw, align=center, anchor=north, inner sep=0}, cross/.style={cross out, draw, minimum size=2*(#1-\pgflinewidth), inner sep=0pt, outer sep=0pt}]
  \node[ball,text width=.18cm,fill,color=black,label=below:{\tiny $v_1$}] at (0,0) (v1) {};  
  \node[ball,text width=.18cm,fill,color=black,right=1.5cm of v1.east,label={[label distance=-.1cm]250:{\tiny $v_2$}}] (v2) {};      
  \node[ball,text width=.18cm,fill,color=black,right=1.5cm of v2.east,label={[label distance=-.1cm]320:{\tiny $v_3$}}] (v3) {};        
  \node[ball,text width=.18cm,fill,color=black,right=1.5cm of v3.east,label={[label distance=-.1cm]320:{\tiny $v_4$}}] (v4) {};        
  \node[ball,text width=.18cm,fill,color=black,above=1cm of v4.north,label={[label distance=-.1cm]320:{\tiny $v_5$}}] (v5) {};        
  \node[ball,text width=.18cm,fill,color=black,above=1cm of v5.north,label=above:{\tiny $v_6$}] (v6) {};          
  \draw[-,thick,color=black] (v1) edge node[text width=.18cm,above=-0.05cm,midway] {\tiny $e_{12}$} (v2);
  \draw[-,thick,color=black] (v1) edge node[very thick,cross=4pt,rotate=0,color=blue,right=0.5cm of v1.east] {} (v2);  
  \draw[-,thick,color=black] (v2) edge node[text width=.18cm,above=-0.05cm,midway] {\tiny $e'_{23}$} (v3);  
  \draw[-,thick,color=black] (v2) edge node[very thick,cross=4pt,rotate=0,color=blue,midway] {} (v3);    
  \draw[-,thick,color=black] (v3) edge node[text width=.18cm,above=-0.05cm,midway] {\tiny $e_{34}$} (v4);    
  \draw[-,thick,color=black] (v3) edge node[very thick,cross=4pt,rotate=0,color=blue,midway] {} (v4);      
  \draw[-,thick,color=black] (v4) edge node[very thick,cross=4pt,rotate=0,color=blue,below=0.2cm of v5.south] {} (v5);
  \draw[thick,color=black] ($(v2)!0.5!(v3)$) circle (.85cm);
  \draw[thick,color=black] ($(v5)!0.5!(v6)$) circle (.6cm);  

  \coordinate (t1) at ($(v2)!0.5!(v3)$);
  \node[color=black,text width=.18cm,above=.825cm of t1.north] {\tiny $e_{23}$}; 
  \node[very thick,cross=4pt,rotate=0,color=blue,above=.05cm of v2.north]{};  
  \node[very thick,cross=4pt,rotate=0,color=blue,above=.05cm of v3.north]{}; 
  \node[color=black,text width=.18cm,below=.825cm of t1.south] {\tiny $e''_{23}$};  
  \node[very thick,cross=4pt,rotate=0,color=blue,below=.76cm of t1.south] {};

  \coordinate (t2) at ($(v5)!0.5!(v6)$);
  \node[color=black,text width=.18cm,left=.7cm of t2.west] {\tiny $e_{56}$}; 
  \node[color=black,text width=.18cm,right=.55cm of t2.east] {\tiny $e'_{56}$};

  \coordinate (a1) at (v2.north);
  \coordinate (a2) at (v3.north);
  \coordinate (a3) at (4.8,0.2);
  \coordinate (a4) at (5.5,-0.2);
  \coordinate (a5) at (3.5,-1);  
  \coordinate (a6) at (2.5,-1.5);
  \coordinate (a7) at (1.5,-.75);

  \draw[thick,red] plot [smooth cycle] coordinates {(a1) (a2) (a3) (a4) (a5) (a6) (a7)};    

  \node[ball,text width=.18cm,fill,color=black,label=below:{\tiny $v_1$},right=1.5cm of v4.east] (x1) {};  
  \node[ball,text width=.18cm,fill,color=black,right=1.5cm of x1.east,label={[label distance=-.1cm]250:{\tiny $v_2$}}] (x2) {};      
  \node[ball,text width=.18cm,fill,color=black,right=1.5cm of x2.east,label={[label distance=-.1cm]320:{\tiny $v_3$}}] (x3) {};        
  \node[ball,text width=.18cm,fill,color=black,right=1.5cm of x3.east,label={[label distance=-.1cm]320:{\tiny $v_4$}}] (x4) {};        
  \node[ball,text width=.18cm,fill,color=black,above=1cm of x4.north,label={[label distance=-.1cm]320:{\tiny $v_5$}}] (x5) {};        
  \node[ball,text width=.18cm,fill,color=black,above=1cm of x5.north,label=above:{\tiny $v_6$}] (x6) {};          
  \draw[-,thick,color=black] (x1) edge node[text width=.18cm,above=-0.05cm,midway] {\tiny $e_{12}$} (x2);
  \draw[-,thick,color=black] (x1) edge node[very thick,cross=4pt,rotate=0,color=blue,right=0.5cm of x1.east] {} (x2);    
  \draw[-,thick,color=black] (x2) edge node[text width=.18cm,above=-0.05cm,midway] {\tiny $e'_{23}$} (x3);  
  \draw[-,thick,color=black] (x2) edge node[very thick,cross=4pt,rotate=0,color=blue,midway] {} (x3);      
  \draw[-,thick,color=black] (x3) edge node[text width=.18cm,above=-0.05cm,midway] {\tiny $e_{34}$} (x4);    
  \draw[-,thick,color=black] (x3) edge node[very thick,cross=4pt,rotate=0,color=blue,midway] {} (x4);   
  \draw[-,thick,color=black] (x4) edge node[very thick,cross=4pt,rotate=0,color=blue,midway] {} (x5);     
  \draw[-,thick,color=black] (x4) -- (x5);
  \draw[thick,color=black] ($(x2)!0.5!(x3)$) circle (.85cm);
  \node[very thick,cross=4pt,rotate=0,color=blue,below=.05cm of x2.south]{};
  \node[very thick,cross=4pt,rotate=0,color=blue,below=.05cm of x3.south]{};  
  \draw[thick,color=black] ($(x5)!0.5!(x6)$) circle (.6cm);  
  \node[very thick,cross=4pt,rotate=0,color=blue,left=.05cm of x5.north west]{};    
  \node[very thick,cross=4pt,rotate=0,color=blue,left=.05cm of x6.south west]{};      

  \coordinate (t3) at ($(x2)!0.5!(x3)$);
  \node[color=black,text width=.18cm,above=.825cm of t3.north] {\tiny $e_{23}$}; 
  \node[very thick,cross=4pt,rotate=0,color=blue,above=.76cm of t3.north] {};
  \node[color=black,text width=.18cm,below=.825cm of t3.south] {\tiny $e''_{23}$};  

  \coordinate (t4) at ($(x5)!0.5!(x6)$);
  \node[color=black,text width=.18cm,left=.7cm of t4.west] {\tiny $e_{56}$}; 
  \node[color=black,text width=.18cm,right=.55cm of t4.east] {\tiny $e'_{56}$};
  \node[very thick,cross=4pt,rotate=0,color=blue,right=.5cm of t4.east] {};

  \coordinate (b1) at (x2.west);
  \coordinate (b2) at (8.8,1);
  \coordinate (b3) at (9.6,1);
  \coordinate (b4) at (10.5,0.2);
  \coordinate (b5) at (11.75,0.2);
  \coordinate (b6) at (11.75,1.2);
  \coordinate (b7) at (12,1.8);
  \coordinate (b8) at (11.75,2.4);
  \coordinate (b9) at (12,2.55);
  \coordinate (b10) at (12.5,1.8);
  \coordinate (b11) at (12,0.2);
  \coordinate (b12) at (11.9,-0.2);
  \coordinate (b13) at (x3.south);
  \coordinate (b14) at (x2.south);
  \draw[thick,red] plot [smooth cycle] coordinates {(b1) (b2) (b3) (b4) (b5) (b6) (b7) (b8) (b9) (b10) (b11) (b12) (b13) (b14)};      
 \end{tikzpicture}  
\end{equation*}
It is easy to see that with this rule no markings can be legally deleted. It is also easy to see that {\it every} allowed marking takes this form. As we have seen, if no middles are marked, we must choose a vertex $v$ and mark all the edges touching it next to $v$, which corresponds to the graph $\mathfrak{g}$ consisting only of $v$. Now, suppose that the middle of some edge $e$ is marked. Then we can always legally delete any markings next to the vertices $e$ attaches to.

If $e$ touches a vertex $v$ which is also touched by another edge $e'$ (and perhaps others) we must have either
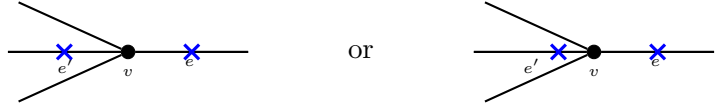
\begin{equation*}
 \begin{tikzpicture}[ball/.style = {circle, draw, align=center, anchor=north, inner sep=0}, cross/.style={cross out, draw, minimum size=2*(#1-\pgflinewidth), inner sep=0pt, outer sep=0pt}]
  \node[ball,text width=.18cm,fill,color=white] at (0,0) (v1) {};    
  \node[ball,text width=.18cm,fill,color=black,right=1.5cm of v1.east,label=below:{\tiny $v$}] (v2) {};
  \node[ball,text width=.18cm,fill,color=white,right=1.5cm of v2.east] (v3) {};  
  \node[ball,text width=.18cm,fill,color=white] at (0.15,0.7) (v4) {};      
  \node[ball,text width=.18cm,fill,color=white] at (0.15,-0.7) (v5) {};        
  \draw[thick,black] (v1) edge node[text width=.18cm,below=-0.05cm,midway] {\tiny $e'$} (v2);
  \draw[thick,black] (v1) edge node[very thick,cross=4pt,rotate=0,color=blue,midway] {} (v2);
  \draw[thick,black] (v2) edge node[text width=.18cm,below=-0.05cm,midway] {\tiny $e$} (v3);  
  \draw[thick,black] (v2) edge node[very thick,cross=4pt,rotate=0,color=blue,midway] {} (v3);  
  \draw[thick,black] (v2) -- (v4);
  \draw[thick,black] (v2) -- (v5);  

  \node[right=1cm of v3.east] (or) {or};

  \node[ball,text width=.18cm,fill,color=white,right=1cm of or.east] (x1) {};    
  \node[ball,text width=.18cm,fill,color=black,right=1.5cm of x1.east,label=below:{\tiny $v$}] (x2) {};
  \node[ball,text width=.18cm,fill,color=white,right=1.5cm of x2.east] (x3) {};  
  \node[ball,text width=.18cm,fill,color=white] at (6.35,0.7) (x4) {};      
  \node[ball,text width=.18cm,fill,color=white] at (6.35,-0.7) (x5) {};        
  \draw[thick,black] (x1) edge node[text width=.18cm,below=-0.05cm,midway] {\tiny $e'$} (x2);
  \draw[thick,black] (x1) edge node[very thick,cross=4pt,rotate=0,color=blue,near end] {} (x2);
  \draw[thick,black] (x2) edge node[text width=.18cm,below=-0.05cm,midway] {\tiny $e$} (x3);  
  \draw[thick,black] (x2) edge node[very thick,cross=4pt,rotate=0,color=blue,midway] {} (x3);  
  \draw[thick,black] (x2) -- (x4);
  \draw[thick,black] (x2) -- (x5); 
 \end{tikzpicture}   
\end{equation*}
In the second case, if {\it all} the other edges touching $v$ are marked next to $v$, then we can leave all the other edges of the graph accessible via $v$ unmarked. If any of the other edges connected to $v$ are marked on the middle, we proceed to the other vertices they connect to, and keep going in this way, forming a connected subgraph of precisely the sort we described above, until we finally encounter vertices $v$ where all the other edges are marked next to $v$, where we stop.

Note that we are after {\it connected} graphs for a simple reason. If we have two or more disconnected graphs, we can always legally set the $\alpha$'s for all but one of the connected components to zero. 

Having chosen the connected subgraph $\mathfrak{g}$ and the external to $e$ ending on it, the vertices belonging to this face are simply $\{{\bf x}_v+{\bf y}_e-{\bf x}_{v'},\,{\bf x}_{v'}+{\bf y}_e-{\bf x}_v\}$ for all the internal edges $e$, and $\displaystyle\left({\bf x}_v+\sum_{\substack{\mbox{\tiny external} \\ {\mbox{\tiny edges } e}}}{\bf y}_e\right)$ for all the ``external'' vertices of $\mathfrak{g}$ that touch external edges. The $\mathcal{W}_I$ associated with this face is then easily seen to be
\begin{equation}\label{eq:WiF2}
 \mathcal{W}_I\:=\:\sum_{v\in\mathfrak{g}}{\bf\tilde{X}}_{v_I}+\sum_{\substack{\mbox{\tiny external} \\ {\mbox{\tiny edges } e} \\ \mbox{\tiny ending on }\mathfrak{g}}}{\bf\tilde{Y}}_{e_I}.
\end{equation}
Note that
\begin{equation}\label{eq:WE}
 \mathcal{W}_I\mathcal{Y}^{I}\:=\:\sum_{v\in\mathfrak{g}}x_v+\sum_{\mbox{\tiny ext. }e}y_e
\end{equation}
which are precisely the ``energy denominators'' we encounter in old-fashioned perturbation theory. Of course, this is expected from the connection between the canonical form $\Omega(\mathcal{Y};\mathcal{P}_{\mathcal{G}})$ for $\Psi_{\mathcal{G}}(\mathcal{Y})$.

The poles of $\Omega$ (and so of $\Psi$) are precisely associated to the faces of $\mathcal{P}_{\mathcal{G}}$, and occur when $\mathcal{W}\cdot\mathcal{Y}\,\longrightarrow\,0$. Since we know precisely which vertices of the polytope lie on any particular codimension-one face, we can determine the structure of the lower-dimensional faces of the polytope as well. For instance, in order to describe the codimension-two boundaries, we simply check whether two of the codimension-one faces, which are $(V+E-2)$-dimensional, share enough vertices to span a $(V+E-3)$-dimensional space, and son on; We can see whether $K$ faces share enough vertices in common to span a $(V+E-1-K)$-dimensional space. Let us illustrate the idea via the star diagram 
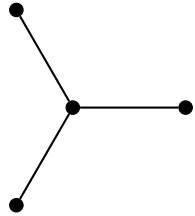
\begin{equation*}
 \begin{tikzpicture}[ball/.style = {circle, draw, align=center, anchor=north, inner sep=0}]
    \node[ball,text width=.18cm,fill,color=black] at (0,0) (x4) {};  
    \node[ball,text width=.18cm,fill,color=black] at ({1.5*cos(0)},{1.5*sin(0)}) (x3) {};    
    \node[ball,text width=.18cm,fill,color=black] at ({1.5*cos(120)},{1.5*sin(120)}) (x1) {};
    \node[ball,text width=.18cm,fill,color=black] at ({1.5*cos(240)},{1.5*sin(240)}) (x2) {};    
    \draw[-,thick,color=black] (x1) -- (x4);    
    \draw[-,thick,color=black] (x2) -- (x4);        
    \draw[-,thick,color=black] (x3) -- (x4);
 \end{tikzpicture}
\end{equation*}
which is a $4+3-1\,=\,6$ dimensional polytope. Let us exhibit two of its five-dimensional faces that {\it do} intersect on a four-dimensional face: the faces
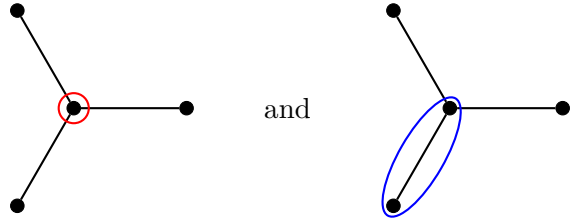
\begin{equation*}
 \begin{tikzpicture}[ball/.style = {circle, draw, align=center, anchor=north, inner sep=0}]
  \begin{scope}
    \node[ball,text width=.18cm,fill,color=black] at (0,0) (x4) {};  
    \node[ball,text width=.18cm,fill,color=black] at ({1.5*cos(0)},{1.5*sin(0)}) (x3) {};    
    \node[ball,text width=.18cm,fill,color=black] at ({1.5*cos(120)},{1.5*sin(120)}) (x1) {};
    \node[ball,text width=.18cm,fill,color=black] at ({1.5*cos(240)},{1.5*sin(240)}) (x2) {};    
    \draw[-,thick,color=black] (x1) -- (x4);    
    \draw[-,thick,color=black] (x2) -- (x4);        
    \draw[-,thick,color=black] (x3) -- (x4);
    \draw[red, thick] (x4) circle (.2cm);
    \node[right=.8cm of x3.east] {and };
  \end{scope}
  \begin{scope}[shift={(5,0)}]
    \node[ball,text width=.18cm,fill,color=black] at (0,0) (x4) {};  
    \node[ball,text width=.18cm,fill,color=black] at ({1.5*cos(0)},{1.5*sin(0)}) (x3) {};    
    \node[ball,text width=.18cm,fill,color=black] at ({1.5*cos(120)},{1.5*sin(120)}) (x1) {};
    \node[ball,text width=.18cm,fill,color=black] at ({1.5*cos(240)},{1.5*sin(240)}) (x2) {};    
    \draw[-,thick,color=black] (x1) -- (x4);    
    \draw[-,thick,color=black] (x2) -- (x4);        
    \draw[-,thick,color=black] (x3) -- (x4);
    \draw[rotate=240,color=blue, thick] ($(x2)!.5!(x4)$) ellipse (.9cm and .3cm);
  \end{scope}
 \end{tikzpicture}
\end{equation*}
Let us use our above notation to denote the vertices {\it not} on each:
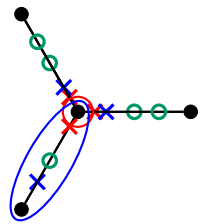
\begin{equation*}
 \begin{tikzpicture}[ball/.style = {circle, draw, align=center, anchor=center, inner sep=0},  cross/.style={cross out, draw, minimum size=2*(#1-\pgflinewidth), inner sep=0pt, outer sep=0pt}]
  \node[ball,text width=.18cm,fill,color=black] at (0,0) (x4) {};  
    \node[ball,text width=.18cm,fill,color=black] at ({1.5*cos(0)},{1.5*sin(0)}) (x3) {};    
    \node[ball,text width=.18cm,fill,color=black] at ({1.5*cos(120)},{1.5*sin(120)}) (x1) {};
    \node[ball,text width=.18cm,fill,color=black] at ({1.5*cos(240)},{1.5*sin(240)}) (x2) {};    
    \draw[-,thick,color=black] (x1) -- (x4);    
    \draw[-,thick,color=black] (x2) edge node[very thick,cross=4pt,rotate=0,color=blue,near start] {} (x4);        
    \draw[-,thick,color=black] (x2) edge node[ball, very thick,text width=.18cm,color=green!60!blue,midway] {} (x4);  
    \draw ($(x2)!.85!(x4)$) node[cross=4pt,very thick,color=red] {};
    \draw ($(x3)!.85!(x4)$) node[cross=4pt,very thick,color=red] {};    
    \draw ($(x3)!.75!(x4)$) node[cross=4pt,very thick,color=blue] {};        
    \draw[-,thick,color=black] (x3) edge node[ball, very thick,text width=.18cm,color=green!60!blue,midway] {} (x4);
    \draw[-,thick,color=black] (x3) edge node[ball, very thick,text width=.18cm,color=green!60!blue,near start] {} (x4);    
    \draw ($(x1)!.85!(x4)$) node[cross=4pt,very thick,color=red] {};    
    \draw ($(x1)!.75!(x4)$) node[cross=4pt,very thick,color=blue] {};        
    \draw[-,thick,color=black] (x1) edge node[ball, very thick,text width=.18cm,color=green!60!blue,midway] {} (x4);
    \draw[-,thick,color=black] (x1) edge node[ball, very thick,text width=.18cm,color=green!60!blue,near start] {} (x4); 
    \draw[-,thick,color=red] (x4) circle (.2cm);
    \draw[rotate=240,color=blue, thick] ($(x2)!.5!(x4)$) ellipse (.9cm and .3cm);
 \end{tikzpicture} 
\end{equation*}
where the vertices not on $R$, $B$ facets are represented by $x$'s above.

We see that there are five vertices left, represented by the open green circles, and it is furthermore easy to see that they are linearly independent. Thus, these five vertices form a five-dimensional simplex in four-dimensions, so these two faces intersect on a four-dimensional face, which is just a simplex. On the other hand, the two faces
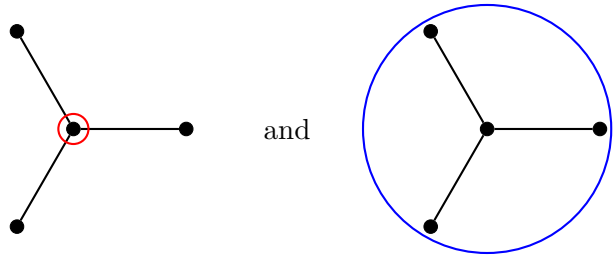
\begin{equation*}
 \begin{tikzpicture}[ball/.style = {circle, draw, align=center, anchor=north, inner sep=0}]
  \begin{scope}
    \node[ball,text width=.18cm,fill,color=black] at (0,0) (x4) {};  
    \node[ball,text width=.18cm,fill,color=black] at ({1.5*cos(0)},{1.5*sin(0)}) (x3) {};    
    \node[ball,text width=.18cm,fill,color=black] at ({1.5*cos(120)},{1.5*sin(120)}) (x1) {};
    \node[ball,text width=.18cm,fill,color=black] at ({1.5*cos(240)},{1.5*sin(240)}) (x2) {};    
    \draw[-,thick,color=black] (x1) -- (x4);    
    \draw[-,thick,color=black] (x2) -- (x4);        
    \draw[-,thick,color=black] (x3) -- (x4);
    \draw[red, thick] (x4) circle (.2cm);
    \node[right=.8cm of x3.east] {and };
  \end{scope}
  \begin{scope}[shift={(5.5,0)}]
    \node[ball,text width=.18cm,fill,color=black] at (0,0) (x4) {};  
    \node[ball,text width=.18cm,fill,color=black] at ({1.5*cos(0)},{1.5*sin(0)}) (x3) {};    
    \node[ball,text width=.18cm,fill,color=black] at ({1.5*cos(120)},{1.5*sin(120)}) (x1) {};
    \node[ball,text width=.18cm,fill,color=black] at ({1.5*cos(240)},{1.5*sin(240)}) (x2) {};    
    \draw[-,thick,color=black] (x1) -- (x4);    
    \draw[-,thick,color=black] (x2) -- (x4);        
    \draw[-,thick,color=black] (x3) -- (x4);
    \draw[thick, color=blue] (x4) circle (1.65cm);
  \end{scope}
 \end{tikzpicture}
\end{equation*}
do not intersect on a four-dimensional face:
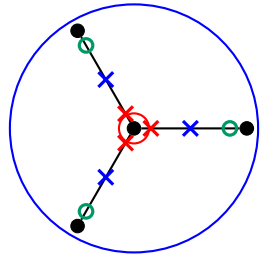
\begin{equation*}
 \begin{tikzpicture}[ball/.style = {circle, draw, align=center, anchor=center, inner sep=0}, cross/.style={cross out, draw, minimum size=2*(#1-\pgflinewidth), inner sep=0pt, outer sep=0pt}]
   \node[ball,text width=.18cm,fill,color=black] at (0,0) (x4) {};  
    \node[ball,text width=.18cm,fill,color=black] at ({1.5*cos(0)},{1.5*sin(0)}) (x3) {};    
    \node[ball,text width=.18cm,fill,color=black] at ({1.5*cos(120)},{1.5*sin(120)}) (x1) {};
    \node[ball,text width=.18cm,fill,color=black] at ({1.5*cos(240)},{1.5*sin(240)}) (x2) {};    
    \draw[-,thick,color=black] (x1) edge node[cross=4pt,very thick,color=blue,midway] {} (x4);
    \draw ($(x1)!.15!(x4)$) node[ball, very thick,text width=.18cm,color=green!60!blue] {};
    \draw ($(x1)!.85!(x4)$) node[cross=4pt, very thick,color=red] {};    
    \draw[-,thick,color=black] (x2) edge node[cross=4pt,very thick,color=blue,midway] {} (x4);        
    \draw ($(x2)!.15!(x4)$) node[ball, very thick,text width=.18cm,color=green!60!blue] {};
    \draw ($(x2)!.85!(x4)$) node[cross=4pt, very thick,color=red] {};  
    \draw[-,thick,color=black] (x3) edge node[cross=4pt,very thick,color=blue,midway] {}  (x4);
    \draw ($(x3)!.15!(x4)$) node[ball, very thick,text width=.18cm,color=green!60!blue] {};
    \draw ($(x3)!.85!(x4)$) node[cross=4pt, very thick,color=red] {}; 
    \draw[thick, color=blue] (x4) circle (1.65cm);
    \draw[thick, color=red] (x4) circle (.2cm);    
 \end{tikzpicture}
\end{equation*}
There are only three vertices common to both faces; These five-dimensional faces intersect only on a two-dimensional triangle.

We can proceed in this way to find all the lower-dimensional faces. Since we can visualize three-dimensional spaces, let us describe some of the three-dimensional facets of the geometry. They range in complexity from a simplex (tetrahedron) to an octahedron. For an example of octahedron, consider the intersection of
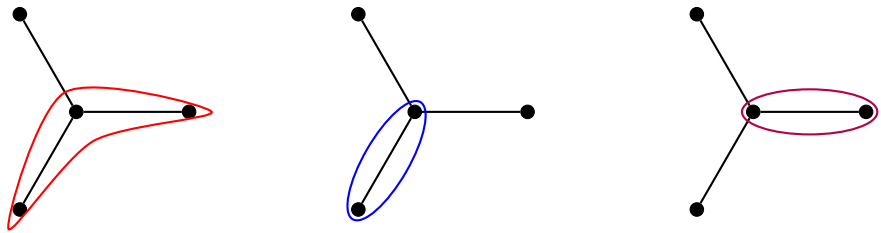
\begin{equation*}
 \begin{tikzpicture}[ball/.style = {circle, draw, align=center, anchor=center, inner sep=0}, cross/.style={cross out, draw, minimum size=2*(#1-\pgflinewidth), inner sep=0pt, outer sep=0pt}]
  \begin{scope}
   \node[ball,text width=.18cm,fill,color=black] at (0,0) (x4) {};  
    \node[ball,text width=.18cm,fill,color=black] at ({1.5*cos(0)},{1.5*sin(0)}) (x3) {};    
    \node[ball,text width=.18cm,fill,color=black] at ({1.5*cos(120)},{1.5*sin(120)}) (x1) {};
    \node[ball,text width=.18cm,fill,color=black] at ({1.5*cos(240)},{1.5*sin(240)}) (x2) {};
    \draw[-,thick,color=black] (x1) -- (x4);
    \draw[-,thick,color=black] (x2) -- (x4);
    \draw[-,thick,color=black] (x3) -- (x4);
    \coordinate (y1) at ($(x4)!.2!(x1)$);
    \coordinate (y2) at ($(x4)!1.2!(x2)$);
    \coordinate (y3) at ($(x4)!-.3!(x1)$);
    \coordinate (y4) at ($(x4)!1.2!(x3)$);
    \draw[thick,red] plot [smooth cycle] coordinates {(y1) (y2) (y3) (y4)};
  \end{scope}
  \begin{scope}[shift={(4.5,0)}]
   \node[ball,text width=.18cm,fill,color=black] at (0,0) (x4) {};  
    \node[ball,text width=.18cm,fill,color=black] at ({1.5*cos(0)},{1.5*sin(0)}) (x3) {};    
    \node[ball,text width=.18cm,fill,color=black] at ({1.5*cos(120)},{1.5*sin(120)}) (x1) {};
    \node[ball,text width=.18cm,fill,color=black] at ({1.5*cos(240)},{1.5*sin(240)}) (x2) {};
    \draw[-,thick,color=black] (x1) -- (x4);
    \draw[-,thick,color=black] (x2) -- (x4);
    \draw[-,thick,color=black] (x3) -- (x4);
    \draw[rotate=240,color=blue, thick] ($(x2)!.5!(x4)$) ellipse (.9cm and .3cm);    
  \end{scope}
  \begin{scope}[shift={(9,0)}]
   \node[ball,text width=.18cm,fill,color=black] at (0,0) (x4) {};  
    \node[ball,text width=.18cm,fill,color=black] at ({1.5*cos(0)},{1.5*sin(0)}) (x3) {};    
    \node[ball,text width=.18cm,fill,color=black] at ({1.5*cos(120)},{1.5*sin(120)}) (x1) {};
    \node[ball,text width=.18cm,fill,color=black] at ({1.5*cos(240)},{1.5*sin(240)}) (x2) {};
    \draw[-,thick,color=black] (x1) -- (x4);
    \draw[-,thick,color=black] (x2) -- (x4);
    \draw[-,thick,color=black] (x3) -- (x4);
    \draw[color=red!70!blue, thick] ($(x3)!.5!(x4)$) ellipse (.9cm and .3cm);    
  \end{scope}
 \end{tikzpicture}
\end{equation*}
which can be represented as 
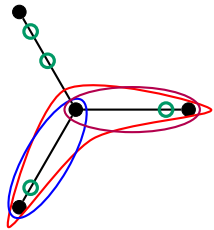
\begin{equation}
 \begin{tikzpicture}[ball/.style = {circle, draw, align=center, anchor=center, inner sep=0}, cross/.style={cross out, draw, minimum size=2*(#1-\pgflinewidth), inner sep=0pt, outer sep=0pt}]
  \begin{scope}
   \node[ball,text width=.18cm,fill,color=black] at (0,0) (x4) {};  
    \node[ball,text width=.18cm,fill,color=black] at ({1.5*cos(0)},{1.5*sin(0)}) (x3) {};    
    \node[ball,text width=.18cm,fill,color=black] at ({1.5*cos(120)},{1.5*sin(120)}) (x1) {};
    \node[ball,text width=.18cm,fill,color=black] at ({1.5*cos(240)},{1.5*sin(240)}) (x2) {};
    \draw[-,thick,color=black] (x1) -- (x4);
    \draw[-,thick,color=black] (x2) -- (x4);
    \draw[-,thick,color=black] (x3) -- (x4);
    \coordinate (y1) at ($(x4)!.2!(x1)$);
    \coordinate (y2) at ($(x4)!1.2!(x2)$);
    \coordinate (y3) at ($(x4)!-.3!(x1)$);
    \coordinate (y4) at ($(x4)!1.2!(x3)$);
    \draw[thick,red] plot [smooth cycle] coordinates {(y1) (y2) (y3) (y4)};
    \draw[rotate=240,color=blue, thick] ($(x2)!.5!(x4)$) ellipse (.9cm and .3cm);    
    \draw[color=red!70!blue, thick] ($(x3)!.5!(x4)$) ellipse (.9cm and .3cm);
    \draw ($(x1)!.2!(x4)$) node[ball, very thick,text width=.18cm,color=green!60!blue] {};
    \draw ($(x1)!.5!(x4)$) node[ball, very thick,text width=.18cm,color=green!60!blue] {};
    \draw ($(x2)!.2!(x4)$) node[ball, very thick,text width=.18cm,color=green!60!blue] {};    
    \draw ($(x3)!.2!(x4)$) node[ball, very thick,text width=.18cm,color=green!60!blue] {};    
  \end{scope}
 \end{tikzpicture}
\end{equation}
where, in order to avoid clutter, we have only shown the vertices common to all the faces. Meanwhile, the intersection of
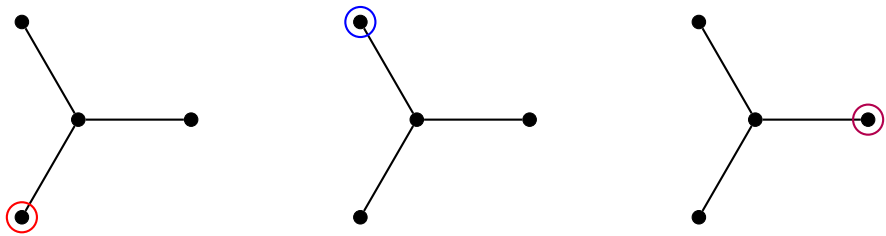
\begin{equation*}
 \begin{tikzpicture}[ball/.style = {circle, draw, align=center, anchor=center, inner sep=0}, cross/.style={cross out, draw, minimum size=2*(#1-\pgflinewidth), inner sep=0pt, outer sep=0pt}]
  \begin{scope}
   \node[ball,text width=.18cm,fill,color=black] at (0,0) (x4) {};  
    \node[ball,text width=.18cm,fill,color=black] at ({1.5*cos(0)},{1.5*sin(0)}) (x3) {};    
    \node[ball,text width=.18cm,fill,color=black] at ({1.5*cos(120)},{1.5*sin(120)}) (x1) {};
    \node[ball,text width=.18cm,fill,color=black] at ({1.5*cos(240)},{1.5*sin(240)}) (x2) {};
    \draw[-,thick,color=black] (x1) -- (x4);
    \draw[-,thick,color=black] (x2) -- (x4);
    \draw[-,thick,color=black] (x3) -- (x4);
    \draw[thick,red] (x2) circle (.2);
   \end{scope}
   \begin{scope}[shift={(4.5,0)}]
   \node[ball,text width=.18cm,fill,color=black] at (0,0) (x4) {};  
    \node[ball,text width=.18cm,fill,color=black] at ({1.5*cos(0)},{1.5*sin(0)}) (x3) {};    
    \node[ball,text width=.18cm,fill,color=black] at ({1.5*cos(120)},{1.5*sin(120)}) (x1) {};
    \node[ball,text width=.18cm,fill,color=black] at ({1.5*cos(240)},{1.5*sin(240)}) (x2) {};
    \draw[-,thick,color=black] (x1) -- (x4);
    \draw[-,thick,color=black] (x2) -- (x4);
    \draw[-,thick,color=black] (x3) -- (x4);
    \draw[thick,blue] (x1) circle (.2);    
  \end{scope}
  \begin{scope}[shift={(9,0)}]
   \node[ball,text width=.18cm,fill,color=black] at (0,0) (x4) {};  
    \node[ball,text width=.18cm,fill,color=black] at ({1.5*cos(0)},{1.5*sin(0)}) (x3) {};    
    \node[ball,text width=.18cm,fill,color=black] at ({1.5*cos(120)},{1.5*sin(120)}) (x1) {};
    \node[ball,text width=.18cm,fill,color=black] at ({1.5*cos(240)},{1.5*sin(240)}) (x2) {};
    \draw[-,thick,color=black] (x1) -- (x4);
    \draw[-,thick,color=black] (x2) -- (x4);
    \draw[-,thick,color=black] (x3) -- (x4);
    \draw[thick,red!70!blue] (x3) circle (.2);
  \end{scope}
 \end{tikzpicture}
\end{equation*}
gives
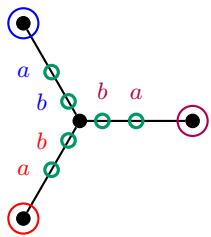
\begin{equation}
 \begin{tikzpicture}[ball/.style = {circle, draw, align=center, anchor=center, inner sep=0}, cross/.style={cross out, draw, minimum size=2*(#1-\pgflinewidth), inner sep=0pt, outer sep=0pt}]
  \begin{scope}
   \node[ball,text width=.18cm,fill,color=black] at (0,0) (x4) {};  
    \node[ball,text width=.18cm,fill,color=black] at ({1.5*cos(0)},{1.5*sin(0)}) (x3) {};    
    \node[ball,text width=.18cm,fill,color=black] at ({1.5*cos(120)},{1.5*sin(120)}) (x1) {};
    \node[ball,text width=.18cm,fill,color=black] at ({1.5*cos(240)},{1.5*sin(240)}) (x2) {};
    \draw[-,thick,color=black] (x1) -- (x4);
    \draw[-,thick,color=black] (x2) -- (x4);
    \draw[-,thick,color=black] (x3) -- (x4);
    \draw[thick,red] (x2) circle (.2);
    \draw[thick,blue] (x1) circle (.2);    
    \draw[thick,red!70!blue] (x3) circle (.2);     
    \coordinate (a1) at ($(x1)!.5!(x4)$);
    \coordinate (b1) at ($(x1)!.8!(x4)$);
    \coordinate (a2) at ($(x2)!.5!(x4)$);
    \coordinate (b2) at ($(x2)!.8!(x4)$);
 
    \coordinate (a3) at ($(x3)!.5!(x4)$);
    \coordinate (b3) at ($(x3)!.8!(x4)$);   
    \draw (a1) node[ball, very thick,text width=.18cm,color=green!60!blue] {};
    \draw (b1) node[ball, very thick,text width=.18cm,color=green!60!blue] {};
    \draw (a2) node[ball, very thick,text width=.18cm,color=green!60!blue] {};    
    \draw (b2) node[ball, very thick,text width=.18cm,color=green!60!blue] {};        
    \draw (a3) node[ball, very thick,text width=.18cm,color=green!60!blue] {};    
    \draw (b3) node[ball, very thick,text width=.18cm,color=green!60!blue] {};        
    \node[left=.15cm of a1.north west,color=blue] {\footnotesize $a$};
    \node[left=.15cm of b1.north west,color=blue] {\footnotesize $b$};    
    \node[left=.15cm of a2.south east,color=red] {\footnotesize $a$};  
    \node[left=.15cm of b2.south east,color=red] {\footnotesize $b$};    
    \node[above=.15cm of a3.north,color=red!70!blue] {\footnotesize $a$};  
    \node[above=.15cm of b3.north,color=red!70!blue] {\footnotesize $b$};
  \end{scope}
 \end{tikzpicture}
\end{equation}
since ${\color{red} a+b}\,=\,{\color{blue} a+b}\,=\,{\color{redb} a+b}$ the convex hull of these six points is an octahedron
\begin{equation}\label{eq:octh}
 \begin{tikzpicture}[line join = round, line cap = round, ball/.style = {circle, draw, align=center, anchor=north, inner sep=0}]
  \begin{scope}
   \pgfmathsetmacro{\factor}{1/sqrt(2)};  
   \coordinate [label=right:{\footnotesize ${\color{redb} b}$}] (X4) at (10.5,-3,-1.5*\factor);
   \coordinate [label=left:{\footnotesize ${\color{red} a}$}] (X3) at (7.5,-3,-1.5*\factor);
   \coordinate [label=right:{\footnotesize ${\color{red} b}$}] (X5) at (10.5,-3.75,1.5*\factor);
   \coordinate [label=left:{\footnotesize ${\color{redb} a}$}] (X6) at (7.5,-3.75,1.5*\factor);  
   \coordinate [label=above:{\footnotesize ${\color{blue} a}$}] (X1) at (9.75,-.65,.75*\factor);
   \coordinate [label=below:{\footnotesize ${\color{blue} b}$}] (X2) at (9.4,-6.05,.75*\factor);


    \draw[-,dashed] (X3) -- (X4) -- (X1) -- cycle;

    \draw[-,dashed] (X3) -- (X4) -- (X2) -- cycle;
    \draw[-,dashed] (X3) -- (X6) -- (X2) -- cycle;

    \draw[-] (X6) -- (X1) -- (X4) -- (X2) --cycle; 
    \draw[-] (X6) -- (X1) -- (X4) -- (X2) --cycle;
    \draw[-,thick] (X6) -- (X5) -- (X2) -- cycle;
    \draw[-,thick] (X4) -- (X5) -- (X2) -- cycle;  
    \draw[-,thick] (X6) -- (X1) -- (X5) -- cycle;   
   \draw[-,thick] (X4) -- (X1) -- (X5) -- cycle;
   \draw[-,thick] (X6) -- (X3) -- (X1) -- cycle;
    \end{scope}
 \end{tikzpicture}
\end{equation}


\subsection{Dual of the Cosmological Polytope}\label{subsec:DCP}

We can think of the faces of $\mathcal{P}_{\mathcal{G}}$ as being the vertices of the dual polytope $\tilde{\mathcal{P}}_{\mathcal{G}}$. The dual polytopes for the two-site chain and the two-site loop are: 
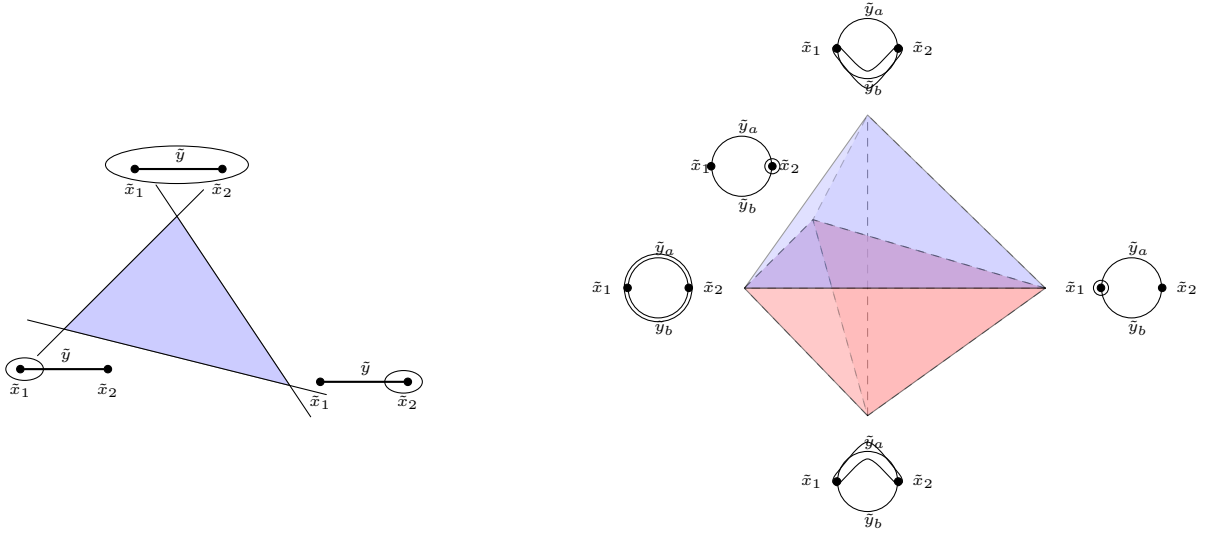
\begin{equation*}
 \begin{tikzpicture}[shift={(1,0)}, line join = round, line cap = round, ball/.style = {circle, draw, align=center, anchor=north, inner sep=0}]
  \begin{scope}
   \coordinate [label=above:{
                \begin{tikzpicture}
                 \node[ball,text width=.1cm,fill,color=black,label=below:{\tiny $\tilde{x}_1$}] at (0,0) (x1b) {};
                 \node[ball,text width=.1cm,fill,color=black,right=1.05cm of x1b.east,label=below:{\tiny $\tilde{x}_2$}] (x2b) {};
                 \draw[-,thick,color=black] (x1b.east) edge node [text width=.1cm,above=-0.05cm,midway] {{\tiny $\tilde{y}$}} (x2b.west);
                 \draw (0.56,0) ellipse (.95cm and 0.25cm);
                \end{tikzpicture}
               }] (A) at (0,0);
   \coordinate [label=below:{
                \begin{tikzpicture}
                 \node[ball,text width=.1cm,fill,color=black,label=below:{\tiny $\tilde{x}_1$}] at (0,0) (x1b) {};
                 \node[ball,text width=.1cm,fill,color=black,right=1.05cm of x1b.east,label=below:{\tiny $\tilde{x}_2$}] (x2b) {};
                 \draw[-,thick,color=black] (x1b.east) edge node [text width=.1cm,above=-0.05cm,midway] {{\tiny $\tilde{y}$}} (x2b.west);
                 \draw (x1b.east) ellipse (.25cm and 0.15cm);
                \end{tikzpicture}
               }] (B) at (-1.5,-1.5);
   \coordinate [label=right:{
                \begin{tikzpicture}
                 \node[ball,text width=.1cm,fill,color=black,label=below:{\tiny $\tilde{x}_1$}] at (0,0) (x1b) {};
                 \node[ball,text width=.1cm,fill,color=black,right=1.05cm of x1b.east,label=below:{\tiny $\tilde{x}_2$}] (x2b) {};
                 \draw[-,thick,color=black] (x1b.east) edge node [text width=.1cm,above=-0.05cm,midway] {{\tiny $\tilde{y}$}} (x2b.west);
                 \draw (x2b.west) ellipse (.25cm and 0.15cm);
                \end{tikzpicture}
               }] (C) at (1.5,-9/4);

   \draw [name path=P1] ($(A)!-0.5cm!(B)$) -- ($(B)!-0.5cm!(A)$);
   \draw [name path=P2] ($(B)!-0.5cm!(C)$) -- ($(C)!-0.5cm!(B)$);
   \draw [name path=P3] ($(C)!-0.5cm!(A)$) -- ($(A)!-0.5cm!(C)$);
   \draw[-,fill=blue!40, opacity=.5] (A) -- (B) -- (C) -- cycle;
  \end{scope}

  \begin{scope}[shift={(9,-1.5)}]
   \pgfmathsetmacro{\factor}{1/sqrt(2)};
   \coordinate [label=right:{
                \begin{tikzpicture}
                 \node[ball,text width=.1cm,fill,color=black,label=left:{\tiny ${\tilde{x}}_1$}] at (0,0) (x1b) {};
                 \node[ball,text width=.1cm,fill,color=black,right=.7cm of x1b.east,label=right:{\tiny ${\tilde{x}}_2$}] (x2b) {};
                 \coordinate (c1) at ($(x1b)!0.5!(x2b)$);
                 \draw (c1) circle (.4cm);
                 \node[text width=.1cm,above=.3cm of c1.north] (ya) {\tiny ${\tilde{y}}_a$};
                 \node[text width=.1cm,below=.3cm of c1.south] (ya) {\tiny ${\tilde{y}}_b$};
                 \draw (x1b) circle (.1cm);
                \end{tikzpicture}                               
               }] (A) at (2,0,-2*\factor);
   \coordinate [label=left:{
                \begin{tikzpicture}
                 \node[ball,text width=.1cm,fill,color=black,label=left:{\tiny ${\tilde{x}}_1$}] at (0,0) (x1b) {};
                 \node[ball,text width=.1cm,fill,color=black,right=.7cm of x1b.east,label=right:{\tiny ${\tilde{x}}_2$}] (x2b) {};
                 \coordinate (c1) at ($(x1b)!0.5!(x2b)$);
                 \draw (c1) circle (.4cm);
                 \node[text width=.1cm,above=.3cm of c1.north] (ya) {\tiny ${\tilde{y}}_a$};
                 \node[text width=.1cm,below=.3cm of c1.south] (ya) {\tiny ${\tilde{y}}_b$};
                 \draw (c1) circle (.45cm);
                \end{tikzpicture}
              }] (B) at (-2,0,-2*\factor);
   \coordinate [label={[label distance=-.2cm]120:
                \begin{tikzpicture}
                 \node[ball,text width=.1cm,fill,color=black,label=left:{\tiny ${\tilde{x}}_1$}] at (0,0) (x1b) {};
                 \node[ball,text width=.1cm,fill,color=black,right=.7cm of x1b.east,label={[label distance=-.1cm]0:{\tiny ${\tilde{x}}_2$}}] (x2b) {};
                 \coordinate (c1) at ($(x1b)!0.5!(x2b)$);
                 \draw (c1) circle (.4cm);
                 \node[text width=.1cm,above=.3cm of c1.north] (ya) {\tiny ${\tilde{y}}_a$};
                 \node[text width=.1cm,below=.3cm of c1.south] (ya) {\tiny ${\tilde{y}}_b$};
                 \draw (x2b) circle (.1cm);
                \end{tikzpicture}
    }] (C) at (0,2,2*\factor);
   \coordinate (Ab) at ($(B)!.5!(C)$); 
   \coordinate (Bb) at ($(A)!.5!(C)$);
   \coordinate (Cb) at ($(A)!.5!(B)$);
   \coordinate (G) at (intersection of A--Ab and B--Bb);
   \coordinate[below=2cm of G.south, label=below:{
                \begin{tikzpicture}
                 \node[ball,text width=.1cm,fill,color=black,label=left:{\tiny $\tilde{x}_1$}] at (0,0) (x1b) {};
                 \node[ball,text width=.1cm,fill,color=black,right=.7cm of x1b.east,label=right:{\tiny ${\tilde{x}}_2$}] (x2b) {};
                 \coordinate (c1) at ($(x1b)!0.5!(x2b)$);
                 \draw (c1) circle (.4cm);
                 \node[text width=.1cm,above=.3cm of c1.north] (ya) {\tiny ${\tilde{y}}_a$};
                 \node[text width=.1cm,below=.3cm of c1.south] (yb) {\tiny ${\tilde{y}}_b$};
                 \coordinate (tmp) at ($(ya.north)!0.5!(ya.south)$);
                 \coordinate (tmp2) at ($(x1b.west)!0.25!(ya.west)$);
                 \coordinate (tmp3) at ($(x2b.east)!0.25!(ya.east)$);                 
                 \draw plot [smooth cycle] coordinates {(x1b.east) (ya.south) (x2b.west) (x2b.south) (x2b.east) (tmp3.east) (tmp) (tmp2.west) (x1b.west) (x1b.south)};
                \end{tikzpicture}
   }] (D);
   \coordinate[above=2cm of G.north, label=above:{
                \begin{tikzpicture}
                 \node[ball,text width=.1cm,fill,color=black,label=left:{\tiny ${\tilde{x}}_1$}] at (0,0) (x1b) {};
                 \node[ball,text width=.1cm,fill,color=black,right=.7cm of x1b.east,label=right:{\tiny ${\tilde{x}}_2$}] (x2b) {};
                 \coordinate (c1) at ($(x1b)!0.5!(x2b)$);
                 \draw (c1) circle (.4cm);
                 \node[text width=.1cm,above=.3cm of c1.north] (ya) {\tiny ${\tilde{y}}_a$};
                 \node[text width=.1cm,below=.3cm of c1.south] (ya) {\tiny ${\tilde{y}}_b$};
                 \coordinate (tmp) at ($(yb.north)!0.5!(yb.south)$);
                 \coordinate (tmp2) at ($(x1b.west)!0.25!(yb.west)$);
                 \coordinate (tmp3) at ($(x2b.east)!0.25!(yb.east)$);                 
                 \draw plot [smooth cycle] coordinates {(x1b.east) (yb.north) (x2b.west) (x2b.north) (x2b.east) (tmp3.east) (tmp) (tmp2.west) (x1b.west) (x1b.north)};
                \end{tikzpicture}
   }] (E);
   \draw[dashed] (D) -- (E);
  
   \draw[-,dashed,fill=blue!35, opacity=.7] (A) --(B) -- (C) --cycle;
   \draw[-,dashed,fill=blue!20, opacity=.4] (A) -- (C) -- (E) -- cycle;
   \draw[-,dashed,fill=red!30, opacity=.4] (B) -- (C) -- (D) -- cycle;
   \draw[-,dashed,fill=red!50, opacity=.45] (A) -- (C) --(D) -- cycle;
   \draw[-,fill=blue!30,opacity=.4] (A) -- (B) -- (E) -- cycle;
   \draw[-,fill=red!30,opacity=.5] (A) -- (B) -- (D) -- cycle;
 
  \end{scope}
 \end{tikzpicture} 
\end{equation*}
For any polytope in a projective space $\mathbb{P}^{N}$ with co-ordinates $\mathcal{Y}$, the canonical form $\Omega(\mathcal{Y}\,;\mathcal{P})$ is related to the volume of the dual polytope via 
\begin{equation}
 \Omega(\mathcal{Y}\,;\mathcal{P}) = \langle\mathcal{Y} d^{N-1}\mathcal{Y} \rangle {\rm Vol}(\mathcal{Y};\tilde{\mathcal{P}})
\end{equation}
where Vol$(\mathcal{Y};\tilde{\mathcal{P}})$ is the dual of the polytope with $\mathcal{Y}$ taken as the hyperplane at infinity. 

Then $\Psi_{\mathcal{G}}(\mathcal{Y})$ is literally the volume of $\tilde{\mathcal{P}}_{\mathcal{G}}$ (with $\mathcal{Y}$ specifying the hyperplane at infinity). As we will see in concrete examples (and can be straightforwardly proven), there is a natural triangulation of $\tilde{\mathcal{P}}_{\mathcal{G}}$ into simplices, such that the volume of each simplex corresponds precisely to a single term in OFPT. There are also nice triangulation of $\mathcal{P}$ itself that reproduce the ``time integral'' expressions, and still others with no pre-existing physical interpretation.


\subsection{The Scattering Amplitude Face}\label{subsec:SAF}

One of the faces of the Cosmological Polytope is of special interest. We have already remarked that the flat-space scattering amplitude is associated with a pole in the wavefunction where $\sum_v{x_v}\,\longrightarrow\,0$. Restricting ourselves to the tree diagrams, the residue of this pole should match the product of all the flat-space propagators $1/P^2_e$ of the graph, both giving the amplitude and manifesting the Lorentz-invariance of the flat-space theory. We will now see how this simple and fundamental fact follows straightforwardly from the geometry of the facet of the cosmological polytope where $\mathcal{W}_I\,=\,\sum_v{{\bf x}_{v\,I}}$. 

We begin with a trivial observation. This facet has exactly $2E$ vertices; Since a tree graph has $V=E+1$, we can write $2E\,=\,E+V-1$ and thus this $(E+V-2)$-dimensional face has $(E+V-1)$ vertices and is therefore just a simplex $\mathcal{S}$!

The residue of the canonical form $\Omega$ on  $\sum_v{x_v}\,\longrightarrow\,0$ is thus just the canonical form of this simplex, which is (trivially) simply the inverse of the product of all the linear factors defining the faces of $\mathcal{S}$.

Now, since $\mathcal{S}$ is a simplex, each face of $\mathcal{S}$ is also a simplex; each face corresponds to leaving out one of the $2E$ vertices of $\mathcal{S}$ and taking the lower-dimensional simplex formed by the others. 

Each vertex of $\mathcal{S}$ is either 
$
 \begin{tikzpicture}[ball/.style = {circle, draw, align=center, anchor=north, inner sep=0}, cross/.style={cross out, draw, minimum size=2*(#1-\pgflinewidth), inner sep=0pt, outer sep=0pt}]
  \node[ball,text width=.18cm,fill,color=black,label=above:{\tiny $v\phantom{'}$}] at (0,0) (v1) {};
  \node[ball,text width=.18cm,fill,color=black,label=above:{\tiny $v'$},right=1.5cm of v1.east] (v2) {};  
  \draw[-,thick,color=black] (v1.east) edge node [text width=.18cm,above=-0.05cm,midway] {\tiny $e$} (v2.west);
  \node[thick, cross=3pt, rotate=0, color=black, right=.25cm of v1.east]{};
 \end{tikzpicture}
$
 ${\bf x}_v+{\bf y}-{\bf x}_{v'}$ or 
$
 \begin{tikzpicture}[ball/.style = {circle, draw, align=center, anchor=north, inner sep=0}, cross/.style={cross out, draw, minimum size=2*(#1-\pgflinewidth), inner sep=0pt, outer sep=0pt}]
  \node[ball,text width=.18cm,fill,color=black,label=above:{\tiny $v\phantom{'}$}] at (0,0) (v1) {};
  \node[ball,text width=.18cm,fill,color=black,label=above:{\tiny $v'$},right=1.5cm of v1.east] (v2) {};  
  \draw[-,thick,color=black] (v1.east) edge node [text width=.18cm,above=-0.05cm,midway] {\tiny $e$} (v2.west);
  \node[thick, cross=3pt, rotate=0, color=black, left=.25cm of v2.west]{};
 \end{tikzpicture}
$ 
 ${\bf x}_{v'}+{\bf y}-{\bf x}_v$. So each face of $\mathcal{S}$ can be represented by decorating the vertex of $\mathcal{S}$ being left off wit an open circle on one or other side of an edge of the graph. For instance, we can have
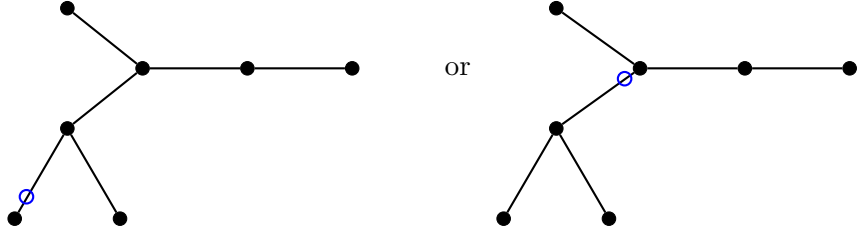
\begin{equation*}
 \begin{tikzpicture}[ball/.style = {circle, draw, align=center, anchor=north, inner sep=0}, cross/.style={cross out, draw, minimum size=2*(#1-\pgflinewidth), inner sep=0pt, outer sep=0pt}]
  \node[ball,text width=.18cm,fill,color=black] at (0,0) (v1) {};  
  \node[ball,text width=.18cm,fill,color=black,right=1.2cm of v1.east] (v2) {};    
  \node[ball,text width=.18cm,fill,color=black,right=1.2cm of v2.east] (v3) {};
  \node[ball,text width=.18cm,fill,color=black] at (-1,.8) (v4) {};    
  \node[ball,text width=.18cm,fill,color=black] at (-1,-.8) (v5) {};    
  \node[ball,text width=.18cm,fill,color=black] at (-1.7,-2) (v6) {};    
  \node[ball,text width=.18cm,fill,color=black] at (-.3,-2) (v7) {};    
  \node[right=1cm of v3.east] (or) {or$\qquad$};
  \node[ball,text width=.18cm,fill,color=black,right=1.25cm of or.east] (x1) {};
  \node[ball,text width=.18cm,fill,color=black,right=1.2cm of x1.east] (x2) {};    
  \node[ball,text width=.18cm,fill,color=black,right=1.2cm of x2.east] (x3) {};
  \node[ball,text width=.18cm,fill,color=black] at (5.5,.8) (x4) {};    
  \node[ball,text width=.18cm,fill,color=black] at (5.5,-.8) (x5) {};    
  \node[ball,text width=.18cm,fill,color=black] at (4.8,-2) (x6) {};    
  \node[ball,text width=.18cm,fill,color=black] at (6.2,-2) (x7) {}; 

  \node[above=.1cm of v6.north] (ref1) {};
  \coordinate (Int1) at (intersection of v5--v6 and ref1--v7);

  \node[above=.35cm of x5.north] (ref2) {};
  \coordinate (Int2) at (intersection of x5--x1 and ref2--x2);  
  
  \draw[-,thick,color=black] (v1) -- (v2) -- (v3); 
  \draw[-,thick,color=black] (v1) -- (v4);
  \draw[-,thick,color=black] (v5) -- (v1);
  \draw[-,thick,color=black] (v5) -- (v7);   
  \draw[-,thick,color=black] (v5) -- (v6); 
  \node[ball,text width=.18cm,thick,color=blue,left=-.09cm of Int1] {};

  \draw[-,thick,color=black] (x1) -- (x2) -- (x3); 
  \draw[-,thick,color=black] (x1) -- (x4);
  \draw[-,thick,color=black] (x5) -- (x1);
  \draw[-,thick,color=black] (x5) -- (x7);   
  \draw[-,thick,color=black] (x5) -- (x6); 
  \node[ball,text width=.18cm,thick,color=blue,left=-.09cm of Int2] {};
 \end{tikzpicture}
\end{equation*}
Since we are dealing with tree diagrams, the circled edge splits the graph into two parts, which we can label ``$\mathcal{C}$'' for close to the circle, and $\mathcal{F}$ far from the circle as in:
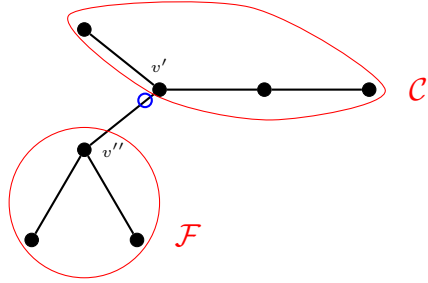
\begin{equation*}
 \begin{tikzpicture}[ball/.style = {circle, draw, align=center, anchor=north, inner sep=0}, cross/.style={cross out, draw, minimum size=2*(#1-\pgflinewidth), inner sep=0pt, outer sep=0pt}]
  \node[ball,text width=.18cm,fill,color=black,label=above:{\tiny $v'$}] at (0,0) (x1) {};    
  \node[ball,text width=.18cm,fill,color=black,right=1.2cm of x1.east] (x2) {};    
  \node[ball,text width=.18cm,fill,color=black,right=1.2cm of x2.east] (x3) {};
  \node[ball,text width=.18cm,fill,color=black] at (-1,.8) (x4) {};    
  \node[ball,text width=.18cm,fill,color=black,label=right:{\tiny $v''$}] at (-1,-.8) (x5) {};    
  \node[ball,text width=.18cm,fill,color=black] at (-1.7,-2) (x6) {};    
  \node[ball,text width=.18cm,fill,color=black] at (-.3,-2) (x7) {};

  \node[above=.35cm of x5.north] (ref2) {};
  \coordinate (Int2) at (intersection of x5--x1 and ref2--x2);  

  \coordinate (t1) at (x3.east);
  \coordinate (t2) at (x4.west);
  \coordinate (t3) at (x1.south west);
  \coordinate (t4) at (x2.south);

  \draw[-,thick,color=black] (x1) -- (x2) -- (x3); 
  \draw[-,thick,color=black] (x1) -- (x4);
  \draw[-,thick,color=black] (x5) -- (x1);
  \draw[-,thick,color=black] (x5) -- (x7);   
  \draw[-,thick,color=black] (x5) -- (x6); 
  \node[ball,text width=.18cm,thick,color=blue,left=-.09cm of Int2] {};
  \draw[red] plot [smooth cycle] coordinates {(3,-.1) (1.2,1) (-1.2,.9) (t3) (1.5,-.5)};
  \node[color=red,right=.3cm of x3.east] {\large $\mathcal{C}$};

  \draw[red] (-1,-1.6) circle (1cm);
  \node[color=red,right=.3cm of x7.north east] {\large $\mathcal{F}$};  
 \end{tikzpicture}  
\end{equation*}
We would now like to explicitly determine the ${\bf \omega}_{I}$ associated with one of those sets of $(2E-1)$ vertices. It is easy to see that 
\begin{equation}\label{eq:wI}
 {\bf\omega}_I\:=\:{\bf\tilde{Y}}_{e_I}+\sum_{v\in\mathcal{C}}{\bf\tilde{X}}_{v_I}.
\end{equation}
Trivially, ${\bf\omega}_I$ annihilates all the vertices of $\mathcal{S}$ that do not involve edges touching the vertices of $e$, including all vertices of $\mathcal{S}$ associated with edges of $\mathcal{F}$. 

Now, consider any edge in $\mathcal{C}$ touching $v'$. Since the vertices of $\mathcal{S}$ involve differences of ${\bf x}_v$'s they are annihilated by ${\bf\omega}_I$. Finally, consider the one vertex of $\mathcal{S}$ associated with the uncircled end of $e$, ${\bf x}_{v''}-{\bf x}_{v}+{\bf y}_e$; Due to the relative $-$ sign between ${\bf y}_e$ and ${\bf x}_{v'}$, this is also annihilated by ${\bf\omega}_I$. Now, ${\bf\omega}\cdot\mathcal{Y}\,=\,y_e+\sum_{v\in\mathcal{C}}x_v$. Furthermore, for each edge we have the two circling of the edge, on one end or the other. Thus, now denoting the two subgraphs bonneted by any edge $e$ to be $\mathcal{L}_e$ and $\mathcal{R}_e$, the canonical form of the simplex $\mathcal{S}$ is proportional to
\begin{equation}
 \prod_{e\in\mathcal{E}_{\mathcal{S}}} \frac{1}{\displaystyle y_e+\sum_{v\,\in\,\mathcal{L}_e} x_v} \,  \frac{1}{\displaystyle y_e+\sum_{v\,\in\,\mathcal{R}_e} x_v} 
\end{equation}
If we recall that $\sum_v x_v\,=\,0$ on this face, we have $\sum_{v\,\in\,\mathcal{L}_e} x_v\,=\,-\sum_{v\,\in\,\mathcal{R}_e} x_v$ and so this expression precisely takes the form of a product of propagators
\begin{equation}
 \prod_{e\in\mathcal{E}_{\mathcal{S}}} \frac{1}{\displaystyle y_e^2-\left(\sum_{v\,\in\,\mathcal{L}_e} x_v \right)^2},
\end{equation}
as we wanted to establish.


\section{Bulk and Boundary Representations From Triangulations of the Polytope}\label{sec:PP}

Given any polytope, it is natural to think about ``triangulating" it in various ways, expressing it the union of a collection of simplices. This is especially useful in the context of determining the canonical form associated with the polytope. The canonical form for a simplex is trivial -- just having poles corresponding to its facets. Given a triangulation of the polytope, the sum of the canonical forms for the simplices give the canonical form for the polytope. This can be even more geometrically understood by the connection between the canonical form and the volume of the {\it dual} polytope. Here it is obvious that the sum of the volumes of simplices in a triangulation of the dual polytope gives the volume of the polytope. It is also clear that a triangulation of the dual polytope can be dualized to the triangulation of the original polytope (though in general it may be an ``inclusion-exclusion" triangulation). 

In the story connecting scattering amplitudes and the canonical form for positive geometries, we have seen that  natural triangulations of positive geometries are associated with natural representations of scattering amplitudes; for instance the BCFW representation of gluon scattering amplitudes is associated with simple triangulations of the amplituhedron. In this section we will see the analogous facts for cosmological polytopes. For the convenience of visualizing some of our examples, we will talk in terms of triangulating the dual of the cosmological polytopes though of course everything can be phrased in terms of the polytope itself as well. 

We will see that old-fashioned perturbation theory corresponds to a very simple triangulation of the dual polytope, which does not introduce any new vertices (and thus does not introduce any spurious poles in the separate pieces that add up to the full canonical form). The time-integral representation is associated with a different type of triangulation, that is forced to use a natural set of points as ``anchors'', these do introduce spurious poles that of course cancel in the sum. There are also many other kinds of triangulation; some corresponding to the recursive computation of tree graphs we discussed in Section 2, and others with no standard physical interpretation. One of these turns out to confine the total energy pole to just a sub-set of terms ({\it i.e.} the related vertex belongs to just a sub-set of the simplices in the triangulation), which thus solely encode the information about the flat-space S-matrix, while the other terms have vanishing residue on the $\sum_v x_v \to 0$ pole and don't contribute to the amplitude, while of course being necessary for the full wavefunction. We will leave a systematic exposition of triangulations of our polytopes, along with the proofs connecting them to old-fashioned perturbation theory and time-integral representations, to future work. In this section we will instead content ourselves for describing these triangulations in various simple examples.

Let us consider some concrete example, starting with the two-site graph at tree level \eqref{eq:2sOFPT}. The graph is characterized by $V\,=\,2$ vertices ($x_1,\,x_2$) and $E\,=\,1$ edge $y$, so that it lives in $\mathbb{P}^2$. Let us choose the line at infinity to be $\mathcal{Y}\,\equiv\,(x_1,\,y,\,x_2)^{\mbox{\tiny $T$}}$. 

The connected subgraphs in which the two-site graph can be divided are three (see  \eqref{eq:2sOFPT}) and thus the polytope associated to the graph is a triangle:

\begin{equation}\label{eq:2sTr}
 \begin{tikzpicture}[shift={(1,0)}, line join = round, line cap = round, ball/.style = {circle, draw, align=center, anchor=north, inner sep=0}]
  \begin{scope}[shift={(0,1)}]
   \coordinate [label=above:$3$] (A) at (0,0);
   \coordinate [label=left:$1$] (B) at (-1.5,-1.5);
   \coordinate [label=right:$2$] (C) at (1.5,-9/4);

   \draw [name path=P1] ($(A)!-0.5cm!(B)$) -- ($(B)!-0.5cm!(A)$);
   \draw [name path=P2] ($(B)!-0.5cm!(C)$) -- ($(C)!-0.5cm!(B)$);
  \draw [name path=P3] ($(C)!-0.5cm!(A)$) -- ($(A)!-0.5cm!(C)$);

  \draw[-,fill=blue!40, opacity=.5] (A) -- (B) -- (C) -- cycle;
\end{scope}
\node (eqt) at (4,0) {$\displaystyle\:=\:\frac{1}{2}\frac{\langle123\rangle}{\langle1,\mathcal{Y}\rangle\langle2,\mathcal{Y}\rangle\langle3,\mathcal{Y}\rangle},$};
\node[right=1cm of eqt.east] (vct) {$\displaystyle
                                        \left\{
                                        \begin{array}{l}
                                         X^{\mbox{\tiny $(1)$}}\,=\,(1,1,0),\\
                                         \phantom{\ldots}\\
                                         X^{\mbox{\tiny $(2)$}}\,=\,(0,1,1),\\
                                         \phantom{\ldots}\\
                                         X^{\mbox{\tiny $(3)$}}\,=\,(1,0,1)
                                        \end{array}
                                        \right.$};
\end{tikzpicture} 
\end{equation}
where $\langle123\rangle\,\equiv\,\varepsilon^{\alpha\beta\gamma}{\bf X}_{\alpha}^{\mbox{\tiny $(1)$}}{\bf X}_{\alpha}^{\mbox{\tiny $(2)$}}{\bf X}_{\alpha}^{\mbox{\tiny $(3)$}}$. The right-hand-side of \eqref{eq:2sTr} is the volume of the triangle, and it coincides with \eqref{eq:23sum}.

\begin{figure}
 \begin{tikzpicture}[shift={(1,0)}, line join = round, line cap = round, ball/.style = {circle, draw, align=center, anchor=north, inner sep=0}]
  \begin{scope}
   \node[ball,text width=.18cm,fill,color=white] at (0,0) (tmp1) {};
   \node[ball,text width=.18cm,fill,color=black,right=1cm of tmp1.east,label=below:$x_1$] (x1) {};
   \node[ball,text width=.18cm,fill,color=black,right=1.5cm of x1.east,label=below:$x_2$] (x2) {};
   \draw[-,thick,color=black] (x1.east) edge node [text width=.18cm,above=-0.05cm,midway] {{$y$}} (x2.west);
  \end{scope}
  \node[right=2cm of x2.west] (a) {};
  \node[right=4cm of x2.west] (b) {};
  \draw[solid,line width=.1cm,shade,preaction={-triangle 90,thin,draw,color=black,shorten >=-1mm}] (a) -- (b);    
  \begin{scope}[shift={(10.5,1)}]
   \coordinate [label=above:{
                \begin{tikzpicture}
                 \node[ball,text width=.1cm,fill,color=black,label=below:{\tiny $x_1$}] at (0,0) (x1b) {};
                 \node[ball,text width=.1cm,fill,color=black,right=1.05cm of x1b.east,label=below:{\tiny $x_2$}] (x2b) {};
                 \draw[-,thick,color=black] (x1b.east) edge node [text width=.1cm,above=-0.05cm,midway] {{\tiny $y$}} (x2b.west);
                 \draw (0.56,0) ellipse (.95cm and 0.25cm);
                \end{tikzpicture}
               }] (A) at (0,0);
   \coordinate [label=below:{
                \begin{tikzpicture}
                 \node[ball,text width=.1cm,fill,color=black,label=below:{\tiny $x_1$}] at (0,0) (x1b) {};
                 \node[ball,text width=.1cm,fill,color=black,right=1.05cm of x1b.east,label=below:{\tiny $x_2$}] (x2b) {};
                 \draw[-,thick,color=black] (x1b.east) edge node [text width=.1cm,above=-0.05cm,midway] {{\tiny $y$}} (x2b.west);
                 \draw (x1b.east) ellipse (.25cm and 0.15cm);
                \end{tikzpicture}
               }] (B) at (-1.5,-1.5);
   \coordinate [label=right:{
                \begin{tikzpicture}
                 \node[ball,text width=.1cm,fill,color=black,label=below:{\tiny $x_1$}] at (0,0) (x1b) {};
                 \node[ball,text width=.1cm,fill,color=black,right=1.05cm of x1b.east,label=below:{\tiny $x_2$}] (x2b) {};
                 \draw[-,thick,color=black] (x1b.east) edge node [text width=.1cm,above=-0.05cm,midway] {{\tiny $y$}} (x2b.west);
                 \draw (x2b.west) ellipse (.25cm and 0.15cm);
                \end{tikzpicture}
               }] (C) at (1.5,-9/4);

   \draw [name path=P1] ($(A)!-0.5cm!(B)$) -- ($(B)!-0.5cm!(A)$);
   \draw [name path=P2] ($(B)!-0.5cm!(C)$) -- ($(C)!-0.5cm!(B)$);
   \draw [name path=P3] ($(C)!-0.5cm!(A)$) -- ($(A)!-0.5cm!(C)$);
   \draw[-,fill=blue!40, opacity=.5] (A) -- (B) -- (C) -- cycle;
  \end{scope}
 \end{tikzpicture}
 \caption{The dual polytope associated to the tree-level two-site graph. Each of its vertices corresponds to a connected subgraph.}\label{fig:triang}
\end{figure}
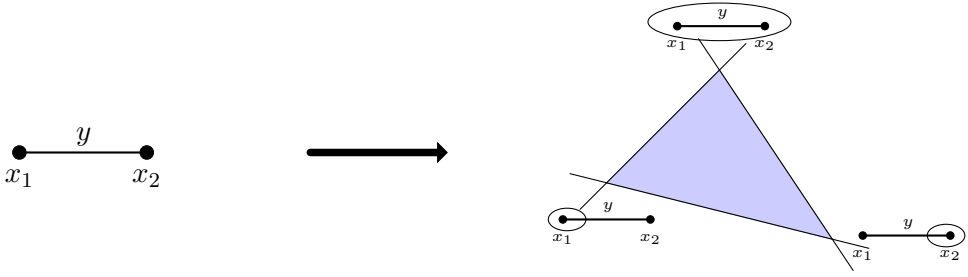
For this graph we discussed two further representations (see Sections \ref{subsec:TimeRep} and \ref{subsec:TreeRR}). They are characterized by being a sum of terms rather than a single term, and by the appearance of spurious poles ($1/y$ in the case of the time representation, and $1/(y-x_1)$ for the recursive one). What is their interpretation in this dual polytope picture? As just pointed out, both of them involve a spurious singularity. Let us introduce two new points corresponding to these singularities, namely ${\bf X}^{\mbox{\tiny $(4)$}}\,\equiv\,(-1,1,0)$ and ${\bf X}^{\mbox{\tiny $(5)$}}\,=\,(0,1,0)$. Note that in the notation of section 3.1 we can also recognize ${\bf X}^{\mbox{\tiny $(4)$}} = \tilde{{\bf Y}} - \tilde{{\bf X}}_1$ and ${\bf X}^{\mbox{\tiny $(5)$}}\,= \tilde{{\bf Y}}$. We then insist on ``triangulating the triangle" (123), enforcing that every triangle in the triangulation has the new point. Thus for instance, we can recognize the interior of $(123)$ as the union of $(513)$ and $(532)$, then subtracting the extra piece $(512)$. This gives us the time-integral representation, which indeed had three terms, with two ``plus" signs and one ``minus" sign. Or, we can recognize the interior of $(123)$ as $(134)$ subtracting $(124)$; here it is important to note that the point $4$ is indeed on the line $(32)$. This gives us the two-term expression for this graph from our recursion relation, which again was expressed as the difference between two terms. 

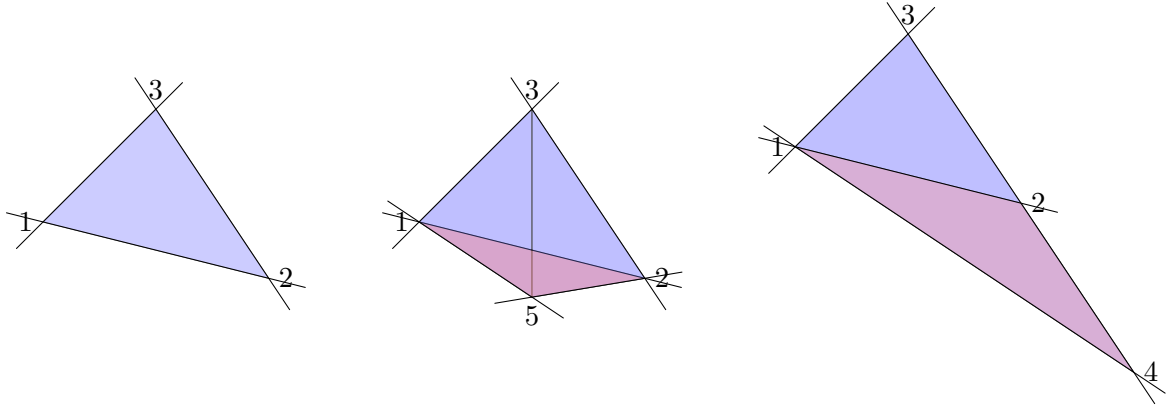
\begin{figure}
 \centering
  \begin{tikzpicture}[line join = round, line cap = round, ball/.style = {circle, draw, align=center, anchor=north, inner sep=0}] 
   \begin{scope}
    \coordinate [label=above:$3$] (A) at (0,0);
    \coordinate [label=left:$1$] (B) at (-1.5,-1.5);
    \coordinate [label=right:$2$] (C) at (1.5,-9/4);

    \draw [name path=P1] ($(A)!-0.5cm!(B)$) -- ($(B)!-0.5cm!(A)$);
    \draw [name path=P2] ($(B)!-0.5cm!(C)$) -- ($(C)!-0.5cm!(B)$);
    \draw [name path=P3] ($(C)!-0.5cm!(A)$) -- ($(A)!-0.5cm!(C)$);
  
    \draw[-,fill=blue!40, opacity=.5] (A) -- (B) -- (C) -- cycle;
   \end{scope}

   \begin{scope}[shift={(10,1)}]
    \coordinate [label=above:$3$] (A) at (0,0);
    \coordinate [label=left:$1$] (B) at (-1.5,-1.5);
    \coordinate [label=right:$2$] (C) at (1.5,-9/4);
    \coordinate (Ab) at ($(B)!.5!(C)$);  
    \coordinate [label=right:$4$] (D) at (3,-9/2);   

    \draw [name path=T1] ($(A)!-0.5cm!(B)$) -- ($(B)!-0.5cm!(A)$);
    \draw [name path=T2] ($(A)!-0.5cm!(D)$) -- ($(D)!-0.5cm!(A)$);
    \draw [name path=T3] ($(B)!-0.5cm!(D)$) -- ($(D)!-0.5cm!(B)$);
    \draw[-,fill=blue!50,opacity=.5] (A) -- (B) -- (D) -- cycle;

    \draw [name path=T5] ($(B)!-0.5cm!(C)$) -- ($(C)!-0.5cm!(B)$);
    \draw[-,fill=red!40,opacity=.4] (B) -- (C) -- (D) -- cycle;
   \end{scope}

   \begin{scope}[shift={(5,0)}]
    \coordinate [label=above:$3$] (A) at (0,0);
    \coordinate [label=left:$1$] (B) at (-1.5,-1.5);
    \coordinate [label=right:$2$] (C) at (1.5,-9/4);
    \coordinate (Ab) at ($(B)!.5!(C)$);  
   \coordinate (D) at (3,-9/2);   
    \coordinate (E) at (-3,-3);

    \coordinate [label=below:$5$] (Y) at (intersection of B--D and C--E);  
 
    \draw [name path=T1] ($(A)!-0.5cm!(B)$) -- ($(B)!-0.5cm!(A)$);
    \draw [name path=T2] ($(A)!-0.5cm!(C)$) -- ($(C)!-0.5cm!(A)$);
    \draw [name path=T3] ($(B)!-0.5cm!(Y)$) -- ($(Y)!-0.5cm!(B)$);
    \draw [name path=T4] ($(C)!-0.5cm!(Y)$) -- ($(Y)!-0.5cm!(C)$);  
    \draw [name path=75] ($(B)!-0.5cm!(C)$) -- ($(C)!-0.5cm!(B)$);  
    \draw[-,fill=blue!50,opacity=.5] (A) -- (B) -- (Y) -- cycle;
    \draw[-,fill=blue!50,opacity=.5] (A) -- (C) -- (Y) -- cycle;
    \draw[-,fill=red!50,opacity=.4] (B) -- (C) -- (Y) -- cycle;
    \end{scope}
   \end{tikzpicture}
\caption{Dual polytope picture for the OFPT, the tree-level recursion relation, and the time integral representations respectively. While the OFPT is given by a single term, the other two representations involve a triangulation through
          a further vertex. This signalize the presence of spurious singularities.}
 \label{Fig:2sTriangD}
\end{figure}


  
   

Let us move on to the next-to-simplest case, provided by the two-site graph at one loop. It is characterized by $V\,=\,2$ vertices and $E\,=\,2$ edges. Hence, the associated polytope lives in $\mathbb{P}^3$, whose hyperplane at infinity is defined as $\mathcal{Y}\,\equiv\,(x_1,\,y_a,\,y_b,\,x_2)^{\mbox{\tiny $T$}}$. The graph can be divided into $5$ connected subgraph, and, consequently, as shown in section 3.2,  the dual polytope is characterized by five vertices:
\begin{equation}\label{eq:2sLoop}
 \begin{tikzpicture}[line join = round, line cap = round, ball/.style = {circle, draw, align=center, anchor=north, inner sep=0}]
  \begin{scope}[shift={(-1.5,-.8)}]
   \pgfmathsetmacro{\factor}{1/sqrt(2)};
   \coordinate [label=right:$1$] (A) at (2,0,-2*\factor);
   \coordinate [label=left:$5$] (B) at (-2,0,-2*\factor);
   \coordinate [label=above:$2$] (C) at (0,2,2*\factor);
   \coordinate (Ab) at ($(B)!.5!(C)$); 
   \coordinate (Bb) at ($(A)!.5!(C)$);
   \coordinate (Cb) at ($(A)!.5!(B)$);
   \coordinate (G) at (intersection of A--Ab and B--Bb);
   \coordinate[below=2cm of G.south, label=below:$4$] (D);
   \coordinate[above=2cm of G.north, label=above:$3$] (E);
   \draw[dashed] (D) -- (E);
 
   \draw[-,dashed,fill=blue!35, opacity=.7] (A) --(B) -- (C) --cycle;
   \draw[-,dashed,fill=blue!20, opacity=.4] (A) -- (C) -- (E) -- cycle;
   \draw[-,dashed,fill=red!30, opacity=.4] (B) -- (C) -- (D) -- cycle;
   \draw[-,dashed,fill=red!50, opacity=.45] (A) -- (C) --(D) -- cycle;
   \draw[-,fill=blue!30,opacity=.4] (A) -- (B) -- (E) -- cycle;
   \draw[-,fill=red!30,opacity=.5] (A) -- (B) -- (D) -- cycle;

  \end{scope}
  \node (eqt) at (5,0) {$\displaystyle\:=\:\frac{1}{\langle1,\mathcal{Y}\rangle\langle2,\mathcal{Y}\rangle\langle5,\mathcal{Y}\rangle}\left[\frac{\langle1235\rangle}{\langle3\mathcal{Y}\rangle}+\frac{\langle1254\rangle}{\langle4\mathcal{Y}\rangle}\right],$};  
 \end{tikzpicture}
\end{equation}
where the vertices are given by ${\bf X}^{\mbox{\tiny $(1)$}}\,=\,(1,1,1,0)$, ${\bf X}^{\mbox{\tiny $(2)$}}\,=\,(0,1,1,1)$, ${\bf X}^{\mbox{\tiny $(3)$}}\,=\,(1,1,0,1)$, ${\bf X}^{\mbox{\tiny $(4)$}}\,=\,(1,0,1,1)$ and 
${\bf X}^{\mbox{\tiny $(5)$}}\,=\,(1,0,0,1)$. The two terms in the right-hand-side of \eqref{eq:2sLoop} correspond to the most obvious triangulation of the polytope on the left-hand-side, simply equating the volume of the polytope with the sum of the volumes of the tetrahedra $\widehat{1253}$ and $\widehat{1254}$. Notice that the vertices are related by ${\bf X}^{\mbox{\tiny $(1)$}}\,+\,{\bf X}^{\mbox{\tiny $(2)$}}\,+\,{\bf X}^{\mbox{\tiny $(5)$}}\,\sim\,{\bf X}^{\mbox{\tiny $(3)$}}\,+\,{\bf X}^{\mbox{\tiny $(4)$}}$, which geometrically translates into the fact that the line identified by the vertices $3$ and $4$ intersects the triangle $\widehat{125}$ in its centroid.

This dual polytope also has another obvious triangulation into simplices which don't introduce any new (spurious) vertices:
\begin{equation}\label{eq:2sLoop2}
 \begin{tikzpicture}[line join = round, line cap = round, ball/.style = {circle, draw, align=center, anchor=north, inner sep=0}]
  \begin{scope}[shift={(-1.5,-.8)}]
   \pgfmathsetmacro{\factor}{1/sqrt(2)};
   \coordinate [label=right:$1$] (A) at (2,0,-2*\factor);
   \coordinate [label=left:$5$] (B) at (-2,0,-2*\factor);
   \coordinate [label=above:$2$] (C) at (0,2,2*\factor);
   \coordinate (Ab) at ($(B)!.5!(C)$); 
   \coordinate (Bb) at ($(A)!.5!(C)$);
   \coordinate (Cb) at ($(A)!.5!(B)$);
   \coordinate (G) at (intersection of A--Ab and B--Bb);
   \coordinate[below=2cm of G.south, label=below:$4$] (D);
   \coordinate[above=2cm of G.north, label=above:$3$] (E);
   \draw[dashed] (D) -- (E);

    \draw[-,dashed,fill=blue!25, opacity=.3] (B) -- (C) -- (E) -- cycle;
    \draw[-,dashed,fill=blue!30, opacity=.3] (C) -- (E) -- (G) -- cycle;
    \draw[-,dashed,fill=blue!20, opacity=.4] (E) -- (G) -- (B) -- cycle;
    \draw[-,dashed,fill=blue!25, opacity=.3] (B) -- (C) -- (D) -- cycle;
    \draw[-,dashed,fill=blue!30, opacity=.3] (C) -- (G) -- (D) -- cycle;
    \draw[-,dashed,fill=blue!30, opacity=.2] (B) -- (G) -- (D) -- cycle;
    \draw[-,dashed,fill=red!40, opacity=.3] (E) -- (C) -- (A) -- cycle;
    \draw[-,dashed,fill=red!45, opacity=.3] (E) -- (G) -- (A) -- cycle;    
    \draw[-,dashed,fill=red!45, opacity=.3] (D) -- (G) -- (A) -- cycle;        
    \draw[-] (A) -- (B);
    \draw[-] (B) -- (E) -- (A) -- (D) -- (B);

  \end{scope}
  \node (eqt) at (6,0) {$\displaystyle\:=\:\frac{1}{\langle3,\mathcal{Y}\rangle\langle4,\mathcal{Y}\rangle}\left[\frac{\langle1234\rangle}{\langle1\mathcal{Y}\rangle\langle2\mathcal{Y}\rangle}+\frac{\langle1534\rangle}{\langle1\mathcal{Y}\rangle\langle5\mathcal{Y}\rangle}+\frac{\langle5234\rangle}{\langle2Y\rangle\langle5Y\rangle}\right],$};  
 \end{tikzpicture}
\end{equation}
where now the polytope is expressed in terms of the sum of the tetrahedra $\widehat{1234}$, $\widehat{1534}$ and $\widehat{5234}$. This represents a new decomposition for the perturbative wavefunction. Its main feature is to separate the terms which have the total energy pole (which in \eqref{eq:2sLoop2} is given by $\langle5Y\rangle$) and, thus, contribute to the flat space scattering amplitude, from those terms which do not show this pole and thus characterize the wavefunction only.

In general, old-fashioned perturbation theory corresponds to a natural triangulation of the dual polytope into simplices formed by it's vertices. By contrast, the ``time-integral" representation is an ``external" triangulation of the (dual) polytope. Note that since every internal propagator has a factor of $1/y_e$, every term in the time-integral representation has a pole $1/(\prod_e y_e)$ corresponding the product of all the edge variables.  This pole is completely absent in the final result, but their presence indicates a triangulation of the dual polytope into simplices, all of which include all the vertices $\tilde{\bf{Y}}_e$. We have already seen an example with the  three-piece ``triangulation of a triangle", a more interesting example is offered by the two-site loop graph: 
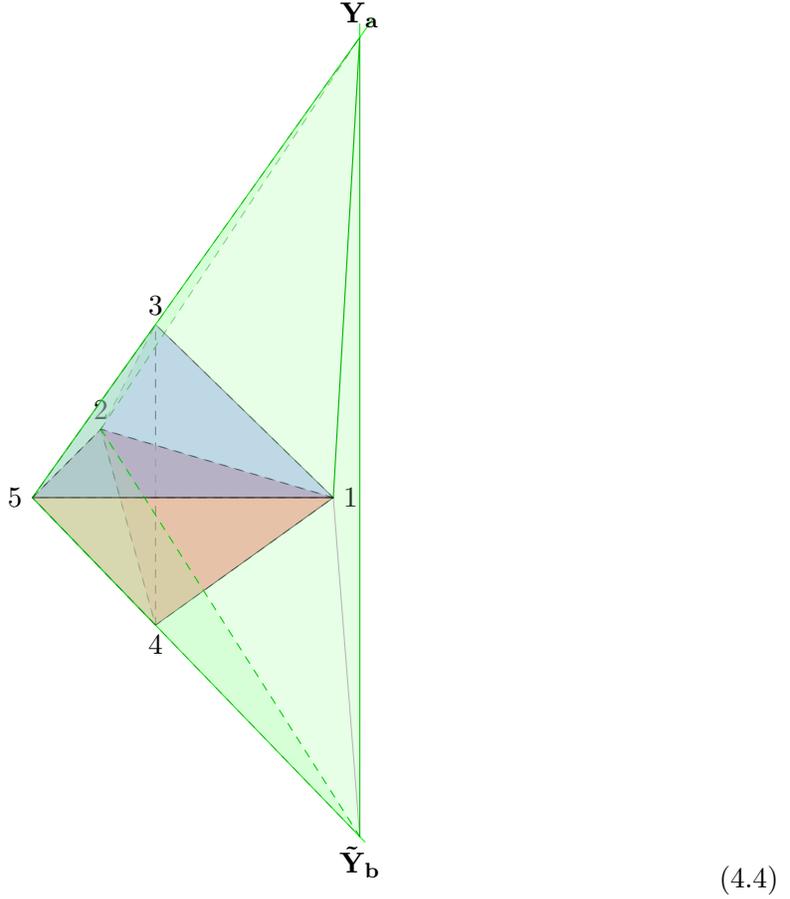
\begin{equation} 
 \begin{tikzpicture}[line join = round, line cap = round, ball/.style = {circle, draw, align=center, anchor=north, inner sep=0}]
  \begin{scope}[shift={(-1.5,-.8)}]
   \pgfmathsetmacro{\factor}{1/sqrt(2)};
   \coordinate [label=right:$1$] (A) at (2,0,-2*\factor);
   \coordinate [label=left:$5$] (B) at (-2,0,-2*\factor);
   \coordinate [label=above:$2$] (C) at (0,2,2*\factor);
   \coordinate (Ab) at ($(B)!.5!(C)$); 
   \coordinate (Bb) at ($(A)!.5!(C)$);
   \coordinate (Cb) at ($(A)!.5!(B)$);
   \coordinate (G) at (intersection of A--Ab and B--Bb);
   \coordinate[below=2cm of G.south, label=below:$4$] (D);
   \coordinate[above=2cm of G.north, label=above:$3$] (E);
   \draw[dashed] (D) -- (E);
  
   \draw[-,dashed,fill=blue!35, opacity=.7] (A) --(B) -- (C) --cycle;
   \draw[-,dashed,fill=blue!20, opacity=.4] (A) -- (C) -- (E) -- cycle;
   \draw[-,dashed,fill=red!30, opacity=.4] (B) -- (C) -- (D) -- cycle;
   \draw[-,dashed,fill=red!50, opacity=.45] (A) -- (C) --(D) -- cycle;
   \draw[-,fill=blue!30,opacity=.4] (A) -- (B) -- (E) -- cycle;
   \draw[-,fill=red!30,opacity=.5] (A) -- (B) -- (D) -- cycle;
 
   \draw [color=green,name path=P1] ($(B)!-0.0cm!(D)$) -- ($(D)!-4cm!(B)$);   
   \draw [color=green,name path=P2] ($(B)!-0.0cm!(E)$) -- ($(E)!-5cm!(B)$);    
   \coordinate [label=below:${\bf\tilde{Y}_b}$] (yb) at ($(D)!-3.9cm!(B)$);
   \coordinate (tmp) at (G -| yb) ;
   \draw[-,color=green,name path=P3] ($(yb)!-0.0cm!(tmp)$) -- ($(tmp)!-6cm!(yb)$);
   \draw[color=green,name intersections={of=P2 and P3,by={yt}}] (yt)--(A) ;
   \coordinate [label=above:${\bf\tilde{Y}_a}$] (ya) at (yt);
   \draw[-,dashed,color=green,opacity=.3] (C) -- (ya);
   \draw[-,dashed,color=green] (C) -- (yb);
   \draw[dashed,fill=green!30,opacity=.2] (C) -- (yb) -- (B) -- cycle;
   \draw[dashed,fill=green!30,opacity=.2] (C) -- (ya) -- (B) -- cycle;   
   \draw[-,fill=green,opacity=.1] (A) -- (yb) -- (B) --cycle; 
   \draw[-,fill=green,opacity=.1] (A) -- (ya) -- (B) --cycle;    
   \draw[-,fill=green!50,opacity=.2] (A) -- (yb) -- (ya) --cycle;    
  \end{scope}
 \end{tikzpicture}
\end{equation}
Note a few non-trivial features of the geometry: the points $(3,5,{\bf\tilde{Y}}_a)$ are collinear, as are $(4,5,{\bf\tilde{Y}}_b)$. Furthermore, $(1,2,3,{\bf\tilde{Y}}_b)$ are coplanar, and $(1,2,4,{\bf\tilde{Y}}_a)$ are also coplanar. There is a natural seven-simplex ``inclusion-exclusion'' triangulation with simplices that all include both $\tilde{\bf{Y}}_a, \tilde{\bf{Y}}_b$. The sum of the two simplices $(5,1,{\bf\tilde{Y}}_a,{\bf\tilde{Y}}_b) + (5,2,{\bf\tilde{Y}}_a,{\bf\tilde{Y}}_b)$ covers the entire space. We then subtract the two square pyramids $({\bf\tilde{Y}}_a;1,2,3,{\bf\tilde{Y}}_b)$ and $({\bf\tilde{Y}_a};1,2,4,{\bf\tilde{Y}_a})$. This oversubtracts the simplex $(1,2,{\bf\tilde{Y}}_a,{\bf\tilde{Y}}_b)$, so we add that back in. Each of the square pyramids $({\bf\tilde{Y}}_a;1,2,3,{\bf\tilde{Y}}_b)$ and $({\bf\tilde{Y}}_b;1,2,4,{\bf\tilde{Y}}_a)$ can be triangulated with two simplices, {\it e.g.} $({\bf\tilde{Y}}_a;1,2,3,{\bf\tilde{Y}}_b) = ({\bf\tilde{Y}}_a,{\bf\tilde{Y}}_b,2,3) + ({\bf\tilde{Y}}_a,{\bf\tilde{Y}}_b,1,3)$. Thus in total we have a triangulation with $2 + 4 + 1\,=\,7$ simplices. Each term has the singularity $1/(y_a y_b)$, and can be identified with the terms in the time integral representation.

As a last example of the dual cosmological polytope, let us consider the one related to the three-site graph, which is characterized by $V\,=\,3$ vertices and $E\,=\,2$ edges. The associated cosmological polytope and its dual live in $\mathbb{P}^{4}$. Interestingly, it has the feature to be self-dual, {\it i.e.} both the cosmological polytope (which has been discussed earlier) and its dual have the same shape of a {\it double square pyramid}. If we choose the hyperplane at infinity to be $\mathcal{Y}\,=\,(x_1,\,y_{12},\,x_2,\,y_{23},\,x_3)^{\mbox{\tiny T}}$, the dual polytope is give by the following six vertices, which, as we saw, correspond to the connected subgraphs in which the three-site graph can be divided: ${\bf X}^{1}\,=\,(1,1,0,0,0)$, ${\bf X}^{2}\,=\,(0,0,0,1,1)$, ${\bf X}^{3}\,=\,(1,0,1,1,0)$, ${\bf X}^{4}\,=\,(0,1,1,1,0)$, ${\bf X}^{5}\,=\,(0,1,1,0,1)$, ${\bf X}^{1}\,=\,(1,0,1,0,1)$. It is easy to see that the vertices $(3,4,5,6)$ are coplanar, with ${\bf X}^{4} + {\bf X}^{6}\,\sim\,{\bf X}^{3} + {\bf X}^{5}$, while the other two vertices live in different spaces and hence also the dual polytope is a {\it double square pyramid}:
\begin{equation}\label{eq:3sDCP}
 \begin{tikzpicture}[line join = round, line cap = round, ball/.style = {circle, draw, align=center, anchor=north, inner sep=0}]
  \begin{scope}
   \pgfmathsetmacro{\factor}{1/sqrt(2)};  
   \coordinate [label=right:{\footnotesize $4$}] (X4) at (10.5,-3,-1.5*\factor);
   \coordinate [label=left:{\footnotesize $3$}] (X3) at (7.5,-3,-1.5*\factor);
   \coordinate [label=right:{\footnotesize $5$}] (X5) at (10.5,-3.75,1.5*\factor);
   \coordinate [label=left:{\footnotesize $6$}] (X6) at (7.5,-3.75,1.5*\factor);  
   \coordinate [label=above:{\footnotesize $1$}] (X1) at (9.75,-.65,.75*\factor);
   \coordinate [label=below:{\footnotesize $2$}] (X2) at (9.4,-6.05,.75*\factor);
   

   \draw[-,dashed,fill=red!30, opacity=.6] (X3) -- (X4) -- (X1) -- cycle;

   \draw[-,dashed,fill=red!30, opacity=.4] (X3) -- (X4) -- (X2) -- cycle;
   \draw[-,dashed,fill=red!40, opacity=.4] (X3) -- (X6) -- (X2) -- cycle;
   

   \draw[-,dashed] (X6) -- (X4);
   \draw[-,dashed,thick,fill=blue!40,opacity=.4] (X6) -- (X5) -- (X2) -- cycle;
   \draw[-,dashed,thick,fill=blue!40,opacity=.7] (X4) -- (X5) -- (X2) -- cycle;   
   \draw[-,thick,fill=blue!35, opacity=.4] (X6) -- (X1) -- (X5) -- cycle;
   \draw[-,thick,fill=blue!35, opacity=.4] (X4) -- (X1) -- (X5) -- cycle;
   \draw[-,thick,fill=red!50, opacity=.4] (X6) -- (X3) -- (X1) -- cycle;

   \coordinate (tt) at ($(X5)!0.25!(X4)$);
   \node[right=1cm of tt.east] (tf) {$\displaystyle\:=\:\frac{1}{\langle6\mathcal{Y}\rangle\langle1\mathcal{Y}\rangle\langle2\mathcal{Y}\rangle\langle4\mathcal{Y}\rangle}
                                      \left[
                                       \frac{\langle61234\rangle}{\langle3\mathcal{Y}\rangle}+\frac{\langle61245\rangle}{\langle5\mathcal{Y}\rangle}
                                      \right]$};
  \end{scope}
 \end{tikzpicture}
\end{equation}
where the colors represent the OFPT triangulation which is returned by dividing the three-site graph into connected components: the first term in the right-hand-side of \eqref{eq:3sDCP} is a simplex $\widehat{12456}$ in $\mathbb{P}^{4}$, highlighted in the picture above in blue, while the second term is given by the simplex $\widehat{12346}$ highlighted in red. (The reader should again remember that this is not a three-dimensional object, these two simplices only touch on their common three-dimensional boundary given by the tetrahedron $\widehat{1246}$.). We can also see a different  two-term representation, which is given by the sum of the simplices $\widehat{12356}$ and $\widehat{12345}$. This new triangulation has the feature to show the total energy pole ({\it i.e.} the vertex $6$) just in one of the two terms, which therefore contains the entire flat space  flat-space scattering amplitude on the pole as $\sum_i x_i \to 0$. 

\section{New Representations of the Wavefunction}\label{sec:NRWF}

We have seen that some familiar representations of the wavefunction can be associated with natural triangulations of the cosmological polytope. Recent work \cite{Arkani-Hamed:2017tmz} has found representations of canonical forms for polytopes not based on any notion of ``triangulation". In this subsection we will apply to of these methods to cosmological polytopes, which will give us new representations for the wavefunction with no existing physical interpretation. One is closely related to the Grassmannian contour-integral formulas familiar from scattering amplitudes. The other is the ``push-forward'' method, which starts with an integer representative of the polytope of interest, and builds from this a ``push-forward'' from a simplex, one-to-one into the polytope.  

\subsection{Contour Integrals}\label{subsec:ContInt}

In \cite{Arkani-Hamed:2017tmz} a new way of generating canonical forms for a convex projective polytope (and related triangulations) was introduced as a contour integral. Let $\mathcal{P}$ be some polytope defined by a set of $\nu$ vertices $\mathcal{X}\,\equiv\,({\bf X}^{\mbox{\tiny $(1)$}},\,\ldots\,{\bf X}^{\mbox{\tiny $(\nu)$}})$ in $\mathbb{P}^{N-1}$, then the canonical rational function $\Omega(\mathcal{P})$ is given by 
\begin{equation}\label{eq:canform}
 \Omega(\mathcal{P})\:=\:\frac{1}{(2\pi i)^{\nu-N}(N-1)!}\int_{\mathbb{R}^N}\frac{d^{\nu}c}{\prod_{i=1}^{\nu}(c_i-i\epsilon_i)}\,
                         \delta^{\mbox{\tiny $(N)$}}\left(\mathcal{Y}-\sum_{i=1}^{\nu}c_i{\bf X}^{\mbox{\tiny $(i)$}}\right),
\end{equation}
which for $\nu\,=\,N$ defines a simplex in $\mathbb{P}^{N-1}$. For the polytopes of interest, this occurs just for the triangle in $\mathbb{P}^2$. As a non-trivial example, let us consider the polytope $\mathcal{P}^{\mbox{\tiny $(2,1)$}}$\footnote{Here the notation $\mathcal{P}^{\mbox{\tiny $(s,L)$}}$ indicates that the polytope is related to a graph with $s$ sites and $L$ loops.} associated with the two-site graph at one loop, which lives in $\mathbb{P}^3$ and has five vertices (see \eqref{eq:2sLoop}). Then
\begin{equation}\label{eq:canform21}
 \Omega(\mathcal{P}^{\mbox{\tiny $(2,1)$}})\:=\:\frac{1}{(2\pi i)\,3!}\int\frac{d^5 c}{\prod_{j=1}^5(c_j-i\epsilon_j)}\,\delta^{\mbox{\tiny $(4)$}}\left(\mathcal{Y}-\sum_{j=1}^5 c_j{\bf X}^{\mbox{\tiny $(j)$}}\right).
\end{equation}
The delta-functions fix four out of the five integration, so that the remaining variable will be integrated over the real axis. The contour of integration can be equivalently closed in the upper-half plane or in the lower one, taking just two different sub-set of the poles in \eqref{eq:canform30}. Geometrically, this translates in two different triangulations. If we choose $c_5$ to remain unfixed by the delta-function, then the latter constrains the other integration variables $c_i$ ($i\,=\,1,\ldots,\,4$) to depend on $c_5$ as follows:
\begin{equation}\label{eq:cf21poles}
 \begin{split}
  &\langle \mathcal{Y}234\rangle\:=\:-c_1\langle4123\rangle+c_5\langle5234\rangle,\\
  &\langle \mathcal{Y}341\rangle\:=\:c_2\langle4123\rangle-c_5\langle5431\rangle,\\
  &\langle \mathcal{Y}412\rangle\:=\:-c_3\langle4123\rangle-c_5\langle5142\rangle,\\
  &\langle \mathcal{Y}123\rangle\:=\:c_4\langle4123\rangle+c_5\langle5123\rangle,
 \end{split}
\end{equation}
where the coefficients of the $c_j$'s are written in such a way that they are positive. In the $c_5$ plane, we can see that the poles $c_j\:=\:i\epsilon_j$ ($j\:=\:5,\,1,\,2$) lie in the upper half-plane, the poles $j\,=\,3,\,4$ in the lower one. 
\begin{figure}
 \centering
 \begin{tikzpicture}[
    scale=5,
    axis/.style={very thick, ->, >=stealth'},
    pile/.style={thick, ->, >=stealth', shorten <=2pt, shorten
    >=2pt},
    every node/.style={color=black}
    ]
  \draw[axis] (-0.6,0) -- (+0.6,0) node(xline)[right]{Re$\{c_5\}$};
  \draw[axis] (0,-0.6) -- (0,+0.6) node(yline)[above]{Im$\{c_5\}$};
  \fill[red] (0,+0.25) circle (.4pt) node(pole5)[right] {$5$};
  \fill[red] (-0.31,+0.25) circle (.4pt) node(pole1)[right] {$1$};
  \fill[red] (+0.33,+0.25) circle (.4pt) node(pole2)[right] {$2$};  
  \fill[red] (-0.34,-0.25) circle (.4pt) node(pole3)[right] {$3$};  
  \fill[red] (+0.31,-0.25) circle (.4pt) node(pole4)[right] {$4$};    
 \end{tikzpicture}
 \caption{Poles location in the $c_5$-plane for the $\Omega(\mathcal{P}^{\mbox{\tiny $(2,1)$}})$ integral. The contour can be closed either in the upper half-plane, enclosing with counter-clockwise orientation the poles
         $1$, $2$, $5$; or in the lower half-plane, enclosing the poles $3$ and $4$ with clockwise orientation.}
 \label{fig:21poles}
\end{figure}
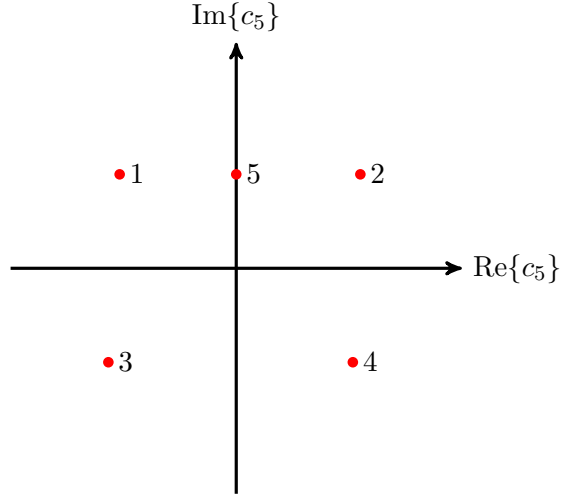
We can thus perform the integration by closing the contour of integration in either of the two half-planes. The different poles enclosed in the two contour provide a different decomposition, {\it i.e.} a different triangulation of the same object:
\begin{equation}\label{eq:cf21triang}
 \begin{split}
  \Omega(\mathcal{P}^{\mbox{\tiny $(2,1)$}})\:&=\:[1524]+[3152]\:=\hspace{3cm}\longleftarrow\quad\mbox{lower half-plane contour}\\
                                              &=\:[3415]+[3452]+[3421].\hspace{1.9cm}\longleftarrow\quad\mbox{upper half-plane contour}
 \end{split}
\end{equation}

As a second example, let us consider the polytope $\mathcal{P}^{\mbox{\tiny $(3,0)$}}$ in $\mathbb{P}^4$ associated to the three-site graph at tree level. If $\mathcal{Y}\,=\,(x_1,y_{12},x_2,y_{23},x_3)^{\mbox{\tiny T}}$ is the  hyperplane at infinity, its vertices are six and are given by ${\bf X}^{\mbox{\tiny $(1)$}}\,=\,(1,1,0,0,0)$, ${\bf X}^{\mbox{\tiny $(2)$}}\,=\,(0,0,0,1,1)$, ${\bf X}^{\mbox{\tiny $(3)$}}\,=\,(1,0,1,1,0)$, ${\bf X}^{\mbox{\tiny $(4)$}}\,=\,(0,1,1,1,0)$, ${\bf X}^{\mbox{\tiny $(5)$}}\,=\,(0,1,1,0,1)$ and ${\bf X}^{\mbox{\tiny $(6)$}}\,=\,(1,0,1,0,1)$. Furthermore, four of these vertices turn out to be coplanar, satisfying the relation ${\bf X}^{\mbox{\tiny $(3)$}}\,+\,{\bf X}^{\mbox{\tiny $(5)$}}\,\sim\,{\bf X}^{\mbox{\tiny $(4)$}}\,+\,{\bf X}^{\mbox{\tiny $(6)$}}$. The canonical rational form $\Omega(\mathcal{P}^{\mbox{\tiny $(3,0)$}})$ is therefore given by

\begin{equation}\label{eq:canform30}
 \Omega(\mathcal{P}^{\mbox{\tiny $(3,0)$}})\:=\:\frac{1}{(2\pi i)\,4!}\int\frac{d^6 c}{\prod_{j=1}^6(c_j-i\epsilon_j)}\,\delta^{\mbox{\tiny $(5)$}}\left(\mathcal{Y}-\sum_{j=1}^5 c_j{\bf X}^{\mbox{\tiny $(j)$}}\right). 
\end{equation}
As in the previous case, the delta functions localize five out of the six integration variables $c_j$. Choosing to leave $c_6$ as unfixed, then the delta function constrains the other variables to 
\begin{equation}\label{eq:cf30poles}
 \begin{split}
  &\langle \mathcal{Y}2345\rangle\:=\:-c_1\langle21345\rangle,\\
  &\langle \mathcal{Y}3451\rangle\:=\:-c_2\langle21345\rangle,\\
  &\langle \mathcal{Y}4512\rangle\:=\:-c_3\langle21345\rangle-c_6\langle21456\rangle,\\
  &\langle \mathcal{Y}5123\rangle\:=\:-c_4\langle21345\rangle+c_6\langle21356\rangle,\\
  &\langle \mathcal{Y}1234\rangle\:=\:-c_5\langle21345\rangle-c_6\langle21645\rangle.
 \end{split}
\end{equation}
Notice that $c_1$ and $c_2$ do not depend on $c_6$, which is a direct consequence of coplanarity of the vertices $3$, $4$, $5$ and $6$. The existence of special hyperplanes is a feature of all tree-level cosmological polytopes and is reflection of the symmetries of the polytope itself and, thus, of the theory, as we will see in Section \ref{sec:SymPol}.

In the $c_6$-plane, there are four poles, two of which lie in the upper half-plane and the other two in the lower one, so that closing the integration contour on the two halves provide again two different triangulations:
\begin{equation}\label{eq:cf30rep}
 \begin{split}
  \Omega(\mathcal{P}^{\mbox{\tiny $(3,0)$}})\:&=\:[12536]+[12354]\:=\hspace{2.7cm}\longleftarrow\quad\mbox{upper half-plane contour}\\
                                             &=\:[12364]+[12465].\hspace{3cm}\longleftarrow\quad\mbox{lower half-plane contour}
 \end{split}
\end{equation}

\begin{figure}
 \centering
 \begin{tikzpicture}[
    scale=5,
    axis/.style={very thick, ->, >=stealth'},
    pile/.style={thick, ->, >=stealth', shorten <=2pt, shorten
    >=2pt},
   every node/.style={color=black}
    ]
  \draw[axis] (-0.6,0) -- (+0.6,0) node(xline)[right]{Re$\{c_6\}$};
  \draw[axis] (0,-0.6) -- (0,+0.6) node(yline)[above]{Im$\{c_6\}$};
  \fill[red] (0,+0.25) circle (.4pt) node(pole5)[right] {$6$};
  \fill[red] (-0.31,+0.25) circle (.4pt) node(pole4)[right] {$4$};
  \fill[red] (+0.35,-0.25) circle (.4pt) node(pole5)[right] {$5$};  
  \fill[red] (+0.20,-0.25) circle (.4pt) node(pole3)[left] {$3$};    
 \end{tikzpicture}
 \caption{Poles location in the $c_6$-plane for the $\Omega(\mathcal{P}^{\mbox{\tiny $(3,0)$}})$ integral. The contour can be closed either in the upper half-plane, enclosing with counter-clockwise orientation the poles $4$ and $6$; or in the lower half-plane, enclosing the poles $3$ and $5$ with clockwise orientation.}
 \label{fig:30poles}
\end{figure}
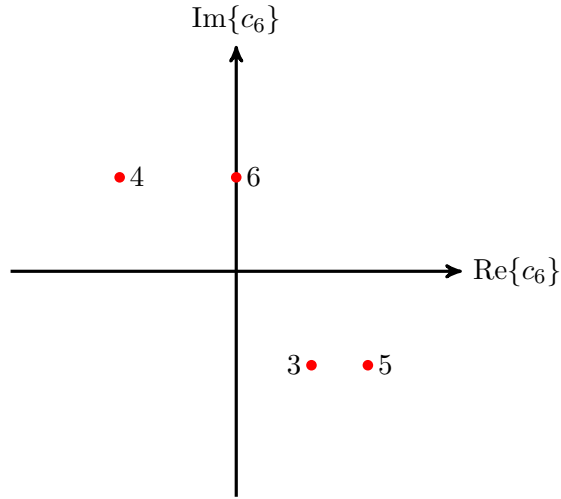

\subsection{Push-Forward}

Any point $\mathcal{Y}$ in an $n$-dimensional polytope with $V$ vertices ${\bf V}_i$ can be written as
\begin{equation}
\mathcal{Y}\:=\:c_i {\bf V}_i, \, {\rm with} \, c_i\,>\,0,
\end{equation}
but this is a highly redundant description; the space of $c_i$ is (projectively) $(V-1)$ dimensional, and $V \geq n + 1$. What we would like is a parameterization of the $c_i(z)$ in terms of $n$ variables $(z_1, \cdots, z_n)$, such that there is diffeomorphism between the points in the $z-$space simplex where $z_a\,>\,0$, and a point in the polytope. Given such a map, quite beautifully the push-forward of the form on the simplex gives us the canonical form on the polytope. In other words, 
\begin{equation} 
 \Omega(\mathcal{Y},\mathcal{P}) = \sum_{\mbox{\tiny $\{\forall\,z\,|\,\mathcal{Y}=c_i(z){\bf V}_i\}$}}\,\prod_{k=1}^{n} \frac{dz_k}{z_k}.
\end{equation}
There is a very simple way of determining the $c_i(z)$. We start with a polytope that has the same shape as the one we are interested in, but with integer-valued vertices. That is, the $i$'th vertex of this integer polytope has co-ordinates $(j_{i;1}, \cdots, j_{i;n})$. Then, 
\begin{equation}
 c_i(z) = z_1^{j_{i;1}} z_2^{j_{i;2}}  \cdots z_n^{j_{i;n}}
\end{equation}
gives us the desired diffeomorphism. An example of this ``Newton polytope" construction for the case of a pentagon is shown below: 
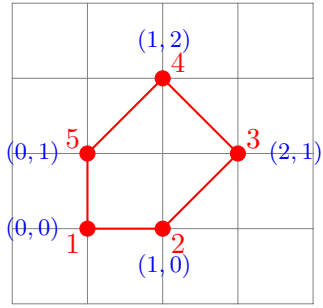
\begin{equation*}
 \begin{tikzpicture}[line join = round, line cap = round, ball/.style = {circle, draw, align=center, anchor=north, inner sep=0}]
   \draw[very thin,color=gray] (-1,-1) grid (3,3);
   \draw[fill=red,color=red]  (1,2) circle (.1cm);
   \draw[fill=red,color=red]  (0,1) circle (.1cm);
   \draw[fill=red,color=red]  (0,0) circle (.1cm);
   \draw[fill=red,color=red]  (1,0) circle (.1cm);
   \draw[fill=red,color=red]  (2,1) circle (.1cm);
   \draw[-,thick,color=red] (1,2) -- (0,1) -- (0,0) -- (1,0) -- (2,1) -- cycle;
   \node[text width=.18cm,color=red] at (1.2,2.2) {$4$};
   \node[text width=.18cm,color=blue] at (.75,2.5) {\footnotesize $(1,2)$};
   \node[text width=.18cm,color=red] at (-.2,1.2) {$5$};
   \node[text width=.18cm,color=blue] at (-1,1) {\footnotesize $(0,1)$};
   \node[text width=.18cm,color=red] at (-.2,-.2) {$1$};
   \node[text width=.18cm,color=blue] at (-1,0) {\footnotesize $(0,0)$};
   \node[text width=.18cm,color=red] at (1.2,-.2) {$2$};
   \node[text width=.18cm,color=blue] at (.75,-0.5) {\footnotesize $(1,0)$};
   \node[text width=.18cm,color=red] at (2.2,1.2) {$3$};
   \node[text width=.18cm,color=blue] at (2.5,1) {\footnotesize $(2,1)$};
   \node[text width=.18cm] at (5,1) {$\displaystyle
                                      \begin{array}{l}
                                       c_1\:=\:z_1^0 z_2^0\\
                                       c_2\:=\:z_1^1 z_2^0\\
                                       c_3\:=\:z_1^2 z_2^1\\
                                       c_4\:=\:z_1^1 z_2^2\\
                                       c_5\:=\:z_1^0 z_2^1                                   
                                      \end{array}
                                     $};
  \node at (1,-2) {$\displaystyle\mathcal{Y}\:=\:{\bf V}_1 + z_1{\bf V}_2 + z_1^2 z_2{\bf V}_3 + z_1 z_2^2{\bf V}_4 + z_2 {\bf V}_5$};
 \end{tikzpicture} 
\end{equation*}
It is especially natural to apply this formalism to cosmological polytopes, since they are defined with  integer vertices to begin with! 

It is natural to associate the positive co-ordinates of the simplex with positive variables $(X,Y)$ associated with the vertices and edges of the graph. The push-forward is a map from these positive co-ordinates into $(x,y)$, given by 
\begin{eqnarray}
 y &=& X Y X^{\prime -1} + X^{\prime} Y X^{-1} - X X^\prime Y^{-1} \nonumber \\ 
 x &=& \sum_{{\rm edges \, touching}\, x} X Y X^{\prime -1} - X^{\prime} Y X^{-1} + X X^{\prime} Y^{-1}
\end{eqnarray}
Then we have 
\begin{equation}
 \sum_{{\rm roots}} \prod \frac{dX}{X} \frac{dY}{Y} = \prod \frac{dx}{x} \frac{dy}{y} \psi(x,y)
\end{equation}

We can write this in a way that makes the connection with simplices more manifest. For every edge $y$ connecting $x,x^\prime$ in the graph we have the three vertices of the polytope, ${\bf x} + {\bf x}^\prime - {\bf y}, {\bf y} + {\bf x} - {\bf x}^\prime, {\bf y}  + {\bf x}^\prime - {\bf x}$, and corresponding weights for each vertex
\begin{equation}
 c_{x + x^\prime - y} = X X^\prime Y^{-1}, \, c_{y + x - x^\prime} = Y X X^{\prime -1}, \, c_{y + x^\prime - x} = Y X^\prime X^{-1}
\end{equation}
Clearly, precisely when two triangles meet at a midpoint, so that the vertices satisfy $V_1 + V_2 = V_1^\prime + V_2^\prime$, the corresponding $c$'s will be related as 
\begin{equation}
c_{V_1} c_{V_2} = c_{V_1^\prime} c_{V_2^\prime} \qquad {\rm if} \quad {\bf V}_1 + {\bf V}_2 = {\bf V}_1^\prime + {\bf V}_2^\prime.
\end{equation}
Trivially translating this in terms of the graph itself, for any vertex $x$ that is shared by two edges $y_1,y_2$, with the vertices connected to $x$ being $x_1,x_2$, we have 
\begin{equation}
 c_{x + x_1 - y_1} c_{x + y_1 - x_1} = c_{x + x_2 - y_2} c_{x + y_2 - x_2}
\end{equation}

Let us illustrate how this works for the case of the three-site chain. In order not to clutter then notation let the vertices of one triangle be $(1,2,3)$ and of the second triangle be $(4,5,6)$. Then we have 
\begin{eqnarray}
 \int dc_1 \cdots dc_6 \frac{1}{c_1 c_2 c_3 c_4 c_5}\times  \frac{1}{\underline{c_6 - c_ 1 c_2/c_5}} \delta^5({\cal Y} - c \cdot {\bf V}) \nonumber \\ =  
 \int dc_1 \cdots dc_6 \frac{1}{c_1 c_2 c_3 c_4 (\underline{c_5 c_6 - c_1 c_2})} \delta^5({\cal Y} - c \cdot {\bf V})
\end{eqnarray}
where the underline is instructing us to sum over all the roots of $c_5 c_6 - c_1 c_2 = 0$. We can relate this formula to a sum over simplices in a simple way. We can use the global residue theorem to write this as a sum over terms where the underline is moved onto $c_1, c_2, c_3, c_4$. Consider the term where the underline is on $c_1$, and note that the quadratic factors simplifies to a product since $c_1 \to 0$:
\begin{equation}
 \int \frac{1}{\underline{c_1} c_2 c_3 c_4 (c_5 c_6 - c_1 c_2)} \delta^5({\cal Y} - c \cdot {\bf V}) = \int \frac{1}{\underline{c_1} c_2 c_3 c_4 c_5 c_6} \delta^5({\cal Y} - c \cdot {\bf V}) = [23456]
\end{equation}
where $[23456]$ is the form for the simplex with vertices $2,3,4,5,6$. In a similar way the term with the underline on $c_2$ forcing the residue at $c_2 \to 0$ simplifies and we get $[13456]$. Now let's look at the term where  $c_3 \to 0$; here there is no simplification of the quadratic factor. Note that setting $c_3 \to 0$, the delta function tells us to express ${\cal E}$ as a linear combination of $1,2,4,5,6$. However, it is precisely the vertices ${\bf V}_1,\,{\bf V}_2,\,{\bf V}_5,\,{\bf V}_6$ that satisfy the relation ${\bf V}_1 + {\bf V}_2 = {\bf V}_5 + {\bf V}_6$, and so $(1256)$ do not span the full $4$ dimensional space; hence the delta function constraint kills this term. The same thing happens with the term where $c_4$ is underlined taking the residue at $c_4 \to 0$. Thus we have found an (in this case quite obvious) triangulation of the form, given by 
\begin{equation}
 \psi = [23456] + [13456].
\end{equation}

\section{Symmetries of Cosmological Polytopes}\label{sec:SymPol}

The wavefunction of the universe can have various symmetries inherited from isometries of the ``bulk", for instance the wavefunction in deSitter space on a flat FRW spatial slice should be (Euclidean) conformally invariant reflecting the bulk deSitter isometries. It is interesting to ask how these symmetries are captured by the geometry of the polytope. Since the canonical form is the volume of the dual polytope, there are seemingly an enormous number of symmetries given by all the ways we can infinitesimally deform the polytope leaving its volume invariant!  But this is too quick. We are looking symmetries corresponding to  differential operators that act of the $x_v,y_e$ variables, which in the polytope picture are acting on the hyperplane at infinity ${\mathcal Y}$. In other words, we are looking for symmetries of the volume of the polytope, where the vertices are fixed but we move only the hyperplane at infinity. This is an interesting mathematical question for general polytopes, but we will focus on finding these symmetries in various examples of cosmological polytopes. We will systematically find first- and second-order differential operators in ${\mathcal Y}$ that annihilate the volume, and then see how these geometric symmetries descend to conformal symmetry in suitable cases. 

To begin with, all tree-level dual polytopes have a rather trivial scaling-like symmetries, whose infinitesimal generators can be written as
\begin{equation}\label{eq:GenSymm}
 \hat{\mathcal{D}}\:=\:\frac{\partial}{\partial \mathcal{Y}^{\alpha}}\mathcal{Y}^{\alpha},\qquad
 \hat{\mathcal{O}}^{\mbox{\tiny $(i)$}}\:=\:\frac{\partial}{\partial\mathcal{Y}^{\alpha}}X^{\alpha}_{i_2\ldots i_k j_{k+1}\ldots j_N}\langle i\mathcal{Y}\rangle,
\end{equation}
where $X^{\alpha}_{i_2\ldots i_k j_{k+1}\ldots j_N}\,\equiv\,\varepsilon^{\alpha\beta_2\ldots\beta_{N}}X^{\mbox{\tiny $(i_2)$}}_{\beta_2}\ldots X^{\mbox{\tiny $(j_N)$}}_{\beta_N}$ indicates a hyperplane identified by the vertices $X^{\mbox{\tiny $(i_2)$}}\,\ldots\,X^{\mbox{\tiny $(j_N)$}}$, which, together with $X^{\mbox{\tiny $(i)$}}$, form a basis of $\mathbb{R}^N$. The different labelling among these vertices come from the fact that all the other vertices of the polytope can be {\it at most} coplanar with the $X^{\mbox{\tiny $(j)$}}$'s. Hence, the volume $\psi_n$ of a cosmological polytope in $\mathbb{P}^{N-1}$ satisfies the following set of first order differential equations:
\begin{equation}\label{eq:DeSymm}
 \hat{\mathcal{D}}\psi_n(X,\mathcal{Y})\:=\:0,\hspace{2cm}
 \hat{\mathcal{O}}^{\mbox{\tiny $(i)$}}\psi_n(X,\mathcal{Y})\:=\:0,\quad (i\,=\,1,\,\ldots k).
\end{equation}

From the graph perspective this structure corresponds to topologies with $k$ external edges: for example, an $n$-site line graph has two symmetries generated by operators such as $\hat{\mathcal{O}}^{\mbox{\tiny $(i)$}}$, while the star graph studied earlier have three.

For the sake of concreteness, let us consider the polytope in $\mathbb{P}^2$. The scaling operators are exactly by \eqref{eq:GenSymm}, however $i$ runs over all the vertex labels:
\begin{equation}\label{eq:TrGenSym}
 \hat{\mathcal{O}}^{\mbox{\tiny $(1)$}}\:=\:\frac{\partial}{\partial\mathcal{Y}^{\alpha}}X^{\alpha}_{\mbox{\tiny $(23)$}}\langle 1\mathcal{Y}\rangle,\qquad
 \hat{\mathcal{O}}^{\mbox{\tiny $(2)$}}\:=\:\frac{\partial}{\partial\mathcal{Y}^{\alpha}}X^{\alpha}_{\mbox{\tiny $(31)$}}\langle 2\mathcal{Y}\rangle,\qquad
 \hat{\mathcal{O}}^{\mbox{\tiny $(3)$}}\:=\:\frac{\partial}{\partial\mathcal{Y}^{\alpha}}X^{\alpha}_{\mbox{\tiny $(12)$}}\langle 3\mathcal{Y}\rangle.
\end{equation}
Notice that if a function is annihilated by any two of these operators as well as by $\hat{\mathcal{D}}$, then it is automatically annihilated by the third one as well. In other words, $\hat{\mathcal{O}}^{\mbox{\tiny $(3)$}}$ can be decomposed in terms of $\hat{\mathcal{O}}^{\mbox{\tiny $(1)$}}$, $\hat{\mathcal{O}}^{\mbox{\tiny $(2)$}}$, and $\hat{\mathcal{D}}$, which constitute a basis in the space of (dimension $0$) operators which annihilate the four-point contribution to the wavefunction:
\begin{equation}\label{eq:O3dec}
 \hat{\mathcal{O}}^{\mbox{\tiny $(3)$}}\:=\:\langle123\rangle\hat{\mathcal{D}}-\hat{\mathcal{O}}^{\mbox{\tiny $(1)$}}-\hat{\mathcal{O}}^{\mbox{\tiny $(2)$}}.
\end{equation}
For this particular case, these operators are a basis for the operators which annihilate the volume of the polytope in $\mathbb{P}^2$ as well as any second order operator can be decomposed on this basis.

Furthermore, the relation \eqref{eq:O3dec} is quite general: given a polytope in $\mathbb{P}^{N-1}$, it is always possible to define a scaling operator $\hat{\mathcal{O}}^{\mbox{\tiny $(I)$}}$ which acts on a special hyperplane only. It is anyhow linearly dependent from the general dilatation generator $\hat{\mathcal{D}}$  and the other operators $\hat{\mathcal{O}}^{\mbox{\tiny $(i)$}}$ which act on the vertices which do not belong to such a  hyperplane. Thus, one can either select the basis $\{\hat{\mathcal{O}}^{\mbox{\tiny $(i)$}};\,\hat{\mathcal{D}},\;i\,=\,1,\ldots,k\}$ or $\{\hat{\mathcal{O}}^{\mbox{\tiny $(i)$}},\,\hat{\mathcal{O}}^{\mbox{\tiny $(I)$}};\;i\,=\,1,\ldots,k\}$. 

The scaling operators such as  $\{\hat{\mathcal{O}}^{\mbox{\tiny $(i)$}};\,\hat{\mathcal{D}},\;i\,=\,1,\ldots,k\}$ account for all the volume preserving transformations just for the cosmological polytope  in $\mathbb{P}^2$, while in $\mathbb{P}^{N-1\,\ge\,3}$ new volume-annihilating operators are allowed. As a prototypical example, let us consider the dual polytope $\mathcal{P}^{\mbox{\tiny $(3,0)$}}$ in $\mathbb{P}^4$ which is defined via six  vertices, four of which (namely $\{3,\,4,\,5,\,6\}$) are coplanar. The space of first order operators annihilating the volume of $\mathcal{P}^{\mbox{\tiny $(3,0)$}}$ turns out to be five-dimensional, with a basis given by 
\begin{equation}\label{eq:P4volgen}
 \begin{split}
  &\hspace{2cm}\hat{\mathcal{D}}\:=\:\frac{\partial}{\partial\mathcal{Y}^{\alpha}}\mathcal{Y}^{\alpha},\hspace{2cm}
   \hat{\mathcal{O}}^{\mbox{\tiny $(i)$}}\:=\:\frac{\partial}{\partial\mathcal{Y}^{\alpha}}X^{\alpha}_{i_2 345}\langle i\mathcal{Y}\rangle,\quad (i\,=\,1,\,2),\\
  &\hat{\mathcal{O}}^{\mbox{\tiny $(35)$}}\:=\:
    \left(
     \frac{\partial}{\partial\mathcal{Y}^{\alpha}}\frac{X^{\alpha}_{1245}}{\langle12456\rangle}-\frac{\partial}{\partial\mathcal{Y}^{\alpha}}\frac{X^{\alpha}_{1243}}{\langle12436\rangle}
    \right) 
    \langle3\mathcal{Y}\rangle\langle5\mathcal{Y}\rangle,
   \qquad
   \hat{\mathcal{O}}^{\mbox{\tiny $(46)$}}\:=\:
   \frac{\partial}{\partial\mathcal{Y}^{\alpha}}X^{\alpha}_{1253}\langle4\mathcal{Y}\rangle\langle6\mathcal{Y}\rangle,
 \end{split}
\end{equation}
where we explicitly used the coplanarity condition $X^{\mbox{\tiny $(4)$}}+X^{\mbox{\tiny $(6)$}}\:\sim\:X^{\mbox{\tiny $(3)$}}+X^{\mbox{\tiny $(5)$}}$ on  the vertex $6$. The transformations generated by $\{\hat{\mathcal{O}}^{\mbox{\tiny $(35)$}},\;\hat{\mathcal{O}}^{\mbox{\tiny $(46)$}}\}$ preserve the volume of the full polytope by preserving the volume of a lower-dimensional hypersurface.


\subsection{The cubic interaction in six dimension}\label{subsec:phi3d6}

As a working example, let us consider the case of the cubic interaction in six-dimension. This theory is conformal: the coupling constant $\lambda_3$ is time-independent and thus the wave-function is purely meromorphic. The wavefunction depends on the momenta just through its energies. It is then useful to know how the generators of the conformal group act in momentum space (and, more concretely, in energy space). In momentum space, the dilatation and the special conformal transformations are given by
\begin{equation}\label{eq:ConfTr}
 \begin{split}
  &\hat{\mathcal{D}}^{\mbox{\tiny $(r)$}}\:=\:-(\Delta_r-d)+p_j^{\mbox{\tiny $(r)$}}\frac{\partial}{\partial p_{j}^{\mbox{\tiny $(r)$}}},\\
  &\hat{\mathcal{K}}^{\mbox{\tiny $(r)$}}_i\:=\:-2(\Delta_r-d)\frac{\partial}{\partial p_{\mbox{\tiny $(r)$}}^{i}}+2p_j^{\mbox{\tiny $(r)$}}\frac{\partial^2}{\partial p_j^{\mbox{\tiny $(r)$}}\partial p_{\mbox{\tiny $(r)$}}^i}
    -p_i^{\mbox{\tiny $(r)$}}\frac{\partial^2}{\partial p_{\mbox{\tiny $(r)$}}^j p^{\mbox{\tiny $(r)$}}_j},
 \end{split}
\end{equation}
where a sum needs to be understood only on the spatial index $j$, while $r$ label the state -- the differential operators above act on a single state. These operators have been written in arbitrary space dimension $d$ and for arbitrary primary of dimension $\Delta_r$. We will be interested just in the six dimensional conformal case, {\it i.e.} $d\,=\,5$ and $\Delta_r\,=\,3\;(\forall r)$.

When we consider the action of the operators \eqref{eq:ConfTr} on the wavefunction $\psi_n$, we will need to consider them as summed over all the states in the $\psi_n$. Their explicit expression in the energy space will depend on the particular $\psi_n$ given that they will depend on the internal energies as well, which parameterize the angles between the momenta. However, it is easy to see that the dilatation operator for a general $\psi_n$ can be written as 
\begin{equation}\label{eq:Dgen}
 \hat{\mathcal{D}}\:=\:\sum_{v\in\mathcal{V}}\frac{\partial}{\partial x_v}x_v+\sum_{e\in\mathcal{E}}\frac{\partial}{\partial y_e}y_e,
\end{equation}
with $\mathcal{V}$ and $\mathcal{E}$ being respectively the set of vertices and edges of the related graph, while $x_v$ and $y_e$ are the energies at the vertex $v$ and edge $e$.


\subsubsection{The four-point contribution}\label{subsubsec:phi3d64pt}

Let us now write explicitly the action of the conformal generators \eqref{eq:ConfTr} in energy space for the four-point contribution to the wave-function:
\begin{equation}\label{eq:4ptwv}
 \psi_2(x,y)\:=\:\frac{1}{(x_1+x_2)(x_1+y)(y+x_2)},\qquad
 \left\{
  \begin{array}{l}
   x_1\:\equiv\:E_1+E_2,\\
   y\:\equiv\:|\overrightarrow{p}^{\mbox{\tiny $(1)$}}+\overrightarrow{p}^{\mbox{\tiny $(2)$}}|,\\
   x_2\:\equiv\:E_3+E_4,
  \end{array}
 \right.
\end{equation}
with $E_r\,=\,|\overrightarrow{p}^{\mbox{\tiny $(r)$}}|$ being the energy of the state $r$. After a little algebra, one finds that
\begin{equation}\label{eq:4ptwvOpt}
 \begin{split}
  &\hat{\mathcal{D}}\:=\:\frac{\partial}{\partial x_1}x_1+\frac{\partial}{\partial x_2}x_2 + \frac{\partial}{\partial y}y,\\
  &\hat{\mathcal{K}}_1\:=\:-\frac{\partial^2}{\partial x_2^2}+\frac{\partial^2}{\partial x_1^2}+2\frac{x_1}{y}\frac{\partial^2}{\partial x_1\partial y}+\frac{\partial^2}{\partial y^2}+\frac{4}{y}\frac{\partial}{\partial y},\\
  &\hat{\mathcal{K}}_2\:=\:-\frac{\partial^2}{\partial x_1^2}+\frac{\partial^2}{\partial x_2^2}+2\frac{x_2}{y}\frac{\partial^2}{\partial x_2\partial y}+\frac{\partial^2}{\partial y^2}+\frac{4}{y}\frac{\partial}{\partial y}.\\  
 \end{split}
\end{equation}
In principle one would expect a further operator for the special conformal transformations. However, it turns out to be reducible to some first order operator acting on the dilatation generator. Notice that also the sum $\hat{\mathcal{K}}_1+\hat{\mathcal{K}_2}$ has this very same structure, so that one can consider independent just one of the two operators.

Furthermore, any of the operator $\hat{\mathcal{K}}_1$ can be decomposed as follows:
\begin{equation}\label{eq:}
 y\,\hat{\mathcal{K}_1}\:=\:
  \left(
   \frac{\partial}{\partial x_2}+\frac{\partial}{\partial y}-\frac{\partial}{\partial x_1}
  \right)
  \hat{\mathcal{D}}+
  \frac{\partial}{\partial x_1} \hat{\mathcal{O}}^{\mbox{\tiny $(1)$}}-
  \frac{\partial}{\partial x_2} \hat{\mathcal{O}}^{\mbox{\tiny $(2)$}},
\end{equation}
where $ \hat{\mathcal{O}}^{\mbox{\tiny $(i)$}}$ are the very same first order operators found for the triangle in $\mathbb{P}^2$ but now written the in the energy space $(x_1,\,y,\,x_2)$. Thus, the action of the special conformal transformations for $\psi_2$ reduces to the action of the first order operators $\hat{\mathcal{O}}^{\mbox{\tiny $(i)$}}$. 

The volume-preserving transformations in Section \ref{sec:SymPol} for the cosmological polytope have been discussed with no reference to the underlying theory the polytope computes the wavefunction of. This means that $\psi_2$, which  from the polytope perspective is computed as volume of a triangle in $\mathbb{P}^2$, {\it always} enjoys the symmetries generated for the operators $\{\hat{\mathcal{D}},\,\hat{\mathcal{O}}^{\mbox{\tiny $(1)$}},\,\hat{\mathcal{O}}^{\mbox{\tiny $(2)$}}\}$, which in the context of cubic interactions in six-dimensions we recognize to be the dilatations and special conformal transformations. For general interactions, the volume of the polytope computes an integrand: to obtain the actual contribution to the wavefunction we would need to integrate over the energies $\varepsilon$ which are related to the time-dependence of the coupling constant, and which in our energy-space computation are implicitly included in the energies $x_v$. Hence, the dilatation and special conformal transformation turns out to be symmetries of the {\it integrand} for any $\psi_2$.


\subsubsection{The five-point contribution}\label{subsubsec:phi3d65pt}

Let us now consider the five-point contribution to the wavefunction, which is given by the three-side graph $\psi_3$ discussed earlier:
\begin{equation}\label{eq:5ptwv}
 \begin{split}
  \psi_{3}(x,y)\:=\:&\frac{1}{(x_1+x_2+x_3)(x_1+y_{12})(y_{12}+x_2+y_{23})(y_{23}+x_3)}\times\\
  &\times
   \left[
    \frac{1}{x_1+x_2+y_{23}}+\frac{1}{y_{12}+x_2+x_3} 
   \right],
 \end{split}
\end{equation}
whose variables, in term of the momenta and energies of each state, are given by
\begin{equation}\label{eq:5ptvar}
 x_1\:\equiv\:E_1+E_2,\quad
 x_2\:\equiv\:E_3,\quad
 x_3\:\equiv\:E_4+E_5,\quad
 y_{12}\:=\:|\overrightarrow{p}^{\mbox{\tiny $(1)$}}+\overrightarrow{p}^{\mbox{\tiny $(2)$}}|,\quad
 y_{23}\:=\:|\overrightarrow{p}^{\mbox{\tiny $(4)$}}+\overrightarrow{p}^{\mbox{\tiny $(5)$}}|.
\end{equation}
The generators of dilatation and special conformal dimensions in energy space can be written as
\begin{equation}\label{eq:5ptgenconf}
 \begin{split}
  &\hat{\mathcal{D}}\:=\:\frac{\partial}{\partial x_1}x_1+\frac{\partial}{\partial x_2}x_2 +\frac{\partial}{\partial x_3}x_3 + \frac{\partial}{\partial y_{12}}y_{12}+\frac{\partial}{\partial y_{23}}y_{23},\\
  &\hat{\mathcal{K}}_1\:=\:-\frac{\partial^2}{\partial x_2^2}+\frac{\partial^2}{\partial x_1^2}+2\frac{x_1}{y_{12}}\frac{\partial^2}{\partial x_1\partial y_{12}}+\frac{\partial^2}{\partial y_{12}^2}+
        \frac{4}{y_{12}}\frac{\partial}{\partial y_{12}},\\
  &\hat{\mathcal{K}}_2\:=\:-\frac{\partial^2}{\partial x_2^2}+\frac{\partial^2}{\partial x_3^2}+2\frac{x_3}{y_{23}}\frac{\partial^2}{\partial x_3\partial y_{23}}+\frac{\partial^2}{\partial y_{23}^2}+
        \frac{4}{y_{23}}\frac{\partial}{\partial y_{23}}.
 \end{split}
\end{equation}
Differently from the four point case, the special conformal generators \eqref{eq:5ptvar} cannot be written purely in terms of the first order operators $\{\hat{\mathcal{D}},\,\hat{\mathcal{O}}^{\mbox{\tiny $(i)$}},\;(i\,=\,1,\,2)\}$, where now the latter operators have the following explicit form
\begin{equation}\label{eq:5ptOs}
 \begin{split}
  &\hat{\mathcal{O}}^{\mbox{\tiny $(1)$}}\:=\:\left(\frac{\partial}{\partial x_1}+\frac{\partial}{\partial y_{12}}-\frac{\partial}{\partial x_2}\right)(x_1+y_{12}),\\
  &\hat{\mathcal{O}}^{\mbox{\tiny $(2)$}}\:=\:\left(\frac{\partial}{\partial x_3}+\frac{\partial}{\partial y_{23}}-\frac{\partial}{\partial x_2}\right)(y_{23}+x_3),
 \end{split}
\end{equation}
which still annihilate $\psi_3$. The special conformal generators therefore turn out not to be reducible, but their form can be simplified using \eqref{eq:5ptOs}:
\begin{equation}\label{eq:5ptconftransf}
 \begin{split}
  &y_{12}(x_1+y_{12})\hat{\mathcal{K}}_1\:=\:
     y_{12}\left(\frac{\partial}{\partial x_1}+\frac{\partial}{\partial y_{12}}+\frac{\partial}{\partial x_2}\right)\hat{\mathcal{O}}^{\mbox{\tiny $(1)$}}-
     2\frac{\partial^2}{\partial x_1\partial y_{12}}\left(y_{12}^2-x_1^2\right),\\
 &y_{23}(x_3+y_{23})\hat{\mathcal{K}}_2\:=\:
     y_{23}\left(\frac{\partial}{\partial x_3}+\frac{\partial}{\partial y_{23}}+\frac{\partial}{\partial x_2}\right)\hat{\mathcal{O}}^{\mbox{\tiny $(2)$}}-
     2\frac{\partial^2}{\partial x_3\partial y_{23}}\left(y_{23}^2-x_3^2\right),
 \end{split}
\end{equation}
where the first term in each line represent purely second order operators. Interestingly, they manifestly annihilate term by term the recursion relation \eqref{eq:TreeRRex2} obtained from the frequency integral. Actually, any tree-level graph is annihilated by such class of operators: this is made manifest exactly by the recursive formula \eqref{eq:TreeRRf}, where the two terms depend on {\it either} the external node energy {\it or} on the related edge energy. These operators, together with the first order scaling-like operators discussed earlier are a feature of {\it any} tree-level $\psi_n$. 

Let us turn to the polytope $\mathcal{P}^{\mbox{\tiny $(3,0)$}}$ in $\mathbb{P}^4$ whose volume compute the $\psi_3$ in question. We can think to classify the space of second order operators which can generate volume-preserving transformations. We are interested in those operators which cannot be reduced to combination of composition of first order operators which annihilate the volume of $\mathcal{P}^{\mbox{\tiny $(3,0)$}}$. This space turns out to be fairly small -- it has dimension $3$ -- and a basis is given by:
\begin{equation}\label{eq:5pt2ndClass}
 \begin{split}
  &\hat{\mathcal{O}}^{\mbox{\tiny $(7)$}}\:\equiv\:\frac{\partial}{\partial\mathcal{Y}^{\alpha}}X^{\alpha}_{\mbox{\tiny $(1245)$}}\frac{\partial}{\partial\mathcal{Y}^{\beta}}X^{\beta}_{\mbox{\tiny $(1236)$}}\langle7\mathcal{Y}\rangle,\qquad
   \hat{\mathcal{O}}^{\mbox{\tiny $(8)$}}\:\equiv\:\frac{\partial}{\partial\mathcal{Y}^{\alpha}}X^{\alpha}_{\mbox{\tiny $(1234)$}}\frac{\partial}{\partial\mathcal{Y}^{\beta}}X^{\beta}_{\mbox{\tiny $(1256)$}}\langle8\mathcal{Y}\rangle,\\
  &\hat{\mathcal{O}}^{\mbox{\tiny $(6)$}}\:\equiv\:\frac{\partial}{\partial\mathcal{Y}^{\alpha}}X^{\alpha}_{\mbox{\tiny $(1245)$}}\frac{\partial}{\partial\mathcal{Y}^{\beta}}X^{\beta}_{\mbox{\tiny $(1234)$}}\langle6\mathcal{Y}\rangle,
 \end{split}
\end{equation}
where the labels $7$ and $8$ correspond to two new (spurious) vertices such that $X^{\mbox{\tiny $(7)$}}\,+\,X^{\mbox{\tiny $(3)$}}\,\sim\,X^{\mbox{\tiny $(4)$}}\,\sim\,X^{\mbox{\tiny $(5)$}}\,+\,X^{\mbox{\tiny $(8)$}}$ (in energy space they correspond to $y_{12}-x_1$ and $y_{23}-x_3$ respectively). The operators $\hat{\mathcal{O}}^{\mbox{\tiny $(7)$}}$ and $\hat{\mathcal{O}}^{\mbox{\tiny $(8)$}}$ annihilate term by term the recursion relations of type \eqref{eq:TreeRRex2}, as the special conformal generators in \eqref{eq:5ptconftransf} do. Not surprisingly, they are actually equivalent -- this is straightforward to show by rewriting \eqref{eq:5pt2ndClass} in energy space\footnote{Saying that the operators \eqref{eq:5pt2ndClass} and the special conformal generators \eqref{eq:5ptconftransf} are equivalent means that they coincide up to terms which can be written as composition of two first order operators $\hat{\mathcal{O}}^{\mbox{\tiny $(L)$}}\,\circ\,\hat{\mathcal{O}}^{\mbox{\tiny $(R)$}}$ with $\hat{\mathcal{O}}^{\mbox{\tiny $(R)$}}$ annihilating the volume of $\mathcal{P}^{\mbox{\tiny $(3,0)$}}$.}.

The operator $\hat{\mathcal{O}}^{\mbox{\tiny $(6)$}}$ not only preserves the volume of the polytope, but it preserves the volume of the simplices of a particular triangulation. This triangulation corresponds to the OFPT representation for the $\psi_3$.

As a last comment, while the relation among the volume preserving transformations and the conformal group have been find for the concrete example of $\lambda_3\phi^3$ theory in six dimension, which is  conformal, the actual treatment on the graphs and on the polytopes do not rely on any concrete feature of the conformal theory, and goes through for {\it any} theory. The main difference between the conformal case and a general theory is that while the in the former graphs and polytopes compute the contribution to the wavefunction $\psi_n$ and thus these operators generates symmetries of the actual theory, in the latter case what we compute are {\it integrands} and thus these are symmetries of the integrand rather than of the integrated wavefunction. 


\section{Time and Transcendentality}\label{sec:TT}

Along the whole discussion, and in particular in the previous section, we stressed how the cosmological polytopes return the {\it integrands} of the wavefunction: We still need to integrate over the energies related to the time-dependent coupling constants. The only situation in which the result is the actual wavefunction is the conformal case, for which the wavefunction is a meromorphic function. This is an important feature: the wavefunction is a meromorphic function just if there is no genuine time-dependence. In other words, the presence of an actual time-dependence is reflected in the wavefunction being represented by transcendental functions. 

In this section, we discuss how the structure of the integrated wavefunction, and therefore its transcendentality, in terms of the symbols is encoded into the cosmological polytopes and can be read-off from them. For the sake of concreteness, let us focus on $\lambda_3 \phi^3$ theory in four-dimensional deSitter space dS$_4$. Here, after the conformal transformation the time-dependent coupling constant is $\lambda_3(\eta) = \lambda/\eta$, and the corresponding Fourier-transform is $\tilde{\lambda}(\varepsilon) = 0$ for $\varepsilon\,<\,0$ and $\tilde{\lambda}(\varepsilon) = \lambda$ for $ \varepsilon\,>\,0$. Consequently, in order to get the final result, we simply have to take our integrand, and integrate the energy variables for each vertex $v$ from some fixed $X_v \to \infty$. 

In this case we have 
\begin{equation}
F(x,y) = 2 y \psi(x,y) = \int_{X_1,X_2}^\infty dx_1 dx_2 \frac{2 y}{(x_1 + y)(x_2 + y)(x_1 + x_2)}
\end{equation}
Let us determine the symbol by finding the differential. We spell out the straightforward detailed steps in this case since they immediately generalize to more complicated cases.  
\begin{eqnarray}
 \frac{\partial F}{\partial X_1} &=& \frac{2 y}{(x_1 + y)}  \int_{X_2}^\infty dx_2 \frac{1}{(x_2 + y)(x_2 + x_1)} \nonumber \\ 
  &=& \frac{2y}{(x_1 + y)(x_1 - y)} \int_{X_2}^{\infty} \left( \frac{1}{x_2 + y} - \frac{1}{x_2 + x_1} \right) \nonumber \\ 
  & = & \left(\frac{1}{x_1 + y} - \frac{1}{x_1 - y} \right) \int_{X_2}^{\infty} \left( \frac{1}{x_2 + y} - \frac{1}{x_2 + x_1} \right)
\end{eqnarray}
The first line is trivial since the differentiation is with respect to the endpoint of the region of integration. In the second line we have expressed the resulting the 1-dimensional integral in terms of pieces normalized to unit residues, and in the third line we have expressed the prefactor in exactly the same way. We can do exactly the same for differentiating w.r.t. $X_2$. Thus we find
\begin{equation} 
 dF = d{\rm log}\frac{x_1 + y}{x_1 - y} G_1 + d{\rm log} \frac{x_2 + y}{x_2 - y} G_2
\end{equation}
where
\begin{equation}
 d G_1 = d{\rm log} \frac{x_2 + y}{x_2 + x_1}, \, dG_2 = d{\rm log} \frac{x_1 + y}{x_1 + x_2}
\end{equation}
and hence we can read off the symbol as 
\begin{equation}\label{eq:S2s}
 {\cal S}(F) = \frac{x_1 + y}{x_1 - y} \otimes \frac{x_2 + y}{x_2 + x_1} +  \frac{x_2 + y}{x_2 - y} \otimes \frac{x_1 + y}{x_1 + x_2}
\end{equation}

Let us now move to the general case. Given a graph $\mathcal{G}$ and $\psi_{\mathcal{G}}(x,y)$, we are interested in the integral
\begin{equation}\label{eq:Intfg}
 f_{\mbox{\tiny $\mathcal{G}$}}(X,Y)\:=\:\prod_{v\in\mathcal{V}}\int_{X_v}^{\infty}dx_v\psi_{\mathcal{G}}(x,Y).
\end{equation}
Now it is a remarkable fact that taking any residue of $\psi_{\mathcal{G}}(x,Y)$ in the $x$ variables always leaves us with $\prod_{e\in\mathcal{E}}1/Y_e$. This has a beautiful interpretation in terms of the geometry of the cosmological polytope, which would take us too far afield to explain here; but it does make it natural to normalize the integral \eqref{eq:Intfg} as

\begin{equation}\label{eq:IntfgN}
 f_{\mbox{\tiny $\mathcal{G}$}}(X,Y)\:=\:\left(\prod_{e\in\mathcal{E}}\frac{1}{Y_e}\right)F_{\mathcal{G}}(X,Y).
\end{equation}
Since  $\psi_{\mathcal{G}}(x,Y)$ has linear factors as poles, and since the residues are $\pm\,\prod_{e\in\mathcal{E}}1/Y_e$, on general grounds we are guaranteed to obtain a sum of ``pure'' polylogarithms of degree $\mathfrak{V}$. We wish to give an expression for the symbol of $F_{\mathcal{G}}(X,Y)$. This has been achieved above by a brute force calculation, but this method becomes rapidly incredibly complicated for more interesting graphs and furthermore gives no insight into the structure of the answer.

However, we now present a ``one-line'' expression for $S[F_{\mathcal{G}}(X,Y)]$ for any graph $\mathcal{G}$. We will see that the symbol is directly a record of the geometry of the cosmological polytope!

The expression we present is a special case of a general investigation into a generalization of ``Aomoto polylogarithms'' \cite{Arkani-Hamed:2017asp}. We will therefore only present the final result here. We need to first introduce some notation. We begin by arbitrarily choosing one of the edges, say $e_1$, and we define a plane $\mathcal{W}_{\mbox{\tiny $(1)\,I$}}^{\mbox{\tiny $Y$}}\,=\,{\bf\tilde{y}}_{e_1\,\mbox{\tiny $I$}}$. We then define for all other edges and vertices of the graph
\begin{equation}\label{eq:WdefXY}
  \begin{split}
  &\left(\mathcal{W}_e^{\mbox{\tiny $Y$}}\right)_{\mbox{\tiny $I$}}\:=\:{\bf Y}_{e_1}{\bf\tilde{y}}_{e\,\mbox{\tiny $I$}}-{\bf Y}_e{\bf\tilde{y}}_{e_1\mbox{\tiny $I$}},\\
  &\left(\mathcal{W}_e^{\mbox{\tiny $X$}}\right)_{\mbox{\tiny $I$}}\:=\:{\bf Y}_{e_1}{\bf\tilde{x}}_{v\,\mbox{\tiny $I$}}-{\bf X}_v{\bf \tilde{y}}_{e_1\mbox{\tiny $I$}}.
 \end{split}
\end{equation}
The motivation for these definitions is that $\left(\mathcal{W}_v^{\mbox{\tiny $X$}}\cdot\mathcal{Y}\right)\,=\,{\bf Y}_{e_1}\left({\bf x}_v-{\bf X}_v\right)$, so that the boundaries of the region of integration over the $x$ variables is $\left(\mathcal{W}_v^{\mbox{\tiny $X$}}\,\cdot\,\mathcal{Y}\right)\,\ge\,0$. Since we are not integrating over the $y$'s, we are localized to $\left(\mathcal{W}_e^{\mbox{\tiny $Y$}}\,\cdot\,\mathcal{Y}\right)\,=\,{\bf Y}_{e_1}\left({\bf y}_e-{\bf Y}_e\right)\,=\,0$.

Now, given a sequence of planes $(\mathcal{W}_1,\,\ldots,\,\mathcal{W}_k)$ and vertices $({\bf V}_1,\,\ldots,\,{\bf V}_k)$, we define 
\begin{equation}\label{eq:WVdef}
 [\mathcal{W}_1,\,\ldots,\,\mathcal{W}_k\; ; \; {\bf V}_1,\,\ldots,\,{\bf V}_k]\:=\:\Det_{\mbox{\tiny $\{a,b\}$}}\left\{\mathcal{W}_a\,\cdot\,{\bf V}_b\right\}.
\end{equation}

We now come to the geometric heart of the story. We look for a sequence of vertices $\{{\bf V}_1,\,{\bf V}_2,\,\ldots,\,{\bf V}_{\mbox{\tiny $E+V-1$}}\}$ of the cosmological polytope $\mathcal{P}_{\mathcal{G}}$ with the following properties. First, $\mathcal{V}_2$ is connected to $\mathcal{V}_1$ by an edge of the cosmological polytope. Now, let us project the cosmological polytope through $\mathcal{V}_1$. We obtain one-lower dimensional polytope $\mathcal{P}_{\mathcal{G}}^{\mbox{\tiny $({\bf V}_1)$}}$, since ${\bf V}_2$ was connected to ${\bf V}_1$ by an edge of $\mathcal{P}_{\mathcal{G}}$, ${\bf V}_1$ will be a vertex of $\mathcal{P}_{\mathcal{G}}^{\mbox{\tiny ${\bf V}_1$}}$ (indeed, in the lower dimensional space where we have quotiented by ${\bf V}_1$, $\mathcal{P}_{\mathcal{G}}$, ${\bf V}_1$ can be thought of as the convex hull of all vertices ${\bf V}$ which were connected to ${\bf V}_1$ by an edge of $\mathcal{P}_{\mathcal{G}}$).

Then ${\bf V}_3$ must be connected to ${\bf V}_2$ by an edge of $\mathcal{P}_{\mathcal{G}}^{\mbox{\tiny ${\bf V}_1$}}$. We then further project through ${\bf V}_2$ to get $\mathcal{P}_{\mathcal{G}}^{\mbox{\tiny $({\bf V}_1{\bf V}_2$})}$; ${\bf V}_3$ is a vertex of  $\mathcal{P}_{\mathcal{G}}^{\mbox{\tiny $({\bf V}_1{\bf V}_2)$}}$ and  ${\bf V}_4$ must be connected to ${\bf V}_3$ by an edge. In general, the sequence $\{{\bf V}_1,\,\ldots,\,{\bf V}_{E+V-1}\}$ has the property that ${\bf V}_j$ is connected to ${\bf V}_{j+1}$ via an edge of the cosmological polytope projected through $\{{\bf V}_1,\,\ldots,\,{\bf V}_{j-1}\}$, $\mathcal{P}_{\mathcal{G}}^{\mbox{\tiny $\{{\bf V}_1,\ldots,{\bf V}_{j-1}$\}}}$.

Let us illustrate what these sequence look like for the two-site loop graph 
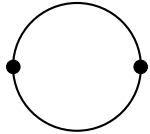
\begin{equation*}
 \begin{tikzpicture}[ball/.style = {circle, draw, align=center, anchor=north, inner sep=0}]
  \node[ball,text width=.18cm,fill,color=black] (x1) at (0,0) {};
  \node[ball,text width=.18cm,fill,color=black,right=1.5cm of x1.east] (x2) {};
  \draw[thick] ($(x1)!0.5!(x2)$) circle (0.85cm);
 \end{tikzpicture} 
\end{equation*}
where the cosmological polytope is:
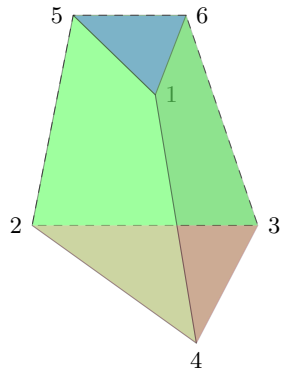
\begin{equation*}
 \begin{tikzpicture}[line join = round, line cap = round, ball/.style = {circle, draw, align=center, anchor=north, inner sep=0}]
 \begin{scope}
  \pgfmathsetmacro{\factor}{1/sqrt(2)};
  \coordinate [label=right:{\footnotesize $6$}] (A) at (.75,0,-.75*\factor);
  \coordinate [label=left:{\footnotesize $5$}] (B) at (-.75,0,-.75*\factor);
  \coordinate [label=right:{\footnotesize $1$}] (C) at (0.75,-.65,.75*\factor);
 
  \coordinate (Ab) at ($(B)!.5!(C)$); 
  \coordinate (Bb) at ($(A)!.5!(C)$);
  \coordinate (G) at (intersection of A--Ab and B--Bb);
 
  \coordinate [label=right:{\footnotesize $3$}] (D) at (1.5,-3,-1.5*\factor);
  \coordinate [label=left:{\footnotesize $2$}] (E) at (-1.5,-3,-1.5*\factor);
  \coordinate [label=below:{\footnotesize $4$}] (F) at (1.5,-3.75,1.5*\factor);

  \draw[-,dashed,fill=green!50,opacity=.6] (A) -- (B) -- (E) -- (D) -- cycle;
  \draw[draw=none,fill=red!60, opacity=.45] (D) -- (E) -- (F) -- cycle;
  \draw[-,fill=blue!,opacity=.3] (A) -- (B) -- (C) -- cycle; 
  \draw[-,fill=green!50,opacity=.4] (B) -- (C) -- (F) -- (E) -- cycle;
  \draw[-,fill=green!45!black,opacity=.2] (A) -- (C) -- (F) -- (D) -- cycle; 
 \end{scope}
 \end{tikzpicture}
\end{equation*}

Let us look at the sequence starting with the vertex $1$. It is connected to vertices $4$, $5$, $6$, so $\mathcal{P}_{\mathcal{G}}^{\mbox{\tiny $(1)$}}$ is the triangle $(456)$. Hence we can have $\{(145),\,(146),\,(154),\,(156),\,(164),\,(165)\}$.

The paths starting from vertex $2$ are $\{(234),\,(235),\,(243),\,(245),\,(253),\,(254)\}$; All other paths are related to these by the obvious up $\longleftrightarrow$ down and left $\longleftrightarrow$ right reflection symmetries of the polytope.

Let us look at the slightly more interesting case of the three-site chain 
$
 
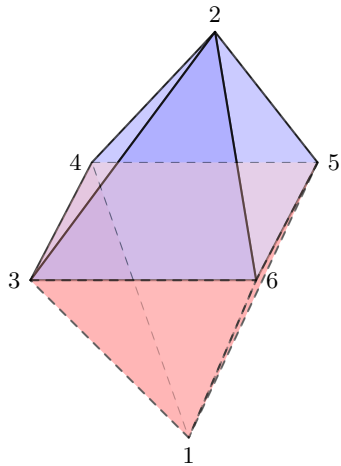
\begin{tikzpicture}[ball/.style = {circle, draw, align=center, anchor=north, inner sep=0}]
  \node[ball,text width=.18cm,fill,color=black] (x1) at (0,0) {};
  \node[ball,text width=.18cm,fill,color=black,right=1.5cm of x1.east] (x2) {};
  \node[ball,text width=.18cm,fill,color=black,right=1.5cm of x2.east] (x3) {};
  \draw[-,thick] (x1) -- (x2) -- (x3);
 \end{tikzpicture}   
$:
\begin{equation*}
 \begin{tikzpicture}[line join = round, line cap = round, ball/.style = {circle, draw, align=center, anchor=north, inner sep=0}]
  \begin{scope}
   \pgfmathsetmacro{\factor}{1/sqrt(2)};  
   \coordinate [label=right:{\footnotesize $5$}] (B2c) at (10.5,-3,-1.5*\factor);
   \coordinate [label=left:{\footnotesize $4$}] (A1c) at (7.5,-3,-1.5*\factor);
   \coordinate [label=right:{\footnotesize $6$}] (B1c) at (10.5,-3.75,1.5*\factor);
   \coordinate [label=left:{\footnotesize $3$}] (A2c) at (7.5,-3.75,1.5*\factor);  
   \coordinate [label=above:{\footnotesize $2$}] (C1c) at (9.75,-.65,.75*\factor);
   \coordinate [label=below:{\footnotesize $1$}] (C2c) at (9.4,-6.05,.75*\factor);
   
   \node at (A1c) (A1d) {};
   \node at (B2c) (B2d) {};
   \node at (B1c) (B1d) {};

   \node at (A2c) (A2d) {};
   \node at (C1c) (C1d) {};
   \node at (C2c) (C2d) {};

   \draw[-,dashed,fill=blue!30, opacity=.7] (A1c) -- (B2c) -- (C1c) -- cycle;
   \draw[-,thick,fill=blue!20, opacity=.7] (A1c) -- (A2c) -- (C1c) -- cycle;
   \draw[-,thick,fill=blue!20, opacity=.7] (B1c) -- (B2c) -- (C1c) -- cycle;
   \draw[-,thick,fill=blue!35, opacity=.7] (A2c) -- (B1c) -- (C1c) -- cycle;

   \draw[-,dashed,fill=red!30, opacity=.3] (A1c) -- (B2c) -- (C2c) -- cycle;
   \draw[-,dashed,thick,fill=red!50, opacity=.5] (B2c) -- (B1c) -- (C2c) -- cycle;
   \draw[-,dashed,fill=red!40, opacity=.3] (A1c) -- (A2c) -- (C2c) -- cycle;
   \draw[-,dashed,thick,fill=red!45, opacity=.5] (A2c) -- (B1c) -- (C2c) -- cycle;
  \end{scope}
 \end{tikzpicture}
\end{equation*}
Let us look at the sequence that start with vertex $1$. This vertex is connected to all five other vertices and looks like a square pyramid
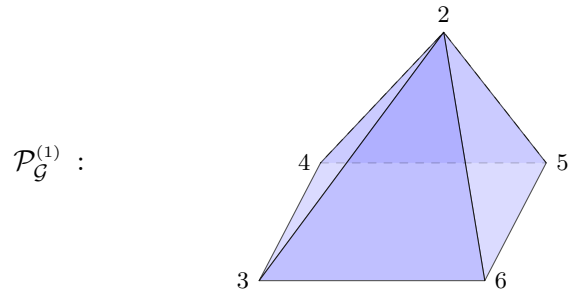
\begin{equation*}
  \begin{tikzpicture}[line join = round, line cap = round, ball/.style = {circle, draw, align=center, anchor=north, inner sep=0}]
  \begin{scope}
   \pgfmathsetmacro{\factor}{1/sqrt(2)};  
   \coordinate [label=right:{\footnotesize $5$}] (B2c) at (10.5,-3,-1.5*\factor);
   \coordinate [label=left:{\footnotesize $4$}] (A1c) at (7.5,-3,-1.5*\factor);
   \coordinate [label=right:{\footnotesize $6$}] (B1c) at (10.5,-3.75,1.5*\factor);
   \coordinate [label=left:{\footnotesize $3$}] (A2c) at (7.5,-3.75,1.5*\factor);  
   \coordinate [label=above:{\footnotesize $2$}] (C1c) at (9.75,-.65,.75*\factor);
   
   \node at (A1c) (A1d) {};
   \node at (B2c) (B2d) {};
   \node at (B1c) (B1d) {};
   \node at (A2c) (A2d) {};
   \node at (C1c) (C1d) {};

   \draw[-,dashed,fill=blue!30, opacity=.7] (A1c) -- (B2c) -- (C1c) -- cycle;
   \draw[-,fill=blue!20, opacity=.7] (A1c) -- (A2c) -- (C1c) -- cycle;
   \draw[-,fill=blue!20, opacity=.7] (B1c) -- (B2c) -- (C1c) -- cycle;
   \draw[-,fill=blue!35, opacity=.7] (A2c) -- (B1c) -- (C1c) -- cycle;

   \node[left=3cm of A1c.west] (lbl) {$\mathcal{P}^{\mbox{\tiny $(1)$}}_{\mathcal{G}}\: :$};
  \end{scope}
 \end{tikzpicture}
\end{equation*}
but we obviously get different shapes for $\mathcal{P}^{\mbox{\tiny $(12)$}}_{\mathcal{G}}$
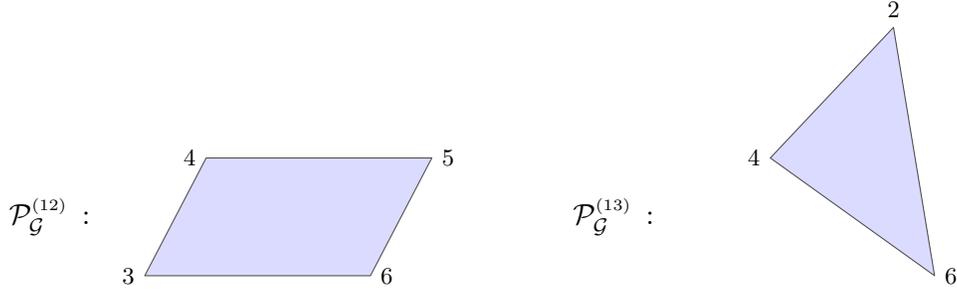
\begin{equation*}
 \begin{tikzpicture}[line join = round, line cap = round, ball/.style = {circle, draw, align=center, anchor=north, inner sep=0}]
  \begin{scope}
   \pgfmathsetmacro{\factor}{1/sqrt(2)};  
   \coordinate [label=right:{\footnotesize $5$}] (B2c) at (10.5,-3,-1.5*\factor);
   \coordinate [label=left:{\footnotesize $4$}] (A1c) at (7.5,-3,-1.5*\factor);
   \coordinate [label=right:{\footnotesize $6$}] (B1c) at (10.5,-3.75,1.5*\factor);
   \coordinate [label=left:{\footnotesize $3$}] (A2c) at (7.5,-3.75,1.5*\factor);  
   
   \node at (A1c) (A1d) {};
   \node at (B2c) (B2d) {};
   \node at (B1c) (B1d) {};
   \node at (A2c) (A2d) {};

   \draw[-,fill=blue!20, opacity=.7] (A2c) -- (A1c) -- (B2c) -- (B1c) -- cycle;
   \coordinate (tmp) at ($(A2c)!0.5!(A1c)$);
   \node[left=1cm of tmp.west] (lbl) {$\mathcal{P}^{\mbox{\tiny $(12)$}}_{\mathcal{G}}\: :$};

   \coordinate [label=left:{\footnotesize $4$}] (A1e) at (15,-3,-1.5*\factor);
   \coordinate [label=right:{\footnotesize $6$}] (B1e) at (18,-3.75,1.5*\factor);
   \coordinate [label=above:{\footnotesize $2$}] (C1e) at (17.25,-.65,.75*\factor);
   \coordinate  (A2e) at (15,-3.75,1.5*\factor);     

   \draw[-,fill=blue!20, opacity=.7] (A1e) -- (B1e) -- (C1e) -- cycle;
   \coordinate (tmp2) at ($(A2e)!0.5!(A1e)$);
   \node[left=1cm of tmp2.west] (lbl) {$\mathcal{P}^{\mbox{\tiny $(13)$}}_{\mathcal{G}}\: :$};
  \end{scope}
 \end{tikzpicture}
\end{equation*}
Thus, the allowed paths starting from vertex $1$ include $(12\,\{34\})$, $(12\,\{45\})$, $(12\,\{56\})$, $(12\,\{63\})$ and $(13\,\{\mbox{choose two of }(246)\})$, where ${\it e.g.}$ $(12\,\{56\})$ means we can have either $(1256)$ or $(1265)$.

Meanwhile, since the vertex $3$ is the only connected to $(1,\,2,\,4,\,6)$, $\mathcal{P}^{\mbox{\tiny $(3)$}}_{\mathcal{G}}$ is just a simplex so that
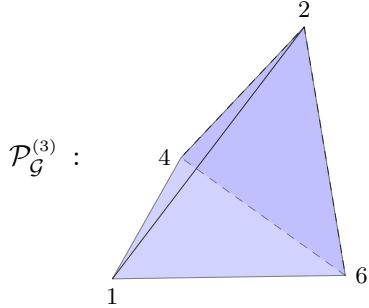
\begin{equation*}
 \begin{tikzpicture}[line join = round, line cap = round, ball/.style = {circle, draw, align=center, anchor=north, inner sep=0}]
  \begin{scope}
   \pgfmathsetmacro{\factor}{1/sqrt(2)};  
   \coordinate [label=left:{\footnotesize $4$}] (A1c) at (7.5,-3,-1.5*\factor);
   \coordinate [label=right:{\footnotesize $6$}] (B1c) at (10.5,-3.75,1.5*\factor);
   \coordinate [label=above:{\footnotesize $2$}] (C1c) at (9.75,-.65,.75*\factor);
   \coordinate [label=below:{\footnotesize $1$}] (C2c) at (7.2,-4,.75*\factor);
   
   \node at (A1c) (A1d) {};
   \node at (B2c) (B2d) {};
   \node at (B1c) (B1d) {};
   \node at (A2c) (A2d) {};
   \node at (C1c) (C1d) {};
   \node at (C2c) (C2d) {};

   \draw[-,dashed,fill=blue!20, opacity=.7] (B1c) -- (A1c) -- (C1c) -- cycle; 
   \draw[-,fill=blue!35, opacity=.5] (A1c) -- (C1c) -- (C2c) -- cycle;
   \draw[-,fill=blue!35, opacity=.5] (B1c) -- (C1c) -- (C2c) -- cycle;   
   \coordinate (tmp) at (A1c);
   \node[left=1.2cm of tmp.west] (lbl) {$\mathcal{P}^{\mbox{\tiny $(3)$}}_{\mathcal{G}}\: :$};
  \end{scope}
 \end{tikzpicture}
\end{equation*}
and all the paths starting from vertex $3$ are $(3,\,\{\mbox{choose three of }(1,\,2,\,4,\,6)\})$.

Note that any allowed sequence $\{{\bf V}_1,\,\ldots,\,{\bf V}_{\mbox{\tiny $E+V-1$}}\}$ can be thought of projecting the entire polytope down to one dimension. A one-dimensional polytope is just an interval, so all of $\{{\bf V}_1,\,\ldots,\,{\bf V}_{\mbox{\tiny $E+V-1$}}\}$ are projected to one boundary. Thus, all the other vertices of the original polytope have either been projected down to the same boundary or are all on the same side of this boundary. In other words, for all the vertices ${\bf V}\,\notin\,\{{\bf V}_1,\,\ldots,\,{\bf V}_{\mbox{\tiny $E+V-1$}}\}$ for which $\langle{\bf V}{\bf V}_1\ldots{\bf V}_{\mbox{\tiny $E+V-1$}}\rangle\,\neq\,0$, all the signs of $\langle{\bf V}{\bf V}_1\ldots{\bf V}_{\mbox{\tiny $E+V-1$}}\rangle$ are the same! We stress that this would clearly not be the case for a randomly chosen collection of vertices $\{{\bf V}_1,\,\ldots,\,{\bf V}_{\mbox{\tiny $E+V-1$}}\}$, it is the ``projective connectivity'' rule that guarantees all non-vanishing $\langle{\bf V}{\bf V}_1\ldots{\bf V}_{\mbox{\tiny $E+V-1$}}\rangle$'s to have the same sign. Let us define $s[{\bf V}_1,\ldots\,{\bf V}_{\mbox{\tiny $E+V-1$}}]$ to be this sign.

There is a similar sign associated with a sequence of planes $\{\mathcal{W}_{a_1},\ldots,\mathcal{W}_{a_{E+V-1}}\}$. Since we have a basis of $(E+V)$ planes, it is simply the sign of the only non-zero $\langle\mathcal{W}\mathcal{W}_{a_1}\ldots\mathcal{W}_{a_{E+V-1}}\rangle$ or, what is the same, the sign of the permutation that ends in $(a_1,\,\ldots,\,a_{E+V-1})$. Let $s\left[\mathcal{W}_{a_1}\ldots\mathcal{W}_{a_{E+V-1}}\right]$ be this sign.

We are now equipped to give the expression for the symbol $\mathcal{S}[F_{\mathcal{G}}(X,Y)]$:
\begin{equation}\label{eq:SFg}
 \begin{split}
  \mathcal{S}[F_{\mathcal{G}}(X,Y)]\:=\:&\sum_{\substack{\{a_1,\ldots,a_{v-1}\}\\ \{b_1,\ldots,b_{E+V-1}\}}}
             s[\mathcal{W}_1^{\mbox{\tiny $Y$}}\ldots\mathcal{W}_E^{\mbox{\tiny $Y$}}\mathcal{W}_{a_1}^{\mbox{\tiny $X$}}\ldots\mathcal{W}_{a_{v-1}}^{\mbox{\tiny $X$}}]\times
             s[{\bf V}_{b_1}\ldots{\bf V}_{b_{E+V-1}}]\times\\
       &\times[\mathcal{W}_1^{\mbox{\tiny $Y$}}\ldots\mathcal{W}_E^{\mbox{\tiny $Y$}};{\bf V}_{b_1}\ldots{\bf V}_{b_{E}}]\otimes
        [\mathcal{W}_1^{\mbox{\tiny $Y$}}\ldots\mathcal{W}_E^{\mbox{\tiny $Y$}}\mathcal{W}_{a_1}^{\mbox{\tiny $X$}};{\bf V}_{b_1}\ldots{\bf V}_{b_{E}}{\bf V}_{b_{E}+1}]\otimes\\
       &\ldots\otimes[\mathcal{W}_1^{\mbox{\tiny $Y$}}\ldots\mathcal{W}_E^{\mbox{\tiny $Y$}}\mathcal{W}_{a_1}^{\mbox{\tiny $X$}}\ldots\mathcal{W}_{a_{v-1}}^{\mbox{\tiny $X$}};
                      {\bf V}_{b_1}\ldots{\bf V}_{b_{E}}{\bf V}_{b_{E+V-1}}],
 \end{split}
\end{equation}
where $\{{\bf V}_{b_1}\ldots{\bf V}_{\mbox{\tiny $b_{E+V-1}$}}$ is a path compatible with our ``projected connectivity'' rule.

For a given $\{a_1,\,\ldots,\,a_{v-1}\}$ and $\{b_1,\,\ldots,\,b_{E+V-1}\}$, we can interpret the symbol as follows. We construct an $(E+V-1)\times(E+V-1)$ matrix of $\left(\mathcal{W}\cdot\mathcal{V}\right)$'s:
\begin{equation}\label{eq:WVmtr}
\bordermatrix{~ & {\bf V}_{b_1}       & \ldots  & {\bf V}_{b_{E+V-1}} \cr
                  \mathcal{W}_{Y_1}       & {}      & {} \cr
                  \vdots                  & {}      & {} \cr
                  \mathcal{W}_{Y_E}       & {}      &  \mathcal{W}_A\cdot{\bf V}_B \cr
                  \mathcal{W}_{a_1}       & {}      & {} \cr
                  \vdots                  & {}      & {} \cr
                  \mathcal{W}_{a_{E+V-1}} & {}      & {} \cr                      
} 
\end{equation}

The symbol entry is given by by $\otimes$'ing together the upper left adjusted minors of this matrix -- $\mathcal{M}_{E\times E}\otimes\mathcal{M}_{(E+1)\times(E+1)}\otimes\ldots\otimes\mathcal{M}_{(E+V-1)\times(E+V-1)}$ -- starting with the $E\times E$ minor.

Note that we have broken the symmetry between the edges in our definition of the $(\mathcal{W}^{\mbox{\tiny $Y$}})$'s, which is a kind of ``gauge-fixing'', but the final symbol will of course not depend on this choice. This provides a strong consistency check on the result.

We see that the symbol is fully determined by the geometry of the cosmological polytope, and indeed can be thought of as a record of all the projection-connected paths we can take through the polytope. The reader can readily verify that the formula reproduces the result for the symbol of the two-site chain \eqref{eq:S2s}. Since there are many more paths through the polytope, the one-line expression for the symbol of the three-site chain turns into a longer explicit formula, the sum of 104 terms:

{\tiny
\begingroup
 \allowdisplaybreaks
 \begin{align*}
  & \mathcal{S}[F_3^{\mbox{\tiny $(0)$}}(X,Y)]\:=\\
  &=\:
    - (X_1 - Y_1) \otimes ( X_2 + Y_1 + Y_2) \otimes ( X_2 + X_3 + Y_1) 
    + ( X_1 - Y_1) \otimes ( X_1 + X_2 + Y_2) \otimes ( X_1 + X_2 + X_3) 
    + ( X_1 - Y_1) \otimes ( X_2 + Y_1 - Y_2) \otimes ( X_2 + X_3 + Y_1) \\ 
  & - ( X_1 - Y_1) \otimes ( X_1 + X_2 - Y_2) \otimes ( X_1 + X_2 + X_3) 
    + ( X_1 - Y_1) \otimes ( X_1 + X_2 - Y_2) \otimes ( X_3 + Y_2)  
    - ( X_1 - Y_1) \otimes ( X_2 + Y_1 - Y_2) \otimes ( X_3 + Y_2) \\
  & + ( X_1 - Y_1) \otimes ( X_2 + Y_1 + Y_2) \otimes ( X_3 + Y_2) 
    - ( X_1 - Y_1) \otimes ( X_1 + X_2 + Y_2) \otimes ( X_3 + Y_2) 
    + ( X_1 + Y_1) \otimes ( X_1 + X_2 - Y_2) \otimes ( X_1 + X_2 + X_3) \\ 
  & - ( X_1 + Y_1) \otimes ( X_1 + X_2 + Y_2) \otimes ( X_1 + X_2 + X_3) 
    - ( X_1 + Y_1) \otimes ( X_1 + X_2 - Y_2) \otimes ( X_3 + Y_2) 
    + ( X_1 + Y_1) \otimes ( X_1 + X_2 + Y_2) \otimes ( X_3 + Y_2) \\
  & - ( X_1 + Y_1) \otimes ( X_2 +Y_1 - Y_2) \otimes ( X_2 + X_3 + Y_1) 
    + ( X_1 + Y_1) \otimes ( X_2 + Y_1 + Y_2) \otimes ( X_2 + X_3 + Y_1) 
    + ( X_1 + Y_1) \otimes ( X_2 + Y_1 - Y_2) \otimes ( X_3 + Y_2) \\ 
  & - ( X_1 + Y_1) \otimes ( X_2 + Y_1 + Y_2) \otimes ( X_3 + Y_2) 
    + ( X_1 - Y_1) \otimes ( X_3 - Y_2) \otimes ( X_2 + X_3 + Y_1) 
    - ( X_1 - Y_1) \otimes ( X_3 - Y_2) \otimes ( X_1 + X_2 + X_3) \\
  & + ( X_1 - Y_1) \otimes ( X_3 - Y_2) \otimes ( X_1 + X_2 + Y_2) 
    - ( X_1 - Y_1) \otimes ( X_3 - Y_2) \otimes ( X_2 + Y_1 + Y_2) 
    - ( X_1 - Y_1) \otimes ( X_3 + Y_2) \otimes ( X_2 + X_3 + Y_1) \\ 
  & + ( X_1 - Y_1) \otimes ( X_3 + Y_2) \otimes ( X_1 + X_2 + X_3) 
    + ( X_1 - Y_1) \otimes ( X_3 + Y_2) \otimes ( X_2 + Y_1 + Y_2) 
    - ( X_1 - Y_1) \otimes ( X_3 + Y_2) \otimes ( X_1 + X_2 + Y_2) \\
  & + ( X_1 + Y_1) \otimes ( X_3 + Y_2) \otimes ( X_1 + X_2 + Y_2) 
    - ( X_1 + Y_1) \otimes ( X_3 - Y_2) \otimes ( X_1 + X_2 + Y_2) 
    - ( X_1 + Y_1) \otimes ( X_3 + Y_2) \otimes ( X_1 + X_2 + X_3) \\ 
  & + ( X_1 + Y_1) \otimes ( X_3 - Y_2) \otimes ( X_1 + X_2 + X_3) 
    - ( X_1 + Y_1) \otimes ( X_3 + Y_2) \otimes ( X_2 + Y_1 + Y_2) 
    + ( X_1 + Y_1) \otimes ( X_3 - Y_2) \otimes ( X_2 + Y_1 + Y_2) \\
  & + ( X_1 + Y_1) \otimes ( X_3 + Y_2) \otimes ( X_2 + X_3 + Y_1) 
    - ( X_1 + Y_1) \otimes ( X_3 - Y_2) \otimes ( X_2 + X_3 + Y_1) 
    - ( X_2 + Y_1 + Y_2) \otimes ( X_1 - Y_1) \otimes ( X_2 + X_3 + Y_1) \\ 
  & + (X_2 + Y_1 + Y_2) \otimes ( X_1 - Y_1) \otimes ( X_1 + X_2 + X_3) 
    - (X_2 + Y_1 + Y_2) \otimes ( X_1 + X_2 + Y_2) \otimes ( X_1 + X_2 + X_3)
    + ( X_2 + Y_1 - Y_2) \otimes ( X_1 - Y_1) \otimes ( X_2 + X_3 + Y_1) \\ 
  & - ( X_2 + Y_1 - Y_2) \otimes ( X_1 - Y_1) \otimes ( X_1 + X_2 + X_3) 
    + ( X_2 + Y_1 - Y_2) \otimes ( X_1 + X_2 -Y_2) \otimes ( X_1 + X_2 + X_3) 
    - ( X_2 + Y_1 - Y_2) \otimes ( X_1 + X_2 - Y_2) \otimes ( X_3 + Y_2) \\ 
  & + ( X_2 + Y_1 + Y_2) \otimes ( X_1 + X_2 + Y_2) \otimes ( X_3 + Y_2) 
    - ( X_2 - Y_1 - Y_2) \otimes ( X_1 + X_2 - Y_2) \otimes ( X_1 + X_2 + X_3) 
    + ( X_2 - Y_1 - Y_2) \otimes ( X_1 + Y_1) \otimes ( X_1 + X_2 + X_3) \\
  & - ( X_2 - Y_1 + Y_2) \otimes ( X_1 + Y_1) \otimes ( X_1 + X_2 + X_3) 
    + ( X_2 - Y_1 + Y_2) \otimes ( X_1 + X_2 + Y_2) \otimes (X_1 + X_2 + X_3) 
    - ( X_2 - Y_1 - Y_2) \otimes ( X_1 + Y_1) \otimes (X_3 + Y_2) \\ 
  & + ( X_2 - Y_1 - Y_2) \otimes ( X_1 + X_2 - Y_2) \otimes ( X_3 + Y_2) 
    + ( X_2 - Y_1 + Y_2) \otimes ( X_1 + Y_1) \otimes ( X_3 + Y_2) 
    - ( X_2 - Y_1 + Y_2) \otimes ( X_1 + X_2 + Y_2) \otimes ( X_3 + Y_2) \\
  & - ( X_2 + Y_1 - Y_2) \otimes ( X_1 + Y_1) \otimes ( X_2 + X_3 + Y_1) 
    + ( X_2 + Y_1 + Y_2) \otimes ( X_1 + Y_1) \otimes ( X_2 + X_3 + Y_1) 
    + ( X_2 + Y_1 - Y_2) \otimes ( X_1 + Y_1) \otimes ( X_3 + Y_2) \\ 
  & - ( X_2 + Y_1 + Y_2) \otimes ( X_1 + Y_1) \otimes ( X_3 + Y_2) 
    - ( X_2 + Y_1 + Y_2) \otimes ( X_2 + X_3 + Y_1) \otimes ( X_1 + X_2 + X_3) 
    + ( X_2 + Y_1 + Y_2) \otimes ( X_3 - Y_2) \otimes ( X_1 + X_2 + X_3) \\
  & - ( X_2 + Y_1 + Y_2) \otimes ( X_3 - Y_2) \otimes ( X_1 + X_2 + Y_2) 
    + ( X_2 + Y_1 - Y_2) \otimes ( X_2 + X_3 + Y_1) \otimes (X_1 + X_2 + X_3) 
    - ( X_2 + Y_1 - Y_2) \otimes ( X_3 + Y_2) \otimes (X_1 + X_2 + X_3) \\ 
  & + ( X_2 + Y_1 + Y_2) \otimes ( X_3 + Y_2) \otimes ( X_1 + X_2 + Y_2) 
    - ( X_2 - Y_1 + Y_2) \otimes ( X_3 + Y_2) \otimes ( X_1 + X_2 + Y_2) 
    + ( X_2 - Y_1 + Y_2) \otimes ( X_3 - Y_2) \otimes ( X_1 + X_2 + Y_2) \\
  & + ( X_2 - Y_1 - Y_2) \otimes ( X_3 + Y_2) \otimes ( X_1 + X_2 + X_3) 
    - ( X_2 - Y_1 - Y_2) \otimes ( X_2 + X_3 - Y_1) \otimes ( X_1 + X_2 + X_3) \\
  & + (X_2 - Y_1 + Y_2) \otimes ( X_2 + X_3 - Y_1) \otimes ( X_1 + X_2 + X_3)
    - ( X_2 - Y_1 + Y_2) \otimes ( X_3 - Y_2) \otimes ( X_1 + X_2 + X_3) 
    - ( X_2 - Y_1 - Y_2) \otimes ( X_3 + Y_2) \otimes ( X_1 + Y_1) \\
  &  + ( X_2 - Y_1 - Y_2) \otimes ( X_2 + X_3 - Y_1) \otimes ( X_1 + Y_1) 
    + ( X_2 - Y_1 + Y_2) \otimes ( X_3 + Y_2) \otimes ( X_1 + Y_1) 
    - ( X_2 - Y_1 + Y_2) \otimes ( X_2 + X_3 -Y_1) \otimes ( X_1 + Y_1) \\
  &  + ( X_2 + Y_1 - Y_2) \otimes ( X_3 + Y_2) \otimes ( X_1 + Y_1) 
    - ( X_2 + Y_1 - Y_2) \otimes ( X_2 + X_3 + Y_1) \otimes ( X_1 + Y_1) 
    - ( X_2 + Y_1 + Y_2) \otimes ( X_3 + Y_2) \otimes ( X_1 + Y_1) \\
  &  + ( X_2 + Y_1 + Y_2) \otimes ( X_2 + X_3 + Y_1) \otimes ( X_1 + Y_1)
    + ( X_3 - Y_2) \otimes ( X_1 - Y_1) \otimes ( X_2 + X_3 + Y_1) 
    - ( X_3 - Y_2) \otimes ( X_1 - Y_1) \otimes ( X_1 + X_2 + X_3) \\
  &  + ( X_3 - Y_2) \otimes ( X_1 - Y_1) \otimes ( X_1 + X_2 + Y_2)  
    - ( X_3 - Y_2) \otimes ( X_1 - Y_1) \otimes ( X_2 + Y_1 + Y_2) 
    - ( X_3 + Y_2) \otimes ( X_1 - Y_1) \otimes ( X_2 + X_3 + Y_1) \\
  &  + ( X_3 + Y_2) \otimes ( X_1 - Y_1) \otimes ( X_1 + X_2 + X_3) 
    + ( X_3 + Y_2) \otimes ( X_1 - Y_1) \otimes ( X_2 + Y_1 + Y_2) 
    - ( X_3 + Y_2) \otimes ( X_1 - Y_1) \otimes ( X_1 + X_2 + Y_2) \\
  &  + ( X_3 + Y_2) \otimes ( X_1 + Y_1) \otimes ( X_1 + X_2 + Y_2)  
    - ( X_3 - Y_2) \otimes ( X_1 + Y_1) \otimes ( X_1 + X_2 + Y_2) 
    - ( X_3 + Y_2) \otimes ( X_1 + Y_1) \otimes ( X_1 + X_2 + X_3) \\
  &  + ( X_3 - Y_2) \otimes ( X_1 + Y_1) \otimes ( X_1 + X_2 + X_3)  
    - ( X_3 + Y_2) \otimes ( X_1 + Y_1) \otimes ( X_2 + Y_1 + Y_2) 
    + ( X_3 - Y_2) \otimes ( X_1 + Y_1) \otimes ( X_2 + Y_1 + Y_2) \\
  &  + ( X_3 + Y_2) \otimes ( X_1 + Y_1) \otimes ( X_2 + X_3 + Y_1)  
    - ( X_3 - Y_2) \otimes ( X_1 + Y_1) \otimes ( X_2 + X_3 + Y_1) 
    + ( X_3 - Y_2) \otimes ( X_2 + X_3 + Y_1) \otimes ( X_1 + X_2 + X_3) \\
  &  - ( X_3 - Y_2) \otimes ( X_2 + Y_1 + Y_2) \otimes ( X_1 + X_2 + Y_2) 
    - ( X_3 + Y_2) \otimes ( X_2 + X_3 + Y_1) \otimes ( X_1 + X_2 + X_3) 
    + ( X_3 + Y_2) \otimes ( X_2 + Y_1 + Y_2) \otimes ( X_1 + X_2 + Y_2) \\
  &   - ( X_3 + Y_2) \otimes ( X_2 - Y_1 + Y_2) \otimes ( X_1 + X_2 + Y_2)  
    + ( X_3 - Y_2) \otimes ( X_2 - Y_1 + Y_2) \otimes ( X_1 + X_2 + Y_2) 
    + ( X_3 + Y_2) \otimes ( X_2 + X_3 - Y_1) \otimes ( X_1 + X_2 + X_3) \\
  &  - ( X_3 - Y_2) \otimes ( X_2 + X_3 - Y_1) \otimes ( X_1 + X_2 + X_3) 
    - ( X_3 + Y_2) \otimes ( X_2 + X_3 - Y_1) \otimes ( X_1 + Y_1) 
    + ( X_3 + Y_2) \otimes ( X_2 - Y_1 + Y_2) \otimes ( X_1 + Y_1) \\
  &  + ( X_3 - Y_2) \otimes ( X_2 + X_3 - Y_1) \otimes ( X_1 + Y_1)  
    - ( X_3 - Y_2) \otimes ( X_2 - Y_1 + Y_2) \otimes ( X_1 + Y_1) 
    + ( X_3 + Y_2) \otimes ( X_2 + X_3 + Y_1) \otimes ( X_1 + Y_1) \\
  &  - ( X_3 + Y_2) \otimes ( X_2 + Y_1 + Y_2) \otimes ( X_1 + Y_1) 
    - ( X_3 - Y_2) \otimes ( X_2 + X_3 + Y_1) \otimes ( X_1 + Y_1) 
    + ( X_3 - Y_2) \otimes ( X_2 + Y_1 +Y_2) \otimes ( X_1 + Y_1)
 \end{align*}
\endgroup
}
The reader can verify that the result has the obvious $Y_1\,\longleftrightarrow\,Y_2$, $X_1\,\longleftrightarrow\,X_3$ symmetry, again a non-trivial consistency check since our ``gauge-fixing'' choice of an edge does not manifest this symmetry.

It would be interesting to completely characterize all the admissible paths $\{{\bf V}_{b_1},\,\ldots,\,{\bf V}_{b_{E+V-1}}\}$ of a cosmological polytope $\mathcal{P}_{\mathcal{G}}$, which may well reduce to a tractable combinatorial problem. We leave these and related explorations for future work.


\section{Conclusion and Outlook}\label{sec:Concl}

In recent years the study of scattering amplitudes has revealed surprising and deep connections between the dynamical processes of particle scattering and new mathematical structures in combinatorics and ``positive geometry'', ranging from the amplituhedron formulation of all-loop-order-planar $\mathcal{N}\,=\,4$ SYM amplitudes, to the more recent discovery of a worldsheet structure common to a wide range of massless scattering amplitudes \cite{Cachazo:2013iaa, Cachazo:2013gna, Cachazo:2013hca, Cachazo:2013iea, Cachazo:2014nsa, Cachazo:2014xea}. The existence of this connection is deeply reliant on the  Lorentzian, dynamical aspect of particle scattering: after all it is the integration of {\it phases} associated with time evolution that is ultimately responsible for the intricate pattern of singularities seen in amplitudes, at least in perturbation theory,  and it is precisely these sharp pattern of singularities that are captured in surprising ways by positive geometries.

This suggests that physical questions that are even more intrinsically ``time-dependent" -- most naturally involving the wavefunction of the universe -- might have an even more fundamental connection to combinatorial and geometric structures. This thought is further encouraged by the observation that wavefunction of the universe {\it contains} scattering amplitudes as a residue of the  $\sum_i E_i \longrightarrow 0$ pole -- if positive geometries play a crucial role on this co-dimension one subspace of the kinematical space, it would seem surprising if their role simply disappeared off this subspace.

In this paper we have seen the first indications for the connection between positive geometries and the wavefunction of the universe, for the computation of the late-time wavefunction of a class of toy models of scalars with non-derivative but time-dependent coupling constants. Conformally coupled scalars in FRW cosmologies, with (non-conformal) polynomial interactions, are a special case. We have seen that the contribution of each Feynman graph to the wavefunction of the universe is identified with the canonical form of a ``cosmological polytope'', which has an entirely canonical and intrinsic definition in its own right, from which the graph structure, and familiar physical properties of the wavefunction, naturally emerge.

The most crucial sense in which these are toy models is that, in the language of Feynman rules, these models only know about ``poles" but no non-trivial ``numerators''. Of course all the extra magical structure in Yang-Mills theory or gravity are precisely about non-trivial numerators, and much of {\it e.g.} the positive structure of the amplituhedron geometrizes not only the singularities associated with poles but also the zeroes  associated with non-trivial numerators. Nonetheless, for precisely this reason the singularity structure of these toy models is also completely universal, valid for any theory, and the cosmological polytope organizes these singularities in a beautifully combinatorial and geometrical way.

Note that since we are working with conformally coupled scalars, we can equally well think in terms of computing Witten diagrams for boundary correlators in Euclidean AdS space. We prefer the Lorentzian way of talking about the physics for two reasons. First,  as explained in our intrinsic description of the cosmological polytope, the structure of the polytope, dictated by the pattern of intersecting triangles on only two of three edges, is most strikingly connected to the way the light-cone divides space-time into $3\,=\,2+1$ regions. Second, the residues of the canonical form on the polytope on lower-dimensional facets are naturally associated with the physics of particle production.  In order to see this, note that the computation of the wavefunction involves an integral over the Fourier transforms of the coupling constants $\tilde{\lambda}_k(\varepsilon)$. Now, for familiar examples such as dS space, we have that $\tilde{\lambda}_k(\varepsilon)=0$ for $\varepsilon\,<0\,$, so the integrand is always manifestly positive and in particular we never hit any of the poles. However if we have $\tilde{\lambda}_k(\varepsilon)\,\neq\,0$ for $\varepsilon\,<\,0$, then we will inevitably hit poles corresponding to the facets, giving the late wavefunction  an oscillatory piece that corresponds to ``particle production''; the coefficient of these terms are precisely the residues on the facets. Thus the information contained in all the facets of the polytope is most fundamentally related to intrinsically ``time-dependent'' physics.

There are clearly a large number of open questions left unanswered by our preliminary investigations. Rather than make a  boring list of all of them, we will simply highlight one {\bf (s)}mall and one {\bf (L)}arge question. 

{\bf (s)} Even sticking with the toy models we have studied here, we would like to have a more systematic understanding of the symmetries of the canonical form for all cosmological polytopes, and the way in which familiar physics symmetries descend from them. We have seen how this works in a number of examples, but it would be highly illuminating to see a more primitive combinatorial origin for {\it e.g.} Lorentz or conformal symmetry from the symmetries of the cosmological polytope.

{\bf (L)} In this paper we have only seen the connection between the geometry and the physics one graph at a time. This is extremely unsatisfying, especially compared to e.g. the way in which the amplituhedron captures entire scattering amplitudes in a single geometry. However there is an obvious obstruction to ``combining the graphs together into a larger geometry''. Even sticking with {\it e.g.} planar cubic graphs, most naively, it would seem that the different cosmological polytopes simply live in different spaces, and that there is no natural way they can be  combined into a bigger object. However some recent progress in the study of scattering amplitudes \cite{Arkani-Hamed:2017vfh,Arkani-Hamed:2017abh} may give us a clue for how to proceed. It has been suggested that amplitudes should most naturally be thought of as differential forms on the space of kinematical data. Furthermore, positive geometries such as the amplituhedron for planar $\mathcal{N}\,=\,4$ SYM \cite{Arkani-Hamed:2017vfh}, and the associahedron for e.g. bi-colored $\phi^3$ theory \cite{Arkani-Hamed:2017abh}, emerge on certain canonical subspaces, again directly in the space of kinematical data. The amplitude differential form is then fully determined by the requirement of matching the canonical form of these positive geometries, when pulled back to the appropriate subspaces. Thus, for instance in bi-colored $\phi^3$ theory, one may be tempted to think of each graph as corresponding to a ``simplex" of some type, but since the variables in different graphs are different, it isn't clear how these ``simplices" could fit together. The key realization is that they {\it do} fit together, indeed they triangulate an associahedron, but that this geometry is seen on certain canonical subspaces of the full kinematical space. Could something analogous happen for cosmological polytopes and the wavefunction of the universe, even in toy examples like bi-colored $\phi^3$ theory, with collections of simplices triangulating the associahedron replaced by cosmological polytopes triangulating some new geometry?

\section*{Acknowledgements}

We are grateful to Hugh Thomas for comments on the manuscript. We thank Yuntao Bai, Freddy Cachazo, Thomas Lam and Hugh Thomas for useful discussion. N.A-H. would like to especially thank Juan Maldacena for many valuable and stimulating comments. P.B. is grateful to the Institute for Advanced Study in Princeton for hospitality during several stages of this work as well as Jos{\'e} Barb{\'o}n and Sebastian Mizera for insightful discussions. P.B. would also like to thank the developers of SageMath \cite{sage}, Maxima \cite{maxima}, and Tikz \cite{tantau:2013a}. Some of the polytope analysis has been performed with the aid of {\tt polymake} \cite{polymake:2000,polymake:ICMS_2006,polymake:FPSAC_2009, polymake_drawing:2010,polymake_XML:ICMS_2016, polymake:2017}.  The work of N.A-H. is supported by the DOE under grant DOE DE-SC0009988. Research at Perimeter Institute is supported by the Government of Canada through the Department of Innovation, Science an Economic Development and by the Province of Ontario through the Ministry of Research, Innovation and Science.

\appendix

\bibliographystyle{JHEP}
\bibliography{cprefs}

\end{document}